\documentclass[11pt]{report}
\usepackage[utf8]{inputenc}
\usepackage{graphicx}
\graphicspath{ {images/} }
\usepackage{caption}
\usepackage{subcaption}
\usepackage{float}
\usepackage[width=150mm,top=35mm,bottom=25mm,bindingoffset=6mm]{geometry}
\usepackage{fancyhdr}
\usepackage{setspace}
\renewcommand{\headrulewidth}{0pt}
\usepackage{tikz}

\usepackage{natbib}

\usepackage{graphicx} 

\usepackage{soul, color, mathtools, amsmath, amsthm, amssymb}
\usepackage{mdframed}
\usepackage{hyperref, soul, color, mathtools, algorithm, algpseudocode, float}
\usepackage{natbib}

\newtheorem{theorem}{Theorem}[section]
\newtheorem{definition}{Definition}[section]
\newtheorem{corollary}{Corollary}[section]
\newtheorem{lemma}[theorem]{Lemma}
\newtheorem{proposition}{Proposition}[section]
\newtheorem{rmk}{Remark}
\newtheorem{example}{Example}

\newtheorem{claim}[theorem]{Claim}
\newtheorem{fact}{Fact}[section]

\newcommand{\gk}{\sqrt{k}}
\newcommand{\gka}[1]{\sqrt{#1}}
\newcommand{\OPT}{OPT}
\newcommand{\polyk}{4k}
\newcommand{\mt}{\frac{\OPT_T}{2c_2}}
\newcommand{\lbt}[1]{\left(  \frac{#1}{T}\right)^2 \mt}

\newcommand{\world}[1]{\mathbf{E}_ #1}
\newcommand{\Lrwf}{\frac{\log(1/\delta_0 - 1)}{\log(p/(1-p))}}
\newcommand{\Lrw}{L}
\newcommand{\alpp}{\left(\frac{1-p}{p}\right)}
\newcommand{\pmin}{p_\text{min}}

\renewcommand{\Pr}[2]{\mathbb{P}_{#1}\left[  #2 \right]}
\newcommand{\E}[2]{\mathbb{E}_{#1}\left[  #2 \right]}

\newcommand{\rmhere}[1]{}

\title{Thesis Title}
\author{Author Name}
\date{Day Month Year}

\begin{document}

\begin{titlepage}
    \begin{center}
        \vspace*{1cm}
        
        \Huge
        \textbf{Algorithmic Approaches to Sequential Decision-Making and Social Epistemology}
        
        \vspace{0.5cm}
        \LARGE
        by\\
    
        \textbf{Kavya Ravichandran}
        
        \vfill
        
        A thesis submitted in partial fulfillment\\
        of the requirements for the degree of\\
        Doctor of Philosophy in Computer Science

        \vspace{1.8cm}

        \Large
        at the\\Toyota Technological Institute at Chicago\\
        Chicago, IL \\
        August 2026\\
        \vspace{1.0cm}
        \begin{flushleft}
        \large
        The dissertation is approved by the following members of the committee: \\
        \setlength{\parindent}{10ex}
        \begin{tabular}{ll}
             Avrim Blum (Thesis Advisor)  & TTIC \\Madhur Tulsiani   & TTIC\\ Alexander Williams Tolbert   & Emory University 
        \end{tabular}
        
        \end{flushleft}
        
    \end{center}
    
\end{titlepage}
\pagenumbering{roman}
\setcounter{page}{2}

\fancyhf{} 
\fancyhead[RO,R]{\thepage} 
\renewcommand{\headrulewidth}{0pt}

\begin{center}
    \Large
    \textbf{Algorithmic Approaches to
Sequential Decision-Making and
Social Epistemology}
    
    
    \vspace{0.4cm}
    \textbf{Kavya Ravichandran}
    
    \vspace{0.9cm}
    \textbf{Abstract}
\end{center}
As humans, we face many decisions that require us to choose between sticking to something and giving up. This thesis uses algorithmic tools to derive insights about such decision-making problems in theoretical models, studying both near-optimal methods and outcomes of social and behavioral influences. Along the way, this thesis sheds light on what we gain and what we lose as we move from a messy and complex real world setting to a very general abstract model by studying various points along this spectrum.

In Part I, we study algorithms for sequential decision-making in the improving multi-armed bandits problem. We provide nearly matching upper and lower bounds in the general case. Then, we then ask what is possible if we have access to similar instances to the one we wish to deploy our algorithm on. To that end, we provide guarantees in the data-driven algorithm design framework, showing that a polynomial number of samples is sufficient for learning good algorithms from a class of algorithms.

In Part II, we study algorithmic approaches for problems in social epistemology. We start by analyzing what role theoretical models can play in the study of social problems. We then study social and behavioral influences in decision-making requiring investment. First, we provide mathematical formalism in which to study the formation of pessimism traps, a phenomenon identified by philosophers in which agents are influenced by their predecessors to engage in less-ambitious goals. We develop financial interventions to sustainably shift communities out of these traps. The second problem we study is the influence of grit as a behavioral trait in ambitious decision-making. Overall, these works seek to theoretically model phenomena in social epistemology and provide a framework for intervening algorithmically.
\clearpage

\vspace*{0.25\textheight}
\begin{center}
\textit{To the women who came before me, who cleared the way and paved the path.}
\end{center}

\chapter*{Acknowledgments}



If it were not for the customary template for thesis acknowledgments, I would have a hard time knowing where to start with the present expression of gratitude. Many people have contributed significantly to my PhD journey, some directly during the PhD and some much earlier in my life by setting me along this path. 

I am lucky to have had weekly access to Avrim's infinite wisdom, creativity, and problem-solving prowess. In the early years, I would bring vague ideas and connections to Avrim, which he let me believe were creative and useful while he steered us toward actionable research ideas. I quickly realized that I should be paying attention to how exactly he went about that, and watching Avrim in action turned out to be one of the greatest treats of the PhD. Avrim helped me build both research skills and confidence through his thoughtful mentoring style. Outside of research, Avrim's wisdom and advice has provided me with a lot of perspective on academic life, and the way he carries himself as a leader -- humble and quick to help -- sets the standard for the rest of us. I couldn't have hoped for a better advisor for my PhD. I will call my academic career a success if I can be a fraction of the researcher and academic that Avrim is. Thank you, Avrim.

Madhur Tulsiani's wisdom is matched by his cheery demeanor. Through working with him as a TA, I  had the privilege of seeing his thinking in action, and I learnt a lot about how to put building blocks together to reach mathematical truths. Madhur's kindness and support, though always present, were especially appreciated at times during the PhD when I really needed it. I have benefitted greatly from his advice on everything from coursework to burnout to advising students. Thank you, Madhur.

I will never be as well-read as Alex Tolbert but  by working with him I both pick up insights from his vast knowledge and am inspired to read more. When I started my PhD, I fully anticipated that I would have to relegate my interest in the social sciences to something I do in my personal time, but thanks to working with Alex, I have been able to incorporate that interest in a meaningful way toward my research work. I am regularly stunned by how our wide-ranging conversations still culminate at research ideas, and I know that I will learn something new each time I pick up a call from Alex. Thank you, Alex, for opening my research trajectory in such a significant way.

Ronitt Rubinfeld has been a source of tremendous inspiration and support since I took her sublinear algorithms class in 2019. Ronitt's insistence on conceptual clarity alongside technical rigor shaped my thinking. Moreover, her kindness and encouragement have buoyed me at many tough points in the PhD. When I faced self-doubt, I often replayed to myself the encouraging things she'd said to me, and it kept me going. Thank you, Ronitt.

I'm grateful that Nati Srebro shared his incisive thinking and research taste during the Machine Learning and Optimization reading group sessions. These meetings and the hours of preparation that went into each one taught me how to read papers critically and understand what a paper claims to be saying vs is really saying. Through this, I learned also how to approach research problems by first understanding what the most basic approach achieves and using that to identify difficulties in the problem. Thank you, Nati.

Emily Diana has repeatedly created opportunities for me and supported me through them. I am grateful that we started working together during her year at TTIC; collaborating with Emily invited me into doing significantly interdisciplinary research, which has been really fruitful for me. I also really appreciate all her wisdom throughout the job search and decision-making. Thank you, Emily.

Maryam Aliakbarpour guided me through my first theory project and proceeded to become a close mentor and friend. Maryam taught me a lot about how to approach theory research and how to turn ideas into math; she has also been a source of wisdom and support through the emotional journey of the PhD. I am grateful to have a mentor with whom I can be so open. Our relationship has taught me a crucial lesson about academia: I will never be able to repay Maryam for the positive influence she has had in my life, so all I can hope to do is pay it forward. Thank you, Maryam.

Zooming out, faculty at both TTIC and MIT have provided me with a lot of advice that has brought me to where I am today and will, I am sure, prove extremely valuable for years to come. I am grateful to Julia Chuzhoy, Zhiyuan Li, Shiry Ginosar, Karen Livescu, Yury Makarychev, and Matthew Turk for interesting classes and valuable advice over the years. Several faculty in that list have been quite candid with me in matters of balancing career and personal life, and I am thankful for their perspectives. I am especially grateful to Greg Shakhnarovich and Matt Walter for their generosity with helping me prepare for interviews during this job market season. Going back further, my academic journey was significantly shaped by many faculty I interacted with in undergrad. An incomplete list: Kai von Fintel, Stefanie Jegelka, Aleskander Madry, Piotr Indyk, Sasha Rakhlin, Anant Madabhushi, George Verghese, Peter Hagelstein, Jacob White, Leslie Kaelbling, Duane Boning, Raul Radovitsky.

My research journey actually started all the way back in high school. It was then that I decided I wanted to do a PhD. Thank you to Mrs. Patty Hunt and Dr. Crystal Miller, as directors of SREP, for finding me amazing opportunities and teaching me so much about structuring research, communicating science, and professionalism. Thank you to Dr. Anirban Sengupta and Dr. Christa Pawlowski for the chance to learn from you -- while I ended up in a field that is quite far away from biomaterials, I still use many skills you taught and modeled for me: breaking down problems, designing complex solutions, and reviewing literature, as well as technical writing and perseverance. \par

I had a great time collaborating with many people from many places during my PhD. Thank you to my coauthors: Maryam Aliakbarpour, Amartya Shankha Biswas, Avrim Blum, Yatin Dandi, Emily Diana, Marten Garicano, Stefani Karp, Francesca Mignacco, Ronitt Rubinfeld, Dravyansh Sharma, Mor Shpigel-Nacson, Daniel Soudry, Nati Srebro, Alex Tolbert. I also learnt a lot from reading papers, discussing research, and working on problems with: Sam Buchanan, Surbhi Goel, Anmol Kabra, Gene Li, Sepideh Mahabadi, Theodore Misiakiewicz, Saeed Sharifi-Malvajerdi, Jiawei Zhou. 

In addition to direct collaborators, I have been lucky to interface regularly with many academic communities. I am always energized by hearing ideas, results, perspectives, and more from people around the world. Some communities I'd especially like to acknowledge: the Les Houches Summer School in July 2022, the NSF-Simons collaboration on Mathematical and Scientific Foundations of Deep Learning, the Simons collaboration on the Theory of Algorithmic Fairness, ALT, FORC, and more.

TTIC has been an incredible place to do a PhD. 
I am constantly inspired by curiosity and rigor with which every researcher here approaches research. This trait often visible at seminars, with people asking thoughtful questions throughout the talk. The environment is also extremely collegial and friendly. I always looked forward to coming into the office and learning something new, whether from class, lunch discussions, or pure gossip. I am grateful for the camaraderie and friendship of: Saba Ahmadi, Idan Attias, Dimitar Chakarov (incl deep chats about how to approach grad school), Lee Cohen, Xiaodan Du,  Melissa Dutz, Mahdi Haghifam, Nirmit Joshi (incl discussions about what constitutes good / useful research), Jiahao Li, Omar Montasser (incl the guidance in the early years of my PhD!), Olga Medrano, Marko Medvedev, Kanishka Mishra,  Amin Mohamadi, Ron Mozenson, Keziah Nagitta, Abhijit Mudigonda, Rachit Nimavat, Ankita Pasad (incl our virtual roommate year!), Adela de Pavia, Donya Saless (incl conferences all over!), Omshi Samal (incl our unrealized intentions to hit up a bakery!), Han Shao (incl job market advice and gossip!), Vaidehi Srinivas, Shashank Srivastava,  Kevin Stangl (incl coffee and stationary discussions!), Kaylene Stocking, Ali Vakilian, Santhoshini Velusamy (how cool to connect again in this phase!), Lingxiao Wang, David Yunis. Special shoutout to the best officemates -- Haochen Wang and Ju-Chieh Chou. You guys have been a constant positive force; I will miss our chats about PhD / research life and learning distilled insights about applied machine learning from you! And I would be remiss to not appreciate the delicious water in the 4th and 5th floor kitchens which motivated me to come to the office in a more significant way than one might think (and the coffee was not too bad, either).

I'm lucky to have gotten to know the TTIC staff through pandemic book club, various internal committees, and all the wonderful community events they put on, including karaoke, summer trips, tea times, and defense celebrations. Thank you to Mary Marre for all the delicious food, fun conversations, and enthusiasm about holidays and celebrations. Thank you to Brandie Jones for infinite support and being a friend on campus. Thank you to Rose Bradford for your energy and passion -- I had such a wonderful time working on outreach activities with you and am inspired by your approach to improving the world around us. Thank you to Randy Landsberg for listening to many versions of my talks and giving me extremely helpful feedback. And thank you to Jessica Jacobson, Chrissy Coleman, Adam Bohlander, Erica Cocom, Celeste Ki, Deree Kobets, Alicia McClarin-DeMuro, and Amy Minick for everything you do to make TTIC such a wonderful workplace. 

My friends have made this journey a memorable and incredibly fun one. I moved to Chicago in 2021 having sort of made virtual friends through zoom, gather.town, and FB messenger with some folks at TTIC and UChicago. Whatever trepidation I had about whether these friendships would hold up ``in the real-world'' went away quickly\footnote{though I must admit it took some time to get used to the fact that these faces on a screen now had height.}. We have spent long nights in West Loop Jeni's, racked up tens of decibels in hearty laughter, and gone on adventures in both Chicago and places we visit for conferences and weddings. Naren, you are always up for an adventure and your energy ensured that alongside work, we also did fun random things together. Kshitij, I learn so much from every conversation with you, whether about research, food, or inane topics, and I admire how carefully you think about each thing that enters your field of view. Anmol, the years we spent together in Chicago were filled with ups and downs, and having you to commiserate with and laugh with made a world of difference. Gene, your taste in memes is truly unparalleled, something I could not have anticipated when you were the responsible, serious counselor in charge of my group at Presidential Scholars 2016. Max, what a privilege it has been to start and finish this journey together. I am grateful for our exploration of Chicago's neighborhoods, the Lula Cafe brunches, walks, and always having you to check in with. Owen, I'm always amazed by how organized you are, and I'm inspired by your commitment to community; I hope to emulate both. Sue, our regular lunchtime conversations give me a lot of perspective on what's important in life and research.
Sudarshan, onnuda kathaigal ku vara siruppu vera kathaigal ku varave varadhu.
Pushkar, your groundedness inspires me, and I enjoy sharing recipes and ideas based on our shared love for food!
Tushant, our conversations on why we want to be in academia always leave me more committed to the cause, and I must, of course, thank you for the post-interview sambar rice!
Naomi, what a blessing to have shared not only these years in Chicago but also the majority of our lives so far. Thank you for being a steadfast friend.
Simran, I will never get over miracle that landed you in Chicago after I started my PhD. Your work ethic inspires me and your support buoys me. I am forever grateful that we have each other as we navigate various life experiences, both personal and professional.
Sarbari, our conversations are equal parts profound and silly, and I always feel happier and more confident after them.
Sushruth, our discussions on math, probability, economics, and philosophy have been so formative for me, both before and during the PhD.

My friends in PhD programs outside of TTIC provided much-needed perspective:
Samyu, though our distance has precluded continuing the shenanigans of our youth, I am so grateful that we can lean on each other during these phases of life and hopefully many more to come.
Josephine, our conversations often land us up talking about the role that theory plays in our respective fields, or what growth and achievement has looked like over the years, or one of a myriad of other topics. I am grateful to have in you a close friend, an ardent supporter, and a role model.
Rohan, I feel lucky that we've been able to compare notes on our PhD journeys, and each time we talk, I am struck by your thoughtful approach to life.

A number of additional shoutouts:
Emma, Katie, Ahona, Prabhav, Jean-Luc, Ying, 
Yamin (who was in Chicago for both my thesis proposal and defense -- incredible and thank you!), Nalini, Kevin, Tam, Sohil,
Gabe, Rene, Karunya, Agni, Haripriya, Aditi, Stuti, Vivek.

During the pandemic period, a group of us had a weekly Advaita reading group. The lessons I learnt from the texts and from our discussions have made me a better person and more equipped to handle the ups and downs of life. The group being enriched with PhDs (and theorists, in fact!) made the lessons from the texts all the more relevant to my daily life. Thank you to Srini, Dheeraj, Suhas, and Aniruddh for engaging with me in this valuable foray.

My curiosity and enthusiasm for learning has been consistently fostered by incredible educators throughout my life. Looking back, I can recognize how each time I asked one too many question or got distracted in class, a teacher made an intentional choice to give me an extra math problem or challenge me to learn a new word, nurturing my desire for academic challenge. I hope to emulate the standard they have set in my own teaching. I could never manage to capture verbally the impact they've had on my life, so for now, my heartfelt thank you to this (highly incomplete) list of former teachers: Mrs. Nirmala Bala, Mrs. Helene Debelak, Mr. Charles Debelak, Mrs. Lorraine Tzeng, Mrs. Connie Miller, Ms. Linda Brown, Mrs. Geetha Vasanth, Mr. William Adler, Ms. Patty Hunt, Mr. Kevin Purpura, Ms. Mary Kay Osredkar, Mr. Jason Habig, and many others.

To have found myself in these rich educational settings was not simply a stroke of luck (though there was some of that, too). It was due to intention and action from the part of my family, who have obviously made too big an impact on my life for me to adequately express my gratitude in this one paragraph. My first educational environment, inside the walls of our home, was filled with books, answers to endless ``but why?''s, and times tables. I cannot overstate the value of the culture of curiosity that my parents cultivated, and they and my extended family matched that with searching for excellent schools and educational opportunities for us. 
Thank you, Amma, for everything, but especially for being my first math teacher, for setting a bar I'll never clear in terms of being an all-around superwoman, for picking every phone call no matter the time or circumstance, and for being not just my mother but also a treasured mentor and friend. Thank you, Appa, for everything, but especially for using those hours driving me to school teaching me about information theory and machine learning, for demonstrating to me the value of intellectual courage and the relentless pursuit of truth, for your attempts\footnote{and fine, I'll admit, successes} in making puns, and for giving me the advice to keep writing my thoughts down while doing research. Shruthi, thanks for being my built-in best friend, for the shenanigans and memes, and for the wisdom and reminding me to not sweat the small stuff. It's been especially cool this year to be on opposite ends of the PhD journey, and I am so excited to see everything you accomplish in yours! A special thank you to Thatha for the stories, wisdom, and constant blessings, and for carrying forward Ammamma's spirit in her absence. For my part, I strive to model her curiosity and care for every person with whom she crossed paths. Halfway through my PhD, I gained a second family. Thank you to Harini Amma and Raghu Appa for your love and consistent checkins and good wishes. Thank you to Maithra and Arun for being the original source of grad school wisdom.

Needless to say, this brings me to the customary final thank you -- to Aniruddh, thank you for always being so supportive in every way, big and small, for constantly working together to build a life that suits us both, for racing\footnote{and usually beating} me to share the memes from ML twitter, for going through the PhD first so I could apply lessons from yours to mine, and for making each day that much more joyful. Thanks for being with me on this crazy journey not just of PhD but of life.

And if you have made it this far, thank you, reader, for your engagement with my exercise in gratitude. Please take one more moment and tell someone how much they mean to you.

\tableofcontents

\listoffigures

\listoftables

\singlespacing
\chapter{Introduction}
\pagenumbering{arabic}

Scholars from a wide range of disciplines engage with questions surrounding decision-making. How do humans weigh various factors and options when making decisions? What are ways for institutions to proceduralize decision-making so they can vet whether the decisions they make align with their goals or not? Why do different people make different decisions in light of similar information? Which decisions can we automate and how do we do so optimally? Decision problems differ in their settings, information environments, objectives, and constraints. Computer science has always been interested in how to automate decisions, and in particular, how to make them efficiently and accurately. In order to apply algorithmic tools to study decision-making, we must start by theoretically modeling the decision problem in an abstract way. Subsequently, we can frame the heuristics used by humans as algorithms and study outcomes of these. This thesis takes this approach to studying decision-making problems where investment is required to succeed. \par

We often find ourselves as individuals in situations where we must decide whether to invest our time and energy or not. Perhaps we are studying a difficult mathematical topic. Or maybe we are deciding whether to pick a challenging college major. Should we do a PhD or go to work immediately? Should we start our own business? Institutions face similar questions, often in the context of resource allocation. Given many technologies we could work on developing, which ones will have the highest payoff and therefore warrant investment? In each of these settings, we can consider myriad factors, including social influences, the impact of personal beliefs and traits, and the effect of the costs of various options and of switching between them. In this thesis, we consider several of these factors in stylized theoretical models. In doing so, a theme that emerges is:
\begin{quote}
    What do we gain, and what do we lose, by abstracting a problem instead of keeping it concrete?
\end{quote}

Our scholarship takes two angles: (1) we provide novel technical results in a number of theoretical settings, and (2) we utilize theoretical models to develop insight about social problems. In synthesizing these two styles of work, we develop understanding of the above question.
Along the abstract to concrete continuum, we have, on the one end, purely theoretical settings and, on the other end, the complex and nuanced real world. 
Our work addresses problems at three points along this contiunuum, which when taken together, span a substantial portion of the range. 
We start with a very general and abstract decision-making problem, proceed to a version of that problem that assumes that the decision-making problem comes from a distribution, and finally develop theoretical models for social phenomena. The lattermost works represent particular instantiations of more general theoretical frameworks intending to capture some of the concreteness of reality.
Before diving into our results and analysis, we specify the framework that unifies the work in this thesis.

\section{Unifying Framework} \label{sec:unif-framework}

We take two perspectives regarding theoretical models for decision-making in this thesis: (1) what I will call the classical ``sequential decision-making'' perspective and (2) what I will call the ``social modeling'' perspective. These two perspectives will echo the ways in which game theory and mechanism design, as fields, view problems; the differences lie in what about the problem setting we hold fix and what we modify. In a decision problem, there are three parts we will need to specify in order to define the abstract version of the problem. First, we delineate the abstract details of the \textbf{setting}. This includes agents, actions available to them, information available to them, and any constraints thereof. Next, we consider the heuristic or \textbf{algorithm} by which agents make decisions. Finally, we must study \textbf{outcomes}. This includes the degree to which the decision objective is optimized, as well as any resulting phenomena or externalities. We will make all of these more precise in the specific settings in which we study them. For now, in the abstract sense, we will discuss what happens to the \textbf{setting}, \textbf{algorithm}, and \textbf{outcomes} in each perspective. \par

In the classical sequential decision-making perspective, we are interested in characterizing optimal algorithms for solving a sequential decision-making task. In our framework, that means that we specify the \textbf{setting}, specify the \textbf{outcome} we wish to achieve, and then identify \textbf{algorithms} and characterize their optimality. As the setting, we would specify the information and actions available to an agent in the decision-making problem, the costs incurred or rewards available thereof, and any resulting changes to the world. As the outcome, we could say that we wish to maximize the accrued reward, or minimize costs. Then, an algorithm or strategy would have to navigate the world, getting information, costs, and reward under the specified goal. This perspective is familiar to theoretical computer scientists. \par

In the ``social modeling'' perspective, we will have two tasks. First, we will fix the setting, fix the algorithm used by agents, and study the outcomes. Usually the goal in this stage will be to ensure our model reproduces outcomes from the real-world setting of interest. Once we are satisfied that the model reflects what we want it to about the real world\footnote{Models reflecting what we want them to about the real world can be subtle! We discuss more on our perspective on what models can be useful for in Chapter~\ref{chap:social-ep-intro}.}, we can think about modifications. We then tweak the setting and explore how that affects the outcome, allowing us to steer outcomes toward desired ones by appropriately intervening on the setting. Further, this tells us what conditions in the setting were sufficient or even necessary for outcomes to look a certain way. Through this exercise, we are able to reason about human behavior and how to make outcomes incentive-compatible. 

For studying decision-making in an abstract sense, it is useful to consider this framework because it allows us to isolate the relative effects of the setting and algorithmic choices. From here, we can characterize optimal decision-making in a way that is familiar to computer science theorists. Additionally, we can develop controlled models in which to study heuristics used by humans in practical decision-making. By comparing these two sides of the coin, we can hope to evaluate outcomes from using heuristics against outcomes that would result from optimal decision-making, and we could even try to incentivize agents to use optimal decision-making strategies. On the other hand, by thinking of the factors/heuristics driving human decision-making as fixed, we also don't take an overly prescriptive position. That is, we never say, ``if people just changed their behavior, things would be better in the world!'' We instead think about what factors in the world might lead decision makers toward ``worse'' decisions and identify ways to change those factors. This perspective is especially useful when studying social problems. \par

We do not claim that this framework is complete or without flaws. It is simply the framework that we use to organize and interpret our results.

\section{Outline of Thesis}

In this thesis, we advance the theoretical study of various aspects of decision-making in contexts where investment is required to witness payoff. We characterize nearly-optimal algorithms in a general version of the problem, study how to go beyond the worst-case paradigm for this problem, and investigate social and personal influences on decision-making that requires ambition. In this section, we explain the structure of the thesis and summarize each chapter. \par

The thesis broadly comprises two parts, each corresponding to one part of the title. In Part \ref{part:imab}, we focus on sequential decision-making, particularly on the improving multi-armed bandits problem. In Part \ref{part:social-ep}, we study social decision-making, focusing on developing theoretical models for a line of work in social epistemology studying social and behavioral influences in setting and working toward ambitious goals. While studying these questions, we also reflect on the value of theoretical models and what we gain and what we lose from abstraction. \par

\subsection{Part I: Improving Multi-Armed Bandits}

\paragraph{Chapter~\ref{chap:imabsetting}} In this chapter, we present preliminaries for the {\em improving multi-armed bandits} (IMAB) problems, first introduced by \cite{heidari_tight_nodate}. The study of IMAB comprises the first half of this thesis. In this setting, there are a set of options to choose between. Each option has some final reward but we are not lucky enough to know the final outcome. Instead, we get rewards along the way that increase toward the final reward in an unpredictable way. We motivate this abstraction through various real-world circumstances it could reflect before formalizing it. Toward that, we first clarify the adversary model and performance comparator. Then, we discuss the measures of quality and optimality under those measures. We also review related work to highlight known results and specify where our work fits into the broader landscape. 

\paragraph{Chapter~\ref{chap:br25}} We characterize decision-making when investment is required via the (IMAB) problem. In this chapter, we take the traditional {\em worst-case} perspective. In particular, we assume that an oblivious adversary can pick a worst-case instance on which to deploy our (randomized) algorithm.  
We provide a lower bound under minimal assumptions and match it up to logarithmic factors with an upper bound. In fact, we show that no algorithm can achieve better than an $\Omega{\sqrt{k}}$ approximation factor against the optimal reward, where $k$ is the number of bandit arms. We do so by constructing a family of bad instances and then applying Yao's principle. Further, we provide a simple randomized algorithm that achieves $O(\sqrt{k} \log k)$ approximation factor.

\paragraph{Chapter~\ref{chap:alg-design-imab}} While the ``worst-case'' setting is familiar and foundational, it often proves pessimistic, in particular when lower-bound instances are pathological or brittle. Thus, it is useful to not only characterize the nature of optimal decision-making in the worst case but also explore the possibility of leveraging structure in practical instances to do better. In this chapter, we apply tools from the {\em data-driven algorithm design} literature to the improving multi-armed bandits problem. We study both cumulative reward and best arm identification objectives. We identify ``niceness'' conditions under which we could hope to do better than the worst-case bounds in the previous chapter, and then we show that by collecting algorithms that are (near-)optimal under these conditions, we can develop learnable families of algorithms. Accordingly, we characterize how many historical instances suffice for learning good algorithms from data. We draw from a wide range of technical tools; we extend the techniques in the previous chapter and utilize complexity characterizations of algorithm families from the data-driven algorithm design literature that allow us to prove uniform convergence bounds in our setting.

\subsection{Part II: Social Epistemology}
\paragraph{Chapter~\ref{chap:social-ep-intro}} We open this section with a meditation on what role mathematical models can play in analyzing real-world systems. We reflect on our own experience developing and working with mathematical models for various phenomena and identify three potential roles, of increasing complexity, models could play in understanding a phenomenon. We map our reflections onto existing literature in the philosophy of science, and we identify recommendations for computer scientists looking to build models, particularly for social phenomena. Along the way, we also explore evaluation and the tradeoffs between abstraction and specificity. Finally, we provide philosophical and sociological context for the specific social problem studied in this latter half of the thesis.

\paragraph{Chapter~\ref{chap:pess-traps}} This chapter studies the role an individual's community plays when they are deciding between an ambitious and a moderate end. Motivated by the work of philosopher Jennifer Morton \cite{morton2022resisting}, we develop a theoretical model for pessimism traps, a phenomenon in which people rationally make less-ambitious choices based on beliefs about their likelihood of success at a risky end. These traps form because taking the less-ambitious choice does not change an agent's perception of the value of the ambitious choice. We formalize this in a sequential decision-making model inspired by the social learning literature: individuals deciding between two options are influenced by those who made the same decision before them and their internal orientation. We then derive interventions to break communities out of these pessimism traps, crucially without forcing individuals to act against their interests. On a technical level, this work develops interventions in an extended information cascade model. In particular, we show how financial incentives drawn from a random process can simultaneously break negative learning cascades in parallel communities. \par

\paragraph{Chapter~\ref{chap:grit}} Having studied the effect of social influences in the previous chapter, in this chapter, we study the role of an individual's behavioral traits on ambition. While folk definitions of grit abound, we focus on the conceptual development of \cite{morton2019grit} and develop mathematical formalism based on their conceptual abstractions. In particular, we develop a specialized instance of the IMAB problem to study the effect of {\em grit} on the strategies and outcomes of different agents. Based on our models, we draw conclusions about the relative benefits of grit and systemic support that corroborate findings in economics and sociology. \par

\subsection{Bibliographic Notes}

This thesis is based on four conference papers and additional analysis. We note the references and collaborators here.
\begin{itemize}
\item \textbf{Chapter~\ref{chap:imabsetting}} draws on material from \cite{blum_nearly-tight_2024}, which was done in collaboration with Avrim Blum.

\item \textbf{Chapter~\ref{chap:br25}} is based on \cite{blum_nearly-tight_2024}, which was done in collaboration with Avrim Blum and appeared in ALT 2025.

\item \textbf{Chapter~\ref{chap:alg-design-imab}} is based on \cite{bgrs2025algorithm}, which was done in collaboration with Avrim Blum, Marten Garicano, and Dravyansh Sharma and will appear in UAI 2026.

\item \textbf{Chapter~\ref{chap:social-ep-intro}} is newly developed for this thesis.

\item \textbf{Chapter~\ref{chap:pess-traps}} is based on \cite{pessimism-traps-alg}, which was done in collaboration with Avrim Blum, Emily Diana, and Alexander Williams Tolbert and appeared in FORC 2025.

\item \textbf{Chapter~\ref{chap:grit}} is based on \cite{blum_theoretical_2025}, which was done in collaboration with Avrim Blum, Emily Diana, and Alexander Williams Tolbert and appeared in AAAI 2026 in the AI for Social Impact Track.
\end{itemize}

\section{Takeaways}

Finally, let us summarize the key contributions and takeaways of this thesis. They are, as mandated by rhetorical best practices, threefold:

\begin{enumerate}
    \item \textbf{Technical results in the IMAB setting:} We provide algorithmic upper bounds and complementary lower bounds for the general version of the problem. We follow this up with a beyond-worst case analysis of the problem in the {\em data-driven algorithm design} paradigm, characterizing the sample complexity of historical instances to learn good algorithms. Our results provide insights for important practical problems like hyperparameter tuning. In general, our algorithms for this problem have an optimistic flavor, and they follow a ``singular evaluation approach'' \cite{klein-sourcesofpower} to decision-making, which involves serially and quickly evaluating whether an option is likely to be the best one and discarding it if not.
    \item \textbf{Incentivizing ambition:} We theoretically characterize where can financial interventions help people act in a way that is in accordance with their beliefs while also being ambition. That is, we ask and answer: can we align ambition with beliefs about self and community? These works show not only that financial incentives can incentivize people to engage in ambition (which is perhaps intuitive) but also that carefully-designed financial incentives can have lasting effects even after the intervention period ends.
    \item \textbf{Applying CS tools to philosophical questions:} This thesis expands our understanding of how to apply computer science tools to problems about belief and epistemology. We, of course, do so through our technical results. We also analyze why this approach is useful and explore limitations of theoretical models. Our reflections on these issues culminate with our  guidelines for developing theoretical CS models for such problems.
\end{enumerate}

\part{Improving Multi-Armed Bandits} \label{part:imab}
\chapter{Setting and Preliminaries} \label{chap:imabsetting}
\section{Improving Multi-Armed Bandit Setting}
Extensive scholarly attention has been devoted to problems where an agent interacting with the world receives some reward that allows them to calibrate their future actions. Often, we are interested in the {\em bandit reward} setting, where the agent receives reward for a particular action taken while in the state in which they find themselves. The structure of this reward affects the achievable performance and algorithms that achieve them. \par

In the improving multi-armed bandits problem, a problem instance consists of $k$ bandit arms (i.e., ``pulling'' the arm reveals the reward) each with reward that increases the more the arm is pulled. In other words, the payoff is not a function of the {\em time} at which an arm is pulled but rather of the {\em number of times it has been pulled so far}, with different arms having (potentially) different increasing functions. Our goal is to maximize the reward we achieve. Some real-world problems captured by this framework include: training multiple learning algorithms, when the performance of an algorithm improves with resources expended and some algorithms are ultimately better than others for the setting at hand \citep{haussler_rigorous_1996, hyperband_2017, li_efficient_2020};
developing new technologies, where investing resources into developing a technology may increase its efficiency and different technologies have different asymptoting utility values \citep{Riv23b};
or even the problem of deciding what research area to work in.
In each of these examples, each algorithm or technology is represented by one bandit arm, and the reward achieved from pulling the arm increases as it is pulled more. The goal, then, is to find a sequence of arms to pull that will maximize some objective. \par

Various objectives might be valid to consider depending on the intended application. Since we are considering rewards (rather than costs), we will always aim to maximize the objective. In some cases, we might want to maximize the total reward accrued (for instance if studying many topics, we might value overall knowledge acquired). In other cases, we may wish to simply identify the best option among the options we are considering, and therefore it suffices to maximize the maximum value of a single pull. For any of these objectives, we can measure the suboptimality of what is achieved by the algorithm relative to the optimal possible additively (regret) or multiplicative (competitive ratio). In this thesis, we consider multiplicative suboptimality in the interest of generality. This is discussed further in Section~\ref{sec:prelims}. \par

Now, we have have two main questions to consider: (1) what algorithm allows us to maximize the objective? (2) what is the best possible performance of any algorithm? These two questions correspond to showing upper and lower bounds, respectively, on the suboptimality (competitive ratio). These results depend on how the instances are chosen, i.e., how much power the adversary has. For a theoretical computer scientist, it is natural to consider a very powerful adversary who can choose a {\em worst-case} instance. While results in this perspective are very strong, they can also be somewhat pessimistic. Thus, in order to impose a little more structure, it might instead make sense to limit an adversary to constructing a distribution from which the instance will be randomly sampled. In such a setting, the algorithm designer has access to a fixed set of samples from this distribution when designing the algorithm. We consider both of these perspectives in this part. 




\paragraph{Summary of Part I} In this section of the thesis, we discuss a setting where the structure of the reward is increasing. We study the {\em improving multi-armed bandits problem} in both the worst-case setting and a setting that makes a distributional assumption on instances we see. In doing so, we unveil nearly-optimal algorithms in the worst case and develop a framework for {\em learning} good algorithms from historical data. In this chapter, we motivate the setting and formally define it. In Chapter~\ref{chap:br25}, we study the problem with the goal of optimizing over worst-case instances. We provide a lower bound and analyze algorithmic building blocks that nearly achieve this lower bound. In Chapter~\ref{chap:alg-design-imab}, we study the problem with the goal of developing algorithms that perform well on average on instances drawn from some distribution. We motivate these two perspectives in the respective chapters. At a high level, these perspectives are complementary and therefore useful, as they help us first isolate what is possible in general and then posit deployment of our methods to settings where we have slightly more information about a specific instance before having to solve the instance.

\section{Formal Preliminaries} \label{sec:prelims}

We follow the problem specification in \cite{patil_mitigating_2023}. In particular, each instance consists of $k$ arms, where each arm $i$ has an associated monotone increasing reward function $f_i\,.$ The reward from pulling arm $i$ for the $t^\text{th}$ time is $f_i(t)\,.$ These functions are not known to the algorithm in advance; the algorithm only gets to know the current reward by interacting with the arm. Further, the argument $t$ of the function is not the time at which the arm is picked, but rather the number of times the arm has been played. Formally:

\begin{definition} \label{defn:imab}
    An instance $I \in \mathcal{I}$ of the improving multi-armed bandits problem consists of $k$ arms, each of which has associated with it a reward function $f_i$ that is nondecreasing 
    as a function of $t_i\,,$ which is the number of times that arm has been pulled so far. 
\end{definition}

\subsection{External and Policy Regret, Adversary Model}

Let us start by discussing the adversary model and how we are running and evaluating our algorithm. We consider {\em oblivious} adversaries and evaluate our algorithm in a {\em policy-competitive} rather than {\em externally-competitive way}\footnote{In the literature, it is standard to discuss external regret and policy regret, which are different measures of regret. We use the same comparators as in that framework but take the ratio of the algorithm's reward with the comparators reward rather than the difference.}. In this section, we flesh out the adversary, comparator, and competitive ratio that we study.

The adversary model in our work is that of an {\em oblivious} adversary. Since we study randomized algorithms, we could consider either adversaries that know the strategy but not the exact instantiation of randomness (oblivious adversaries) or adversaries that know both the strategy and the instantiation of randomness (adaptive adversaries). Once we give the adversary access to the instantiation of randomness, there is very little we can do to outperform it \cite{bbkg-powerofrandomization-94}, and so we restrict our attention to oblivious adversaries. 

We now describe the pattern of interaction in our setting involving randomized algorithms. First, the algorithm is defined. The pseudocode for the algorithm is published, but the specific randomization is kept private. Next, the adversary starts by simulating a run of the algorithm, keeping in mind that there will eventually be randomness. The adversary picks rewards (in our case, functions to assign to the arms) based on this simulation. Once the adversary has picked all the rewards, the game can begin. The algorithm makes its first move, outputting some action $x_1$ and the adversary responds with a reward function $r_1(\cdot) \in \mathcal{R}\,.$ (In the improving multi-armed bandits case, $x_1$ is the identity of the first arm pulled, and $r_1(x_1)$ is $f_{x_1}(1)\,.$) This continues for $T$ time steps. 

Now, let us consider the comparator. We can think of the comparator as what kind of optimal strategy we use to evaluate an adversary's choices. The standard comparator in much of online learning, including in the very first chapter of \cite{hazan2023introductiononlineconvexoptimization} is the best single action in hindsight. In this case, the comparator reward is accrued as follows. Once the interaction is complete, we consider all the reward functions $\{r_t(\cdot)\}_{t=1}^T\,.$ We pick the $x$ that maximizes the sum of the known reward functions over all time. Formally, the accumulated rewards of the algorithm and the comparator are defined below, where the $r_t$s are chosen to be worst case:


\begin{align}
    \text{reward of comparator} &: \max_{x \in \mathcal{X}}\sum_{t = 1}^T r_t(x) \\
    \text{reward of algorithm} &: \sum_{t = 1}^T r_t(x_t^{\mathcal{A}})
\end{align}

Typically, in online learning, we study {\em external regret}, which quantifies the difference between the reward of the algorithm and the comparator, as defined below, following \cite{hazan2023introductiononlineconvexoptimization}:

\begin{equation} \label{eqn:external-regret}
    R_T(\mathcal{A}) \coloneqq 
    \sup_{\{r_t\}_{t=1}^T \in \mathcal{R}^T}\left \{ 
    \max_{x \in \mathcal{X}}\sum_{t = 1}^T r_t(x) - \sum_{t = 1}^T r_t(x_t^{\mathcal{A}}) 
    \right \} \,.
\end{equation}
While this can be a reasonable thing to study in general settings, in the improving multi-armed bandits setting, this comparator is not very sensible. Consider a setting in which we are attempting to select hyperparameters that will perform well when we train to convergence. The comparator in external regret asks which single hyperparameter, if chosen, would maximize the {\em witnessed} training accuracy. However, when we deploy the hyperparameter setting, we are not going to train it for only the number of steps for which we ran it in the selection phase; we will train it to convergence. Then, the sequence of rewards will result in a much higher final reward, and in fact the chosen arm might not be the best arm after all. To address this, we define an alternate comparator, the one that appears in {\em policy} regret. \par

Now, instead of considering a comparator that receives the set of reward functions $r_t(\cdot)$ and chooses the best $x\,,$ suppose the comparator can actually interact with the adversary. The comparator sets its algorithm, and, as described above, the adversary chooses a set of reward functions $r_t^\star(\cdot)$ according to the comparator algorithm. Then, we measure how well this comparator algorithm does. Now, we pick the algorithm that maximizes its own reward under this procedure, and the reward of that algorithm, $\mathcal{A}^\star$ that is the value we to which compare. Formally:

\begin{align}
    \text{reward of comparator} &: \sum_{t = 1}^T r^\star_t(x_t^{\mathcal{A}^\star}) \\
    \text{reward of algorithm} &: \sum_{t = 1}^T r_t(x_t^{\mathcal{A}})
\end{align}

The regret metric associated with this comparator, policy regret is defined as follows:

\begin{equation} \label{eqn:policy-regret}
        R_T(\mathcal{A}) \coloneqq  \sup_{\{r^\star_t\}_{t=1}^T \in \mathcal{R}^T}\sum_{t = 1}^T r^\star_t(x_t^{\mathcal{A}^\star}) - \inf_{\{ r_t\}_{t = 1}^T \in \mathcal{R}^T}\sum_{t = 1}^T r_t(x_t^{\mathcal{A}})\,.
\end{equation}


Now, we reproduce Example 1 from \cite{heidari_tight_nodate} to illustrate the difference between external and policy regret in improving multi-armed bandits.

\begin{example}{(Example 1 in \cite{heidari_tight_nodate})}
    Consider a setting in which there are two arms
and time horizon $T>>10$. Arm 1 returns a reward of $i/T$ when pulled for the $i$th time, and arm 2 always returns a reward of 0.1. Consider the algorithm that always pulls arm 2. The external regret of this algorithm is zero because at every time step, it pulls the arm that would give it the largest reward on that time step. But the policy regret of this algorithm grows linearly with $T$, as the best policy in hindsight indeed pulls arm 1 at every time step.
\end{example}

Thus, we can see that policy regret is a stronger notion of regret than external regret. In general, we will thus use the comparator arising from an optimal {\em policy} (algorithm) rather than the best fixed action in hindsight.

\subsection{Optimality and Competitive Ratio}
Having established exactly what we are considering an ``optimal comparator,'' let us now characterize it. Notice that the best strategy in hindsight is to just identify and pull the arm $i^\star = \arg \max_i \sum_{t=1}^T f_i(t)$ the entire time. (Proposition 1 in \cite{heidari_tight_nodate}\footnote{We emphasize that this is {\em still} different from the comparator in external competitive ratio / regret, because the {\em policy} of playing a single arm for all time induces a different (and stronger) sequence of rewards than the sequence of rewards the comparator is evaluated on in the {\em external} setting.}). We will often refer to how things compare to the ``optimal'' arm; when we say that, we are referring to that arm. For simplicity of notation, we elide the $i$ and refer to this arm as $f^\star(\cdot)\,.$ The reward of playing the optimal arm for the whole time horizon $T$ is $\OPT_T\,,$ which we may refer to as $\OPT\,,$ and we say playing the best arm for $T'$ steps accrues reward $\OPT_{T'}\,.$

Our goal is to maximize the expected reward $ALG_T$ achieved by our algorithm, i.e., find a sequence of pulls which gives us the maximum cumulative reward over the $T$ time steps (in expectation over internal randomness in the algorithm)\footnote{As is standard in the bandits literature, we define the objective as maximizing cumulative reward. The reader may notice that in some of the motivating applications, the natural goal instead might be to maximize the largest single pull. In Appendix~\ref{appendix:maxreward}, we show how all the results from the paper translate to this slightly different objective function.}. 

Next, we can show that it is impossible to achieve sublinear additive regret.

\begin{example}
Suppose one arm linearly increases from 0 to $1/2$ until time $T/2\,,$ and then flattens out, and the other arm increases with the same slope throughout until time $T\,.$ The algorithm has no way of distinguishing which arm it is playing until time $T/2 + 1\,.$ At this point, if it finds it is playing the first arm, switching to the other arm is worse than continuing to play the first arm. Since there is an $\Omega(T)$ gap between the rewards of the two arms, no algorithm can achieve $o(T)$ regret. 
\end{example}
Accordingly, we aim to minimize the approximation factor, which we define as follows.
\begin{definition}
    Suppose the expected reward achieved by the algorithm is $ALG$ and the optimal reward achievable is $OPT\,.$ Then, we say that our algorithm achieves a \textit{$g$-approximation} to the optimal reward if $ALG \ge OPT/g\,.$ 
\end{definition}

Not only is approximation factor a natural objective to study (it is fundamental, as defined in the textbook of \cite{borodin_online_2005}), but also it is the metric on which we could hope to understand the problem better. As discussed earlier, achieving sublinear regret in general is impossible, and indeed achieving any non-trivial guarantee with high probability is impossible. To see this, consider a distribution over instances in which there are $k$ arms: $k-1$ of them increase until time $2T/k$ and then flatten, and one arm (chosen randomly) increases linearly throughout. For any algorithm, with probability at least $1/2\,,$ the algorithm will never receive reward better than 
$2T/k\,,$ because the algorithm can choose at most $k/2$ arms to play more than $2T/k$ times. Thus, achieving a non-trivial guarantee with high probability is not possible.

\subsection{Diminishing Returns}
If the rewards are {\em arbitrarily} increasing, then we cannot guarantee much: we could have one arm that gives 0 reward for the first $T/2$ pulls and reward 1 after that, and $k-1$ arms that are 0 regardless of how many times they've been pulled; the good arm and the bad arms in this case are indistinguishable until it is too late. 
Thus, prior papers on this problem \citep{heidari_tight_nodate, patil_mitigating_2023} consider reward functions that have diminishing returns, i.e., where the difference between two consecutive rewards is non-increasing. The continuous equivalent of this property is concavity. 
Following \cite{heidari_tight_nodate, patil_mitigating_2023}, we assume that the reward function for each arm follows the diminishing returns property. Formally,

\begin{definition}
    A function $f$ is said to have {\em diminishing returns} if the following holds for all $t \ge 1$:
    $$
    f(t+1) - f(t) \le f(t) - f(t-1)\,.
    $$
\end{definition}

Finally, we assume that $f(0) = 0$ for all of the arms.







\section{Related Work}

Our work directly follows up on \cite{heidari_tight_nodate,patil_mitigating_2023}. \cite{heidari_tight_nodate} define the setting and problem and shows how to achieve asymptotically sublinear regret\footnote{More specifically, \cite{heidari_tight_nodate} use the fact that for any $k$ bounded and non-decreasing functions $f_i(t)$ and any $\epsilon>0$, there must exist some $T_\epsilon$ after which each function is within $\epsilon$ of its asymptotic value to achieve asymptotically sublinear regret.  In contrast, we will be interested in the case that the adversary can choose the functions $f_i(t)$ after $T$ is fixed.}. \cite{patil_mitigating_2023} extends the model in \cite{heidari_tight_nodate} to the notion of approximation we consider in our paper. They show that deterministic algorithms must incur an approximation factor of $k\,,$ where $k$ is the number of arms; they also match this with an algorithm that can achieve an $O(k)$ approximation factor. We will show that randomized algorithms can overcome the $\Omega(k)$ lower bound deterministic algorithms suffer. \par

A separate line of work that includes \cite{metelli_stochastic_2022, mussi2024best} has studied regret guarantees for the {\em rested rising bandits} problem. The structure of the received rewards are as in our setting. This work has largely focused on developing regret guarantees and solving the best arm identification task. In order to do these things, works in this vein develop guarantees that are a function of the ``niceness'' of the instance, and then they instantiate those guarantees for specific structures of niceness. While in Chapter~\ref{chap:br25} we make no assumptions on the structure of the reward functions beyond diminishing returns, in Chapter~\ref{chap:alg-design-imab}, we study various conditions under which algorithms in the families we define are optimal. In Table~\ref{tab:imab-related-results}, we summarize objectives, conditions, and results from both the competitive ratio and regret lines of work on this problem.







\begin{table}[]
    \centering
    \begin{tabular}{p{2cm}|p{4cm}|p{4cm}|p{3cm}}
        citation & objective (regret/CR) & condition & result \\
        \hline \hline &&& \\
         \cite{heidari_tight_nodate} & regret with different order of quantifiers (fix instance, send $T\rightarrow \infty$ ) & - & sublinear as $T\rightarrow 0$\\
         \hline &&& \\
         \cite{metelli_stochastic_2022} & regret (regular order of quantifiers) & $\Gamma(q)$ measure of how quickly the instance converges (smaller implies easier instance) & $2k + kT^q\, \Gamma(q)$ \\
          \hline && &\\
         \cite{metelli_stochastic_2022} & regret & $f(t) - f(t-1) \le t^{-c}$ & $\begin{cases}
             T & c<1 \\
             kT^{1/c} & c \ge 1
         \end{cases}$  (up to $\log$ factors)\\
         \hline &&& \\
         \cite{patil_mitigating_2023} & competitive ratio & - & $\Theta(k)$\\
         \hline &&& \\
         \textbf{\cite{blum_nearly-tight_2024}} & competitive ratio & - & $\Omega(\sqrt{k})$ lower bound, $O(\sqrt{k}\log k)$ upper bound\\
         \hline &&& \\
        \textbf{\cite{bgrs2025algorithm}} & competitive ratio & $\frac{f(t)}{f(T)} \ge \left( \frac tT \right)^\beta$, $0 < \beta \le 1$ & $\Omega(k^{\beta/(\beta+1)})$ lower bound, $O(k^{\beta/(\beta+1)} \log k)$ upper bound
         \\
         \hline &&& \\
    \end{tabular}
    \caption{This table summarizes results in reward maximization for IMAB (equivalently, rested rising bandits). We collect results for both regret and competitive ratio, as the spirit of both is reward maximization. For brevity, we omit best arm identification results. All works assume diminishing returns. The ``conditions'' column only includes any additional conditions necessary for the results.}
    \label{tab:imab-related-results}
\end{table}


Another area in which a similar problem has been studied is in the area of online algorithms. Here, the problem is framed as one of a class of {\em searching} problems \citep{noauthor_theory_2003, gal_search_2011}. \cite{fiat_oil_2009} consider the problem of a reward (oil, say) present at some depth at one of $k$ locations. Their goal is to minimize the amount of time spent searching; in contrast, in the setup we study, the search time is fixed, and the rewards are gradual, making the objective to maximize reward. \par

More generally, our work follows in a rich line of work regarding multi-armed bandits \citep{thompson_likelihood_1933, lattimore_bandit_2020, slivkins_introduction_2022}.
Multi-armed bandits have been well-studied in many different settings. The main idea is that there are many actions an agent could take, but the utility of taking any action is only known (possibly only partially) after taking it. In particular, the nature of the reward function associated with each arm could be stochastic, adversarial, or dependent on some world context. In this work, we study one such setting with rewards that are adversarial up to conforming to the improving, concave structure.
\chapter{Nearly-Tight Approximation Guarantees in the Worst Case} \label{chap:br25}

\section{Worst-Case Setting}

Having defined the problem we study in this part, let us turn our attention toward solving this problem in the worst case. It is standard in theoretical computer science to study {\em worst-case guarantees}. Typically, we have an algorithm playing against an adversary. The role of the adversary is to pick the instance on which the algorithm must perform, and the algorithm's goal is to solve some optimization problem on that instance. As indicated by the name, in what we will call the {\em worst-case} perspective, the adversary could, if they wished, choose the worst possible instance for that algorithm. That is, the algorithm is playing against an adversary who has access to its details (think pseudocode), but not any specific randomness used. Thus, if the algorithm $A \in \mathcal{A}$ (the class of all randomized algorithms) is solving an optimization problem with (maximization) objective $\mathcal{V}$ on instance $I \in \mathcal{I}\,,$ the goal of the algorithm designer is to design the algorithm that solves the following minimax objective:
\begin{equation} \label{minimax-objective}
\max_{A \in \mathcal{A}} \min_{I \in \mathcal{I}} \mathcal{V}(A(I))
\end{equation}
Once again, note that in this objective, the algorithm designer must commit to the algorithm before the adversary chooses the instance $I\,.$ Equivalently, the adversary has knowledge of the algorithm before choosing the instance and so can pick the worst possible instance for that algorithm. Thus, the algorithm designer must design an algorithm such that its performance on the worst instance for it is better than any other algorithm's performance on its respective worst instance. \par

In our improving multi-armed bandits setting, an instance is a collections of arms. An algorithm identifies a sequence of arms to pull that attempts to maximize the total reward (or the maximum reward seen). In this chapter, we study this problem in the worst case perspective. In doing so, we identify an instance on which no randomized algorithm can surpass a certain performance. Thus, if the adversary chooses that instance as the one on which to evaluate our algorithm, we cannot hope to do better than that performance. We then devise an algorithm that achieves nearly that performance on arbitrary instances. Recall that the performance metric we study is competitive ratio (as opposed to regret as is standard in the bandits literature).

In \cite{patil_mitigating_2023}, the authors study deterministic algorithms and show tight upper and lower bounds of $\Theta(k)$ for the problem. We show that with randomization, we can, in fact, achieve an approximation ratio of $O(\sqrt{k})$ if the algorithm knows the maximum value achieved by the best arm and $O(\sqrt{k}\log k)$ if it does not.  We also provide an $\Omega(\sqrt{k})$ lower bound, nearly completely characterizing the achievable approximation guarantees for this problem.  Our upper bound is achieved through careful analysis of the recursive structure induced by a natural randomized method for searching through bandit arms. Finally, in Appendix~\ref{appendix:maxreward}, we extend our results to the case where the objective is to maximize the maximum reward achieved rather than the total reward achieved, and in Appendix~\ref{sec:noisyrewards}, we show that even if reward functions are only approximately concave (i.e., concave + noise), we can get nearly the same guarantees.

In the rest of this chapter, we start by presenting a lower bound, establishing that no randomized algorithm can achieve a better approximation factor than $\Omega(\sqrt{k})\,.$ Then, we show that an algorithm that knows the maximum value of the best arm can exactly achieve $O(\sqrt{k})$ approximation factor. Finally, we show that the algorithm can estimate the maximum value of the best arm and then run the aforementioned algorithm and still achieve $O(\sqrt{k} \log k)$ approximation factor.

\section{Lower Bound} \label{sec:LB}

First, we argue that no randomized algorithm can achieve an $o(\sqrt{k})$ approximation to the optimal reward. In order to show this, we apply Yao's principle \citep{yao-principle}, i.e., we focus on a distribution over instances that is hard for deterministic algorithms. 
Theorem~\ref{thm:lb} presents the distribution over instances and argues the $\sqrt{k}$ lower bound for deterministic algorithms over this distribution. Corollary~\ref{cor:lb-randalg} uses Yao's principle to translate this result to randomized algorithms. 

\begin{theorem} \label{thm:lb}
    There exists a distribution over instances of the increasing bandits problem where no deterministic algorithm can achieve expected reward greater than $3 \, \OPT/\sqrt{k}$.
\end{theorem}

\begin{proof}
We begin by defining the following functions. Let $f^\star(t) = t/T \; \forall \; t \in [1, T]\,.$ Let $f(t)$ be the following reward function:
    $$
    f(t) = \begin{cases}
        \frac tT & 1 \le t \le \frac{T}{\sqrt{k}} \\
        \frac{1}{\sqrt{k}} & \frac{T}{\sqrt{k}} < t \le T
    \end{cases}\,.
    $$
Let us describe a game played by the universe and the algorithm with the following stages.
    \begin{enumerate}
        \item \textbf{Universe} sets each of $k$ arms to have reward $f(t)\,.$ 
        \item \textbf{Algorithm} runs for $T$ time steps.
        \item \textbf{Universe} chooses one arm $i^\star$ uniformly at random and changes its reward function to $f^\star(t)\,.$
        \item \textbf{Algorithm} gains $\OPT$ reward if it had played the arm replaced with $f^\star$ for at least $T/\sqrt{k}$ steps. Otherwise it keeps the reward it gained originally.
    \end{enumerate}

The distribution over instances is exactly specified by the (uniform) distribution over choices for the arm that gets $f^\star(t)$ as its reward function in step (3) above.

First, we explain how the game detailed above is strictly more generous to the algorithm than the original game in which the algorithm has to maximize reward when playing the instance consisting of $k$ arms. Because the sequence of arms \textbf{Algorithm} pulls is deterministic, if it has not played arm $i^\star$ more than $T/\sqrt{k}$ times, it cannot change its sequence of pulls based on where $f^\star$ is located. 
Thus, provided we give the algorithm any extra reward for playing $i^\star$ long enough to identify it, which is done in step 4, we can fix its run before revealing $i^\star$ in step 3. 

Now, we argue that the above game ensures that the expected reward of the algorithm, over the randomness in the instance, is at most $3\OPT/\sqrt{k}$. We do so by upper bounding the amount an algorithm can achieve while playing this game and lower bounding the value the optimal arm achieves by $T/2$. 

Observe that in the $T$ time steps that \textbf{Algorithm} runs, at most $\sqrt{k}$ of the arms could have been played for more than $T/\sqrt{k}$ steps. Let us consider the two complementary events: 
\begin{align*}
E &= \mathbb{I}\left( f^\star \text{ assigned to an arm \textbf{Algorithm} played for } > T/\sqrt{k} \text{ steps}\right) \\
E^c &= \mathbb{I}\left( f^\star \text{ assigned to an arm \textbf{Algorithm} played for } \leq T/\sqrt{k} \text{ steps}\right)
\end{align*}

Notice that the maximum reward the algorithm can achieve {\em conditioned} on event $E^c$ is by playing a single non-optimal arm for all $T$ time steps.  We thus upper bound \textbf{Algorithm}'s expected reward as follows:
\begin{align*}
    ALG &= \mathbb{P}\left[E\right] \cdot \text{Reward under event }E + \mathbb{P}\left[E^c\right] \cdot \text{Reward under event }E^c \\
    &\le \frac{\sqrt{k}}{k} \OPT + 1 \cdot \left( \frac12 \frac{T}{T\sqrt{k}}\left(\frac{T}{\sqrt{k}} + 1 \right)+ \left(T-\frac{T}{\sqrt{k}}\right) \frac{1}{\sqrt{k}} \right)  \\
    &= \frac{\OPT}{\sqrt{k}} + \frac{T}{k} + \frac{T}{\sqrt{k}} - \frac{T}{k} \le \frac{3\,\OPT}{\sqrt{k}}\,,
\end{align*}

since we know that $\OPT \ge T/2\,.$ 
%
%
\end{proof}

\begin{corollary} \label{cor:lb-randalg}
    For any randomized algorithm, there exists an instance for which its approximation factor is at least  $\sqrt{k}/3$.
\end{corollary}
\section{Upper Bound} \label{sec:UB}

In this section, we present an algorithm with approximation factor $O(\sqrt{k})$, matching the $\Omega(\sqrt{k})$ lower bound in Section \ref{sec:LB}, when the value of $f^*(T)$ and the value of $T$ itself are given to the algorithm in advance.  In Section~\ref{sec:remassump}, we show how to remove these assumptions with an $O(\log k)$ factor loss in approximation. 
To describe the algorithm, we use $m$ to denote the value given to the algorithm that represents $f^*(T)$; we begin by assuming $m=\Theta(f^*(T))$ and then later show how to relax, and then remove, this assumption. Also, we show that in this setting, we can achieve the same approximation factor in terms of $k$ even when the functions do not exactly follow diminishing returns but are close to functions that do (Appendix~\ref{sec:noisyrewards}).

Note that a natural (upper confidence bound-style) algorithm that consistently chooses the arm with the highest slope or optimistic estimate (similar to the algorithm in \cite{patil_mitigating_2023}) could perform suboptimally on instances where some arms increase quickly and then flatten while the best arm increases steadily but less quickly. Indeed, this is evident from the fact that such an algorithm would be deterministic and therefore subject to the $\Omega(k)$ lower bound of \cite{patil_mitigating_2023}.

\paragraph{Algorithm} The algorithm chooses an arm $i$ uniformly at random, pulls it so long as its current reward $f_i(t_i)$ is at least $m t_i / T$, where $t_i$ is the number of pulls of arm $i$ so far, and then switches to a different uniformly random arm. (Notice that if $m \leq f^*(T)$ then we will never switch away from the optimal arm.)  See Algorithm \ref{alg:rrr}.
In Theorem~\ref{thm:alg1approx}, we show that if $m$ is within a constant factor of $f^*(T)$, then the algorithm achieves approximation factor $O(\sqrt{k})$, matching the lower bound from Section~\ref{sec:LB}. We then consider algorithms that aim to adaptively learn a good value of $m$ in Section~\ref{sec:remassump} and give approximation guarantees for the case that $f^*(T)$ and $T$ are not known in advance.

\begin{algorithm}
     \caption{Random round robin}
     \label{alg:rrr}
\begin{algorithmic}
    \State Parameters $m, T$
    \State $t \leftarrow 0$
    \State $R \leftarrow 0$ 
    \While{time not yet expired}
        \State $i \leftarrow$ Randomly choose arm that has not been chosen so far
    \State $t_i \leftarrow 0$
     \While{$f_{i}(t_i) > m \, t_i/ T  $}
     \State $t_i \leftarrow t_i + 1$ \;
    
     \State $t \leftarrow t + 1$ \;
     
     \State pull arm $i$ \;
     
     \State $R \leftarrow R + f_{i}(t_i)$ \;

     \EndWhile
     \EndWhile \\
     \Return{$R$}
 \end{algorithmic}
\end{algorithm}

    
    
    
    

         
    
     
     


To analyze the performance of this algorithm, we begin with the following helpful fact.

\begin{claim} \label{cl:armreward}
    If for some constant $c_2 > 1\,,$ $m \in [\frac{1}{c_2} f^*(T), f^*(T)]$ and Algorithm~\ref{alg:rrr} plays arm $i$ for $t_i$ steps, then, from that arm, the algorithm receives total reward at least $\lbt{t_i-1}$.
\end{claim}
\begin{proof}
    We know that $i$ is played until $f_{i}(t_i) < m \, t_i/ T$. This gives us that the algorithm's reward is at least:
    \begin{align}
         \sum_{\tau = 1}^{t_i-1} f_{i}(\tau) &\ge \frac mT \sum_{\tau = 1}^{t_i-1} \tau \ge \frac{(t_i-1)^2}{2} \frac mT.
    \end{align}
    Since $\OPT_T \leq f^*(T)T \leq c_2 mT$, this is at least $\lbt{t_i-1}$ as desired.
\end{proof}

Next, define $V(T', k') = \mathbb{E}[$ reward from Algorithm \ref{alg:rrr} when run for $T'+k'$ steps]. Note that this is in the worst case over subsets $\mathcal{A}$ of $k'$ of the original $k$ arms, subject to $\mathcal{A}$ containing the optimal arm. The additional $k'$ steps we give the algorithm in this calculation are simply an analysis tool to help account for the last step on an arm that does not gain sufficient reward but is necessary for the algorithm to know it needs to switch. We will later discuss how to use this recurrence to calculate the quantity we are actually interested in, the expected reward after $T$ steps.
We show that $V(T',k')$ satisfies a natural recurrence and then solve that recurrence.

\begin{lemma} \label{lem:recurrence}
    If $m \in [\frac{1}{c_2} f^*(T), f^*(T)]$ then the quantity $V(T', k') = \mathbb{E}[$ reward from Algorithm \ref{alg:rrr} when there are a total of $k'$ arms and the algorithm is run for time $T'+k']$ satisfies the recurrence below:
      \begin{align*}
        V(T', k') 
        &\ge \frac {1}{k'} \OPT_{T'+k'} + \left( 1 - \frac{1}{k'} \right) \min_{0 \le t \le T'+k'-1} \left\{ \lbt{t} + V(T'-t, k'-1)  \right\}
    \end{align*}
    where we define $V(T',k')=0$ for $T' \leq 0\,, \forall \, k'$. 
\end{lemma}
\begin{proof}
With probability $\frac{1}{k'}$, Algorithm~\ref{alg:rrr} finds the optimal arm in its first random choice, in which case it receives reward $\OPT_{T'+k'}$. If the arm chosen was not the best arm, let the time duration for which reward is received from it be $t + 1$. By Claim~\ref{cl:armreward}, the algorithm receives reward at least $\lbt{t}$ while playing that arm, after which the algorithm recurses in a game with one fewer arm (so $k'\leftarrow k'-1$) and with $T' + k' - t - 1$ time steps to go (so $T' \leftarrow T' - t$). Here, we use the fact that we never discard the optimal arm, which follows from $m \leq f^*(T)$.

Since the value of $t$ depends on the adversarially-chosen function $f_i$, we take a worst case view and lower bound it by the worst possible value of $t \in [0,T'+k'-1]$, 
giving us the recurrence shown.
 \end{proof}

 Now, we analyze this recurrence to get a closed form bound on its value in terms of the amount of reward received by the optimal strategy played for $T$ steps.

\begin{lemma}  \label{lemma:recur-value}
    Suppose we run Algorithm~\ref{alg:rrr} with a parameter $m \in [\frac{1}{c_2} f^*(T), f^*(T)]$.
    Then the recurrence given in Lemma~\ref{lem:recurrence} evaluates as follows: for all $T'\,, V(T', k') \ge \lbt{T'} \cdot \frac{1}{\gka{k'}}\,.$ In particular, $V(T, k) \ge  \frac{\OPT_T}{2c_2 \cdot \gk}\,.$ 
\end{lemma}

\begin{proof}
We prove the desired statement by induction on $T'$ and $k'$. 

\paragraph{Base Case:} We start by considering the base cases $V(1, k')$ and $V(T',1)$.  For the first case, notice that if the very first pull of an arm produces reward less than $m/T$, then the algorithm will immediately choose a new arm for its next pull.  Therefore, the algorithm is guaranteed to at least once pull an arm with reward at least $m/T \ge \frac{\OPT_T}{c_2 T^2} \ge \lbt{1}$.  For the second case, the instance only has one arm, which must be the optimal arm, so it clearly receives $\OPT_{T' + 1} \ge \OPT_{T'} \geq \lbt{T'}$ reward, where the last inequality follows from diminishing returns and $2c_2 > 1\,$.

\paragraph{Inductive Assumption:} Assume that $\forall \, T'' < T'\,$ and $k'' < k', V(T'', k'') \ge \lbt{T''} \cdot \frac{1}{\gka{k''}}\,.$

\paragraph{Induction:} 
First, observe that the right-hand side of the  recurrence in Lemma~\ref{lem:recurrence} can be lower bounded by replacing $\OPT_{T'+k'}$ with $\OPT_{T'}$: 
\begin{align}
    V(T', k') \ge \frac{1}{k'} \OPT_{T'} + \left( 1 - \frac{1}{k'} \right) \min_t \left\{ \lbt{t} + V(T'-t, k'-1)  \right\}\,. 
\end{align}
From the inductive assumption, we have: 
\begin{align}
    V(T', k') 
    &\ge \frac{1}{k'} \OPT_{T'} + \left( 1 - \frac{1}{k'} \right) \min_t \left\{ \lbt{t} + \frac{\lbt{T'-t}}{\gka{k'-1}}  \right\} \\
    &\ge \frac{1}{k'} \OPT_{T'} + \left( 1 - \frac{1}{k'} \right) \min_t \left\{ \lbt{t} + \frac{1}{\gka{k'}} \lbt{T'-t}\right\} \label{eqn:plugin}
\end{align}
Next, let us compute the minimum desired. We take the derivative with respect to $t\,$ and set it to 0 (note that at $t=0\,,$ the derivative is negative, and at $t=T\,,$ the derivative is positive, so the minimum must lie in the middle): 
\begin{align}
    \frac{2t}{T^2} \mt + \mt \frac{1}{\gka{k'}} \cdot 2 \left(\frac{T'-t}{T} \right) \frac{-1}{T} &= 0 \\
    t &= \frac{1}{\gka{k'}} (T' - t ) \Leftrightarrow
    t = \frac{T' }{\gka{k'} + 1}
\end{align}

Here, note that $\frac{T'}{\gka{k'} + 1} < T'$. 
Plugging this back in, we get:

\begin{align}
    \min_t \left\{ \left(\frac tT \right)^2 \mt + \frac{1}{\gka{k'}} \left( \frac{T'-t}{T} \right)^2 \mt \right\} &= \mt \left(  \frac{T'}{T}  \right)^2 \frac{1}{\gka{k'} + 1}
\end{align}

Finally, we can plug this back to get the guarantee:
\begin{multline*}
    \frac{1}{k'} \OPT_{T'} + \left(  1 - \frac{1}{k'} \right) \frac{\OPT_T}{2c_2} \left( \frac{T'}{T} \right)^2 \frac{1}{\gka{k'} + 1} \\
\hspace{45mm}
\end{multline*}

\vspace*{-10mm}

\begin{align*}
    & \ge \frac{\OPT_T}{2c_2} \left( \frac{T'}{T} \right)^2\left( \frac{1}{k'}  + \left(  1 - \frac{1}{k'} \right) \frac{1}{\gka{k'} + 1} \right)\\  
     &= \frac{\OPT_T}{2c_2} \left( \frac{T'}{T} \right)^2 \left(  \frac{\gka{k'} + k'}{k'(\gka{k'} + 1)}  \right) = \frac{\OPT_T}{2c_2} \left( \frac{T'}{T} \right)^2 \frac{1}{\sqrt{k'}}\,.
\end{align*}
\end{proof}
With this fact in hand, let us return our attention to computing the reward for the algorithm run for $T$ steps. For this, we must consider first the value of $V(T-k, k)\,,$ which we get in terms of $OPT_{T-k}.$ Thus, we next compare that to $OPT_T\,,$ which is the actual optimal value the algorithm must compete with.

\begin{theorem} \label{thm:alg1approx}
    If $m \in [\frac{1}{c_2} f^*(T-k), f^*(T-k)]$ and $T \geq 2k$, then Algorithm~\ref{alg:rrr}, run with $T-k$ as its parameter ``$T$'', receives expected reward at least $\frac{\OPT_T}{8 c_2\gk}$ in $T$ steps.  
\end{theorem}
\begin{proof}
    Lemma \ref{lemma:recur-value} shows that Algorithm \ref{alg:rrr}, run for $T+k$ steps, receives expected reward at least $\frac{\OPT_T}{2c_2 \cdot \gk}$.  Therefore, if we run the algorithm with $T-k$ as its value of ``$T$'', then in $T$ steps, where $T \ge 2k\,,$ it will receive expected reward at least 
    $$\frac{\OPT_{T-k}}{2c_2 \cdot \gk} \; \geq \left(\frac{T-k}{T}\right)^2 \frac{\OPT_{T}}{2c_2 \cdot \gk} \; \geq \; \frac{\OPT_{T}}{8c_2 \cdot \gk}\,. $$
\end{proof}

We briefly remark here about an interesting practical property of our algorithm. This algorithm chooses a single arm, evaluates it until it fails and then discards it before moving on to another arm. Further, the algorithm performs best in instances where it can discard many or most arms quickly (we exploit this and discuss it more in the next chapter). This serialized approach to decision making is termed a ``singular evaluation approach'' \cite{klein-sourcesofpower}, and empirical evidence shows that many experienced decision makers use this approach in high-stakes decision-making.

\section{Removing Assumptions} \label{sec:remassump}

In the previous section, we assumed knowledge of $f^\star(T)\,,$ the maximum value attained by the best arm, to within a constant factor. In this section, we give an extension of our previous algorithm that essentially makes an educated guess of this value based on some initial exploration. In order to do so, we now require that $T > 4k\,.$ This is so that we can conduct an initial exploration phase for the first half of the time steps and then conduct our previous algorithm for the second half based on our ``learned'' guess for $m\,,$ which we shall call $\hat{m}\,.$

\paragraph{Main Ideas For Learning $\pmb{\hat{m}}$} We spend half of the time available to us ``learning'' the parameter $m$ with which we will run  Algorithm~\ref{alg:rrr}, as detailed in Algorithm~\ref{alg:get-mhat}. In the other half of the time available, we accrue reward through running Algorithm~\ref{alg:rrr}. In order to learn the parameter, we follow the following main steps: first, we pull each arm $T/(2k)$ times. We use the last two values to determine an upper and lower bound for the maximum possible value achievable by {\em that} arm, $f_i(T)\,,$ (lines 1-3 of Algorithm~\ref{alg:get-mhat}, analyzed in Lemma~\ref{lemma:maxintbytstar}). We then narrow down a range inside which the maximum of the {\em best} arm ($f^\star(T)$) must lie (lines 4-5 of Algorithm~\ref{alg:get-mhat}, analyzed in Lemma~\ref{lemma:intervalbdd}). Finally, we randomly choose a value in that interval that, with probability $\Omega(\frac{1}{\log k})$, is within a factor of 2 of $f^\star(T)\,$ (lines 6-7 of Algorithm~\ref{alg:get-mhat}, analyzed in Theorem~\ref{thm:generalresult}). Once we have that, we can use the guarantee from Theorem~\ref{thm:alg1approx} to analyze the output of Algorithm~\ref{alg:rrr} run with the learned parameter $m\,.$\par

Technically, we will learn an estimate of $f^*(\frac{T}{2} - k)$ (rather than $f^*(T)$) since that will be the value of ``$T$'' given to Algorithm \ref{alg:rrr}; for this reason, we provide a parameter $T_{pred}$ to the algorithm below.  Note that $f^*(T)$ and $f^*(\frac{T}{2} - k)$ differ by only a constant factor.

\begin{algorithm}
\caption{Find $\hat{m}$} \label{alg:get-mhat}
    \begin{algorithmic}
    \State \textbf{Input:} parameter $T_\text{pred}$ 
    
    \For{each arm $i$}
        \State pull arm $T/(2k)$ times 
        
        \State $\hat{m}_L^{(i)} \leftarrow f_i\left(\frac{T}{2k}\right)$ 
        
        \State $\hat{m}_U^{(i)} \leftarrow f_i\left(\frac{T}{2k}\right) + (f_i\left(\frac{T}{2k}\right) - f_i\left(\frac{T}{2k} -1 \right)) \left(T - \frac{T}{2k}\right)$ \;
        
     \EndFor
    \State $L \leftarrow \frac 12 \max_i \hat{m}_L^{(i)}$ \;
    
    \State $U \leftarrow \max_i \hat{m}^{(i)}_U$ \;
    
    \State Define uniform probability distribution $p$ over $\{-\log (T/T_\text{pred}), \hdots, \, 0,  1, 2, \hdots\,, \log(U/L) \}$ 
    
    \State Draw $j \sim p$ \\
    
     \Return{$L \cdot 2^j$}
    \end{algorithmic}
\end{algorithm}

    
        
        
        
    
    
    
    

\subsection{Computing Range for a Fixed Arm}
Here, we describe how we compute the range in which the maximum lies for a fixed arm, i.e., we compute upper and lower bounds on $f_i(T)$ based on an exploration phase on $f_i\,.$


        

\begin{lemma} \label{lemma:maxintbytstar}
    The procedure detailed in lines 1-3 of Algorithm~\ref{alg:get-mhat} provides, for a fixed arm with reward function $f$, a range $\mathcal{R} \coloneqq [\hat{m}_L, \hat{m}_U]$ such that $f(T) \in \mathcal{R}$ and $\hat{m}_U \le  2k \, \hat{m}_L\,.$ 
\end{lemma}
\begin{proof}
    We use the diminishing returns property to show both parts. First, by the fact that the reward functions are increasing, it is clear that $f(T) \ge f(T/(2k))\,.$ Then, due to the diminishing returns property:
    \begin{align}
        f(T) 
        &= f\left(\frac{T}{2k}\right) + \sum_{n = 1}^{T-\frac{T}{2k}} \left \{f\left(\frac{T}{2k} + n\right) - f\left(\frac{T}{2k} + n -1\right) \right \}\\
        &\le f\left(\frac{T}{2k}\right) + \left(T - \frac{T}{2k}\right) \left( f\left(\frac{T}{2k} + 1\right) - f\left(\frac{T}{2k}\right)  \right) \\
        &\le f\left(\frac{T}{2k}\right) + \left(T - \frac{T}{2k}\right) \left( f\left(\frac{T}{2k}\right) - f\left(\frac{T}{2k} - 1\right)  \right)\,.
    \end{align}

Next, we show that $\hat{m}_U \le  2k \hat{m}_L$, which follows immediately from the diminishing returns property. 
 In particular, $\hat{m}_U =  f(\frac{T}{2k}) + (f(\frac{T}{2k}) - f(\frac{T}{2k}-1)) (T-\frac{T}{2k}) \leq f(\frac{T}{2k}) + \frac{2k}{T}f(\frac{T}{2k})(T-\frac{T}{2k}) = f(\frac{T}{2k}) + (2k-1)f(\frac{T}{2k}) = 2k \, \hat{m}_L.$
 \end{proof}

\subsection{Upper and Lower Bounds on Final Value of Best Arm From Exploration}
Now, for each arm, we have a range within which its maximum must lie. Next, we combine these bounds to get a relatively small interval in which we can be sure $ f^\star(T)$ lies. 

\begin{lemma}\label{lemma:intervalbdd}
    Define $L \coloneqq \frac 12 \max_{i} \hat{m}^{(i)}_L$ and $U \coloneqq \max_{i} \hat{m}^{(i)}_U\,.$ Then, $f^\star(T) \in [L, U]\,,$ and $\frac{U}{L} \le 4k.$
\end{lemma}
\begin{proof}
First, observe that by diminishing returns, we have that: 
    \begin{align}
        f^\star(T) \, T  \ge \sum_{t = 1}^{T} f^\star(t) &= \max_i \sum_{t = 1}^{T} f_i(t) \ge
        \sum_{t=1}^{T} f_p(t) \quad \text{ where } p \coloneqq \arg \max_i f_i\left(\frac{T}{2k}  \right) \\
        &\ge f_p\left(\frac{T}{2k}  \right) \frac {T}{2} = \hat{m}^{(p)}_L \frac{T}{2}
        \Leftrightarrow f^\star(T) \ge \frac{\hat{m}^{(p)}_L}{2} = L\,.
    \end{align}
    
Next, since $f^\star(T) \le f^\star(\frac{T}{2k}) + (T-\frac{T}{2k}) ( f^\star(\frac{T}{2k}) - f^\star(\frac{T}{2k} - 1))$ (due to diminishing returns), we also have that $f^\star(T) \le \max_i f^{(i)}(\frac{T}{2k}) + (T-\frac{T}{2k}) ( f^{(i)}(\frac{T}{2k}) - f^{(i)}(\frac{T}{2k} - 1)) = \max_i \hat{m}^{(i)}_U = U\,.$

Now that we have that $m \in [L, U]\,,$ we verify the size of the interval. In particular, for some index $i'$ we have $U = \hat{m}^{(i')}_U \leq 2k \hat{m}^{(i')}_L \leq 4kL$, where the first inequality follows from Lemma \ref{lemma:maxintbytstar}.
%
\end{proof}

\subsection{Reward Approximation Guarantee With Learned Parameter}
Putting the pieces together, we show that choosing $\hat{m}$ randomly according to the defined uniform distribution in line 6 of Algorithm~\ref{alg:get-mhat} allows us to get reward as before with the loss of a factor only logarithmic in $k$.

\begin{theorem} \label{thm:generalresult}
When $T > 4k\,,$ picking the parameter $m$ required by Algorithm~\ref{alg:rrr} using the procedure detailed in Algorithm~\ref{alg:get-mhat} with parameter $T_\text{pred} = \frac{T}{2} - k$ and running Algorithm~\ref{alg:rrr} with that value $\hat{m}$ achieves expected reward $ALG =\Omega(\OPT/(\sqrt{k} \log k))\,.$
\end{theorem}
\begin{proof}
    To show this, we analyze the three parts of the algorithm. First, we analyze lines 1-3 of Algorithm~\ref{alg:get-mhat} in Lemma~\ref{lemma:maxintbytstar}. Then, we analyze the range $\mathcal{R} = [L, U]$ in Lemma~\ref{lemma:intervalbdd}. 
    With these in hand, we show that we can find a constant factor approximation to $f^\star(T)$ with good enough probability that the overall expected reward only gets worse by a logarithmic factor.

    First, let us untangle the relationship between $f^\star\left( \frac{T}{2} - k\right)\,,$ which we must use in calling Algorithm~\ref{alg:rrr} to get the guarantee as per Lemma~\ref{lemma:recur-value}, and $f^\star(T)\,,$ which we know we can approximate from the previous two lemmas. Observe that due to increasing nature of $f^\star\,, f^\star(T) \ge f^\star\left( \frac{T}{2} - k\right) \ge f^\star\left( \frac{T}{4} \right)\,.$ Further, due to the diminishing returns property, we have $f^\star\left( \frac{T}{4}\right) \ge f^\star(T) \frac{T}{4}/T  = \frac{f^\star(T)}{4}\,.$ Thus, since $U \ge f^\star(T) \ge L\,, U \ge f^\star(T/4) \ge L/4\,.$ Note that $\log(T/T_\text{pred}) \le \log(4).$

    We focus on the selection of $\hat{m}\,.$ We divide up the interval $\mathcal{R}$ by doubling $L\,,$ i.e., the $i^{th}$ interval endpoint is $L \cdot 2^{i-1}\,,$ for $ i \in \{-2 \le -\log(T/T_\text{pred}), \hdots, 1, 2, \hdots, \log\left(\frac{U}{L}\right) \le \log(\polyk)  \}\,.$ 
    We know that $f^\star(T/4)$ must lie in one of these intervals, so if we pick the endpoint, we will be within a factor of 2 as desired above. In particular, let us choose $L \cdot 2^i$ uniformly over the set. 
    The probability of choosing $L \cdot 2^i$ is exactly $\frac{1}{3 + \log(U/L)}$ which is at least $\frac{1}{3 + \log(\polyk)}\,.$ Thus, with probability $\ge \frac{1}{3 + \log(\polyk)}\,,$ we choose $\hat{m}$ such that $\hat{m} \le f^\star(T/2 - k) \le 2\hat{m}\,.$
    When $\hat{m} \le f^\star(T/2 - k) \le 2 \hat{m}\,,$ we have that $\hat{m} \le f^\star(T) \le 8 \hat{m}\,,$ since $f^\star(T)/4 \le  f^\star(T/2 - k) \le f^\star(T)$
    Thus, we can apply Lemma~\ref{lemma:recur-value} with $c_2 = 8\,,$ and $T' = T/2 - k$ to get that:
    $$
    ALG_{T/2} = V\left(\frac T2 - k, k\right) \ge \frac{\OPT_{T/2-k}}{2c_2 \gk} \ge \left( \frac{\frac T2 - k}{T} \right)^2 \frac{\OPT_T}{2c_2 \gk} \ge  \frac{\OPT_T}{256 \sqrt{k}}\,.
    $$


    Putting these together, the expected reward is at least 
    $\frac{1}{3 + \log(\polyk)}$
    times the reward as described above, giving us that the reward here is at least:
    $$
    \frac{1}{3 + \log(\polyk)} \cdot \frac{OPT_T}{256 \sqrt{k}} \ge \Omega \left( \frac{OPT}{\sqrt{k} \log(k)}\right)\,.
    $$
\end{proof}

\subsection{Removing Dependence on Having To Know T}

Finally, we describe how to use a standard doubling trick to remove dependence on knowing $T,$ the time horizon, a priori. At a high level, we start with a guess, $T' = T_0 = 4k\,,$ pretend that is all the time we have, and run the combined exploration-exploitation algorithm with it. Then, if it turns out we have not run out of time, we set $T' \leftarrow 2 \cdot T'\,$ and repeat the same process. At a high level, our argument will show that if we have a ``good'' $T'\,,$ i.e., one that is within a constant factor of the true $T\,,$ then the reward received when 
running $T'/2$ steps of Algorithm~\ref{alg:get-mhat} followed by $T'/2$ steps of Algorithm~\ref{alg:rrr} will be within a constant factor of the reward received if we were to play until time $T\,.$
Overall, in the exploration phase, we use the algorithm given in Algorithm~\ref{alg:get-mhat}, but a key change is that in line 5, we instead set $U = 4 \max_i \hat{m}_U^{(i)}\,.$ We present the algorithm and result here and defer the proof to Appendix~\ref{appendix:unknownTproof}.

\begin{algorithm}
\caption{Unknown Time Wrapper} \label{alg:Tuk}
\begin{algorithmic}
    \State $R \gets 0$ \;
    
    \State Guess $T' \leftarrow T_0$ \;

    \While{True}
        \State Try: \;
        
        \hspace{5mm} $T_\text{pred} = T'/2 - k$ \;
        
        \hspace{5mm} $\hat{m} \gets $ modified Algorithm~\ref{alg:get-mhat} for $T'/2$ steps with $T_\text{pred}$ as above \;

        \hspace{8mm} \texttt{// in line 5 of Algorithm~\ref{alg:get-mhat}, $U = 4 \cdot \max_i \hat{m}_U^{(i)}$} 

        \hspace{5mm} $R \gets R \, + $ reward from Algorithm~\ref{alg:rrr} for $T'/2-k$ steps \;

        \State Except: \;
        
        \hspace{5mm} \Return{R} \;

        \State $T' \gets 2 \cdot T'$
    
    \EndWhile
    \end{algorithmic}
\end{algorithm}

    

        
        



        

    

\begin{lemma} \label{lemma:unknownT}
    Suppose we run the procedure described in Algorithm~\ref{alg:Tuk}. If $T > 4k\,,$ then the reward achieved is at least $\frac{\OPT_T}{8192 \gk \log(128k)}$.
\end{lemma}

\section{Conclusion}


In this work, we provided nearly-tight guarantees for the improving multi-armed bandits problem. In particular, we showed that no randomized algorithm could hope to achieve better than an $\Omega({\sqrt{k}})$ approximation factor, and we provided an algorithm that achieves an $O(\sqrt{k}\log k)$ approximation factor. (If the algorithm has access to an additional piece of information, then the approximation factor is $O(\sqrt{k})\,.$) \par

This work provides a theoretical result for a stylized decision-making setting. However, the algorithmic techniques are interesting for their simplicity and practicality. The core idea of the algorithm is to play an arm as long as it {\em could} still be the best arm and give up as soon as we have evidence that it cannot be. This is a very practical strategy. Notably, compared to many algorithms, the worst-case number of arm switches is much smaller. Also it serializes evaluation of the current option, reflecting a decision-making strategy called ``singular evaluation'' \cite{klein-sourcesofpower}. For those of us that started working on this project in search of life advice, there is some satisfaction that we developed such a simple strategy as an answer.

\chapter{Beyond Worst-Case: Data-Driven Algorithm Design} \label{chap:alg-design-imab}


The previous chapter considered the worst-case analysis setting familiar to theoretical computer scientists. In that setting, we managed to get strong algorithmic guarantees that are known to be nearly optimal for very general instances. Indeed, we made very few assumptions on the instances. On the other hand, the lower bounds are, as a result, somewhat pessimistic. While the strong adversary in that setting could indeed choose to cause us great trouble, we might also hope the world is not quite so antagonistic. \par

One natural way to overcome pessimistic lower bounds would be to strengthen the conditions satisfied by instances and then derive algorithms that are optimal for those. However, then we are stuck with algorithms that only work if we know those conditions hold. In this chapter, we go beyond last chapter's analysis in a different way. We propose to learn good algorithms from data, following the data-driven algorithm design paradigm (Chapter 29 \cite{balcan2020data} of \cite{Roughgarden_2021}). Instead of assuming our instances satisfy further conditions, we instead assume we have access to offline instances that are {\em similar} to the instances we expect to see in the future. Thus, we can adapt to more specific settings. However, the tradeoff is that we must have access to historical instances. We explore this perspective in this chapter.

\section{Data-Driven Algorithm Design Setting}

As before, we have an algorithm playing against an adversary. The role of the adversary is to pick the instance on which the algorithm must perform, and the algorithm's goal is to solve some optimization problem on that instance. In contrast to the previous chapter, we now will study a situation where the adversary can only define a distribution over instances. Then, a neutral party chooses the specific instance randomly, and the goal of the algorithm will be to do well on average over instances drawn from the distribution.
Thus, if the algorithm $A \in \mathcal{A}$ (the class of all randomized algorithms) is solving an optimization problem with (maximization) objective $\mathcal{V}$ on instance $I \sim \mathcal{D}$ over domain $\mathcal{I}\,,$ the goal of the algorithm designer is to design the algorithm that solves the following {\em expected} objective:
\begin{equation} \label{avg-objective}
\max_{A \in \mathcal{A}} \mathbb{E}_{I \sim \mathcal{D}}{ \mathcal{V}(A(I))}
\end{equation}

Note that if we can get guarantees that hold for any distribution, then that guarantee must, in particular, hold for the worst distribution the adversary could pick. \par

In the improving multi-armed bandits setting, an instance comprises a set of $k$ arms, each of which has an increasing reward function. Thus, when we talk about a distribution over instances, we are considering distributions that are supported on $\mathcal{I}$, the space of all $k$ tuples of arms. (Note that this is richer than assuming a distribution over arms and then taking its product $k$ times to get a product distribution over instances.)

In the previous chapter, we developed nearly-tight approximation guarantees for this problem. Work on the problem has typically assumed that the reward functions are concave (i.e., satisfy diminishing returns), and lower bounds  involve at least some arms that have minimally concave (i.e., linear) reward functions. However, in many practical settings, we would expect the reward functions to not only be concave but also satisfy some stronger condition on the growth rate. In this chapter, we study the problem of designing algorithms for the improving bandits problem that achieve stronger performance guarantees on more benign problem instances. In particular, we design families of algorithms parameterized by some parameter(s) $\alpha\,,$ and we wish to learn the ``best'' algorithms from this family.\looseness-1 \par

Consider the example of tuning hyperparameters such as learning rates for neural networks. Here each bandit arm corresponds to a value of the hyperparameter, an arm pull corresponds to running an additional training epoch for the corresponding value of the hyperparameter, and the reward function corresponds to the learning curves that capture the training accuracy as a function of the epochs (for the different hyperparameter values). Often the number of arms is very large (e.g.\ a grid of multi-dimensional hyperparameters), so existing approximation factors may be too pessimistic here. Furthermore, in many practical settings, we have access to historical data consisting of learning curves for training similar models on related tasks or datasets (similarly, we may have access to past clinical records, or data regarding click-through-rates for online advertising and recommendation). We can use these related previously-seen tasks to design our algorithm for the current instance. Formally, we assume we have access to multiple IMAB instances drawn from an unknown distribution. We seek to perform as well as the best parameter $\alpha$ in the defined algorithm family, on average over the distribution.\looseness-1

In order to do this, we first develop stronger approximation guarantees that depend on the strength of concavity of the reward functions. Our guarantees are achieved by a family of algorithms, parameterized by a parameter $\alpha$ that corresponds to the strength of concavity. We show that by setting $\alpha$ appropriately, we achieve the optimal approximation guarantees for every strength of concavity. In other words, if we know how ``nice'' our improving bandits instance is, we can use the appropriate algorithm from our family. On the other hand, if we do not have this information, we resort to a data-driven approach to learn the best algorithm parameter from the data. We obtain bounds on the sample complexity, that is, the number of IMAB instances sampled from the distribution that suffice to learn the best algorithm. \par

Next, we turn our attention to the best-arm identification (BAI) task. Approaches from prior work either guarantee that the best arm is exactly recovered on a sufficiently ``nice'' instance (using a notion of niceness that is different from the strength of concavity~\cite{mussi2024best}), or select an arm such that the reward of the selected arm is a good approximation to the reward of the best arm on any (worst-case) instance~\cite{blum_nearly-tight_2024}. However, the former may suffer from sub-optimal approximation ratios on more challenging instances, while the latter may fail to recover the exact best arm on the nicer instances. We propose a hybrid algorithm which resolves this gap in the literature and obtains a best-of-both-worlds guarantee. That is, on a nice instance, our algorithm will recover the best arm, while still guaranteeing the optimal approximation factor (up to constants) on the worst-case instance. We further show how to tune the parameters of our hybrid algorithm to obtain the near-optimal parameters on typical instances for a fixed problem domain, by giving bounds on the sample complexity of data-driven algorithm design.

\subsection{Our Contributions}
\begin{enumerate}
    \item We introduce a parameter $\beta$ to measure the ``strength'' of concavity of reward functions and develop algorithms with optimal approximation ratios for each $\beta$ (Sections~\ref{subsec:algfam-ptrr} and \ref{sec:sharper CR}). For parameter $\beta\in(0,1]$, we show that the optimal approximation ratio is $O(k^{\beta/(1+\beta)})$, which gives an improvement over the worst-case optimal bounds of $O(\sqrt{k})$ from prior work whenever $\beta<1$.\looseness-1
    \item Our algorithms that achieve the optimal guarantees constitute a parameterized family. In Section~\ref{sec:sample-complexity}, we show how to tune the algorithm parameter, given access to similar instances drawn from a fixed but unknown distribution over IMAB instances. We employ techniques from data-driven algorithm design to bound the sample complexity of configuring our algorithm given reward curves of ``training'' instances. 
    As in typical data-driven algorithm design settings, our results are optimal on average over the distribution. However, we can additionally reason about instances on which we can get strong per-instance guarantees due to the relationship between our algorithm family and sufficient conditions for stronger guarantees.\looseness-1 
    
    Our approach of designing a parameterized family of assumptions (that includes worst-case instances for a fixed value of the parameter) and a corresponding tunable family of algorithms that give optimal algorithms for each value of the assumption parameter is novel in the context of data-driven algorithm design, and may be useful to design algorithms with powerful expected-case as well as per-instance worst-case guarantees in the context of other algorithm design problems.
    \item In Section~\ref{sec:bothworlds}, we study best-arm identification (BAI) for improving bandits. Previous literature has a gap for this problem. Namely, there are algorithms that achieve exact BAI on certain ``nice'' instances but with sub-optimal worst-case performance~\cite{mussi2024best}, and the algorithms that achieve the near-optimal worst-case approximation~\cite{blum_nearly-tight_2024} fail to identify the best arm on these easier instances. In Sections~\ref{sec:hybriddef} and \ref{sec:hybridguarantees}, we propose a hybrid approach that switches between an algorithm that explores many arms and our parameterized algorithm family above and obtains best-of-both-worlds guarantees. We also  bound the sample complexity of simultaneously tuning the switching time of our hybrid approach and the concavity-strength  parameter.
\end{enumerate}

\textbf{Related Work.} We improve the tight worst-case bounds on $\tilde{\Theta}(\sqrt{k})$ on the competitive ratio of IMAB by exploiting the strength of concavity of instances. Prior work \cite{metelli_stochastic_2022,mussi2024best} also gives regret bounds for more benign IMAB instances, and we discuss their notions of benign-ness in Chapter~\ref{chap:imabsetting}. We develop algorithms that achieve a best-of-both-worlds guarantee by simultaneously achieving low regret on benign instances and asymptotically optimal competitive ratios on worst-case instances. We also extend techniques from the data-driven algorithm design framework~\cite{balcan2020data,sharma2025offline} to the IMAB setting to bound the sample complexity to tuning the parameters in our algorithms. See Appendix \ref{appendix:additional-related} for a more detailed discussion on the related prior literature.

\section{Data-Driven Algorithm Design Perspective}
\label{sec:data-drivenpreliminaries}

In Sections~\ref{sec:sample-complexity} and \ref{sec:hybridcomplexity}, we take a {\em data-driven algorithm design} perspective, i.e., we show that we can use samples from a distribution over instances to {\em adapt} to the setting at hand and get better guarantees. First, we describe the context and motivation for this perspective. 

\begin{figure}
    \centering
    \includegraphics[width=0.9\linewidth]{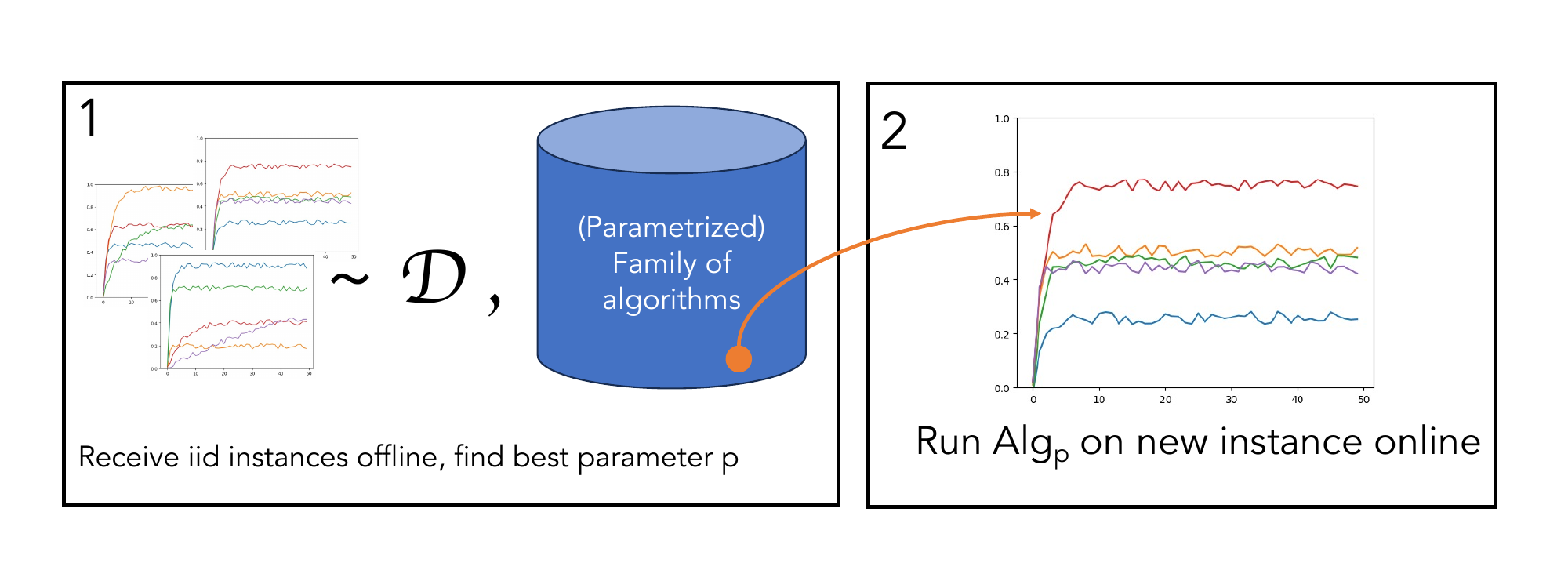}
    \caption{This figure summarizes the framework studied in this chapter. In phase 1 (left), the learning algorithm receives instances sampled iid from some distribution. It uses its {\em offline} access to these instances to select the best algorithm from a parameterized family of algorithms. This algorithm is the version with parameter $p\,.$ In the phase 2 (right), the algorithm with the value of the parameter set to $p$ is run {\em online} on a new instance. Thus, we can see why we call this framework ``offline-to-online transfer.''}
    \label{fig:offline-online}
\end{figure}

\paragraph{Framework.}
Suppose that we are in a setting where we must train and deploy a new machine learning model for a prediction problem each month. Each month, we must select hyperparameters and then train a deployable model with those hyperparameters. How can we leverage historical data to solve the hyperparameter selection problem? We neither want to assume that the best model on a given day is the best model on another day, nor even that the best hyperparameters for one day are the best hyperparameters for another day. Instead, let us simply assume that the algorithm we use for hyperparameter selection should be similar across days, and we wish to learn the best such algorithm. Now suppose we have historical data from, January to June, and we wish to use that {\em offline} data to learn which hyperparameter selection algorithm is most appropriate to deploy in July. Further, assume we can inspect the available data {\em completely}, i.e., for each model and hyperparameter setting trained in January through June, we have access to the {\em full} learning curves. Then, our goal will be to learn a good algorithm from a family of algorithms. In this work, we identify relevant families of algorithms to solve the problem of {\em identifying the best bandit algorithm for future hyperparameter tuning} and study what happens when we choose an algorithm from this family by using empirical risk minimization (ERM) with respect to the offline instances. Finally, with this ``best'' algorithm identified, we deploy it on future hyperparameter tuning problems. We conclude that if future instances are drawn from the same distribution as the training instances, then in expectation over the distribution, we will do well on the future instances.\looseness-1

Now, we formally define the preliminaries for our study of the improving multi-armed bandits problem in the data-driven algorithm design framework following prior work on stochastic bandits~\cite{sharma2025offline}.
We consider loss functions of an algorithm in an algorithm family on an instance, defining $l(I, \pmb{\alpha})$ as the loss of the algorithm with the parameters fixed to $\pmb{\alpha} \in \mathcal{P}$ on instance $I\,.$ We also define the dual of a loss function.\looseness-1


\begin{definition}
    A loss function $\ell: \mathcal{I} \times \mathcal{P} \rightarrow \mathbb{R}$ is {\em piecewise-$H$-bounded} if it is piecewise-constant and has a  bounded range $[0, H]$.
    We fix the instance and consider the loss on a fixed instance as a function of the algorithm. Since the algorithm arises from a parameterized family, the loss is  a function of the parameter vector. In particular, the {\em dual of the loss function} is given by: $l_T^{I}(\pmb{\alpha}) = \ell_T(I, \pmb{\alpha})\,.$ 
\end{definition}

\noindent Finally, we formally define the problem we study. 

\begin{definition}[Hyperparameter Transfer Setting] \label{defn:hypertransf}
    Suppose we have a distribution $\mathcal{D}$ over $\mathcal{I}\,,$ the space of instances $I$ of the improving multi-armed bandits problem as defined in Definition~\ref{defn:imab}. Consider a family of algorithms parameterized by a vector of parameters $\pmb{\alpha} \in \mathcal{P}$. Finally, consider a piecewise-$H$-bounded loss, $l.$ We achieve sample complexity $N(\epsilon,\delta)$ in the {\em Hyperparameter Transfer Setting} if, for any $\epsilon,\delta\in(0,1)$, given $N(\epsilon,\delta)$ instances sampled iid from $\mathcal{D}\,,$ we identify 
    $
    \hat{\pmb{\alpha}}$ such that with probability $1-\delta$,  
    \begin{equation} \label{eqn:goal} 
    \left |\E{I \sim \mathcal{D}}{l_T(I, \hat{\pmb{\alpha}})} - \min_{\pmb{\alpha} \in \mathcal{P}} \E{I \sim \mathcal{D}}{l_T(I, \pmb{\alpha})} \right | < \epsilon\,.
    \end{equation}
\end{definition}

\section{Using Strength of Concavity
}\label{sec:intro3}

In this section, we introduce a family of algorithms designed to address the question of providing sharper bounds for the improving multi-armed bandits problem. The design of this family is motivated by two factors: (1) the family must contain the algorithm that is known to provide the worst-case near-optimal guarantee of \cite{blum_nearly-tight_2024}; (2) there must exist an algorithm in the family that performs better than the general worst-case if instances satisfy a stronger regularity condition. In Section~\ref{subsec:algfam-ptrr}, we define the family. Then, in Section~\ref{sec:sharper CR}, we define a strengthening of concavity and show that a variant of \cite{blum_nearly-tight_2024} achieves optimal competitive ratio guarantees on instances that satisfy the strengthened property. In Section~\ref{sec:sample-complexity}, we show that we can learn the best algorithm from this family for instances arising from a distribution with polynomially-many samples. Finally, in Section ~\ref{sec:empirical}, we present empirical evidence on real learning-curve data that different instances prefer different values of $\alpha$, corroborating the intuition that adapting the choice of this parameter to data is valuable.\par

The data-driven perspective we take in this section differs from standard worst case guarantees in several ways. On the one hand, we do not have to make stronger assumptions to get the guarantees and can instead {\em adapt} to the distribution of data. On the other hand, we must be in a setting where such a distributional assumption holds and we have access to other instances from the distribution. \looseness-1

\subsection{Algorithm Family}
\label{subsec:algfam-ptrr}

\noindent 
 Each algorithm in the family, $\textit{PTRR}_\alpha$ for fixed $\alpha$, is a slightly modified version of Algorithm $1$ in \cite{blum_nearly-tight_2024}. While playing arm $i$, we keep pulling it as long as 
$
f_i(t_i)\geq m\Bigl(\frac{t_i}{\tau}\Bigr)^\alpha,
$
where $\tau$ is an internal horizon parameter 
(set to $T-k$ when $T$ is known), $m$ is an approximation of the maximum final pull, and $t_i$ denotes the number of pulls of $i$ so far. 
When the inequality first fails, we abandon $i$ and move to a uniformly random new arm, repeating until time $T$. In this section, we focus on the cumulative reward $R$ accumulated by the algorithm. We will study the estimated best arm $\hat{i}$ in Section~\ref{sec:bothworlds}.\looseness-1

\begin{algorithm}[H]
\caption{$\textit{PTRR}_\alpha$} \label{alg:ptrr}
\begin{algorithmic}[1]
\State estimated max final pull $m$,  internal horizon  $\tau$
\State $t \gets 0$, $R \gets 0$, $S \gets \{1, \ldots, k\}$
\While{$t < T$}
  \State \textbf{sample} $i$ uniformly from $S$
  \State $S \gets S \setminus \{i\}$,  $t_i \gets 0$
  \While{$f_i(t_i) \geq m\Bigl(\frac{t_i}{\tau}\Bigr)^\alpha$}
    \State \textbf{pull} arm $i$
    \State $t_i \gets t_i + 1$, $t \gets t + 1$, $R \gets R + f_i(t_i)$
  \EndWhile
\EndWhile \\
\Return $R, \hat i\gets \text{argmax}_i f_i(t_i)$ 
\end{algorithmic}
\end{algorithm}


\begin{figure}
    \centering
    \begin{subfigure}{0.45\textwidth}
    \includegraphics[width=0.9\textwidth]{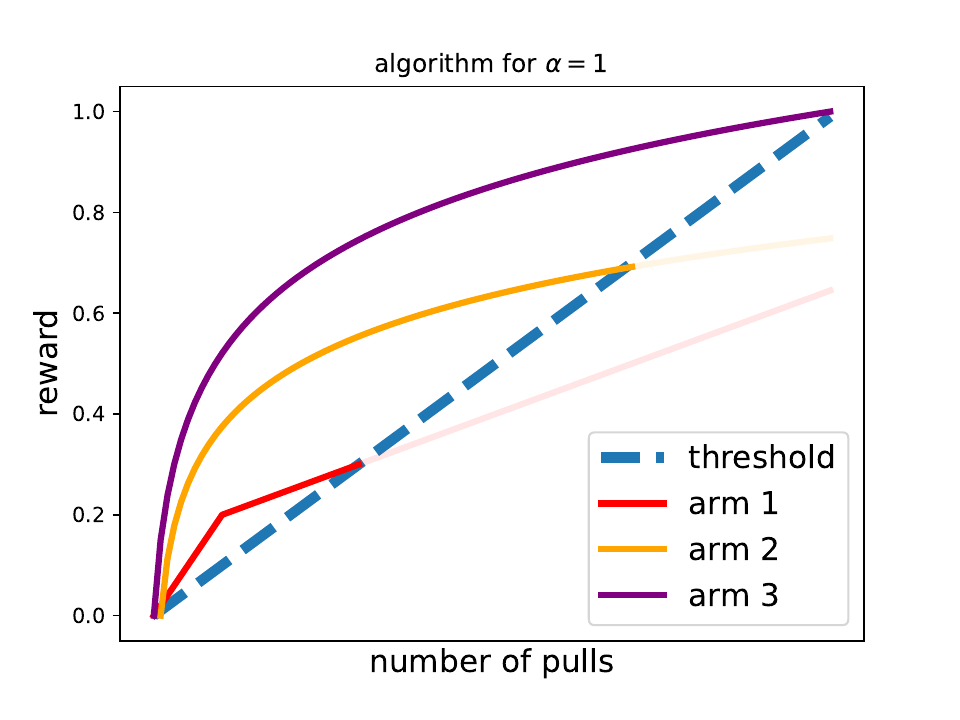} 
    \end{subfigure}
    \begin{subfigure}{0.45\textwidth}
        \includegraphics[width=0.9\textwidth]{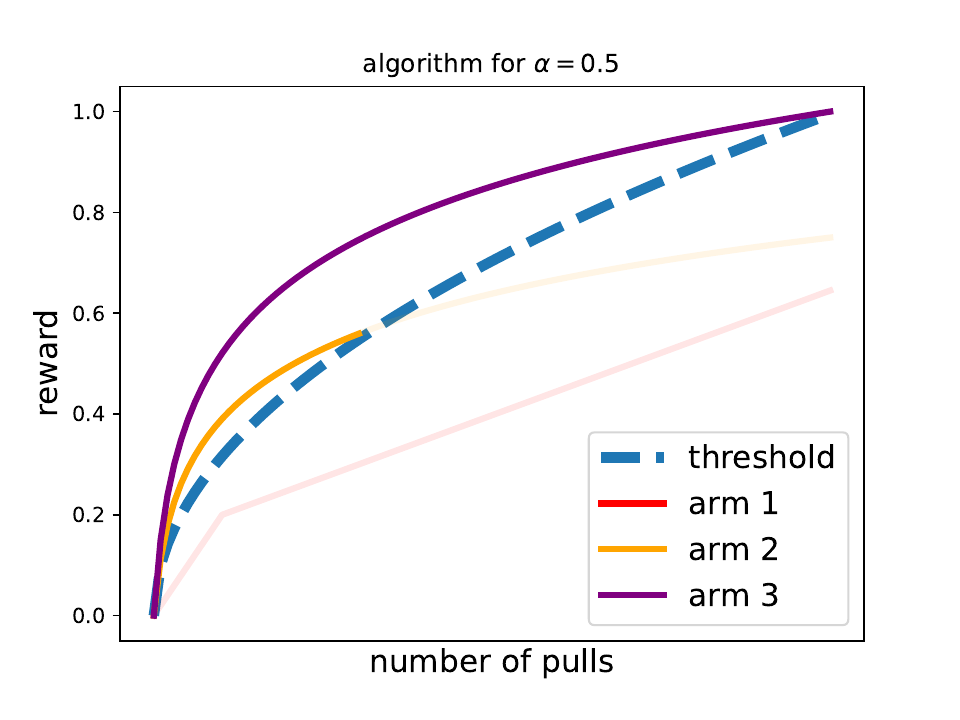} 
    \end{subfigure}
    \caption{This figure shows a snapshot of running PTRR with differing $\alpha$s on the same instance. In the case where $\alpha = 1\,,$ recovering the algorithm of \cite{blum_nearly-tight_2024}, we see that each arm is discarded once it crosses the linear lower bound. Note that the best arm (purple) is never discarded. On the right, we see the run of the algorithm when $\alpha=0.5$. Here, since the best arm satisfies the CEE with $\beta=0.5\,,$ we never discard the best arm. However, we discard worse arms faster. Thus, we can see the motivation for using $PTRR_\alpha$ for the largest possible $\alpha$ such that we still ensure we do not discard the best arm.}
    \label{fig:alg-snapshot}
\end{figure}

\begin{definition} \label{defn:alg-fam}
    Define the family of algorithms $\textit{PTRR}$ as $ \{ \textit{PTRR}_\alpha: \alpha \in (0,1] \},$ where $\textit{PTRR}_\alpha$ ($\alpha$-Power-Thresholded Round Robin) is Algorithm~\ref{alg:ptrr}.
\end{definition}

\subsection{Sharper Competitive Ratio
}\label{sec:sharper CR}

We will now argue that it is natural to consider the simple one–parameter family \textit{PTRR} from  Definition~\ref{defn:alg-fam}.
First, we observe that our family contains the algorithm from \cite{blum_nearly-tight_2024} that is known to be optimal in the unrestricted case. We then show that there are algorithms in \textit{PTRR} that improve the competitive ratio on every instance that satisfies a mild growth condition.
A matching lower bound shows that these algorithms are optimal on such instances.

For simplicity, we assume that both $T$ and $f^*(T)$ are known to the algorithm. In reality, our analysis---mirroring \cite{blum_nearly-tight_2024}---only requires that $m$ lies within a constant factor of $f^*(T-k)$, which we achieve by setting $m:=\frac{T-k}{T} \cdot f^*(T)$. We can avoid this requirement altogether by spending half our time learning $m$, thereby incurring only an extra $O(\log k)$ factor in our competitive ratio. This method is described in detail in section $5$ of \cite{blum_nearly-tight_2024}. When $T$ is also unknown, we embed the same half-explore/half-exploit method into a standard doubling schedule, preserving this $O(\log k)$ overhead. A full proof is included in Appendix \ref{appendix:unknownT}.\looseness-1

We first define a slightly stronger version of concavity.\looseness-1

\begin{definition}[Per-arm $\beta$-Lower Envelope, LE($\beta$)]\label{def:per-arm floor}
For any $\beta \in (0,1]$, we say that arm $i$ satisfies LE($\beta)$ if\looseness-1 $$f_i(t) \geq f_i(T) \left(\frac{t}{T}\right)^\beta \quad \text{for all } t \leq T.$$
\end{definition}

\begin{definition}[Concavity Envelope Exponent, $\beta_I$]\label{def:inst}
For every instance $I$,  define its Concavity Envelope Exponent $\beta_{I}$ as\looseness-1
$$
\beta_{I} \coloneqq \inf \left\{ \beta \in (0,1] : \text{ every arm in $I$ satisfies } \mathrm{LE}(\beta) \right\}.
$$
\end{definition}

\begin{rmk}
    Note that smaller $\beta_I$ indicate larger early rewards, making learning easier. Moreover, note that every non-decreasing concave function with $f(0) \geq 0$ satisfies LE$(1)$, which implies that $1$ is an upper bound on the CEE. When $\beta_I$ approaches $1$ (an instance contains near‑linear arms), the problem reverts to the $\Theta(\sqrt{k})$ regime.
\end{rmk}

\noindent We will now evaluate the performance of our family. 
Note that setting $\alpha=1$ recovers the Random Round Robin algorithm from \cite{blum_nearly-tight_2024}, which ensures that $\textit{PTRR}$ preserves the $O(\sqrt{k})$ guarantee in the unrestricted case.

Now suppose that an instance has $\beta_I < 1$, which occurs whenever it has no (arbitrarily close to) linear arms. Under this assumption, we will prove that choosing any $\alpha\in (\beta_I,1)$ yields the strictly smaller competitive ratio $O\!\bigl(k^{\alpha/(\alpha+1)}\bigr)=o(\sqrt{k})$. Letting $\alpha$ approach $\beta_I$ gives $O\!\bigl(k^{\beta_I/(\beta_I+1)}\bigr)$.

\begin{theorem} Given an IMAB instance $I$ with Concavity Envelope Exponent $\beta_I$. If $T \ge 2k$, then $PTRR_\alpha$ for $\alpha \in (\beta_I, 1)$  with $\tau =T-k$, $m = \frac{\tau}{T} f^*(T)$  achieves
$$
\mathbb{E}[\text{reward from PTRR}_\alpha] \ge \frac{1}{2^{\alpha+3}(\alpha + 1)} \cdot \frac{\text{OPT}_T}{(k + 1)^{\frac{\alpha}{\alpha + 1}}}, 
$$
Equivalently, the competitive ratio is $O(k^{\alpha/(\alpha+1)})$. Setting $\alpha = 1$ recovers the $O(\sqrt{k})$ bound.\label{thm:ptrr-alpha}
\end{theorem}

\begin{proof}[Proof Sketch]
Two simple facts drive the analysis of our algorithm. First, the optimal arm is never abandoned when $\alpha > \beta_I$.  Second, if we ever abandon a non‑optimal arm at time $t$, the cumulative reward collected on that arm is at least the “area” under the lower envelope up to $t$.\looseness-1

These two facts yield a recurrence that trades off (i) the chance we picked the optimal arm now, versus (ii) the value we get after spending $t$ pulls and moving on. At any state $(\tau', k')$, we either draw the best arm now (probability $1/k'$) and then keep it forever, earning $\mathrm{OPT}_{\tau'+k'}$, or we draw a bad arm. If it is bad and we abandon it at a pessimistic time $t$, we have already earned at least the threshold-area up to $t$ and we continue from the smaller state $(\tau' - t, k' - 1)$. Induction on $(\tau', k')$ yields the desired result. A full proof is included in Appendix~\ref{appendix:upperbd-proof} (see Theorem \ref{thm:upper}).
\end{proof}

Our updated keep-test affects exactly three steps of the proof in \cite{blum_nearly-tight_2024}. (i) ``Never drop \(f^\star\)'' now needs the explicit safety range \(m \le f^\star(T)(\tau/T)^\alpha\). (ii) The area-before-abandonment bound scales with \((t - 1)^{\alpha+1}\) (not quadratic). (iii) The recurrence minimization shifts from a quadratic to minimizing \(u\,y^{\alpha+1} + v(1 - y)^{\alpha+1}\), which requires a new balancing inequality that yields the factor \((k+1)^{-\gamma}\) with \(\gamma = \alpha/(\alpha + 1)\).

We next provide a matching lower bound by showing that there is a distribution on instances with the same $\beta_I$ on which no algorithm beats $\Omega(k^{\beta_I/(\beta_I+1)})$. It follows that the exponent is optimal. \textit{PTRR} attains this when $\alpha = \beta_I$.

\begin{figure}
    \centering
    \begin{subfigure}{0.45\textwidth}
    \includegraphics[width=0.9\textwidth]{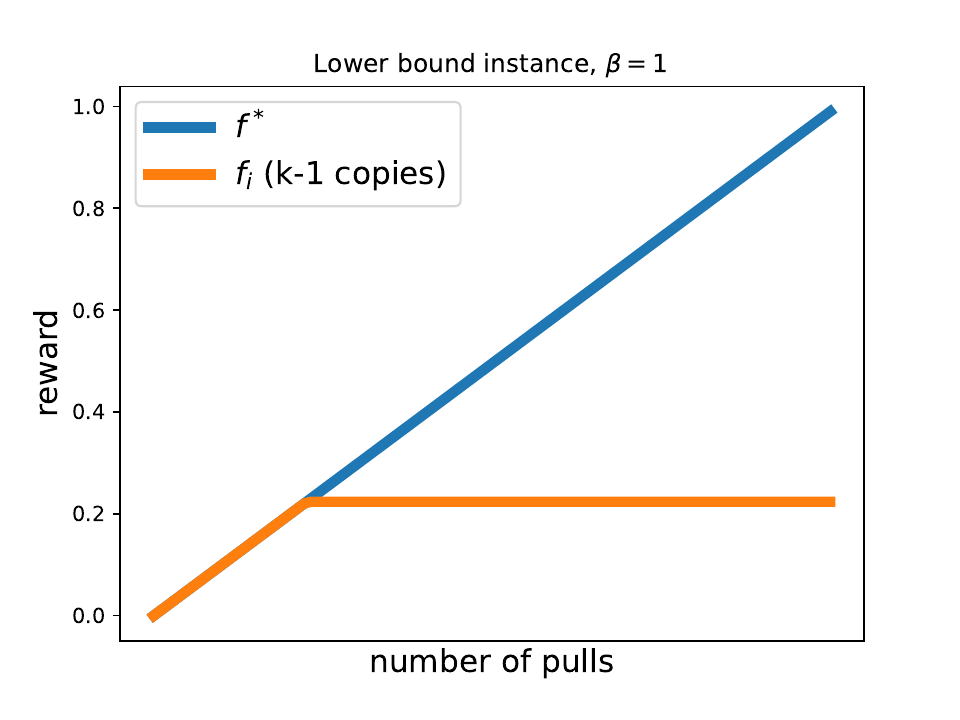} 
    \end{subfigure}
    \begin{subfigure}{0.45\textwidth}
        \includegraphics[width=0.9\textwidth]{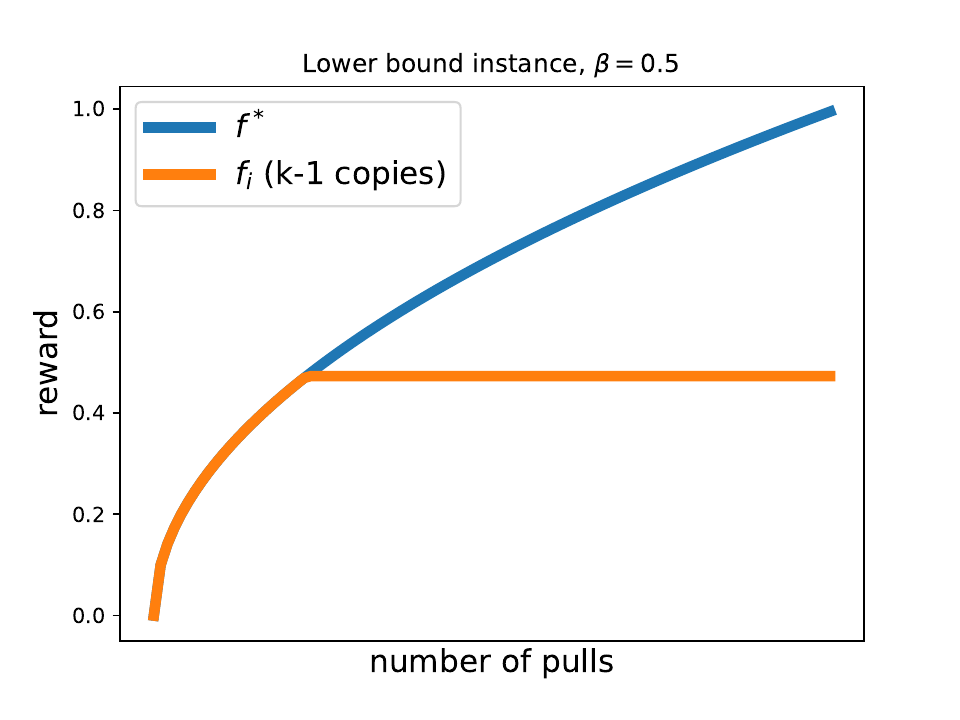} 
    \end{subfigure}
    \caption{\textbf{Lower bound instances:} The spirit of the lower bound instances is that most arms flatten after $T/\sqrt{k}$ pulls but one arm keeps increasing. However, since the algorithm needs to play any arm for a while before figuring out whether it is the good arm or one of the regular arms, the expected reward of the algorithm cannot exceed a certain amount. On the left, we reproduce the instance for $\beta = 1\,,$ recovering the lower bound instance of \cite{blum_nearly-tight_2024}, and on the right we show instance for $\beta = 0.5\,.$}
    \label{fig:lower-bd}
\end{figure}

\begin{theorem}[Lower Bound]
Fix $\beta \in (0, 1]$ and $k \ge 2$. 
For every (possibly randomized) algorithm, for $T$ sufficiently large, there exists an instance with Concavity Envelope Exponent $\beta_I = \beta$ such that
$
\frac{\mathbb{E}[\text{ALG}_T]}{\text{OPT}_T} \le C_\beta k^{-\beta/(\beta+1)},
$
with $C_\beta = \frac{3}{2}(\beta + 1)^2 \left[\beta(\beta + 1)\right]^{-\beta/(\beta+1)}$. Equivalently, any such algorithm has an $\Omega(k^{\beta/(\beta+1)})$ competitive ratio. 
\end{theorem}

\begin{proof}[Proof Sketch] A full proof is in Appendix \ref{appendix:lowerbd-proof}.
We construct an instance similar  to \cite{blum_nearly-tight_2024}. Define a ``good'' arm as the power curve $g(t) = m(t/T)^\beta$, choose this arm at random, and let the other $k - 1$ arms match $g$ exactly for the first $t'$ pulls and then flatten at $g(t')$. Every arm satisfies $\text{LE}(\beta)$, and the good arm violates any stricter floor, so $\beta_I = \beta$. 
At a high-level, we carefully pick $t'$ such that any deterministic algorithm must suffer bad competitive ratio on the distribution over instances. By Yao's principle, this results in a lower bound for randomized algorithms.
\end{proof}

Again, we will briefly contextualize our proof. The hard instance in \cite{blum_nearly-tight_2024} uses a linear ``good'' arm, while ours uses a power curve \(g(t) = m(t/T)^\beta\). This change is propagated in two places. (i) The generous upper bound becomes \(h(x) = 1/(kx) + (\beta + 1)x^\beta\) (instead of \(1/(kx) + 2x\)). (ii) the calculus minimizer moves to \(x^\star = [k\beta(\beta + 1)]^{-1/(\beta+1)}\), giving the exponent \(k^{-\beta/(\beta+1)}\) instead of \(k^{-1/2}\). The rounding step and use of Yao's principle are unchanged.

\subsection{Sample Complexity of Learning the Curvature Parameter}
\label{sec:sample-complexity}

In this section, we analyze the number of samples required to learn a near-optional algorithm from the algorithm family \textit{PTRR} (Definition~\ref{defn:alg-fam}). Previously, we considered a fixed $\alpha$ and showed its optimality for a fixed $\beta.$ However, in practice, we may not know the true $\beta$ and therefore cannot pick the optimal $\alpha.$ Now, suppose we have access to historical examples of improving multi-armed bandits problems (e.g., learning curves for the hyperparameter tuning problem on previous time periods of data). 
Then, we could hope to {\em learn} the best possible $\alpha$ for this distribution, provided we have sufficiently many samples. To analyze the sample complexity of this task, we extend techniques developed by \cite{sharma2025offline} for stochastic multi-armed bandits.
\par

\begin{rmk}
    A related approach for meta-learning bandits is the {\em corralling} framework of \cite{agarwal2017corralling, arora2021corralling, luo2022corralling}. However, existing corralling techniques can only handle a finite number of hyperparameters. We further show that meta-learning even with a finite band of algorithms may not be possible without further assumptions in the improving bandits setting (see Appendix~\ref{appendix:corralling}).
\end{rmk}

\paragraph{Derandomized Dual Complexity and result of \cite{sharma2025offline}.}

We start by explaining how to use results from \cite{sharma2025offline}. In their work, they defined a relevant quantity, the {\em derandomized dual complexity}, $Q_\mathcal{D}$, which accounted for randomness in data sampling and any additional randomness after fixing the instance. Since our family of algorithms includes randomized algorithms, we must also derandomize this additional source of randomness. We do this by defining $\mathcal{D'}$ as an extension of the original distribution. For each element in the support of the original distribution, we define $k!$ copies, each with a different permutation $\pi_k$ of $[k]$ associated with it. Each new augmented instance $(I, \pi_k)$ has probability $\mathcal{D'}_I \coloneqq \mathcal{D}_{I}/k!$ associated with it, where $\mathcal{D}_I$ is the probability associated with that instance originally. 
Wherever the results of \cite{sharma2025offline} refer to the distribution $\mathcal{D}$ over instances, we instead consider the distribution over instances and permutations as defined above. By derandomizing in this way, we can immediately apply their results\footnote{For our purposes, it suffices to handle randomness in the algorithm in this way. It would be interesting and valuable to extend their results to general randomized algorithms in future work.\looseness-1}.

From the results of \cite{sharma2025offline} (reproduced in Appendix~\ref{appendix:ss25results}), we know that if we have sufficiently many samples, as a function of $Q_\mathcal{D}, H$ and the accuracy and success probability parameters, then with high probability the empirical loss will be close to the population loss. 
We know that if we have enough samples to achieve uniform convergence, we will find a hyperparameter value that achieves near-optimal performance on future instances from the distribution (see Appendix~\ref{appendix:ucpoploss}). Thus, if we pick the value of $\alpha$ with the smallest empirical loss (i.e., run empirical risk minimization (ERM)), we can guarantee that the objective in Eqn.~\ref{eqn:goal} is satisfied. It remains to (1) bound $Q_\mathcal{D}$ for our problem and (2) investigate various reasonable piecewise-$H$-bounded losses.\looseness-1 

\paragraph{Bounding derandomized dual complexity in our setting.}
Now, we compute a bound on the value of $Q_\mathcal{D}\,,$ by bounding the number of possible behaviors of algorithms from $\mathcal{A}\,.$ To do so, we identify a sufficient statistic for the loss of a fixed algorithm on an instance then count the number of possible values of the statistic.

\begin{lemma} \label{lemma:bdqd}
    For the family $\mathcal{A}$ defined in Defn.~\ref{defn:alg-fam}, the improving multi-armed bandits problem, and {\em any} piecewise-constant loss function, $Q_\mathcal{D} \le kT$\,.
\end{lemma}



\newcommand{\algalpha}{\mathcal{A}_\alpha}
\begin{proof}
    We show this lemma by defining a instance\footnote{again, augmented instance}- and algorithm- specific tuple $R_\alpha\,.$ We define $R_\alpha$ such that it contains sufficient information to evaluate the loss of the algorithm on the fixed (augmented) instance. That is, since one value of the tuple gives rise at most one value of loss, we can bound the number of possible values of loss by counting the number of such tuples. \par

    First, fix an algorithm $\algalpha\,.$ This algorithm proceeds by first generating a curve to which any chosen arm is compared, namely $c_\alpha(t) \coloneq\left(\frac tT \right)^\alpha f^\star(T).$ Next, call the first arm chosen by the algorithm $f_1(t).$ The algorithm plays $f_1$ until the first time $t^{(1)}_\text{stop} \in \{ 1, 2, \hdots \, T \}$ such that $f_1(t^{(1)}_\text{stop}) < \left( \frac{t^{(1)}_\text{stop}}{T} \right)^\alpha f^*(T).$
    Similarly, we can calculate a $t^{(2)}_\text{stop}$ for the second arm played by the algorithm and so on. This provides us a tuple $R_\alpha \coloneqq (t^{(1)}_\text{stop}, t^{(2)}_\text{stop}, \hdots, t^{(k)}_\text{stop})$ that provides enough information to compute the loss of the algorithm on the instance.

    Now, we observe that as we vary $\alpha\,,$ the number of possible tuples that can be generated is upper bounded by the total number of possible tuples. Thus, we simply need to count the number of possible such tuples $R_\alpha\,.$ Since each $t^{(i)}_\text{stop}$ takes on a discrete value in $[T]\,,$ there are at most $T$ values an element in the tuple could take on, and since there are $k$ elements in the tuple, the total number of such tuples is {\em at most} $kT\,.$
\end{proof}

\paragraph{Sample complexity.}
We present sample complexity results for generic piecewise-$H$-bounded loss functions by extending Theorem~\ref{thm:ss25main} and Lemma~\ref{lemma:bdqd}. In Appendix~\ref{appendix:avgregret}, we instantiate this for a concrete loss function.

\begin{theorem}
    The sample complexity of Hyperparameter Transfer in family $\mathcal{A}$ for the improving multi-armed bandits problem with generic piecewise-$H$-bounded loss is $N = O\left( \left(\frac{H}{\epsilon}\right)^2 (\log kT + \log \frac 1 \delta \right).$
\end{theorem}

While the above result guarantees sample efficiency for learning the best $\alpha,$ it is also interesting to consider the question of computational efficiency. In Appendix~\ref{appendix:erm-alg}, we provide a framework for implementing ERM 
that is efficient if the number of arms is a small constant. In Appendix~\ref{appendix:approx-learning-alpha}, we propose and analyze a potentially more practical algorithm to select a value of $\alpha$ that provides a good approximation for most instances in the distribution.\looseness-1

\subsection{Empirical Evaluation} 
\label{sec:empirical}

\begin{figure}[h]
    \centering
    \includegraphics[width=0.5\linewidth]{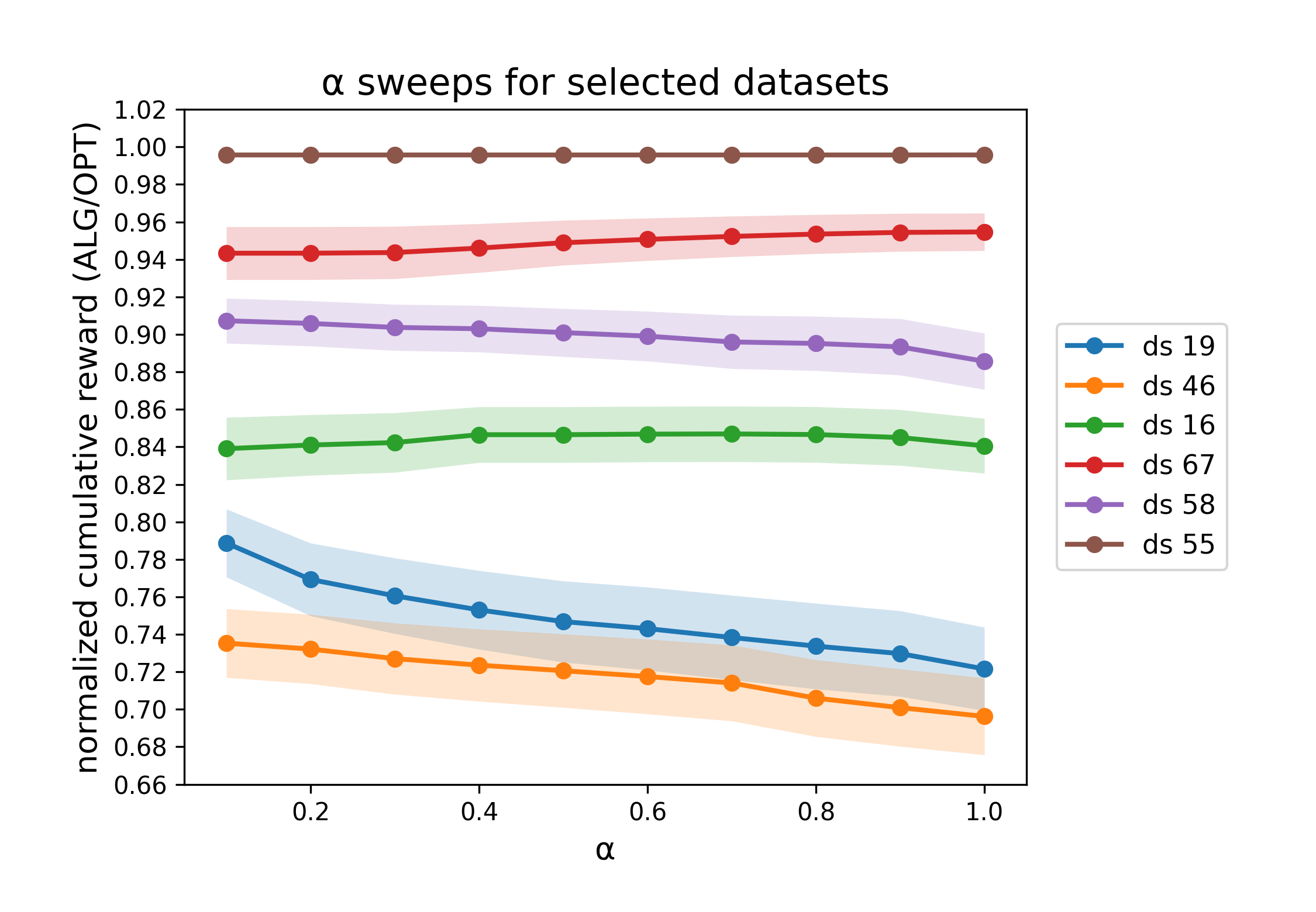}
    \caption{\textbf{Sensitivity of $\mathrm{PTRR}_\alpha$ to $\alpha$ on selected LCDB instances ($T=44$, $k=22$).}
Each curve corresponds to one CC-18 dataset $d$ from LCDB~1.1 and reports the normalized cumulative reward
$\mathbb{E}[\mathrm{PTRR}_\alpha(d)]/\mathrm{OPT}(d)$ as a function of $\alpha\in\{0.1,0.2,\dots,1.0\}$, where
$\mathrm{OPT}(d)=\max_i\sum_{t=1}^T r_{i,d}(t)$ is the cumulative reward of the best fixed-arm policy in hindsight under the same horizon (which is not necessarily the optimal policy due to absence of monotonicity, but it is a useful proxy.) 
and
$r_{i,d}(t)=1-\mathrm{err}_{i,d}(t)$ is the mean (over cross-validation) reward at anchor $t$ for arm $i$.
For each $(d,\alpha)$, $\mathbb{E}[\mathrm{PTRR}_\alpha(d)]$ is estimated by averaging over $200$ random arm orderings. The shaded regions are pointwise 95\% Student-$t$ confidence intervals across the 200 runs (mean $\pm\, t_{0.975,199}\cdot \mathrm{sd}/\sqrt{200}$).
The displayed datasets are selected to illustrate the range of sweep shapes observed across the full benchmark (near-flat curves, monotone trends, and interior maxima). For most datasets, performance differences across $\alpha$ are small relative to the confidence intervals, while a minority show a significant trend across $\alpha$ on this grid. Complete sweeps over all $27$ usable datasets are included in Appendix~\ref{appendix:empirical}.}
    \label{fig:6 sweeps}
\end{figure}

\begin{figure*}[h]
    \centering
    \includegraphics[width=1\linewidth]{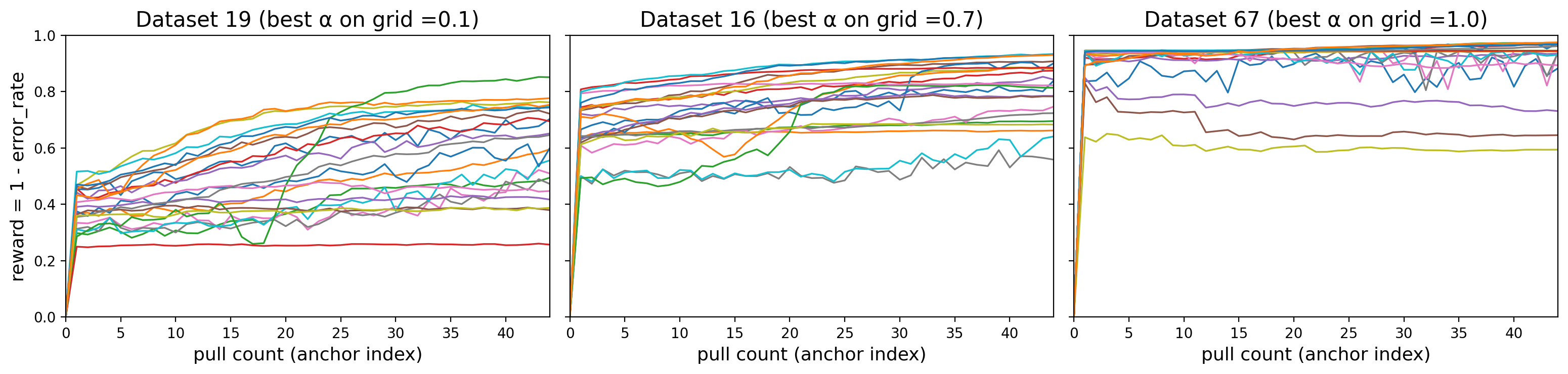}
    \caption{\textbf{Mean LCDB reward curves for three datasets with distinct best $\alpha$ values on the grid.} Each panel overlays the mean reward curves $r_{i,d}(t)=1-\mathrm{err}_{i,d}(t)$ across anchors $t$ for all $k=22$ arms on a single CC-18 dataset $d$. The title of each panel reports the value of $\alpha\in\{0.1,0.2,\dots,1.0\}$ that maximizes the estimated normalized cumulative reward $\mathbb{E}[\mathrm{PTRR}_\alpha(d)]/\mathrm{OPT}(d)$ at horizon $T=44$ on that dataset. 
    We selected these datasets to illustrate diverse best $\alpha$ values on the grid. Qualitatively, these plots suggest a mechanism consistent with the influence of $\alpha$ on $\mathrm{PTRR}_\alpha$, where datasets preferring smaller $\alpha$ tend to exhibit early separation between `good' and `bad' arms (making aggressive abandonment beneficial).\looseness-1
    \label{fig:learning curves}}
\end{figure*}

We provide empirical evidence that the value of $\alpha$ used in running $PTRR$ can affect the cumulative reward using data from a learning curve dataset, LCDB~1.1 (CC-18 benchmarks,~\cite{yanlcdb}). We map each dataset in the LCDB dataset to an IMAB instance as follows: the $k$ arms are the LCDB learners, the time index
corresponds to the number of samples seen, and the rewards arise from 
inverting the error rates through $r_{i,d}(t)=1-\mathrm{err}_{i,d}(t)$.
We average over cross-validation and retain only datasets for which the curve is defined for all $t \in \{1, \dots, T\}\,.$ This yields 27 datasets with $k = 22$ arms each and a time horizon of $T=44.$ We run $PTRR_\alpha$ for each $\alpha \in \{0.1, 0.2, \ldots, 1.0\}$ and report normalized cumulative reward $\mathbb{E}[\mathrm{PTRR}_{\alpha}(d)] / \mathrm{OPT}(d)$ 
(the reciprocal of the competitive ratio in Definition~2.2) averaged over 200 seeds. Figure \ref{fig:6 sweeps} shows that the $\alpha$ value maximizing this quantity varies across datasets.
Figure \ref{fig:learning curves} plots mean reward curves for three datasets to help visualize the underlying reward dynamics. Full details of the  setup and results are provided in Appendix~\ref{appendix:empirical}.\looseness-1

\section{Achieving Best-of-both-worlds Best Arm Identification}\label{sec:bothworlds}

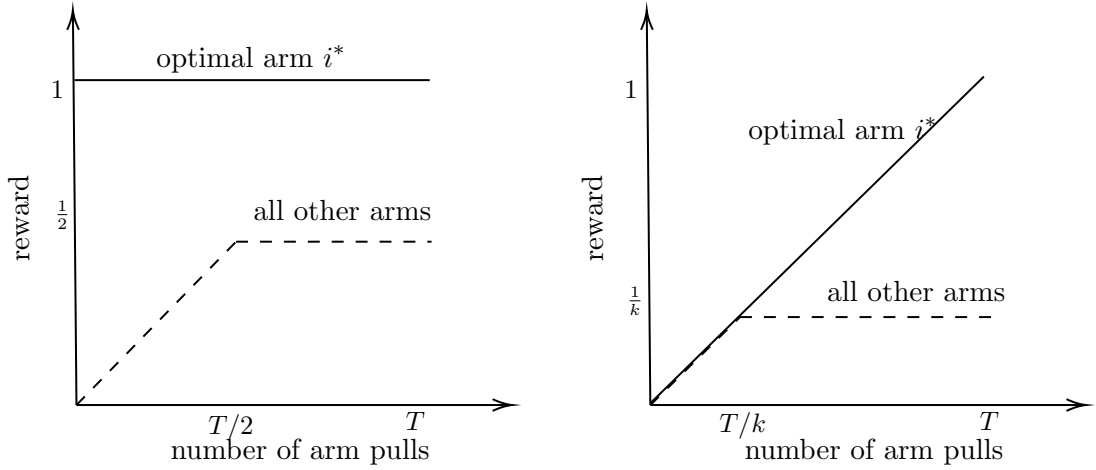
\begin{figure}[t]
    \centering

\tikzset{every picture/.style={line width=0.75pt}} 

\begin{tikzpicture}[x=0.75pt,y=0.75pt,yscale=-0.9,xscale=0.9]

\draw    (71,600.46) -- (312,600.46) ;
\draw [shift={(314,600.46)}, rotate = 180] [color={rgb, 255:red, 0; green, 0; blue, 0 }  ][line width=0.75]    (10.93,-3.29) .. controls (6.95,-1.4) and (3.31,-0.3) .. (0,0) .. controls (3.31,0.3) and (6.95,1.4) .. (10.93,3.29)   ;
\draw    (71,600.46) -- (69.02,382.46) ;
\draw [shift={(69,380.46)}, rotate = 89.48] [color={rgb, 255:red, 0; green, 0; blue, 0 }  ][line width=0.75]    (10.93,-3.29) .. controls (6.95,-1.4) and (3.31,-0.3) .. (0,0) .. controls (3.31,0.3) and (6.95,1.4) .. (10.93,3.29)   ;
\draw    (70,419.46) -- (268,419.46) ;
\draw  [dash pattern={on 4.5pt off 4.5pt}]  (71,600.46) -- (160,509.46) ;
\draw  [dash pattern={on 4.5pt off 4.5pt}]  (160,509.46) -- (269,509.46) ;
\draw    (391,600.46) -- (632,600.46) ;
\draw [shift={(634,600.46)}, rotate = 180] [color={rgb, 255:red, 0; green, 0; blue, 0 }  ][line width=0.75]    (10.93,-3.29) .. controls (6.95,-1.4) and (3.31,-0.3) .. (0,0) .. controls (3.31,0.3) and (6.95,1.4) .. (10.93,3.29)   ;
\draw    (391,600.46) -- (389.02,382.46) ;
\draw [shift={(389,380.46)}, rotate = 89.48] [color={rgb, 255:red, 0; green, 0; blue, 0 }  ][line width=0.75]    (10.93,-3.29) .. controls (6.95,-1.4) and (3.31,-0.3) .. (0,0) .. controls (3.31,0.3) and (6.95,1.4) .. (10.93,3.29)   ;
\draw    (391,599.46) -- (577,417.46) ;
\draw  [dash pattern={on 4.5pt off 4.5pt}]  (391,600.46) -- (441,551.46) ;
\draw  [dash pattern={on 4.5pt off 4.5pt}]  (441,551.46) -- (581,551.46) ;

\draw (122,617.46) node [anchor=north west][inner sep=0.75pt]   [align=left] {number of arm pulls};
\draw (31.88,522.09) node [anchor=north west][inner sep=0.75pt]  [rotate=-269.09] [align=left] {reward};
\draw (114,398.46) node [anchor=north west][inner sep=0.75pt]   [align=left] {optimal arm $\displaystyle i^{*}$};
\draw (55,418.46) node [anchor=north west][inner sep=0.75pt]  [font=\small] [align=left] {$\displaystyle 1$};
\draw (253,602.86) node [anchor=north west][inner sep=0.75pt]  [font=\small]  {$T$};
\draw (143,603.86) node [anchor=north west][inner sep=0.75pt]  [font=\small]  {$T/2$};
\draw (57,484.86) node [anchor=north west][inner sep=0.75pt]  [font=\footnotesize]  {$\frac{1}{2}$};
\draw (168,484.46) node [anchor=north west][inner sep=0.75pt]   [align=left] {all other arms};
\draw (31,642.46) node [anchor=north west][inner sep=0.75pt]   [align=left] {(a) \cite{blum_nearly-tight_2024} may be suboptimal by a factor of\\two on some ``nice" instances};
\draw (442,617.46) node [anchor=north west][inner sep=0.75pt]   [align=left] {number of arm pulls};
\draw (351.88,522.09) node [anchor=north west][inner sep=0.75pt]  [rotate=-269.09] [align=left] {reward};
\draw (444,438.46) node [anchor=north west][inner sep=0.75pt]   [align=left] {optimal arm $\displaystyle i^{*}$};
\draw (375,418.46) node [anchor=north west][inner sep=0.75pt]  [font=\small] [align=left] {$\displaystyle 1$};
\draw (573,602.86) node [anchor=north west][inner sep=0.75pt]  [font=\small]  {$T$};
\draw (428,602.86) node [anchor=north west][inner sep=0.75pt]  [font=\small]  {$T/k$};
\draw (376,532.86) node [anchor=north west][inner sep=0.75pt]  [font=\footnotesize]  {$\frac{1}{k}$};
\draw (488,529.46) node [anchor=north west][inner sep=0.75pt]   [align=left] {all other arms};
\draw (351,642.46) node [anchor=north west][inner sep=0.75pt]   [align=left] {(b) \cite{mussi2024best} may be suboptimal by a factor of\\$\displaystyle \sqrt{k}$ on worst-case instances};

\end{tikzpicture}
    
    \caption{Examples demonstrating the need for a best-of-both-worlds approach.}
    \label{fig:motivating-examples}
\end{figure}

In this section, we bridge a gap in the literature: so far, algorithms can either identify the best arm in an instance that is sufficiently benign or identify an approximate best arm but not {\em both}, i.e., identify the best arm if possible and achieve the approximate goal if the exact goal cannot be achieved. To this end,
we introduce a family of hybrid algorithms that address the task of identifying the arm with the highest final reward, the best arm identification (BAI) task\footnote{While the reader may be more familiar with BAI in the sense of maximum mean in stochastic bandits, we note that in that case, we could equivalently study cumulative reward from a single arm.}. Each algorithm in this family
guarantees best arm identification whenever an instance is sufficiently benign, while simultaneously achieving optimal (up to constants) multiplicative guarantees on worst-case instances. 
We first provide examples that show that no existing algorithm achieves near-optimal results on both nice and worst-case instances. We define our family in Section~\ref{sec:hybriddef}. In Section~\ref{sec:hybridguarantees}, we formalize a `niceness' condition under which it is possible to sample every suboptimal arm enough to safely discard it by the switch time. We then prove that $\text{Hybrid}_{1,T/2}$ is guaranteed to identify the best arm on instances that satisfy this condition and, on all other instances, returns an arm whose expected maximum pull at time $T$ competes optimally with the highest single pull in the instance. We also develop results for choosing $\alpha, B$ in a data-driven way.\looseness-1 


\textbf{Motivating Examples.} 
Prior work gives two distinct types of guarantees for improving bandits---optimal approximation factors for worst-case instances \cite{patil_mitigating_2023,blum_nearly-tight_2024}, and better guarantees (sublinear policy regret, or small sample complexity of best arm identification) for nicer instances \cite{heidari_tight_nodate,metelli_stochastic_2022,mussi2024best}.
It is natural to ask whether we can achieve the best-of-both-worlds, that is, get almost optimal results on nice instances, while being within the optimal approximation factors (up to constants) in the worst-case. It turns out that the algorithms proposed in prior literature fail to achieve such a guarantee. 

The following example shows that the algorithm of \cite{blum_nearly-tight_2024} has sub-optimal guarantees on nice instances.\looseness-1

\begin{example}
{\it The randomized algorithm of \cite{blum_nearly-tight_2024} may fail to identify the best-arm on instances where the UCB-style algorithms of \cite{metelli_stochastic_2022,mussi2024best} find the best-arm.} We set the reward function for the best arm as $f_{i^*}(t)=1$ for all $t$, and for any other arm $i\ne i^*$ as $f_{i}(t)=\min\{\frac{t}{T},\frac{1}{2}\}$, where $T$ is the time horizon. We assume that $k$ is large (in particular, $k\ge 4$). Now, the randomized round-robin algorithm selects the optimal arm as its first or second arm with probability at most $2/k$. Otherwise, it keeps playing a sub-optimal arm till time $T/2$, and a different sub-optimal arm for the rest of the time horizon. Thus, with probability at least $1-\frac{2}{k}$, the best arm identified by the algorithm is sub-optimal by at least a factor of 2 (with respect to both its cumulative reward and its final pull). On the other hand, for the UCB-based algorithms, we can upper bound the number of exploratory pulls of the sub-optimal arms, and the algorithm succeeds in identifying the best arm $i^*$.
\end{example}

\noindent The following example shows that the UCB-variant (R-UCBE) developed for the improving bandits BAI problem by~\cite{mussi2024best}  has sub-optimal worst-case performance.

\begin{example}
{\it The UCB-based algorithm of ~\cite{mussi2024best} for best-arm identification may output an arm with $\Omega(k)$ sub-optimal reward compared to the actual best arm on some instances.} We adapt the example used by \cite{blum_nearly-tight_2024} in their lower bound construction. Set $f_{i^*}(t)=t/T$ for all $t$ for the optimal arm $i^*$, and for any other arm $i\ne i^*$ as $f_{i}(t)=\min\left\{\frac{t}{T},\frac{1}{k}\right\}$. Due to the exploration term in UCB, while the arm rewards are identical, each arm gets pulled an equal number of times. By time $T$, each arm gets pulled $T/k$ times, and all the arms appear identical to the algorithm. Thus, the best arm learned by the algorithm may be sub-optimal (with respect to both its cumulative reward and its final pull) by a factor of $\Omega(k)$. In contrast, the PTRR algorithm family guarantees a worst-case competitive ratio of $O(\sqrt{k})$ or smaller.\looseness-1
\end{example}

\subsection{Hybrid Algorithm Family}\label{sec:hybriddef}
We now define the hybrid algorithm family. As before, we work in the standard improving‑bandits setting with $k$ arms, a known horizon $T$, and non‑decreasing concave reward functions $f_i$. Each algorithm Hybrid$_{\alpha,B}$ has two stages. Stage 1 uses a UCB-style envelope: At each step, the algorithm computes a lower bound $L_i(n)$ and terminal upper bound $U_i(n)$ on the final reward $f_i(T)$ of every arm and pulls the arm with the largest optimistic estimate $U_i$. If the lower bound of one arm dominates the terminal upper bound of every other arm, Hybrid$_{\alpha,B}$ commits to this arm. 
If no commit occurs by time $B$, Stage 2 runs PTRR$_\alpha$ and finds an arm whose expected terminal reward is at least a substantial fraction of the best arm’s. 

For the terminal envelope, we define $L_i(t) := f_i(t),
\triangle_i(t) :=  (T-t) \gamma_i(t - 1),$ and $
U_i(t) := L_i(t) + \triangle_i(t), 
$
where $\gamma_i (t-1) := f_i(t) - f_i (t-1)$. We set $U_i(0) : = \infty$ to ensure first pulls. Using concavity and monotonicity, it is straightforward to prove that $L_i(t) \leq f_i(T) \leq U_i(t)$ for all $i,t$ (see Lemma \ref{lem:envelope} in Appendix~\ref{appendix:hybridproofs}).

\begin{definition} \label{def:fullyhybridfam}
Define the family of algorithms $\textit{Hybrid} := \{\textit{Hybrid}_{\alpha, B} (Algorithm \ref{alg:hybrid}) : \alpha \in (0,1]$ and $B \in [T]$\}.
\end{definition}

\begin{algorithm}[t]
\caption{$\textit{Hybrid}_{\alpha, B}$ BAI}\label{alg:hybrid}
\begin{algorithmic}[1]
\State \textbf{Require:} maximum final pull $m$, horizon $\tau$
\State \textbf{Stage 1}: $t\gets0$ 
\For{each arm $i$}
\State $t_i\gets0$, $L_i\gets0$, $U_i\gets+\infty$
\EndFor
\While{$t<B$}
  \For{each $i$ with $t_i\ge1$}
     \State $L_i \gets f_i(t_i)$ \hfill 
     \State $\gamma_i \gets f_i(t_i)-f_i(t_i-1)$ \hfill 
     \State $U_i \gets L_i + (\tau-t)\,\cdot \gamma_i$ 
  \EndFor
  \State $\hat i \gets \arg\max_i L_i$,\quad $U_{\mathrm{next}}\gets \max_{j\ne \hat i} U_j$
  \If{$L_{\hat i} > U_{\mathrm{next}}$}
     \State \textbf{Return} $\hat i$.
  \EndIf
  \State $i' \gets \arg\max_i (U_i - L_i)$
  \State \textbf{pull} $i'$;\; $t_{i'}\gets t_{i'}+1$, $t\gets t+1$
\EndWhile
\State \textbf{Stage 2}:
 $\tau' \gets (\tau - B) - k$, $m' \gets \left(\frac{\tau'}{T}\right) \cdot m$
\For{each $i$}
\State $g_i(s)\gets f_i(t_i + s)$
\EndFor \\
\Return $\hat i \gets \text{arm returned by } \textit{PTRR}_\alpha$ with parameters $(m', \tau')$ on $\{g_i\}$ for $T-B$ steps.

\end{algorithmic}
\end{algorithm}


\subsection{Best-of-Both-Worlds BAI guarantees}
\label{sec:hybridguarantees}

In this section, we argue that it is meaningful to study the two-parameter family of \textit{Hybrid} algorithms from Definition \ref{def:fullyhybridfam}. We will do this by showing that \textit{Hybrid} contains algorithms that simultaneously (i) guarantee best arm identification on sufficiently benign instances, and (ii) preserve tight (up to constants) multiplicative bounds for approximating the best arm
on adversarial instances.\footnote{With additional information about stronger concavity, we achieve sharper bounds by using PTRR$_\alpha$ with appropriate $\alpha$.}

We start by defining a class of `sufficiently benign' instances and prove that our algorithm is guaranteed to return the best arm on all members of this class. If an instance is not in this class, Stage~2 pursues {\em approximate} BAI and runs $PTRR_\alpha$ with $\alpha$ dependent on the strength of concavity ($\alpha = 1$ works for {\em all} instances). Having reverted to an approximate goal, we identify an arm $\hat{i}$ whose final reward satisfies
$\mathbb{E}[f_{\hat{i}}(T)] \ge \Omega\left(k^{\frac{-\alpha}{1+\alpha}}\right) f^{\star}(T)$. We assume that both $T$ and $f^*(T)$ are known to the algorithm, which we input as $\tau$ and $m$.\looseness-1 

We will now define a condition (Definition \ref{def:GCC}) under which we can guarantee best arm identification. For any arm $i$ and $\epsilon>0$, the {\it terminal budget} $h_i$ of arm $i$ is defined as
$
h_i(\epsilon) := \min\{\,n \in \{2,\ldots,T\} : \triangle_i(n) \le \epsilon\,\},
$
where $\triangle_i(t) := (T-t) \, \gamma_i(t - 1)$.  For any instance $I$, its {\it Best Arm Gap} $\Delta_I$ is defined as
 $\Delta_I := f^\star (T) - \max_{j \ne i^*} f_j(T).$

\begin{definition}[Gap Clearance Condition, $GCC(B)$]\label{def:GCC}
For any $B\le T$, we say that an instance satisfies the Gap Clearance Condition condition $GCC(B)$ if $\Delta_I > 0$ and
$
\sum_{i=1}^K h_i(\Delta_I / 3) \;\le\; B.
$
\end{definition}

Under concavity, $\triangle_i(n)$ is an upper bound on the remaining possible increase in terminal value of arm $i$ after $n$ pulls: it denotes the most additional reward an adversary can ``hide in the tail''  by continuing with the last observed slope. The per-arm budget $h_i(\epsilon)$ is therefore the minimal number of pulls needed to ensure its optimistic terminal value is close to current lower envelope. 
Taking $\epsilon = \Delta_I/3$, once a suboptimal arm has been pulled this many times, its best possible continuation still loses to the best arm’s lower envelope. 
Thus $\mathrm{GCC}(B)$ ensures that the total work needed to certify the best arm fits within budget $B$. 
If the sum exceeds $B$, concavity allows at least one suboptimal arm to remain plausibly optimal by time $B$, so no sound mid-horizon certificate can be guaranteed.
For a comparison of this requirement to other `niceness' conditions from the literature and a proof of Theorem \ref{thm:hybrid-single2}, see Appendix~\ref{appendix:PR}.

\begin{theorem}[best-of-both-worlds guarantees] \label{thm:hybrid-single2}
Suppose an instance $I\in\mathcal{I}$ has Concavity Envelope Exponent $\beta_I\in(0,1]$. Algorithm \ref{alg:hybrid} with $\alpha \in(\beta_I, 1]$  satisfies the following:
\begin{enumerate}
\item[(1)] If the instance further satisfies $\Delta_I>0$, and $GCC(\theta)$ holds for $\theta\le T/2$, then the algorithm with $B\in (\theta,T/2]$  identifies and commits to the best arm $i^\star$ in Stage~1.
    \item[(2)] If Stage~1 does not certify a best arm by time $B$, then Stage~2 finds an approximate best arm $\hat i$ such that 
    $\mathbb{E}\big[f_{\hat i}(T)\big]\ \ge\ \Omega{\left(k^{\frac{-\alpha}{1+\alpha}}\right)}\; f^\star(T),$ where the expectation is  over the randomness of the algorithm. 
\end{enumerate}
\end{theorem}

\begin{proof}[Proof Overview]
(1) $\mathrm{GCC}(\theta)$ implies that the per–arm budgets $h_i(\Delta_I/3)$ sum to at most $\theta$. Since the algorithm pulls the arm with the largest slack, we know that these budgets are met within $B\ge\theta$ pulls. Then every suboptimal arm has $U_j\le f^*(T)-2\Delta_I/3$ and the best arm has $L_{i^\star}\ge f^*(T)-\Delta_I/3$, which implies that $L_{i^\star}\ge\max_{j\ne i^\star}U_j$. We commit to $i^\star$.
(2) If no certificate fires by $B\le T/2$, running $\textit{PTRR}_\alpha$ for $T_{\mathrm{rem}}$ steps yields an average reward within a $k^{\alpha/(1+\alpha)}$ factor of the residual optimum (up to constants). Using the fact that the best single pull is at least the average, and $\mathrm{OPT}^{\mathrm{res}}$ is at least a constant times $g^\star(T_{\mathrm{rem}})\,T_{\mathrm{rem}}$ by concavity, we get $\mathbb{E}[f_{\hat i}(T)]\ge \Omega\!\big(k^{-\alpha/(1+\alpha)}\big)\,f^\star(T)$. 
\end{proof}

\subsection{Sample Complexity for Tuning Curvature and Switch Parameters}
\label{sec:hybridcomplexity}

In the previous section, we described a hybrid algorithms which proceeds as follows: first, it spends a portion of the time attempting to solve exact best-arm identification (BAI) and then it reverts to approximate best-arm identification if it is not successful\footnote{In Appendix~\ref{appendix:cum-reward-bai}, we also discuss a cumulative reward version of the BAI task and provide Algorithm~\ref{alg:cumhybrid} to solve it. The discuss in this section holds for that, as well. These algorithms differ only in the implementation of lines 6,7, and 8, i.e., in how the tracked variables are defined.}. In general, instances could lie anywhere along the spectrum of easy-to-identify to worst-case. Is there a better way to decide how {\em much} time to allocate to attempting exact BAI and when to switch to worst-case approximate BAI? Suppose the instances we see have some stationarity. In particular, if we see historical examples of instances arising from a certain task, and we are expected to complete that task with a good algorithm in the future, we might model the instances as having been drawn from from a distribution. In that case, we could hope to {\em learn} to what degree it is worth attempting exact BAI. In particular, we could hope to learn the time after which we should switch from the exact goal to the relaxed goal. We show that we can pick a switch time and a PTRR parameter with polynomially many samples from the distribution over instances that is almost as good as the optimal such time and parameter {\em on average} over the instances. Here, unlike in the previous sections, we require a distributional assumption to hold and access to samples from the distribution, and we achieve an {\em on average} rather than per instance guarantee. However, we adapt to the data distribution. \par

We study the fully hybrid algorithm family defined in Definition~\ref{def:fullyhybridfam}. Each algorithm in this family has two parameters: the first is the time $B$ at which it switches from the exact BAI goal to the approximate BAI goal, and the second specifies which $PTRR_\alpha$ algorithm it runs {\em after} switching to the approximate BAI goal. Our goal is to learn from a set of offline instances which value of $B$ and $\alpha$ gets the best performance on a new instance from the same distribution.
In order to analyze this, we can apply the same set of data-driven algorithm design tools as before. Once again, we need to understand for a fixed instance, how many possible behaviors an algorithm from the family could have in terms of loss. To this end, it remains to (1) bound $Q_\mathcal{D}$ for this problem and (2) investigate various reasonable $H$-bounded losses. We do this next.

\vspace{-4mm}

\paragraph{Bounding derandomized dual complexity in our setting.}
We bound $Q_\mathcal{D}\,$ by computing the number of possible behaviors of algorithms from $\mathcal{H}\,.$ 

\begin{lemma} \label{lemma:bdqd-hybrid}
    For the family $\mathcal{B}$ defined in Defn.~\ref{defn:alg-fam}, the improving multi-armed bandits problem, and {\em any} piecewise-constant loss function, $Q_\mathcal{D} \le k\,T^2$\,.
\end{lemma}

\vspace{-4mm}

\paragraph{Sample complexity results.}
As in Section~\ref{sec:sample-complexity}, we combine Thm.~\ref{thm:ss25main} and Lemma~\ref{lemma:bdqd-hybrid} to derive sample complexity results for generic $H$-bounded loss functions. We then instantiate it for both BAI and regret loss functions. 

\begin{theorem} \label{thm:hybrid-alg-sc}
    For the Hyperparameter Transfer setting for tuning $\alpha,B$ in Algorithm \ref{alg:hybrid} 
    with generic $H$-bounded loss, $O\left( \left(\frac{H}{\epsilon}\right)^2 (\log kT + \log \frac 1 \delta )  \right)$ instances drawn from $\mathcal{D}$ suffice to get the uniform convergence guarantee in Theorem~\ref{thm:ss25main}.\looseness-1
\end{theorem}

Finally, we instantiate the loss function for BAI. We can study ``maximum pull regret,'' and this loss is $H$-bounded for $H = m\,.$ Note that since $m$ will affect the sample complexity, we need an upper bound on it to know how many samples to draw.

\begin{definition}
    Define  ``maximum pull regret'' as $R_{mp}(T) \coloneqq \max_{i \in [k]} f_i(T) - \max_{i\in [k]} f_i(t_i)\,,$ where $t_i$ is the number of rounds for which the algorithm played arm $i\,.$\looseness-1
\end{definition}

Thus, with knowledge of an upper bound for the best pull of the best arm, we have that:

\begin{corollary}
    For the Hyperparameter Transfer setting for the improving multi-armed bandits problem optimizing for averaged regret, $N = O\left( \left(\frac{m}{\epsilon}\right)^2 (\log kT + \log \frac 1 \delta )  \right)$ instances drawn from $\mathcal{D}$ suffice to get the uniform convergence guarantee in Theorem~\ref{thm:ss25main} for the $Hybrid_{\alpha, B}$ algorithm family.\looseness-1
\end{corollary}

Thus, with polynomially many samples from a distribution over instances, we can achieve near-optimal loss on a new instance drawn from the same distribution.\looseness-1

\section{Conclusion}
We provide beyond worst-case guarantees for the improving multi-armed bandits problem. 
Employing the lens of data-driven algorithm design, we show that using sampled instances from a distribution, we can find algorithms that are near-optimal for instances arising from {\em that} distribution. We introduce two novel and interesting families of algorithms. Our first family achieves stronger competitive ratios that depend on the strength of concavity of the reward curves, and the second achieves a best-of-both-worlds guarantee on benign as well as worst-case instances which prior techniques in the literature fail to achieve. \looseness-1

Our work opens up several directions for future research. It would be interesting to extend our notion of strength of concavity to more refined arm-dependent and piecewise structures with unknown switching times, where techniques for ``tracking regret'' \cite{herbster2001tracking,sharma2020learning} might be helpful. Our best-of-both-worlds results mirror known ``benign'' regimes where sublinear regret is possible, but it would interesting to determine the precise necessary and sufficient conditions for achieving sub-linear regret in improving bandits. Computationally efficient implementation and sample complexity lower bounds are  concrete questions raised by our offline-to-online hyperparameter transfer results. Finally, it would be interesting to develop similar guarantees for variants of the improving bandits model studied in the literature, including stochastic (arm reward means are concave non-decreasing, with sub-Gaussian variance) and restless (arms evolve with time, not pulls) versions.

\part{Social Epistemology} \label{part:social-ep}
\chapter{Why Algorithmic Approaches?} \label{chap:social-ep-intro}

In this chapter, we introduce our perspective on applying algorithmic tools to studying problems in social epistemology. We start by laying out our view of theoretical modeling and the role it can (and cannot) play in the study of social problems. We engage with literature on the philosophy of models and draw on our own experiences to present our argument, and we include recommendations for how to approach social modeling. Then, we introduce the broad social epistemological problem we consider in this part.

\section{On The Use of Models}

\subsection{Introduction}

When I first started reading papers in theoretical computer science (TCS), particularly in algorithms, I found myself puzzled for more reasons than one. You see, I found my way into TCS, like many, through competitive math, where the problem statement was given, and we had to find the solution. Thus, I expected the papers to read the same way: here's the problem, and here's the algorithm that solves it. Instead, I was met with theorem statements that read something like:

\begin{theorem}
    There exists an algorithm that solves the {\em What-I-Want-To-Solve} problem in $poly(n, k)$ time and succeeds with probability 0.9.
\end{theorem}

Why was the main result framed as {\em existence} of an algorithm, and not highlighting the cool algorithm that solved the problem? Why was the specific algorithm secondary? At first, I thought it might be one particular research group that cared significantly about algorithm existence. But when I kept encountering this format in paper after paper, I realized I had to look more carefully at the practice. 

I reflected on the role that theorems play in our understanding of the systems we study, and as I underwent this realization many years ago, I wrote down, ``If a theorem is supposed to be a truth about the universe, then this makes sense. How we design such an algorithm is almost irrelevant, since it depends on what our finite minds can come up with, but that it can be solved is the more important point.'' Indeed, in addition to abstracting real-world computing problems into frameworks where we could study them theoretically, theoretical computer science also abstracts the {\em solution} to these problems. Algorithmic techniques come and go, but for a fixed computing model, knowledge of what is and is not computable remains firm. Slightly restating my realization as a neophyte, the exact description of the method is less relevant than the fact that we have characterized the realm of the possible. \par

Results in TCS are presented this way to highlight the complexity theoretic question that lies at the heart of TCS: can this problem be solved in polynomial time (or perhaps samples)? or will it take exponentially long? And in studying this question for a variety of problems, TCS gets to the heart of something even more fundamental: under what conditions, settings, assumptions about our world can we solve problems easily, and under which conditions is it hard? This perspective is further borne out by the fact that in theoretical computer science, we make an effort to interrogate exactly which of those conditions is necessary for our result and which is just sufficient. We construct instances that don't satisfy an assumption of interest and show that any algorithm must fail hopelessly on such instances. \par

This perspective, as it turns out, is remarkably valuable for studying the social world. Though TCS arose from a desire to model and study computing systems, i.e., to gain an understanding of what is and isn't computable in reasonable models of computation, the machinery developed allows us to isolate aspects of reality that we might not be able to reproduce in the real world. 
In the social sciences, this is especially useful, as we often cannot develop perfectly controlled experiments involving human subjects. However, we can often identify properties of the behavior of individuals or groups of individuals. \par

This abstraction-upon-abstraction perspective familiar to theoretical computer scientists but jarring to a new entrant to the field provides a powerful tool with which to characterize systems well-beyond those arising from computers or collections of computers. The early part of this essay will focus on what models can do in general, while the latter part will engage with what is special about the abstraction-upon-abstraction perspective. \par

\vspace{5mm}



\noindent I wish to situate this essay as a theoretical computer scientist examining and naming our practices. Over years of study, we, as researchers, conceive of the epistemology of TCS by drawing from textbooks, research papers, lectures, seminar talks, and conversations with senior members of the field. We find ourselves inducted into a worldview that we, no doubt, find compelling -- after all, that is why we stay in the field rather than wandering toward other areas. However, it can take interaction with the ``outside'' to really articulate differences in epistemology. For instance, when taking an algorithmic statistics class, I took for granted the desire to have finite sample guarantees (since that is what my professor told me!); when working on a research project in the area and looking to classical statistics literature for background and inspiration, I was surprised to find guarantees that held as the number of samples asympotically approached infinity. \par

Within theoretical computer science, it is extremely common for us to {\em inherit} our models. We often study models that someone decades ago, or at least a senior member of the field a few years ago, codified. As a result, we instinctively have particular ways of simplifying settings, abstracting and idealizing considerations, setting objectives, and evaluating outcomes. As we think about developing novel models, we bring much of that perspective by default, though it is worth articulating some of these considerations. As we study problems in the social world, namely problems that are studied by a vast range of scholars and practitioners, it becomes all the more useful to articulate the perspective we come from. \par

When I started working on building TCS-inspired models for social problems a few years ago, I was initially afraid that it would be easy to fall into producing {\em ad hoc} models that behave exactly as we want them to because we've baked everything about what we want to study into them. 
As I engaged critically with the task of modeling and drawing insight from new models, I developed for myself, a framework for thinking about models. In this essay, I present that framework and show how the epistemic purpose associated with a given model provides a way to evaluate it. Along the way, I map my reflections to arguments in the rich literature about models in philosophy of science. Finally, I conclude with procedural recommendations for computer scientists applying our methods to modeling social problems.

\subsection{Taxonomy of TCS-based Social Models} \label{sec:taxonomy}
In this essay, I reflect on three epistemic purposes a theoretical model might serve. First, a satisfactory model will reproduce a phenomenon observed in the real world. A richer model would propose an explanation for why this phenomenon occurs. Finally, the richest model is one that suggests an intervention that would allow us to achieved a desired real-world outcome. Typically, a model should achieve at least one of these epistemic purposes. Certainly, I never claim that a model will be correct in an absolute sense; the processes of abstraction and idealization inherent to modeling necessarily excise the complexity, and often messiness, of the real world. Thus, when considering the adage ``All models are wrong, but some are useful,'' frequently attributed to statistician George Box, I seek to argue about the different ways in which models can be useful. I will then discuss how to evaluate models against those intended purposes. \par 

The idea that models should be designed for and then assessed with respect to a certain purpose is what, in her 2020 paper, Wendy Parker calls an ``adequacy-for-purpose view'' \cite{Parker_2020_adequacypurpose}. We will take a similar perspective to hers, arguing that rather than study how accurately a model reflects reality, we will assess whether it fits its purpose well or not. Parker further notes that, ``for a model to be adequate-for-purpose, it must stand in a
suitable relationship not just with a representational target but with a target,
user, methodology, and background circumstances jointly.’’ \cite[pg. 460]{Parker_2020_adequacypurpose}. We will engage with all of these aspect as we think about models and evaluation.

While ontological and semantic questions abound about these models, I will focus narrowly on theoretical models, which I am defining as models that include formal mathematical definitions of any agents, their action and reward spaces, and ``update rules'' for both agent behavior and reward provided by the universe. The model specification should also include any constraints on when the model and regimes in which it is / isn't relevant to the real-world task at hand. An agent is anyone that has some ability to take actions in the world. In the most general sense, actions might not even be adaptive; they could just be a collection (for instance, we can think of training data for a machine learning model as the set of ``actions'' in this framework). Reward comes from playing the action in context (this is clear in the decision-making setting; in the machine learning setting this is the loss). Update rules define the dynamics and could either comprise how agents choose their next action / interaction and/or what changes about the environment (in the machine learning example, the change in the weights during gradient descent could be the change in the environment). This is a sufficiently general framework for the kinds of models we will discuss here. 
When I refer to the ``setting,'' I mean  the tuple that specifies the things defined so far. The ``phenomenon'' is the outcome of the model under certain conditions. The ``outcome'' could be the forward evolution of initial conditions, or it could be a comparison of two different parameter regimes in the setting. My construction draws on Michael Weisberg's conception of a theoretical model \cite{Weisberg-whoismodeler}. He writes that:

\begin{quote}
The construal of a model is composed of four parts: an assignment, the modeler’s intended scope, and two kinds of fidelity criteria.
The assignment and scope determine and help us evaluate the relationship between parts of the model and parts of the real-world phenomenon. The
fidelity criteria are the standards theorists use to evaluate a model’s ability to
represent real phenomena.
\end{quote}

The assignment,  ``which is the
specification of the phenomenon in the world to be studied and the explicit
coordination of parts of the model with parts of the real-world phenomenon,'' (\cite[pg. 219]{Weisberg-whoismodeler}) arises through the specification of the agents and rewards and formalizing the action spaces. The scope, which Weisberg writes, ``specif[ies] which parts of the model are to be ignored,'' is handled by making explicit the regimes in which the model is meant to be used and not used and by specifying the constraints on the model. Finally, the fidelity criterion arise from the purposes for which we develop the models. We discuss these in the respective sections. \par

Note that when we talk about the epistemic purpose of the model, it is (almost always) with respect to {\em outcomes}. That is, a model is useful with respect to modeling some outcome, or some set of outcomes, in the setting, not just the setting. Again, this corresponds to Parker's view that models should be evaluated with respect to a purpose, not just with respect to their ability to reproduce reality. Parker  further details the distinction between adequacy for {\em instance} of use and {\em context} of use \cite{Parker_2020_adequacypurpose}, the latter being a generalization of the former. Finally, Parker argues that, ``for a model to be adequate-for-purpose, it must stand in a
suitable relationship not just with a representational target but with a target,
user, methodology, and background circumstances jointly’’ \cite[pg. 460]{Parker_2020_adequacypurpose}. That is, they must be developed and evaluated with the user, the ways in which they are deployed, and broader conditions in which they are deployed. The taxonomy of epistemic purposes I develop here can be viewed as an adaptation of Parker's framework for the specific class of models I focus on, namely TCS-style models for social problems. 
 \par 

\subsubsection{Running Example: Schelling Model}


Throughout this essay, we will refer to a ``Schelling model,'' and so we briefly introduce it. The Schelling model \cite{Schelling01071971} is a computational model that aims to answer the question, ``can cities end up segregated through {\em only} the preference of people to live close to those similar to them?'' In order to study this, the model instantiates a grid in which each cell represents a household. At each time step, an agent evaluates whether or not at least $\alpha$ fraction of their neighbors are like them. If this condition is satisfied, then the agent stays put. Else, the agent moves to the nearest empty location. \par

Suppose we instantiate this model with a random initial distribution of agents according to types, and $\alpha = 0.3.$ Then computer simulations on a 51 by 51 grid result in almost 70\% similar neighbors. Thus, the model answers in the affirmative that a mechanism {\em purely} driven by agent preference can lead to segregation\footnote{It is important to already note the following. The model set out to explore whether segregation could result from pure preferences and confirmed that it was possible. This ought not to be interpreted as ``segregation in real cities {\em does} result purely from preferences.'' The purpose of this model was to explore whether such a phenomenon could exist, and the model only shows that it {\em could} exist.}.



\subsection{Reproduce Phenomenon} \label{sec:reproduce}
The most basic purpose of a model is to reproduce a phenomenon observed in the world. In the natural sciences, this observation would arise in the natural world. However, in general such an observation could exist at any level of abstraction, including in constructed objects. 
The model is consistent with observations as long as there are some initial conditions for which the real world outcome and the model outcome match. 

Models that achieve this goal allow us to hold a phenomenon in our head, or at least in a compact way with a small number of formal definitions and specified equations. We then can strip away all the complexity and nuance of the real world to tell a simple but self-contained story about what is happening. Such a mathematical characterization induces a short description that suffices for isolating exactly what it is we wish to study. This is useful in its own right, even if subsequent goals (explanation and intervention) are not satisfied; such a model proposes a {\em sufficient} set of conditions under which the phenomenon of interest occurs, which is valuable even without establishing {\em necessity} of these conditions. Models that are capable of predicting behavior are related; reproduction is ``prediction'' of a known circumstance, while prediction is more generally associated with new but related circumstances. \par

Catherine Elgin, in \cite{elgin2017} (as quoted in \cite{sep-models-science}), writes that understanding a setting means to have:
\begin{quote}
    an epistemic commitment to a comprehensive, systematically linked body of information that is grounded in fact, is duly responsive to reasons or evidence, and enables nontrivial inference, argument, and perhaps action regarding the topic the information pertains to.
\end{quote}

Our description of a model that reproduces a phenomenon maps well onto Elgin's framework. In particular, our model is grounded in fact as it is able to reproduce observations. It is ``responsive to reasons or evidence'' as its details can be adapted to better fit new evidence. Finally, ``enabl[ing] nontrivial inference'' corresponds to prediction, which is, as discussed previously, an extension of the reproduction goal. The intents of ``argument'' and ``action'' are higher-order epistemic purposes that we discuss in the rest of our argument in Sections~\ref{sec:explain} and \ref{sec:intervene}.
\vspace{5mm}

\noindent A few years ago, for a number of different reasons{\footnote{the primary practical reason was that I was to attend a summer school on applying statistical mechanics techniques to analyzing deep learning phenomena, but I also felt that I finally had the mathematical maturity to attempt another physics class after a failed attempt to double major in physics during my undergraduate studies.}, I decided to take a graduate statistical mechanics class. I frequently found myself puzzled by approximations that came as second nature to the physicists (i.e., everyone else) in the classroom. The math that we did in the course often felt, to me, pretty far from math and pretty close to magic. As I bristled up against the epistemology of this field that had previously eluded me, but which I was now committed to develop a stronger understanding of, I spent significant time asking basic questions to the professor\rmhere{, Professor Wendy Zhang}. At that point in my career, I had finally developed a reasonable understanding for how theoretical computer science framed and solved problems. In an attempt to build up a parallel, albeit significantly less deep, understanding of other such fields, I asked the professor one day, ``what do you think is the point of theory in your field?'' ``Theory is what makes physics human,'' they replied\rmhere{\footnote{\hl{not direct quote, check with her}}}. I argue that the same can be said here for theoretical models. A good theoretical model should highlight the key aspects of a setting and a phenomenon and marry them in such a way that humans can interpret and analyze them easily. Paring down the complexity of the real world and focusing on specific concepts of interest allows us to deepen our engagement with and understanding of what we aim to study. It bears noting here that ``making human'' necessitates the specification of a user (referring back to Parker's adequacy framework \cite{Parker_2020_adequacypurpose}), since what is interpretable by people with differing backgrounds can differ greatly.


\subsubsection{What This Implies for Evaluation of Model}

Let us briefly reflect on what this framing implies for evaluating models. At the most basic level, if a model does not align with observations from which it is developed, then it is not a good model. Of course, ``aligning with observations'' is itself somewhat subtle, as there may be issues such as noise or measurement error. Thus, perhaps a better way to phrase this criteria is to say that we require our models to align closely with observations and in cases where they do not, offer explanations for why not. In Weisberg's terms\cite{Weisberg-whoismodeler}, these are our {\em dynamic fidelity criteria}. Angela Potochnik formulates a closely related criterion at the level of individual posits within a model: ``A posit is epistemically acceptable when its divergence from truth is insignificant, taking into account (a) the posit's role in the representation and (b) the epistemic purpose to which that representation is put'' \cite[pg. 100]{potochnik-idealization-aims}.
Overall, it is necessary for a good model to satisfy this question but this claim does not identify conditions that are sufficient to meet to be deemed a good model. \par

Further, per Parker's development \cite{Parker_2020_adequacypurpose}, we must also ensure that the model must be appropriate for not only reproduction of the phenomenon but also for the user, methodology, and background circumstances. In our development, the models we devise are most likely to be accessible to fellow theoretical computer scientists, and if we wish them to be useful to other users, we must devise them accordingly. Models meant to reproduce phenomena but not intended for deployment in action-oriented pipelines are more safe from methodology and background circumstance considerations. These points become more stark in Section~\ref{sec:intervene}, where we discuss models that propose interventions.

This goal is the weakest of the three we discuss, since we could, in theory, build a model to reflect nearly any kind of behavior. However, constraining our model to be mathematically-sound and simple protects against ``overfitting'' the real world.  So far, the discussed evaluation conditions leave open a critical question about prediction or generalization. Namely, what does the model say about situations that weren't ``baked'' into it? If the model happens to capture  other phenomena that occur in the setting, and not just the phenomenon it was designed for, then the model is more robust\footnote{We use the term ``robust'' informally, but an interested reader may refer to \cite{Weisberg2018-validating-idealized} for a discussion of robustness, clarifying the differences between parameter, structural, and representational robustness.}.

\subsection{Explain Phenomenon} \label{sec:explain}
The next kind of model goes beyond just reproducing a phenomenon. It posits an explanation for why this outcome occurred. The explanation often modularizes the muddled impacts of various abstract pieces. This explanation can then be tested in the real world by intervening on the posited mechanism, i.e., on the module that has significant impact. In this way, such a model provides probes for future inquiry\footnote{as a field, physics makes tremendous use of this perspective}. The posited mechanism should, ideally, be falsifiable. Abstraction and modularity are familiar terms to computer scientists and are key parts of the design of computer systems. In this section, we show how they are relevant for building theoretical models for phenomena of interest. \par

The philosophy of modeling literature distinguishes between two different kinds of explanations, how-possibly and how-actually explanations (\cite{dray1964philosophy, dray1979laws, FORBER-howpossible, VERREAULTJULIEN-2019-howpossibly}, among others). How-actually explanations are meant to explain, as their name suggests, {\em how actually} some phenomenon arises in the real world. On the other hand, how-possibly explanations essentially show that an outcome is not impossible. They provide a possible mechanism without necessarily providing a complete explanation\footnote{See \cite{Ylikoski02012014-hpe, VERREAULTJULIEN-2019-howpossibly} for thorough ontological accounts of how {\em how-possibly} explanations could possibly explain. For our purposes, it suffices to accept that they can.}. In our development, we will defend how-possibly explanations as valuable in specific circumstances, especially when we argue in Section~\ref{sec:can-cannot} that TCS brings to modeling a unique perspective on encapsulation as a topic of study.

A satisfactory explanation must address two puzzles: why the outcome occurs and what would change about the outcome if things changed in the setting. Often, causal considerations come into play, and the latter part described above is indeed a counterfactual that a model would be able to predict. To this end, James Woodward writes in \cite{woodward2003} that the relationships and dependence between ``explanans'' (factors) and ``explanandum'' (things to be explained):
\begin{quote}
    enable us to see what sort of difference it would have made for the explanandum if the factors cited in the explanans had been different in various possible ways.
\end{quote}

Thus, in the terms used by Woodward, a model allows us to {\em isolate} the explanans and and the explanandum in such a way that we can modify the explanans in order to see the effect on the explanandum. \par

Once explanans and explanands are isolated, we have modules that comprise our model. Each module reflects some or several aspects of reality in some simplified way, by removing either ``noise'' or excess complexity. These modules correspond to the components Weisberg describes as comprising an ``abstraction hierarchy'' \cite{Weisberg2020AbstractionAR}. This perspective is valuable because it also, as suggested by Weisberg, draws out the question of where in the hierarchy {\em representation} occurs (in fact, it is up to the modeler to decide the answer to this question through the way in which they choose to organize the modules):

\begin{quote}
    Parts of computational structure becomes
representational when theorists have the appropriate construals. As a
consequence, construals will help us answer the question of which parts of
the hierarchy are represented. For our purposes, the most important part of a
modeler’s construal is her intended scope and her assignment. Intended scopes
specify which aspects of a target phenomenon are intended to be represented
by a model, while assignments tell us which parts of the model are intended
to represent which parts of a real or imagined target. Taken together, they
allow the modeler to specify which part of the model should be understood
as representational and which is just part of the non- representational infrastructure
of the model. \cite[pg.219]{Weisberg2020AbstractionAR}
\end{quote}

Thus, we may conclude that as modelers, our intents in framing the target and defining modules to model it are crucial parts in the modeling task. This, then, provides us the freedom to choose modules that isolate explanations. Naturally, we gravitate toward simplifying the real-world when devising modules. Indeed, a rich line of work argues that not only do models explain {\em despite} their simplicity, but also they only explain {\em because} of their simplicity. Among them, Batterman and Rice 2014 argue that, ``an explanatory story devoid of representation relations is {\em required} to understand
how these models can be used to such great effect'' (emphasis mine) \cite{battermanrice2014}. They go on to argue that such models and {\em many} real-world phenomena lie in the same universality class.
Potochnik develops this idea further, arguing that ``idealizations themselves play a positive representational role. Idealized models and other scientific products do not represent despite their idealizations but partly in virtue of those idealizations. Moreover, idealizations aid in representation not simply by what they eliminate, such as noise or non-central influences, but in virtue of what they add, that is, their positive representational content'' \cite[pg. 50]{potochnik-idealization-aims}.

\vspace{5mm}

\noindent During a wonderful month spent in the French Alps,
pondering what aspects of statistical physics could be applied to machine learning \footnote{This experience raised for me many questions of how we should shape the epistemology of deep learning ``theory'' inspired by physics. Time spent considering this, no doubt, must've influenced my contemporaneous reflections on models and universality}, I found myself marveling at how the erosion patterns of the solid rock comprising the mountains so closely resembled the erosion patterns of the ice making up the glaciers that crowned them. Indeed, at the level of {\em erosion}, it did not (significantly) matter whether the underlying substance was rock or solid water. 
Similarly, returning to the Schelling computational model, Elliott-Graves and Weisberg argue:

\begin{quote}
    Imagine if we replaced
the highly idealized utility function with a more psychologically realistic one. Imagine further
that we included a realistic city map and realistic barriers to moving house. Such a model
may well exhibit the Schelling dynamic, but now it would be impossible to tell which aspects
of the model were responsible for the dynamic. This would result in a loss of explanatory
power, even while we might gain representational realism. In cases like these, idealizations
are justified and will remain. \cite[pg. 180]{eg-w-idealization}
\end{quote}

The desire for simple components that provide explanatory clarity, then, raises the question of what it means to ``isolate'' these explanatory aspects.  To answer the question of what it means to isolate the modules, we have to consider exactly what about the real world we are ``ignoring'' in order to develop the model.
We can separate the kinds of simplifications we make into two categories: abstractions and idealizations. The distinctions between then two and the resulting relative roles have been studied extensively in the literature, including in \cite{valentini2012, eg-w-idealization}. 
On the one hand, abstraction involves ``black boxing,'' assuming that item has some input-output behavior or certain characteristics; an abstraction is not dependent on exactly how the inputs cause the outputs or why the characteristics arise. Weisberg develops a thorough account in \cite{Weisberg2020AbstractionAR} of abstraction and {\em encapsulation}. He writes: ``By encapsulate, I mean that we
hide details of how things work, only knowing that a certain kind of input
is expected, a procedure will be applied...,
and then the transformed output will arrive in a particular form.'' We use the term ``abstraction'' to refer to this process.

On the other hand, idealization involves taking some limit or stylizing behavior in some fictional way. Elliot-Graves and Weisberg develop a rich account of idealization in \cite{eg-w-idealization}. They identify three different kinds of idealizations: Galilean, minimal, and multiple-models. Galilean idealization \cite[pg. 177-8]{eg-w-idealization} are ones made due to lack of capacity to consider the full version, and we expect to be able to de-idealize them with availability of better data or computational power. Minimal idealizations \cite[pg. 178]{eg-w-idealization} entail the modeler choosing factors they deem relevant to the purpose of their model. Finally, multiple-model idealizations \cite[pg. 178]{eg-w-idealization} occur when several parallel, related, but incompatible models are developed to address different aspects of the problem. \par

In the TCS-style models we study, we typically utilize abstraction\footnote{of course! we are computer scientists, after all.} and focus on minimal and multiple-models idealizations. The distinctions between these different approaches affect the kinds of explanations that arise from models that use them. Abstraction allows us to focus on the role components with certain inputs and outputs play in a system without getting distracted by the particularities of the map between the inputs and outputs. We idealize concepts exactly so we can focus on the idealized versions without having to deal with the complexity of the de-idealized concept. A minimal model explains {\em because} of its simplicity, and attempting to add complexity back would destroy the explanation. Models employing multiple-model idealizations each explain isolated components of a system but do not explicitly combine these factors.


\vspace{5mm}
\noindent It is not always easy to go beyond idealizations in social modeling. 
In my own work, 
I have been guilty of assuming ``rationality'' of human actors\footnote{This is one of Potochnik's paradigm cases: ``Many people are familiar with the common assumption in physics of frictionless planes and with the common assumption in economics that humans are perfectly rational agents. These are both idealizations: every plane has friction, and no human is a perfectly rational actor'' \cite[pg. 42-3]{potochnik-idealization-aims}.
}. While it is convenient to argue that we are simply following the assumptions of the philosophers on whose work we base ours, there are significant reasons to deeply question the simplification (and particularly the idealization) of ``humans are rational actors.'' On the one hand, we can talk about all kinds of reasons for which people might not behave rationally, ranging from inability to reason about uncertainty to unexpected influences. However, harping on humans' inability to be rational belies a deeper issue with idealizing the setting in this way: when we formalize behavior, we impose that an agent must be able to specify a quantitative objective such that optimizing that objective yields rational behavior. In practice, as people, we do not, and we actually ought not, make most of our decisions according to optimizing a quantitative objective. In Weisberg's terms, by making a rationality assumption, we are defining a procedure followed by agents in our computational model. He notes that in the Schelling model, ``The model will explain the segregation
in a real city if an analogue to that procedure characterizes the real
agents in that city'' \cite[pg. 213]{Weisberg2020AbstractionAR}. Thus, the possibility that humans' decision-making processes are unlikely to be similar to the procedure induced by the rationality assumption is quite troubling for the purposes of deriving an explanatory model. \par 

In our work assuming rational human behavior, we often made these assumptions partially due to the emphasis on rationality in the philosophy source material and partially with the aim of mathematical tractability. For instance, it can interesting to note that bad outcomes occur even when agents behave rationally. Thus, in this case, the use of idealization allows us to show that simply being rational cannot itself protect against negative outcomes. 
However, it does make it harder to evaluate whether our interventions might be relevant for the real world. 

\vspace{5mm}
Appropriately modularizing components of a system, whether through abstraction or idealization, allows us to examine the real world system, absent subtleties and noise that we have decided are okay to set aside for now. Assuming the first goal (reproduction of the phenomenon) is fulfilled, an explanation will consist of rules for how one module affect other modules and/or how changing one module will affect another one. 


\subsubsection{What This Implies for Evaluation}

Per Weisberg's account in \cite{Weisberg2018-validating-idealized}, we are actually not validating the model itself, but rather we are validating hypotheses about the model. In particular, he writes, ``
One way to develop such an account is to formulate hypotheses about a
model’s similarity to its target. \dots we can formulate them as follows:
`Model M is similar to target T in respects X, Y, and Z.''' \cite[pg. 250]{Weisberg2018-validating-idealized}. If we are simply interested in reproduction, as in the previous section, $X,Y,Z$ would essentially capture first-order measurements that are easy to evaluate. On the other hand, in the case of explanations, $X,Y,Z$ could be posited mechanisms that can then must evaluated in the target real-world system. \par

In particular, if the posited mechanism is falsifiable, then we could hope to design an experiment to test whether this mechanism is reflected in the real world. The modeling exercise involves isolating modules that effect outcomes, and such a module could represents a probe in the real world. For example, if a model suggests that the age of individuals with whom someone interacts should affect outcomes, we could either passively measure whether there are differences in the data between individuals that have different age communities or actively rearrange social groups and observe effects. \par

It is especially important here over other places that we keep the {\em user} of the model in mind, as Parker encourages \cite{Parker_2020_adequacypurpose}. If our explanation is in the language of our TCS model, the natural user base would be individuals who have the appropriate background and language to interpret the resulting theorems. An explanation that appeals to, say, convergence of an iterated dynamic to a Nash equilibrium is adequate-as-explanation only for users who can make sense of that conceptual apparatus. \par

Thus, if the goal of the model is producing a mechanistic explanation, we  end up with a natural way to evaluate the model: if the mechanism is true, then the model is good; if not, the model is bad. But is it still useful to identify false mechanisms? Suppose by ``false,'' we mean a mechanism that is not at all at play in the real world. This could still be useful if it allows us to rule bad explanations out to make way for the true explanation. It could also help posit a counterfactual world; in particular, if the false explanation has attractive properties, then we could think about modifying the world so that that explanation takes root. This brings us toward the idea of using a model to identify interventions. We cover this in detail in the next section.

\subsection{Suggest Intervention} \label{sec:intervene}
In the sequence we have been developing, the grandest purpose of a model is to suggest how to change the real-world system so we may achieve desirable outcomes. It could be the case that the posited explanation justifies why an action taken in practice is a wise idea. The model might suggest a mechanism that we could hope to intervene on. Alternatively, the modularized model might evince a ``location'' (in terms of a module, or a few) where changing something would appreciably change outcomes under the same dynamics. In these ways, having a model clarifies the key parts of the dynamics and setting that lead to the observed outcomes and allows us to identify which to interact with to change the outcomes. \par

Elliot-Graves writes of a distinction between ``confirmatory prediction'' and ``applied prediction'' \cite{Elliott-Graves2026-pointofprediction}. The former is meant for testing theories and evaluated by accuracy and variance, while the latter is meant for intervention and evaluated for its utility even in resource- and information-poor settings \cite{Elliott-Graves2026-pointofprediction}. Thus, in this section, we shift our focus from confirmatory to applied prediction. These aspects of models and interventions leads us to an additional, crucial, perspective arising from the literature on causal inference. 
Per Woodward's interventionist account of causal modeling, to claim a causal relationship between variable $A$ and variable $B$ is to claim that an intervention on variable $A$ always results in a predictable change in variable $B$ \cite{WOODWARD2007-causalmodelssocial}. Thus, interventions provide an extremely natural framework for evaluation of the causal structure ina model. We return to this argument and perspective when we address evaluation.  \par

First of all, interventions do not need to be novel and grand. If there is some existing input into the system, positing an explanation for why that input effects its outcome can justify an existing intervention. Thus, depending on how we break up the system and inputs to it, just solving the second goal (explanation) might already have implications for the intervention goal. For instance, if a subsidy program is in place in the real-world, a model that explains why that is useful justifies the use of that subsidy scheme as an intervention. \par

A helpful perspective to take in this discourse involves setting up a parallel between physics and engineering, and in particular, control theory. When it comes to the quality of the model, it can turn out that a model that suffices for explaining most behaviors in the prediction or reproduction sense is not sufficient enough for control. For instance, we might have a non-linear system for which a first-order model mostly covers the observed behavior. The non-linear correction might be ignorable when we are just comparing trajectories. However, if the control input brings the system into a problematic regime, the error might blow up badly. 
\vspace{5mm}

\noindent In fact, I can give this example with conviction, as this happened in my final project for a control theory class. My lab partner and I built a LEGO elevator and characterized its behavior in open loop. A simple model mostly fit observations and definitely incurred noise, but we figured that is what happens in the real world. Thus, we proceeded to derive the controller and implement it. It failed miserably. We redid everything several times before whispering to each other, ``should we try a more complex model?'' We did, rederived the controller, and had great success. \par

To frame this anecdote in the language of the philosophy of modeling literature, the real-world target remained consistent throughout. For the initial purpose of reproduction, the {\em dynamic fidelity criterion} \cite{Weisberg-whoismodeler} was okay to incur some noise. However, when our purpose shifted to intervention, what we asked of the model changed, and with that, so did the dynamical fidelity criterion. Likewise, in the language of Wendy Parker's adequacy-for-purpose framework\cite{Parker_2020_adequacypurpose}, the simpler model was adequate for the purpose of reproduction but not for the purpose of intervention. Thus, this anecdote really highlights a core argument of this essay, that models must be constructed and evaluated according to the epistemic purpose they serve, even when they are meant to reflect the same target.

To think about interventions, it is important to have not only a good understanding of the world but also access to some way of controlling the system. That is, we need some sort of way of accessing and modifying the relevant part of the system. This access point, in the engineering setting, could be electrical, mechanical, or digital. Note that we need to be careful about the control environment in the model. For instance, it is nonsensical to suggest changing the laws of physics, but it might be sensible to add a motor to an item in order to change its position. Likewise, in a social model, we must carefully consider who carries out the intervention, what kinds of resources or influences they have, and what information they have access to. Sure, we can change outcomes if individuals automatically do different things, but this is not a realistic, implementable intervention. Similarly, assuming an intervener has access to information they do not have (such as access to a relevant probability distribution, or agents' private information) is silly. The concept is familiar to computer scientists as ``lexical scoping.'' Weisberg shows that in the Schelling model, the information available to each of the agents can be viewed as scoping:

\begin{quote}
Lexical scoping can also be a representational resource in more scientifically
interesting contexts. Consider an agent-based model such as Schelling’s
model. In this model, each agent is very simple but still has an identity. In
this case, an identity means that it is numerically distinct, has a location and
a utility function, and has a current level of utility. All of this information
should not be available to all of the other agents. In other words, the scope
of these agent variables is restricted to the agent and sometimes the agent’s
neighbors. In a slightly more sophisticated version of the model that had
strategic interactions, this would be even more relevant. One agent may need
to “guess” what another is going to do, but if it had access to the internal state
of all the agents, there would be no need to guess; all the information would
be available. \cite[pgs. 226-7]{Weisberg2020AbstractionAR}
\end{quote}


Could interventions that use privileged information still have value? Possibly, in that they could guide us to consider how to get them access to information or develop incentives, but we must be aware that we are not telling a complete story yet.


Finally, we must ask: is an intervention useful if it hasn't been implemented in the real world? To that I say: absolutely! Studying interventions, even without implementing them, provides additional insight about the setting in the model: it shows us what is possible and what heuristic aspects of the world could be affecting our outcomes. We can think of proposed interventions as experiments we can run on our model to, on the one hand, identify cause and effect and, on the other hand, stress test which aspects of the model are robust.




\subsubsection{What This Implies for Evaluation of Model}

Having made the shift from confirmatory to applied prediction (as distinguished by Elliott-Graves in \cite{Elliott-Graves2026-pointofprediction}), it is important to note that the standards for evaluating the two families of models are different. Thus, it would be inappropriate, and possibly even harmful, to only consider metrics like accuracy of the predictions of the model on real-world targets.

Following up on our claim in the beginning of this section, by developing interventions in our models, we may evaluate the causal structure in our modeling. An {\em interventionist} perspective on causality, of which Woodward is a prominent proponent, argues that a model represents causal relationships between its variables if the effect of an intervention on a variable in the model is reflected in a corresponding change in the target when the real-world variable is intervened upon.
In particular, following Woodward's terminology, a model is said to be {\em invariant} under an intervention when it correct predicts the outcome of an intervention in the target. Importantly, he notes that: ``Invariance under interventions is
thus a relative matter: a generalization may be invariant under some interventions
(and thus qualify as causal) but not under others'' \cite[pg. 162]{WOODWARD2007-causalmodelssocial}. Accordingly, we can see that {\em intervention} poses a stronger criterion for model evaluation than reproduction.\footnote{If we can pose explanation as intervention (i.e., devise an experiment to intervene on the explanatory mechanism), then explanation also allows for this criterion.} A model that is invariant under intervention is a strong model. \par




Practically speaking, implementing versions or variations of our intervention could help us understand, in the positive case, how robust our intervention is, and in the negative case, whether the model had fatal flaw or the lack of success of the initial intervention was just a result of simplification (and therefore a version that accounts for noise works in the real-world case).

\subsection{What Models Can and Cannot Do} \label{sec:can-cannot}

\begin{quote}
    ``Science is a human tool. It is a remarkably powerful tool, and it is surely our most important epistemic tool. Once we properly appreciate the aims of science and the contexts in which those aims are pursued, features of science that appear to be shortcomings are instead revealed to be strengths. '' \cite[pg. 22]{potochnik-idealization-aims} \par

\end{quote}

Now we turn our attention to an equally important question: what can models {\em not} do? It is important to explore the limitations of theoretical models to temper our optimism about what models {\em can} do. That way, we can use them toward their intended purposes, not for more or for less. \par

The first caveat to our optimism has been stated already: models can only really be evaluated with respect to a particular epistemic purpose. When developing a model, it is important to be clear about what we want it to be able to do, both at the level of the overall goal and in terms of the specific real-world phenomenon we aim to study. On the one hand, this is disappointing because we would like to have a grand unified theory of everything we touch. On the other hand, it is freeing, because by accepting the limitations of the model, we can use it to its fullest in its legitimate setting. \par

In our development, we have focused on realistic models, or models that reflect the real world. There are situations in which we develop what seem like completely unrealistic models, making seemingly absurd assumptions. However, those assumptions can evince structure in a problem that gets to the heart of what is difficult about a situation. This is a specialty of theoretical computer science; a familiar example is assuming the availability of oracles. An oracle is an object that is able to receive queries of a certain type and return the perfect answer to the query for free in terms of computational cost. 
For instance, in machine learning theory, often works assume access to empirical risk minimization (ERM) oracles. We know that in many cases, solving the ERM problem is computationally hard. Yet, it is often valuable to assume access to an ERM oracle. If an otherwise-intractable problem is computationally feasible given access to an ERM oracle, we know that the difficulty must lie in ERM. \par

More generally, identifying the right ``oracles'' for a problem can help us isolate where the difficulty lies. Returning to our discussion of representation, an interesting and powerful aspect of the theoretical computer science approach is that we study encapsulation itself. We spend significant time constructing various oracles and understanding complexities of problems under these oracles. Once we have found oracles under which the problem is (usually polynomially) tractable, 
we have identified the source of difficulty. The process allows us to make wildly unrealistic assumptions via encapsulation while still resulting in very useful models. On a more abstract level, TCS's focus on studying encapsulation itself is a novel modeling perspective not covered by standard frameworks thus far. \par

Could we hope for powerful theoretical models that solve many epistemic purposes in parallel? Possibly, though I am neither optimistic about finding such models nor convinced they are necessary. Several years ago, I worked on deep learning theory. My dream was that when I was sixty and had grey hair, I would prove the grand theorem that tells us why neural networks generalize. Now, at my current age and with more grey hair than I expected to have at this point, I no longer dream of this. The cynical reader might ask if I have given up on theory ever catching up to practice. No, not at all. I have simply broadened my view of what it means to model complex real-world phenomena. For most problems in the world for which theoretical models have provided insight, they are far from the only tools used to study the problem. I now firmly believe that ``understanding'' is a much messier prospect than what stylized and rigorous models can hope to provide. 

At best, our theoretical models will work in conjunction with experimental, statistical, and qualitative analysis. Paraphrasing \rmhere{Andrea Montanari} something I heard once at a discussion regarding the directions in which to steer deep learning theory, ``no one talks of a unified theory of a dishwasher. Why should we expect there to be one for deep learning?'' 
Particularly when models are deployed in social settings, with the intent of informing decisions that have massive impacts on individuals, we must have an abundance of evidence in favor of our intended actions from a vast range of epistemic tools.
This perspective is quite related to the one taken by proponents of the multiple-models idealization framework \cite{eg-w-idealization}. 
 Especially when it comes to selecting potential interventions, it could be wise to develop a collection of models, each isolating particular aspects of the target. Then, we can develop interventions and select those which are successful in multiple of these models. One limitation of this approach is that it might not allow for studying how different aspects of the system interact with one another. However, it would still help us understand the interactions between the intervention and various aspects of the target\footnote{A complementary limitation identified by Kathleen Creel points out that {\em transparency} in algorithmic systems and models could entail functional (knowledge of the algorithm), structural (knowledge of its implementation in code), and run (knowledge of its execution on particular hardware and data) transparency\cite{Creel_2020}. In TCS models, we automatically have functional transparency, but the other two are essential for real-world deployment, and other epistemic approaches to evaluation will play significant roles in providing run transparency, in particular.}. Potochnik addresses something related to this tension, writing ``rampant and unchecked idealization does not mean unprincipled idealization. Idealizations reflect researchers' interests, and they serve those interests in the face of specifiable cognitive, computational, and other limitations'' \cite[pg. 59]{potochnik-idealization-aims}. \par

At worst, theoretical models are cute exercises for our own self-satisfaction. This is roughly the position taken by Ariel Rubinstein in his book, {\em Economic Fables} \cite{rubinstein-fables}. He writes:

\begin{quote}
    If the models we develop in yellow notepads or on blackboards constitute a basis for predicting human behavior, it would be miraculous in my eyes. There are no miracles in economics, but there are wonders. In my studies in the Department of Mathematics in Jerusalem, I learned to see wonders in the world of formalities. I sometimes also see them in economic theory. I approach economics as someone with a sense of curiosity who is trying to understand the logic of human interaction a bit better. This may not be much, but perhaps it is not so little either.
\end{quote}

Indeed, we do not need our models to predict the grandest of human, social, or natural outcomes. I will be satisfied if we can provide slightly more conceptual clarity and inject slightly more human ingenuity into our pursuit of truth.

\subsection{On Process} \label{sec:on-process}


Now that I have developed a framework for thinking about and evaluating models, the natural next question is: how do we operationalize this? In this section, I overview a seven-step process for developing theoretical models for social problems that is consistent with the discussed framework. \par

First, we must ground ourselves in social science scholarship: as computer scientists, we are not the first to study the social problem of interest. As we do so, we internalize the argument or arguments made in the respective papers and identify the crucial parts of it. This distillation process provides a checklist for whether our model has adequately modeled the phenomenon or not. Much like how in a scientific experiment, we must define the hypothesis we wish to test before designing and running the experiment, here, we must define what we want our model to show before building it. Otherwise it is too easy to move the goal posts. \par

Next, we should find general theoretical frameworks that are useful. These could be relevant because others have used them as coarse models for human behavior, or they could model decision problems in an interesting way to which we can add social considerations. From there, we set up the rules specific to the setting. Then, we can ``evolve'' the system forward to investigate the outcomes. Do they match what existing scholarship says or not? Which parts of the setting lead to those outcomes? Can we simplify the model and get the same results? These are important questions to engage with in the first phase of developing such a model. At this stage, it is critical to explicitly articulate assumptions and conditions in order to legitimize the construction of the model \cite{Parker_2020_adequacypurpose}. \par

Now, let us pause and think about the modules that comprise our model. Approaching this question from a TCS perspective, we can carefully devise and redevise the modules in an attempt to isolate properties of interest (earlier, we spoke about how oracles allow us to isolate computational complexity, but we can more broadly consider ``oracles'' that encapsulate other properties of interest). We must start by articulating the mapping between the parts of the model and the respect aspects of the target. Next, we must ask which are abstractions and which are idealizations. Articulating these explicitly will help us understand the relationship between our model and the world we are modeling, and it clarifies for us the different levers we have within the model. \par

We can next try to use the modules to develop explanations. The initial step is to once again make explicit which aspects of our model relate to which aspects of the target \cite{mohseni2024methods}. Then, we explicitly state a hypothesized explanation. This should then lead us to which module should be affected according to that explanation. We may try changing it as a thought experiment. As we vary things within a module of the model, we will gain intuition for what effect that module has. \par

Next, we think about interventions. What do we want to be different about the world? What does the explanation suggest as a potential location to intervene? What kind of intervention access is possible in the real world setting, and what informational constraints does it come with? We can use these questions to guide the kind of intervention we develop. \par

Now, suppose we have an intervention that solves the desired problem. We must now critically engage with the proposed intervention. What properties does it have? What are the benefits, drawbacks? Are there ethical concerns with implementing such an intervention in the real world? \par

Finally, we can consider practical evaluations. These can comprise experiments on synthetic data or real world data. If we are ambitious, we might even consider real-world experiments. Our essay, however, is focused on theoretical models, and so we leave details of this important stage to experts in designing and running experiments for social problems.


\section{Modeling Ambition}

Having articulated our position on the role of theoretical models in the study of social problems, we turn our attention to the specific social problem we study in this thesis. At a high level, we investigate how individuals make decisions when investment is required to succeed, and in particular, we focus on social influences (Chapter~\ref{chap:pess-traps}) and behavior traits (Chapter~\ref{chap:grit}). We base our inquiry on a series of works in social epistemology \cite{morton2019grit, morton2022resisting, morton_moving_2021}.

Myriad reasons contribute to communities facing generational poverty and disadvantagement, most systemic. 
A key question is what role individuals can play in lifting their communities out of poverty and other challenging circumstances.
Often, people take the neoliberalist view that assigns blame to individuals for not empowering themselves to exit their negative circumstances \cite{azevedo_neoliberal_2019}. 
The logic goes something like this: if only individuals were willing to work harder, use their money more wisely, or make bigger investments in their future, they could lift themselves out of dire circumstances, bring more income and status to their communities, influence others by being positive role models, and overall improve the community's standing. 

On the other hand, scholars including Jennifer Morton argue that it is morally indefensible to obligate individuals to lift their communities out of difficult circumstances, often caused by generations of systemic oppression. She argues in her work, and particularly in her book \cite{morton_moving_2021}, that not only should we not hold an individual  responsible for lifting a community out of poverty, but also we must provide individuals engaging in ambitious ends with a balanced and ethical accounting of the costs of their choices. Morton moreover argues for structuring conversations and empowering interventions around communities rather than just individuals. 
\par

Thus, an important question to study theoretically, and the motivation of the works described in Chapter~\ref{chap:pess-traps} and Chapter~\ref{chap:grit}, is:
\begin{quote}
\centering
    \textit{What role do individuals with high amounts of grit play in their communities? How can we design empowering interventions to benefit entire communities?}
\end{quote}

In the rest of this part of the thesis, we present our progress toward this question. Our results in Chapter~\ref{chap:pess-traps} show how to help communities break out of pessimism traps without forcing anyone to take actions that are suboptimal for them. The intervention ideas in this section are also, to the best of our knowledge, technically novel in the opinion dynamics literature. Our results in Chapter~\ref{chap:grit} help theoretically formalize what having grit even means in the context of decision-making. We conclude with what our results mean taken together and a description of what other theoretical inquiry would be valuable toward this grand question.

\chapter{Pessimism Traps and Algorithmic Interventions} \label{chap:pess-traps}
\section{Introduction}

Studying decision-making under uncertainty has interested scholars in several disciplines, such as psychology, economics, and philosophy. In economics, specifically in statistical discrimination, scholars such as \cite{coateLoury} have studied how individuals make decisions, particularly in marginalized communities where societal structures and perceived barriers can impact their choices. This often leads to self-censorship or a withdrawal from available opportunities due to a pessimistic outlook. The concept of pessimism has also been explored in philosophy, specifically in recent work by \cite{morton2022resisting}. \cite{morton2022resisting}'s research examines the dilemmas faced by individuals belonging to marginalized groups when presented with evidence that diminishes their chances of success. As a result, they may rationally choose to redirect their efforts elsewhere to pursue other less valuable alternatives. However, this decision only reinforces the pessimistic mindset that influenced it, perpetuating a cycle of limited aspirations and self-doubt. \cite{morton2022resisting} criticizes the simplistic solution of turning towards optimism and proposes a more critical and situational application of optimistic beliefs.
\cite{morton2022resisting} characterizes \textit{pessimism traps} as follows: 
\begin{quote}
A \textbf{pessimism trap} \dots is
meant to capture how negative beliefs about one’s likelihood of success can play a role in agents pursuing less risky, modest ends instead of ambitious ones, thereby further entrenching the negative evidence that the agent herself and other agents in a similar position face. The central idea, however, has broader applicability.\dots
for this paper, I will focus on those cases in which poverty, prejudice, and discrimination play a role in providing agents with the sort of evidence that would \textit{prima facie} make it \textbf{rational} for them to arrive at the pessimistic beliefs that play a role in thwarting their ambition.
\end{quote}

Our work extends Morton's qualitative description of pessimism traps (summarized in Appendix~\ref{appendix:features}) with mathematical formalism and studies algorithmic decision-making within this formalism.  We explore the theoretical model of \textit{information cascades} (also referred to as ``herding'' in the opinion dynamics literature \cite{mossel1, pathological}) to shed light on decision-making under uncertainty and pessimism in such contexts. Using this model, we propose interventions to break pessimism traps and redirect the population into favorable states.

\subsection{Our Contributions}

In this chapter, we study pessimism traps as conceptualized by Morton and offer interventions to shift communities out of these traps. Our first major contribution is to link the concept of a pessimism trap to the opinion dynamics literature. We do so by interpreting a single-dimensional information cascade model as a decision sequence in a community. We extend this in Section~\ref{sec:k-group} to consider $k$ parallel information cascades representing the decision processes of $k$ independent communities.

Our next major contribution is an intervention for the single community setting that sustainably shifts communities out of pessimism. We consider a subsidy enacted by the government to incentivize certain behavior in an agent. Two natural ideas fail: first, suppose we subsidize the ambitious action by a large fixed amount. While this intervention incentivizes agents to act ambitiously in the short term, once the subsidy program ends, subsequent agents will ascribe prior agents' ambition entirely to the subsidy they received and fall back into pessimism traps. Secondly, subsidizing by too small an amount fails to incentivize the ambitious action over the moderate one at all. 
Thus, we derive the precise size of the subsidy that will incentivize those who were already leaning toward ambition while not moving those who were anyway not considering the ambitious choice. Herein lies our first surprising insight -- the non-monotonicity associated with the effect of the size of subsidy.

Next, we study $k$ communities, each behaving as above. The government still intervenes but must act impartially with respect to community membership. 
Each of these groups may have a different optimal choice among the two options, and the government (a) does not know which choice is optimal for each group and (b) must be blind to  community membership in providing subsidies. We show that in this case, we can construct a distribution $\mathcal{D}$ such that if the government draws the subsidy randomly from that distribution, then eventually each community will shift toward what is the optimal end for them. This intervention relies crucially on the previous insight: if the provided subsidy is in the ``just right'' range discussed above for an agent, the agent's action will shift their community toward optimal action. However, if it is too large or too small (outside of that range), the community will remain in the same state. Thus, provided the distribution assigns at least a minimum probability to the ``just right'' range for a fixed community, that group will eventually settle into their optimal action. Therefore, our second main insight is that since we can guarantee (1) the existence of a range of subsidy values that shift the community toward the correct action for that community and (2) a single community faces no \textit{negative} effect when provided a subsidy outside of that range, the government can use randomization to shift \textit{all} groups into optimal behaviors, even without knowing what is optimal for any given community.

Finally, we verify the effectiveness of our proposed intervention in the one-group case via experiments. Our experiments show that the intervention we develop is indeed successful in shifting sequential decisions into optimism. Further, we confirm that the required budget for the intervention is tractable. Our experiments provide additional confirmation of the benefits of our intervention in the simple model that we study.

We view our work as a step forward in providing a theoretical model in which to study pessimism traps and, more broadly, in developing empowering interventions for marginalized communities. While our 
model is necessarily stylized, our major insights could help guide interventions in more complex settings. 

In summary, our contributions are:
\begin{enumerate}
    \item Linking the opinion dynamics and pessimism traps literatures by studying pessimism trap formation in the information cascade model.
    \item Identifying an intervention that sustainably shifts a community of agents away from pessimism traps. 
    \item Extending this result into a setting where there are $k$ different groups and where the government knows neither which the correct action is for a group, nor how close they are to escaping the trap and showing the power of randomization in this setting.
    \item Corroborating the success of our theoretical interventions experimentally.
\end{enumerate}

The remainder of this chapter is organized as follows. First, we briefly survey work from a diverse array of fields related to our work.
Next, we present the formal mathematical model in which we study pessimism traps along with some preliminary facts about them, and we reflect on the strengths and weaknesses of this modeling approach. In Section~\ref{sec:subsidy}, we introduce and analyze our intervention to shift a single group toward optimism. We then extend this in Section~\ref{sec:k-group} to the setting with $k$ groups each of whom has an unknown optimal action.  Finally, we provide experiments that show the success of our proposed interventions, substantiating our theoretical results.

\subsection{Related Work}

Here, we provide a brief survey of related works that contextualize our approach to studying pessimism traps and interventions aimed at promoting optimism.  

As quoted above, \cite{morton2022resisting} formalizes the notion of pessimism traps, building upon several empirical characterizations from scholars studying the phenomenon in the field of education \cite{cohen_2006_intervention}. Morton highlights the importance of \textit{belief}, which is why we focus our modeling and interventions on shifting beliefs in a sustainable manner.

Much work in the literature studies the connection between individual behavior and community-wide effects. This includes the literature on opinion dynamics and social learning, which are surveyed in \cite{sirbu_opinion_2017} and \cite{chamley_rational_2004}. They discuss both simple and more complex models for how people's beliefs vary in relation to the beliefs of those with whom they interact. \cite{acemoglu_opinion_2011} studies both Bayesian and non-Bayesian models for how agents update their beliefs, investigating consensus and asymptotic learning of state (i.e., what are the true beliefs in the world). On the behavioral economics side, \cite{kahneman1974judgment} study the psychological reasons behind why individuals might act in a way that disregard their private information.

Since individual behavior can impact community-wide outcomes, as studied in the works above, it is reasonable to consider ``nudging'' or intervening at the individual level. Works that have studied this include \cite{thaler2008nudge} in behavioral economics, showing how minor policy adjustments can realign individual decision-making with optimal outcomes.
 Similarly, work in theoretical computer science and game theory, such as \cite{Balcan2013CircumventingTP}, study ``nudging'' specifically in the context of equilibria, where the goal is to redirect a population from a less desirable to a more desirable equilibrium.
 
Additionally, empirical studies have supported the application of nudges in various domains. For example, \cite{johnson2012beyond} extends the discussion on the efficacy of nudges in real-world settings. Integrating nudging into public policy, particularly in health and environmental strategies, as discussed by \cite{benartzi2017should}, demonstrates how these concepts have evolved beyond theoretical discussions to practical implementations. 

Therefore, we utilize information cascade models to develop ``nudging''-style interventions aimed at combating the pessimism traps that frequently arise in marginalized communities. 
\cite{bikhchandani1992theory} initially framed the classic information cascade model, demonstrating how individuals, despite possessing private information, often conform to the erroneous actions of predecessors due to the strong influence of prior actions. This model has served as a baseline for exploring various dimensions of information processing within groups. Building on this foundational work,  \cite{anderson1997information} conducted laboratory experiments to observe cascade behavior in controlled settings, adding empirical evidence to the theoretical predictions. We mainly focus on the former.

\section{Preliminaries and Basic Model}
\label{sec:pess-traps-prelims}
We study two settings in this chapter: first, we look at a single sequence of $T$ agents deciding whether to take action  $A$ or $B$; second, we will expand this to $k$ parallel sequences of agents, where the best action (between $A$ and $B$) differs group-to-group. Here, we define the signal and sequence dynamics for a single sequence; in Section~\ref{sec:k-group}, we flesh out the setting with $k$ parallel groups. We take $T$ to be finite, but our analysis extends to infinite $T$. 
\subsection{Single Sequence Model}
First, for the case of a single community (sequence) 
we assume that $A$ is uniformly the better option (that is, it is best for all agents), though this fact is unknown to the agents. When we consider $k$ parallel sequences, then each sequence will have its own better option. We let $\world{A}$ represent the event that action $A$ is correct and $\world{B}$ represent the event that action $B$ is correct. When referring to the set of these possible events, we define $\mathcal{E} \triangleq \{\world{A}, \world{B}\}$ to indicate the set of possibilities. In particular, we assume that \textit{a priori}, agents have no bias toward either action, which we model by saying that they have a common, uniform prior over $\mathcal{E}$. Formally, $\Pr{\world{A}}=\frac{1}{2}$ and $\Pr{\world{B}} = \frac{1}{2}$. Further, one action has associated with it reward 1 and the other reward 0. 
In addition to the public prior, each agent $t$ receives a private signal $s_t$ indicating that one of $A$ or $B$ is the correct action.  This signal is correct with probability $ p > \frac{1}{2}$ and incorrect with probability $1-p$. Therefore, we have $\Pr{s_t|\world{A}} = p^{\delta_{s_t,A}} (1-p)^{\delta_{s_i,B}}$ and $\Pr{s_t|\world{B}} = p^{\delta_{s_t,B}} (1-p)^{\delta_{s_i,A}}$, where \( \delta \) is the Kronecker delta function.
\newcommand{\barH}[1]{\bar{H}_{#1}}

Given this signal and the actions of preceding agents, the agent decides whether to take action $A$ or $B$. Let $h_t$ indicate the action of agent $t$, where an agent is identified by the time they act and $H_{t-1}$ be the history of actions for the first $t-1$ agents. Agent $t$ will take an action if its expected reward exceeds the expected reward of the other action.  For concreteness, we adopt the tie-breaking convention that if an agent is indifferent between the two options based on the calculated posteriors, they follow their private signal. Now, we define an information cascade.
\begin{definition}[Information Cascade]
An information cascade occurs when an agent's action does not depend on their private signal. This means that one of the following occurs:

\begin{enumerate}
\item $\Pr{\world{A}|s_t=1,H_{t-1}} > \frac{1}{2} \text{ and } \Pr{\world{A}|s_t=0,H_{t-1}} > \frac{1}{2}$

\item $\Pr{\world{A}|s_t=1,H_{t-1}} < \frac{1}{2} \text{ and }\Pr{\world{A}|s_t=0,H_{t-1}} < \frac{1}{2}$
\end{enumerate}

\end{definition}
\begin{rmk}
A cascade begins when the observed history \( H_{t-1} \) becomes sufficiently skewed towards either adoption or rejection. Therefore, enough previous agents have taken the same action to make it rational for subsequent agents to follow the trend, based on a simple Bayesian updating procedure, even if the history contradicts their private signal.
\end{rmk}

To formally capture the concept of a pessimism trap for a sequence setting, we define the conditions under which agents, influenced by the actions of their predecessors, consistently choose the inferior option $B$. This phenomenon occurs when the aggregated evidence from prior decisions leads to a bias that overrides the agents' private signals. We represent this situation mathematically as follows:

\begin{definition}[Pessimism Trap]
A pessimism trap occurs when an information cascade leads agents to consistently choose the inferior action $B$ due to an overwhelming influence of prior agents' incorrect actions. Formally, a pessimism trap is defined by the following condition:
$$\Pr{\world{B}|s_t=1, H_{t-1}} > \frac{1}{2} \text{ and }\Pr{\world{B}|s_t=0, H_{t-1}} > \frac{1}{2}$$

This indicates that despite the private signal $s_t$ suggesting action $A$ is better, history $H_{t-1}$ has led agent $t$ to believe that $B$ is better. Subsequently, agents' decisions are based purely on observed history rather than private signals. 
\end{definition}

\begin{rmk}
When computing posteriors, we assume that agents only consider actions of those up until a cascade began, since if agents are rational, they realize that no further inferential information can be obtained from individuals who do not use their signals. 
\end{rmk}

As a result, we will often be interested in $\barH{t-1},$ the history of actions taken by agents who chose an action if and only if it was their signal. We express mathematically the Bayesian updating procedure through which agents come to their posterior belief about the best action based on the history of actions before a cascade begins and then formalize the kinds of governmental / central interventions we study:
\begin{equation}
\label{eq:post}
\Pr{}{\world{A}|s_t, H_{t-1}} = \Pr{}{\world{A}|s_t, \barH{t-1}} = \frac{\Pr{}{s_t|\world{A}}\Pr{}{\barH{t-1}|\world{A}}}{\sum_{w \in \{A,B\}} \Pr{}{s_t|\world{w}} \Pr{}{\barH{t-1}|\world{w}}}
\end{equation}

\begin{definition}
    A subsidy of size $r$ toward action $a$ is a benefit provided to an agent taking said action independent of the true world. Thus, in the world where $a$ is the correct action, instead of receiving just $R$ reward for taking action $a\,,$ the agent receives $R+r\,$ reward, and in a world in which it is the incorrect action, the agent still receives $r$ reward.
\end{definition}

\textbf{Formulation as Random Walk.} The main technical tools used for analysis of the cascade and subsidy come from framing the process as a random walk. We isolate the probability with which action $A$ is taken and the probability with which action $B$ is taken, and we analyze the information cascade as taking steps on a random walk. This walk in general terminates when either an up or a down cascade has been reached. Once we introduce a subsidy, the stopping point of interest will be the up cascade.
In Appendix~\ref{appendix:random-walk}, we show how we can use the random walk formalism to derive the length of time for which the subsidy must be in place and the required budget.

\subsection{On Our Modeling Choices}
A significant contribution of our work is that we propose using the information cascade model to study pessimism traps and associated interventions. In this section, we justify our use of this model and discuss the trade-offs made.

\paragraph*{Justification}
A natural question is: why choose this particular level of abstraction, i.e., why model the pessimism trap as hinging in this way on decisions made by previous agents? Several works have empirically characterized this phenomenon in the context of education, building on which Morton gives an epistemic characterization. At its core, as conceptualized by Morton, a pessimism trap occurs as a result of people's beliefs (about the world, about the rationality of others, etc), realized as herding behavior. Thus, in order to model this faithfully, we require a model in which we may quantitatively update the \textit{beliefs}. A Bayesian formulation is natural for this. Further, though sequentialness is not inherent to Morton's characterization of pessimism traps, there is a natural sense in which agents consider the context of those who went \textit{before} them when making decisions. Accordingly, a sequential model is a good choice for organizing the manner in which agents are influenced.

\paragraph*{Strengths} In some sense, the Bayesian posterior update represents the ``optimal'' rational decision. By modeling pessimism traps this way, we are showing that even with perfect rationality, a trap can form, corroborating Morton's point that pessimism traps do not occur due to lack of rationality. Further, as summarized in the appendix, two key features Morton identifies regarding pessimism traps are: (1) there is evidence of similar people not succeeding at the ambitious end and (2) not pursuing the ambitious end will not change the agent's view of its value. Both of these are well-represented in the information cascade model. Particularly once we extend to multiple groups, agents are looking to the history of actions taken by people like them (i.e., in the same group), and witnessing several people taking a certain action would indicate to them that taking the opposite action does not tend to be beneficial for people like them. Likewise, in the information cascade model, there is no feedback for whether a different action would have been correct, and therefore there is no reason for an agent to change their pessimistic view about it.

\paragraph*{What this model misses}
On the other hand, since the reward from the taken action is received in one step, this model does not reflect the fact that the ambitious choice requires investment and is contrasted with a choice that has a reasonable payoff throughout. Similarly, the model does not capture risk associated with the ambitious choice. 
These are important modeling considerations for future work.

\section{Time-Varying Subsidy} \label{sec:subsidy}
Without external intervention, once a cascade begins, it persists by definition. However, is it possible to derive an intervention from an external entity, such as the government, that can lift a population out of an incorrect cascade and redirect it toward a correct one? Importantly, can this subsidy be designed so that the correct cascade remains stable once the subsidy is removed? Note that the na\"{i}ve strategy of subsidising the ``correct'' action fails the sustainability criterion, because upon removal, agents have no reason to believe that that action was correct. They would simply believe that agents who took it during the subsidized period did so due to said subsidy. In this section, we focus on the question of how to design a good intervention for a single group / sequence that has \textit{sustainable} effects. 

Since a cascade occurs when an agent makes the same decision regardless of their signal, breaking a cascade requires influencing at least some agents to act according to their respective private signals. Indeed, this is our approach to designing a subsidy.
We consider subsidies for the ``correct'' action, which, in this section, we assume to be action $A$. The net reward for choosing the correct action is $R$.

Let the subsidy the government provides at time $t$ be $r_t$. Recall the two possible states of the world: world $A$, where $A$ is the correct action, and world $B$, where $B$ is the correct action. Let $|A|$ indicate the number of choices for $A$ outside of a cascade state, and similarly for $|B|$. Finally, we assume that the entity providing this subsidy is not trusted by the agents, and so the agents cannot infer from the direction of the subsidy which action is correct. 

Algorithm \ref{alg:subsidy} implements this subsidy scheme. We assume that this algorithm is applied after an incorrect cascade has already begun, and the purpose is to strategically provide a subsidy $r_t$ to the agent acting at time $t$ to break the community out of the pessimism trap. When agents are acting according to their own signals, or when the correct cascade has been reached, the subsidy need no longer be applied. The guarantees and derivations for Algorithm \ref{alg:subsidy} are provided in Theorems \ref{thm:subsidy} and~\ref{thm:subsidy_budget}, with complete proofs in the appendix. The main idea is that once agents act in accordance with their signals, after we see enough agents reveal their signal, a simple majority vote will, in expectation, reveal the correct action. At this point, even if the subsidy is removed, rational agents will act optimistically.
\begin{algorithm}
\caption{Redirecting Pessimism Traps}
\begin{algorithmic}[1]
\Require Start of incorrect cascade $t'$, history $H_{t''}$ for $t'' \geq t'$, $(T - t'')>> \frac{4}{2p-1}$, correct action $A$, incorrect action $B$, private signals $\{s\}_{t=1}^{T}$, signal strength $p$, reward $R$ for correct action
\State $|A| \gets$ choices for $A$ in $\barH{t'-1}$
\State $|B| \gets$ choices for $B$ in $\barH{t'-1}$
\State $ t \gets t''$
\While{$t < T$}
\If{$|A| - |B| \leq - 2$ (In incorrect cascade)}
\State $\gamma_t \gets \left( \frac{1-p}{p} \right)^{2|A|-t} \frac{1-p}{p}$ ; $r_t \gets R \left( \frac{\gamma_t -1 }{1 + \frac{1- p}{p} \gamma_t}\right)$
\Else
\State $r_t \gets 0$ 
\EndIf
\If{$r_t + R \cdot \Pr{}{\world{A} \mid H_{t-1}, s_t} > R \cdot \Pr{}{\world{B} \mid H_{t-1}, s_t}$}
\State Agent chooses action $A$
\If{$|A| - |B| \leq 1$ (Not in cascade)}
\State $|A| = |A| + 1$
\EndIf
\Else
\State Agent chooses action $B$
\If{$|A| - |B| \leq 1$ (Not in cascade)}
\State $|B| = |B| + 1$
\EndIf
\EndIf
\State Update history and increment $t$
\EndWhile
\end{algorithmic}
\label{alg:subsidy}
\end{algorithm}

\begin{theorem}
\label{thm:subsidy}
The subsidy scheme used in Algorithm 1 causes all agents to act according to their signals until the population falls into the correct cascade. The subsidy value is
$
r_t = R \left( \frac{\gamma_t -1 }{1 + \frac{1- p}{p} \gamma_t}\right)\,,
$
where $\gamma_t = \left( \frac{1-p}{p} \right)^{2|A|-t} \frac{1-p}{p}\,.$
\end{theorem}

\textbf{Proof Sketch} We want the subsidy to incentivize taking action $A$ only if it is already aligned with the agent's signal.
We seek to determine $r_t$ such that it ensures action $A$ is at least as preferable when $S_t = A$ and less preferable when $S_t = B$. That is, we want $r_t$ to satisfy the following conditions, so we simply solve:
\begin{align*} \text{(1) } r_t + &R \cdot \Pr{}{\world{A} \mid H_{t-1}, s_t = A} \geq R \cdot \Pr{}{\world{B} \mid H_{t-1}, s_t = A} \\ \text{(2) } r_t + &R \cdot \Pr{}{\world{A} \mid H_{t-1}, s_t = B} \leq R \cdot \Pr{}{\world{B} \mid H_{t-1}, s_t = B}
\end{align*}

Due to the fact that the subsidy induces signal-revealing, in order to compute the posterior, we can view the history as a series of revealed signals and consider the likelihood of seeing that stream in each of the worlds. $\hfill \square$

\begin{theorem}
\label{thm:subsidy_budget}
In expectation, the subsidy is provided over fewer than $\frac{4}{2p-1}$ rounds and totals no more than $R \frac{4}{2p-1}$.
\end{theorem}
\textbf{Proof Sketch} To show this, we appeal to the random walk formulation discussed above. If the subsidy value is too small, the agent will decide based on the history, if the subsidy value is too large, the agent will choose the subsidized action, and if the subsidy is ``just right,'' the agent will act according to their signal. The first two cases correspond to not taking any steps on the random walk, and the last case corresponds to taking a step in the direction of the signal on the random walk. Thus, we can analyze the number of total steps that need to be taken in order to net sufficiently many steps to the correct directions. We apply Wald's equation to do this. 
A detailed computation can be found in  Appendix ~\ref{appendix:proof-subsidylen}.  $\hfill \square$

\section{Extension to multiple groups} \label{sec:k-group}
In this section, we remove the strong assumptions of the previous section, namely that (1) the government knows the correct action and (2) everyone has the same correct action.
    Perhaps a more natural assumption would be that there are $k$ groups, where each group has its own correct action (between the two, A and B). 
    In this section, we show how to develop a subsidy scheme in which \textit{all groups} end up in their respective correct cascade, even if the government is not privy to the respective correct actions.
    
    In this setting, an agent in a group only sees other agents from their group. For example, consider a small town where the decisions of a student are primarily, if not completely, affected by those in the same town. 
    Suppose the government knows the strength of signals provided to agents, but it does not know the agent's town nor which action they take. At each time $t$, the government may give a subsidy $r_t$ with the goal that, eventually, all groups will reach a stable up cascade on the correct action for that group.

    In this section, we show how the government can construct a distribution $\mathcal{D}$ without knowledge of anything more than described above, such that if at each time $r_t \sim \mathcal{D}\,,$ after sufficiently many steps, with high probability, all groups will have converged to their respective correct cascades.
    Importantly, the key idea here is the same as in Algorithm \ref{alg:subsidy}: one kind of subsidy that will help each group make decisions that are right for them in the long run is one that encourages an agent to reveal their signal.
    Further, recall that we derived exactly a signal-revealing subsidy in the previous section. Thus, we know that for each value of $|A|-|B|\,,$ i.e., for each location along the random walk, there exists a value of the subsidy that incentivizes the agent to reveal their signal. Provided this subsidy value is chosen with at least some fixed minimum probability, we can use a similar random walk analysis to before, only this time slowed down by that probability factor.
    In this section, we flesh this argument out. 
    Proofs can be found in the Appendix in the full version. 

\subsection{Formally Defining The Setting}
We define the setting formally below.

\begin{definition} \label{defn:multigroup-game}
    Suppose that in the world, there are two potential actions $A$ and $B\,.$ Each agent in the world has an index in $[k]$ associated with them which we call their group. For each group in $[k]\,$, one of the actions is the ``right'' one and the other is the ``wrong'' one. Notably, which action is correct differs between groups. Formally, we may assume the correct action for a group is $A$ with probability 1/2 and $B$ with probability 1/2. \par
    At time $t\,,$ the ``universe'' draws an index $j \sim \mathcal{G}\,,$ where $\mathcal{G}$ is a distribution supported on $[k]\,,$ with $j$ signifying the group index, and the minimum value of the probability mass function is $g_\text{min}$. It is a new agent's turn to make a decision, and they are a member of group $j\,.$ This agent sees the subsidy history for all agents and need rationally only consider the action history of those in group $j$ who have gone before them. They receive a private signal $s_t\,,$ which is the correct action for the group to which they belong with probability $p_j > 1/2$ and incorrect with probability $1-p_j\,.$ They consider their group history and private signal to update their posterior belief as shown in Equation~\ref{eq:post}, and then incorporate the present subsidy to make their final decision.
\end{definition}

Next, we define the subsidy scheme followed by the government. In fact, it suffices to define a uniform distribution over the possible values the subsidy would need to take, i.e., all values that would encourage signal-revealing in an agent before that agent's group hits a cascade. 

\begin{definition} \label{defn:government-subsidy}
    Let $v_{x, p}$ be the size of subsidy that causes signal revelation by the current agent when there are $x$ more $A$ actions than $B$ ones.
    Define distribution $\mathcal{D}$ as supported on $\mathcal{V}$, all possible values of $v_{x, p}\,,$ and having probability mass associated with any $v_{x, p}$ as $1/|\mathcal{V}|\,.$ 
    The government draws the subsidy they provide from this distribution $\mathcal{D}\,.$ 
\end{definition}

\subsection{Main Result}

In this section, we show that the random process described in Definition~\ref{defn:government-subsidy} applied to the setting described in Definition~\ref{defn:multigroup-game} for a reasonable amount of time shifts groups into correct cascades with high probability.

\begin{theorem} \label{thm:kgroup}
    Suppose the government provides the subsidy detailed in Definition~\ref{defn:government-subsidy} for the game in Definition~\ref{defn:multigroup-game}. That is, at each time step $t > 0 \,,$ the government (without knowledge of the group of the current agent or the history) draws a subsidy at random $r_t \sim \mathcal{D}\,,$
    where the probability of any element in the support of $\mathcal{D}$ is at least $p_\text{min}\,.$ 
    Then, for all $\delta > 0,$ after $\frac{2\, k}{g_\text{min}} \, \left( \frac{ 2 \frac{\log(3k/\delta - 1)}{\log(p/(1-p))}}{\pmin \, (2p - 1)} + \frac{2 \log(3k/\delta)}{\pmin^2 (2p-1)^2}   + \log(3k/\delta)\right)$ steps, with probability at least $1-\delta\,,$ 
    all $k$ groups will end up in what is for them an up-cascade. 
\end{theorem}

\textbf{Proof Sketch} At a high level, the proof proceeds as follows: (1) We fix a group and show that the subsidy behaves, as before, as a random walk that group takes on the number line. (2) We show how long it takes to achieve with high probability a sufficient condition for the walk to finish in a cascade on the correct action. (3) We union bound over the failure probability and appropriately scale the time required to achieve the stated result. 

\textbf{(1)} Let us first describe a random walk $\mathcal{R}_\text{gvt}$ that models the setting in Definition~\ref{defn:multigroup-game} with the government subsidy described in Definition~\ref{defn:government-subsidy}. See the appendix for a proof. Next, we define a related but simpler random walk, $\mathcal{R}$. In both cases, the reverse walk describes the same for a group for whom the correct direction is the left.

\begin{lemma} \label{lemma:random-walk-gvt}
    The following random walk, which we shall call $\mathcal{R}_\text{gvt}\,,$ describes the walk taken by a group for whom the correct action is to the right when the government provides the subsidy described in Definition~\ref{defn:government-subsidy}:
        $\Pr{i \text{ to } i-1} = \alpha_i \cdot (1-p) \, ; \,
        \Pr{i \text{ to } i} = 1-\alpha_i \, ; \,
        \Pr{i \text{ to } i+1} = \alpha_i\cdot p\,,$
    where $\alpha_i \coloneqq \mathbb{P}_\text{gvt choice}\left[\text{signal is revealed in this state}\right]\,.$ 
\end{lemma}

\begin{definition} \label{defn:rw-aug}
    Let us call the random walk with the following transition probabilities $\mathcal{R}$:
        $\Pr{i \text{ to } i-1} = p_\text{min} \cdot (1-p) \, ; \,
        \Pr{i \text{ to } i} = 1-p_\text{min} \, ; \,
        \Pr{i \text{ to } i+1} = p_\text{min}\cdot p\,.$ 
\end{definition}

We can show that we can analyze $\mathcal{R}$ instead of $\mathcal{R}_\text{gvt}\,$ (details in full version).

\textbf{(2)} Next, we consider three modes of failure: first, if there are not sufficiently many signals aligned to the correct direction, the majority vote will not align with the correct action for the group; second, since the government only encourages signal-revealing but does not behave differently depending on which action is correct for a group, we may accidentally hit a bad cascade before hitting a good one just due to a bad ordering in the sequence of signals; third, we may not see enough people from this group to take sufficiently many steps on the random walk. We reason formally about each of these, computing first a number of steps after which with probability $1-\delta/(3k)$ the walk is sufficiently far to one side and then arguing that the side to which the walk is shifted is the \textit{correct} one for that group with probability $1-\delta/(3k)$.
Finally, we apply the Hoeffding bound with failure probability $\delta/(3k)$ to ensure we see sufficiently many people. With that, the analysis is complete for a single fixed group. 

\textbf{(3)} Finally, we union bound over the failure probability so that the result holds and report after how long of the government providing such a subsidy, with high probability {\em all} groups stabilize to optimism cascades. Detailed proofs for this theorem and its constituent lemmas are in the appendix of the full version. $\hfill \square$

\begin{rmk}
    For sake of generality, we present the result in Theorem~\ref{thm:kgroup} in terms of $\pmin\,,$ and $g_\text{min}\,.$ However, let us plug these in and discuss the scaling for intuition. 
    First, the support of the distribution $\mathcal{D}$ has size at most $2\Lrw$, so $p_\text{min} \le 1/(2L) \le 1/(2 \Lrwf) \,.$ Plugging this in, upper bounding, and ignoring constants, we get that the number of total required steps scales like $\frac{k\,L^2}{g_\text{min}} \, \cdot \, \frac{\log(3k/\delta)}{\left( p - \frac 12 \right)^2}\,,$ where $L = \frac{\log(3k/\delta - 1)}{\log(p/(1-p))}\,.$ We can see that the number of steps scales like $1/(p-1/2)^2\,,$ implying that the closer the signal strength to $1/2\,,$ the longer it takes to drift far enough in the walk. This is standard for problems where we must distinguish whether a ``coin flip'' has bias $1/2+\epsilon$ or $1/2-\epsilon\,.$ Next, we see the standard dependence on the failure probability, $\log(1/\delta)\,,$ and union bound, $\log 3k\,$. Finally, we see the inverse dependence on $g_\text{min}\,,$ meaning that the lower the minimum probability of seeing a group, the longer this subsidy needs to be in place. 
    Now, if the probability of an agent belonging to a group is uniform across the $k$ groups, then $g_\text{min} = 1/k\,,$ and so the scaling is like $ L^2 \, k^2 \, \cdot \, \frac{\log(3k/\delta)}{\left( p - \frac 12 \right)^2}\,.$ From this, we can see that the dominant dependence of this bound on the number of groups is through the prevalence of the lowest-prevalence group and making sure each group takes sufficiently many steps. 
\end{rmk}
 
\section{Experiments}

Finally, we conducted simulations to assess the effectiveness of financial supplements in overcoming pessimism traps in our theoretical model. We study the first setting, with a single group and correct action known to government. The primary objectives of our experiments were to evaluate the impact of financial supplements on the proportion of correct cascades, analyze the magnitude of the subsidies, and understand the scalability of interventions across population sizes.

\subsection{Data Generation and Procedure}

The data for our experiments were generated through simulations that model the sequential decision-making process of agents. Each agent must choose between two actions, $A$ and $B$, where $A$ is uniformly the better option, though this is unknown to the agents \textit{a priori}. The agents receive private signals indicating the correctness of their choice, with a probability $p$ of being correct. 
Each simulation was conducted as follows:
    We initialized a population of $N \in \{10, 100, 1000\}$ agents to explore the effects of population size on eventual cascade behavior.
    Each agent received a private signal with strength $p \in [0.51, 0.99]$.
    For each pair of values for $N$ and $p$, we repeat the experiment 100 times and average the results. For the proportion of correct cascades, we repeat each such experiment 10 times and report one standard deviation/$\sqrt{10}$ as the error bars. For the subsidy size, we report the standard deviation over the 100 trials / $\sqrt{100}$ as the error bars. 
    The agents sequentially made their decisions after observing the actions of all preceding agents.
    Financial supplements as derived above were introduced to influence the agents' decisions.

\subsection{Results}

\subsubsection{Impact of Financial Supplement on Correct Cascades}

\begin{figure}[ht!]
    \centering
    \includegraphics[width=0.8\textwidth]{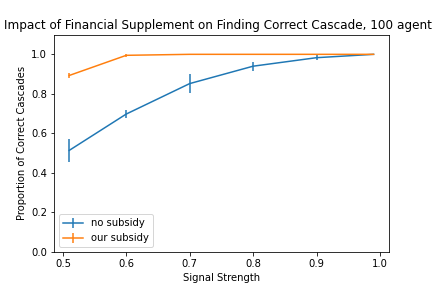}
    \caption{Impact of Financial Supplement on Finding Correct Cascade with Supplement, 100 agents.}
    \label{fig:using_supplement}
\end{figure}

\begin{figure}[ht!]
    \centering
    \includegraphics[width=0.8\textwidth]{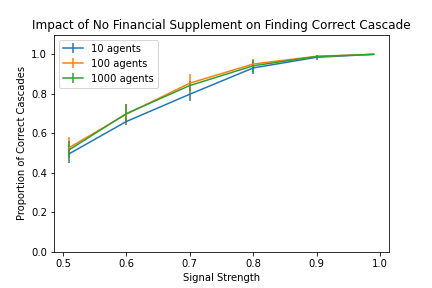}
    \caption{Probability of Finding Correct Cascade without Supplement}
    \label{fig:without_supplement}
\end{figure}

Figure \ref{fig:using_supplement} shows the proportion of correct cascades when a financial supplement is provided (orange line) to 100 agents compared to when it is not (blue line). The results indicate a significant improvement in the proportion of correct cascades, especially at lower signal strengths. For example, with a signal strength of 0.6, the proportion of correct cascades increases markedly (including across different population sizes (10, 100, and 1000 agents), see appendix for plots) when the supplement is used. This demonstrates the benefit of the proposed financial incentive in decision-making.

In contrast, Figure \ref{fig:without_supplement} shows the proportion of correct cascades without the supplement for varying numbers of agents. The performance is notably low, particularly for weaker signals. This highlights the promise of financial interventions in overcoming pessimism traps and steering agents toward optimal decisions. Also observe that the number of agents in the sequence has very little effect -- once a cascade forms, there is no new information to switch out of it even when there are new agents.

\subsubsection{Average Subsidy Progression}

To understand the dynamics of the financial supplement, we analyzed the progression of subsidies over time for different signal strengths. We present results for 100-agent cascades here, and further experiments can be found in the appendix in the full version.

\begin{figure}[ht!]
    \centering
    \includegraphics[width=0.80\textwidth]{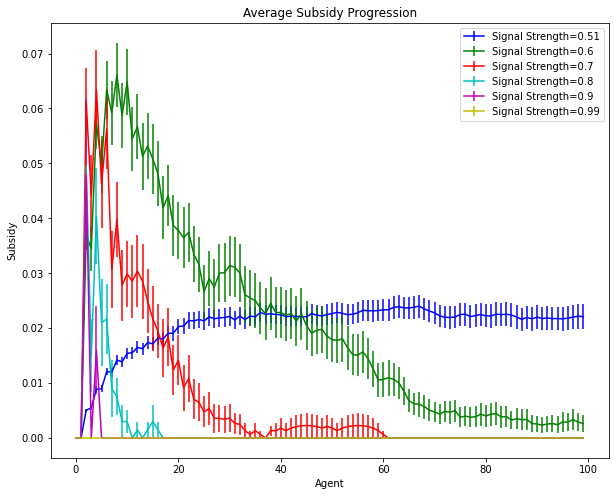}
    \caption{Average Subsidy Progression for 100 Agents \vspace{-3mm}}
    \label{fig:subsidy_100_agents}
\end{figure}

Figure \ref{fig:subsidy_100_agents} presents the average subsidy progression for a population of 100 agents. The required subsidy stabilizes after a few agents have made their decisions, indicating that once the initial agents are guided correctly, the need for subsequent subsidies diminishes, a benefit for allocation of resources. The line for the weakest signal levels off -- based on our theoretical results, we know that it takes a long time to stabilize when the signal margin is narrow.

Interestingly, the required supplement was smallest among the first few agents in the setting with the weakest private signal. 
We observe a spike in the supplement needed at the beginning, yet the necessary supplement typically returns to zero quite quickly. All three plots, however, show that in the case where $p=0.51$, the subsidy stays consistent across the population. This is likely because the signal is weak enough that it takes too long on average for the best action to be detected compared to the number of agents. It is also interesting to observe the average starting times for the subsidies, which align with the beginning of the pessimism trap and almost always begin within the first 10 rounds.

\subsection{Discussion and Implications}

The experiments validate our theoretical predictions regarding the impact of financial supplements on decision-making under uncertainty. The simulations provide several key insights. First, the introduction of financial supplements significantly improves the proportion of correct cascades, particularly in scenarios with lower signal strength and smaller populations. This underscores the importance of targeted interventions in guiding agents towards optimal decisions. Second, the progression of subsidies indicates that early interventions are critical. Once the initial agents are influenced correctly, the necessity for continued subsidies diminishes, suggesting an efficient use of resources. Lastly, the results suggest that the scalability of financial interventions is feasible. Larger populations benefit from reduced per-capita subsidies, making such policies more practical for broader applications.

\section{Discussion and Conclusion}

In this work, we developed a theoretical model for a social epistemological phenomenon of pessimism traps. We reproduce the phenomenon in our theoretical model, and the development of these traps in our model are due to limitations in the evidence perceived by the agent, mirroring the Morton's characterization in \cite{morton2022resisting}. From there, we show how a centralized agency can intervene to shift a community out of pessimism traps if they know what action is good for that community. We then generalize our model to a setting where there are $k$ parallel communities, each with its own ``good'' action between the two. Here, we show that the centralized agency can intervene effectively {\em without} information about group membership or correct action for an agent. \par

Our results hold in this stylized model that reflects important parts of Morton's description of pessimism traps. Yet, we cannot say that our model accurately reflects reality for reasons discussed earlier in this part. Thus, it is hard to know what implications our proposed interventions have in the real-world in the absence of practical experiments. However, our results are still interesting for the insight they provide about the setting. Our results show that indeed availability of trustworthy evidence can shift communities out of pessimism traps. Moreover, our proposed intervention utilizes randomness strategically and therefore is able to shift multiple communities into optimism while being fair and non-paternalistic. These properties of the intervention we develop make this style of intervention interesting to consider practically. On a technical level, it would be interesting to study such interventions in a broader range of social learning settings.
\chapter{A Theoretical Model for Grit} \label{chap:grit}

\section{Introduction}

Scholars across various fields have long been interested in understanding how humans make decisions between immediate and long-term rewards in the face of uncertainty. Costs associated with a given action can also greatly influence how agents make these decisions, and so we wish to understand how to support exploratory and ambitious behavior, particularly in groups who have historically faced lack of access to such opportunities. One important factor that has received much attention in recent years is {\em grit}, with researchers across various fields studying the role resilience and optimism play not only in an individual's success but also in lifting disadvantaged communities out of bad circumstances. \par

While it is difficult to give a single, unifying definition of grit, philosophers Jennifer Morton and Sarah Paul provide a thesis that serves as our guiding qualitative description. They highlight that grit is rational (distinguishing it from delusional optimism) and that it is an outcome of {\em beliefs} an agent holds about their circumstances. In more detail:

\begin{quote}
    ``Grit is not simply the ability to withstand the pain of effort and setbacks, or to resist the siren song of easier rewards; it is a trait or capacity that consists partly in a kind of epistemic resilience. \par

    This is a descriptive rather than a normative claim, and it has not gone unnoticed by psychologists who study perseverance. Angela Duckworth emphasizes the relevance of hope in underwriting the capacity for grit, where hope is defined as the expectation that one’s efforts will pay off. And Martin Seligman touts the importance of optimism, which involves a distinctive style of explaining to oneself why good and bad events happen.'' \cite{morton2019grit}
\end{quote}

Individuals with grit are able to persevere despite the lack of immediate reward: for instance, they are more willing to practice a math or sport skill until it is mastered, or they are willing to tolerate a significant degree of financial austerity while starting a company until revenue is accrued.

For a long time, prevailing guidance in educational settings focused heavily on personal characteristics such as grit, discipline, and resilience as the way to succeed in ambitious ends \cite{Gibbon_seligman_2020}, often even at the protest of the scholars whose work was used to justify this perspective \cite{kamenetz_key_2015}. Subsequent longitudinal work paints a more ambivalent picture, studying outcomes for students from disadvantaged backgrounds who follow those lessons in attempting to succeed at ambitious long-term ends such as college; these works find that simply pushing grit can often have negative effects \cite{morton2019grit, wooten_precarious_2022}. 
These findings highlight an empirical tension, that grit is useful in pursuing long-term ends that require investment but can make those who cannot cushion early losses susceptible to the precarity associated with such ends. The puzzle posed by these findings posit that grit is not {\em uniformly} positive but perhaps instead {\em conditionally} useful. What, then, are the situations in which grit is productive? Harmful?

In this work, we attempt to bridge this analytic gap by isolating where additional grit helps and where it hurts via a simple theoretical quantitative model. We also investigate how outcomes are influenced by having financial supplements. Our goal is to study decision-making dynamics in a controlled, quantitatively-defined setting. By characterizing the landscape of grit and its interplay with external support, we unify the diverse empirical observations in this area and provide a model in which further quantitative analysis can take place. Our work (1) provides further understanding of the relationship between grit and financial support in succeeding when pursuing ambitious outcomes and while doing so, (2) introduces a simple two-armed bandit theoretical model that shows promise as a formalization of a decision-making problem that juxtaposes stable reward against ambition. Our analysis isolates three crucial parameters -- level of grit, amount of external support, willingness to tolerate discomfort -- and develops an understanding of outcomes as a function of these, reproducing with proof several patterns observed by qualitative scholars.

\subsection{Our Approach To Studying Grit}

Morton and Paul emphasize that grit leads agents to {\em rationally} stick with an option that others are not willing to stick with. We consider two formal models of rationality in decision-making. First, we study agents who maximize their competitive ratio. Measuring the ratio between achieved outcome and best possible outcome in hindsight, the competitive ratio being maximized implies that the agent has minimal multiplicative ``regret'' about the past. Thus, this model of rationality is a backward-looking model. Secondly, we study agents who take a Bayesian perspective to uncertainty quantification. In this view, an agent has an explicit prior probability distribution over the possible outcomes and updates their posterior each time they receive new information. In constructing a prior, the agent explicitly quantifies what they think the future holds, making this a forward-looking model of rationality. In this chapter, we will compare and contrast the aspects of grit we are able to study in each of these formal models of rationality and see how the conclusions we draw from each on similar formal models relate.

In order to understand the impact of grit, we study two things -- first, we consider the effect of grit on the ``policy'' or ``strategy'' that the agent follows, i.e., the actions they take; second, we study what the impact of that ``policy'' is on the reward they achieve. This modular breakdown will become especially useful when we consider the impact of a trust fund on the actions of an agent, allowing us to disentangle the effect of grit as a trait and other interventions that lead agents to behave similarly to those who are gritty.

\subsection{Our Contributions}

In this work, we propose studying grit in the {\em improving multi-armed bandits} framework. We develop an instance that allows us to satisfactorily investigate the impact of grit as a characteristic the agent has, and we show that the instance we propose is the simplest instance in which the strategy is non-trivial. In order to understand strategies resulting from grit, which as discussed above is rational, in this model, we must formalize what we mean by ``rational,'' and we do so by appealing to two notions of rationality well-studied in computer science. In particular, we first consider the competitive ratio, a standard notion in the analysis of online algorithms, which is optimized when a strategy minimizes multiplicative regret in hindsight; we also study a Bayesian notion of rationality in which an agent has a prior that helps quantify their uncertainty about the future, which they update to a posterior based on evidence from their environment. Between these two notions, we study how gritty behavior is reflected in both forward- and backward- looking formalizations of rationality. \par

With the model and notions of rationality in hand, we study how grit affects an agent's strategy and how that in turn affects their outcomes. When grit is related to the optimistic outlook of an agent, we can quantitatively derive cases in which grit helps and hurts the agent, showing that an excess of grit harms the agent by causing them to net less reward than their less-gritty counterparts. When grit is reflected in how willing an agent is to tolerate discomfort, we conclude that though agents who require comfort can minimize their multiplicative regret in hindsight pretty easily, agents who can tolerate more discomfort can explore for longer. This section culminates with studying how financial support changes agents' behavior, essentially showing that financial support allows agents to expand their exploration horizon while allowing them to still receive comparable reward to their less gritty counterparts. 
Finally, we investigate similar questions in the Bayesian setting, and we find that grit when framed as uncertainty tolerance has a similar effect on an agent's policy, namely that more grit causes an agent to strive for longer.

The rest of the chapter is structured as follows. First, we survey related work from several fields, including philosophy, sociology, and computer science. Then, we formally introduce the multi-armed bandits instance we study. In Section~\ref{sec:cr}, we study the competitive ratio-based notion of rationality, following which in Section~\ref{sec:trust-fund} we introduce models for financial support and study the effects of the financial safety net on behavior and reward. Finally, in the Appendix,
we investigate the Bayesian notion of rationality, studying uncertainty tolerance in this setting.

\subsection{Related Work}

First, we discuss work from the social sciences that observes grit in people and analyzes its role and impact in society.
One of the most popular studies of grit is psychologist Angela Duckworth's book \cite{duckworth2016grit}.
Relatedly, psychologist Martin Seligman has a large body of work encouraging positive attitudes as a path to success and good outcomes \cite{seligman_authentic_2002, seligman_learned_2006}.
More recently, sociologist Tom Wooten studied the impact of the ``no excuses'' educational approach, which draws on the above ideas, in his dissertation \cite{wooten_precarious_2022}, particularly exploring the mechanisms by which these educational systems perpetuate poverty. Abstracting findings from several such observational studies, Morton and Paul develop a philosophical theory of grit \cite{morton2019grit}. We build heavily on the abstractions derived in this work. \par

Next, we survey computer science literature that we draw on in order to quantitatively study the question of decision-making with grit. Our formalizations of rationality are inspired by objectives that are commonly optimized in the computer science literature, including the competitive ratio studied in online learning \cite{borodin_online_2005} and the Bayesian approach to uncertainty quantification  \cite{bernardosmith_bayesian_2000}. The framework we use to represent the decision problem is an instance of the multi-armed bandits (MAB) problem, a well-studied framework for making decisions when the payoff is unknown \cite{slivkins_introduction_2022}. In particular, we suppose the bandit arms have structured reward, and the structure of interest is improving, first formalized by \cite{heidari_tight_nodate} and studied by \cite{patil_mitigating_2023, blum_nearly-tight_2024}.

\subsubsection{Key Features of Grit According to Morton and Paul}

Since we base our development of a theory of grit on the account of \cite{morton2019grit}, it will be helpful to summarize the key points from their work. In their work, Morton and Paul describe gritty behavior as displaying a form of ``epistemic resilience.'' An important part of this is how an agent redefines their goal in the presence of encouraging or discouraging evidence. They also reason that grit is {\em rational}, and therefore agents displaying grit cannot {\em ignore} evidence but rather should be sensitive to failure. Citing works from psychologists Angela Duckworth and Martin Seligman, Morton and Paul further highlight the importance of hope and optimism. They culminate by providing a description of an ``Evidential Threshold'' that captures the decision-making of a gritty agent. The Evidential Threshold as conceptualized by them asks how compelling evidence must be to change the actions of an agent; for a gritty agent, the Evidential Threshold is higher than that of an ``impartial observer.'' Since this threshold hinges on how compelling the agent finds the evidence, Morton and Paul argue that Permissivism applies, and different agents can witness the same evidence but come to different conclusions about what implications that evidence should have on their actions. 
Throughout the chapter, we will connect back to these various facets described by Morton and Paul, including by providing a quantitative analog of the Evidential Threshold.

\section{Formal Setting: Improving MAB}

We propose studying grit in the improving multi-armed bandits framework \cite{heidari_tight_nodate}. Overall, we follow the framework in Section~\ref{sec:prelims}. We detail the specific instance we study now.

In our models for grit, it is natural to think of options having different payoffs, some of which are static over time and some of which take a while to start paying off but then pay off well once they do. Thus, we propose a two-armed bandit instance in which to study gritty behavior. All of our models will include a stable arm, $f_1(t) = 1 \, \forall \, t$ that represents an option that starts paying off immediately and consistently rewards the agent the same amount (i.e., little scope for growth). The other arm is one in which there is no reward at first (or in certain cases where specified, there is actually a {\em cost} to striving), after which the arm starts providing non-negative reward.
We refer to this arm as the ``striving'' arm, and it provides a reward of 0 units for the first $\theta$ time steps that it is played ($\theta$ being unknown to the agent\footnote{for instance, during a PhD or while starting a business}), following which it increases linearly at a slope of $\alpha\,$ (agents will have beliefs about $\alpha$).
Formally, the two bandit arms are:

\begin{equation} \label{eqn:main-instance}
f_1(t_1) = 1 \, \forall \, t \qquad f_2(t_2) = \begin{cases}
    0 & t_2 < \theta \\
    \alpha(t_2-\theta) & t_2 \ge \theta 
\end{cases}\,,
\end{equation}
where $t$ represents the amount of time for which that arm has been played.

In most of the chapter, we use this model, though in some sections we set $\alpha = 1.$ Later on, in Section~\ref{subsec:discomfort}, we also define a notion of ``comfort'' which places a restriction on how often $f_2$ can be played, requiring that $f_1$ be played frequently enough to build up a buffer.

This choice of model is natural: the improving multi-armed bandits problem is well-studied, and there is a clear understanding of what we could hope to achieve in the general case \cite{patil_mitigating_2023, blum_nearly-tight_2024}. The structure in the reward function allows us to capture the fact that {\em investing} time into an option may change its payoff. Finally, the instance described is abstract and flexible, allowing us to model a wide range of real-world settings. On the other hand, a limitation is that this instance only allows for studying an agent's decision between two options. The stable option is arguably over-simplified, since stable options can also lead to growth in the real world. Overall, however, we believe this is a good starting point for formally modelling the decision problem of interest. Further, within the improving two-armed bandit setting, this is the simplest model in which we see non-trivial behavior: if the second arm's payoff were flat instead of linear, the trivial strategy of playing $f_1$ all along would suffice for optimizing the competitive ratio. This is discussed in detail in the Appendix.

\section{Rationality in Terms of Competitive Ratio} \label{sec:cr}

\subsection{Rationality in This Model}

The first formal model for rationality we study is one in which an agent minimizes regret in hindsight by optimizing the competitive ratio. Competitive ratio measures the relationship between the achieved reward and the optimal reward. This is a commonly-studied notion in theoretical computer science, and particularly in online algorithms \cite{borodin_online_2005}. Achieving a competitive ratio of 1 ensures that the agent did as well as they could hope. Formally, we define the competitive ratio as follows:

\begin{definition}
    An algorithm achieves {\em competitive ratio} $g$ if the ratio of its reward $ALG$ to the optimal achievable reward $OPT$ is at least $g\,,$ i.e., if $ALG/OPT \ge g\,.$
\end{definition}

In certain situations where we are interested in the accumulated reward up to a certain time point $t$ of the algorithm, we will refer to it as $ALG_t\,.$

We will consider agents who have beliefs about $\alpha\,,$ the potential for payoff, and who optimize their competitive ratio over an unknown $\theta\,,$ the time after which the payoff begins. Suppose an agent has a deterministic algorithm to choose how to play arms $f_1$ and $f_2$ as defined above. Each time the agent plays $f_1$ and then switches back to arm $f_2\,,$ they could instead have played $f_2$ followed by $f_1$ while gaining the same reward but possibly witnessing the increase sooner. Thus, there is no benefit to interweaving steps of the arms, and any deterministic strategy for playing this instance can be boiled down to the point at which it switches from $f_2$ to $f_1\,.$ We formalize this in the following lemma.

\begin{lemma} \label{lemma:switchpt}
    Any strategy that interweaves plays of $f_1$ and $f_2$ can be converted into a strategy that plays only $f_2$ followed by only $f_1$ that achieves at least as much reward.
\end{lemma}
\begin{proof}
    Suppose the interweaving strategy plays a total of $t_1$ steps on $f_1$ and $s$ steps on $f_2\,.$ Consider two cases: in the first case, $s < \theta\,.$ Then, the total reward of the policy is $t_1\,,$ and playing $s$ steps of 0-reward-accruing $f_2$ followed by $t_1$ steps of $f_1$ achieves this reward, the same as any interweaving version. In the second case, $s \ge \theta\,.$ Now, by the previous argument, the policy that plays $\theta$ steps of $f_2$ followed by $t_1$ steps of $f_1$ still receives $t_1$ reward. However, after $\theta$ steps on $f_2\,,$ the agent witnesses the increase and therefore has no incentive to switch to the stable arm. The reward of playing the remaining $s-\theta + t_1$ steps on the striving arm is $\frac 12 (s-\theta + t_1)^2\,,$ which is greater than $t_1.$ Thus, we have shown that for each interweaved strategy, there is a non-interweaved one that accrues at least as much reward.
\end{proof}

As a result of this lemma, it suffices to study strategies that play $f_2$ for a while and then permanently switch to $f_1.$ Thus, we will extensively study this switch point, and it will be an interesting quantity to study as a proxy for gritty and non-gritty strategies.  

Morton and Paul discuss an ``Evidential Threshold,'' writing ``In a given context, how much evidence is required -- that is, how compelling must the evidence be --before the thinker comes to a conclusion about what to believe or revises her current beliefs?'' \cite{morton2019grit}. In our proposed framework of analysis, the switch point reflects this evidential threshold. The agent's strategy can be summarized as ``if I don't see evidence that striving is going to pay off until time $s$, I will give up.'' The threshold $s$ is different for different agents, and so the policy the agent follows exactly corresponds to their evidential threshold as described by Morton and Paul.

For a fixed switch point $s\,,$ there are two ``worst'' cases -- the first is when the arms are such that playing the stable arm the whole time would provide the optimal reward, and so the longer the agent spends exploring before switching, the worse the competitive ratio gets. On the other hand, if the striving arm pays off right after the agent switches, then staying on the arm just a little longer would have paid off, so this competitive ratio is increasing with $s\,.$ In order for our strategy to minimize overall regret, we pick $s$ such that the ratio is the same regardless of which extreme case we are in, i.e., we solve for $s$ when the two extreme cases are equal.

\newcommand{\alphat}{\tilde{\alpha}}

\subsection{Modelling Grit: Optimism} \label{subsec:optimism}

Now, let us discuss the relationship between the grittiness of an agent and their approach to the multi-armed bandit problem above. A gritty agent is optimistic about the potential for reward of the risky action, or they would not persevere in taking it.
Accordingly, for this setting, we consider an agent with higher guess for the slope of the increasing portion of arm $f_2$ to be more gritty. Based on this, we can investigate consequences (in terms of both strategy and reward) of demonstrating grit and offer mechanistic insight as to why these consequences exist. The guess for the slope
affects how long the agent is willing to play the striving arm (details in Appendix~\ref{appendix:proof-alphat-switch}). Formally:

\begin{definition}
    In the ``grit-as-optimism'' setting, an agent is $\alphat$-gritty if they guess that the slope of the increasing portion of the striving arm is $\alphat\,.$
\end{definition}


\begin{lemma} \label{lemma:alphat-switch}
    Suppose an agent guesses a value for $\alpha$ that we call $\alphat\,,$ i.e., is $\alphat-$gritty. Assume their goal is to maximize the competitive ratio. Then, they play $f_2$ for $T- \sqrt{\frac{2T}{\alphat}}$ steps, following which they switch to $f_1$ permanently.
\end{lemma}

\subsubsection{Results}

Now, let us consider A, an $\alphat_A$-gritty agent. A is not particularly gritty, so $\alphat_A$ is small. On the other hand, B is somehow privy to perfect information, so $\alphat_B = \alpha\,.$ Finally, consider C, an $\alphat_C$-gritty agent. C is very gritty, and so $\alphat_A < \alphat_B = \alpha < \alphat_C\,.$ (We are simply instantiating the agents in this way to study ``high'' and ``low'' grit as they compare to perfect information.) 
In order to understand the impact the grit-induced strategy has on the reward the agent accrues, we are interested in understanding (1) what level of grit witnesses the striving arm paying off; (2) what level of grit results in good stable reward.

\paragraph*{Observation 1: Duration of Attempt} 
Applying Lemma~\ref{lemma:alphat-switch}, we have that agent A switches at time $s_A = T- \sqrt{\frac{2T}{\alphat_A}}\,,$ agent B at $s_B = T- \sqrt{\frac{2T}{\alpha}}\,,$ and agent C at $s_C = T- \sqrt{\frac{2T}{\alphat_C}}\,.$ Note that $s_C > s_B > s_A\,.$ Our first conclusion, therefore, is that the duration for which an agent explores is longer for a grittier person.

\paragraph*{Observation 2: When Does Increased Grit Benefit the Agent?} Let us now study under what conditions each agent comes out on top. The reward achieved by any agent depends on the relationship between $\theta\,,$ the threshold beyond which the striving arm starts increasing, and $s\,,$ the agent's switch point as stated in the below proposition.

\newcommand{\optreward}{\frac{\alpha}{2} (T-\theta)^2}
\newcommand{\smreward}[1]{\sqrt{\frac{2T}{\alphat_{#1}}}}

\begin{proposition}
    If $\theta \le s\,,$ then an agent switching at $s$ receives $\optreward$ reward, but if $\theta > s\,,$ then the agent receives $T-s$ reward.
\end{proposition}

\begin{proof}
    If $\theta \le s\,,$ then the arm starts paying off while the agent is still playing it. This means that the agent accrues reward starting at time $\theta$ up until time $T$ as the function increases linearly. Hence, the reward is $\frac \alpha2 (T-\theta)^2\,.$ On the other hand, if $\theta > s\,,$ then the agent gains no reward from the striving arm. They gain reward from the stable arm from time $t=s$ to time $t=T\,,$ which is $T-s$ units of reward.
\end{proof}

Let us consider how different levels of grit affect reward (Figure~\ref{fig:reward-visual} may help visualize the relative rewards.):

\begin{enumerate}
    \item \textbf{Case 1:} $\pmb{\theta < s_A\,.}$ Everyone's reward is the same in this case, since all agents receive $\frac \alpha 2 (T-\theta)^2$ reward.
    \item \textbf{Case 2:} $\pmb{s_A < \theta < s_B\,.}$ In this case, A has given up and switched to the stable arm. As a result, they receive $\smreward{A}$ reward. However, agents B and C stay on the striving arm long enough to witness $\theta\,,$ and so they receive $\optreward$ reward. We see that this is a situation where a lack of grit fares worse than being rather gritty.
    \item \textbf{Case 3:} $\pmb{s_B < \theta < s_C\,.}$ In this case, A and B have both given up and switched to the stable arm, but C valiantly perseveres. Here, A receives $\smreward{A}$ reward, B receives $\smreward{B}$ reward, and C receives $\optreward\,.$ C outshines even B, who had perfect information about the rate of reward increase. In this case, curiously, the least gritty person actually fares better than someone with perfect knowledge of the payoff. We can understand this as follows: there are two reasons why the received reward would be small -- one is location of $\theta$ and one is size of $\alpha$. In this case, someone that is pessimistic about the value of the reward magnitude due to pessimism about $\alpha$ ends up reaping the side benefit when the reward is indeed small, but it is because $\theta$ is large i.e., they are right about the reward being small but for the wrong reasons\footnote{This is an interesting consideration for further modeling -- in this particular setting, the model does not disentangle between these reasons for the reward to be small. More broadly, a competitive ratio-based model that studies reward in general rather than particular kinds of reward may not be able to disentangle this at all.}. 
    \item \textbf{Case 4:} $\pmb{s_C < \theta\,.}$ In this case, no one receives the reward from the striving arm. However, since C has stuck around for so long, they actually also receive less reward overall from the stable arm. This indicates that there is a failure mode when an agent is {\em too} optimistic. Further, the resulting strategy of sticking it out for a long time can fail when $\theta$ is quite large. This reflects what \cite{wooten_precarious_2022} calls the ``effort paradox,'' where students encouraged to be gritty often end up burnt out.
\end{enumerate}

\begin{figure}[h!]
    \centering
    \includegraphics[width=0.85\linewidth]{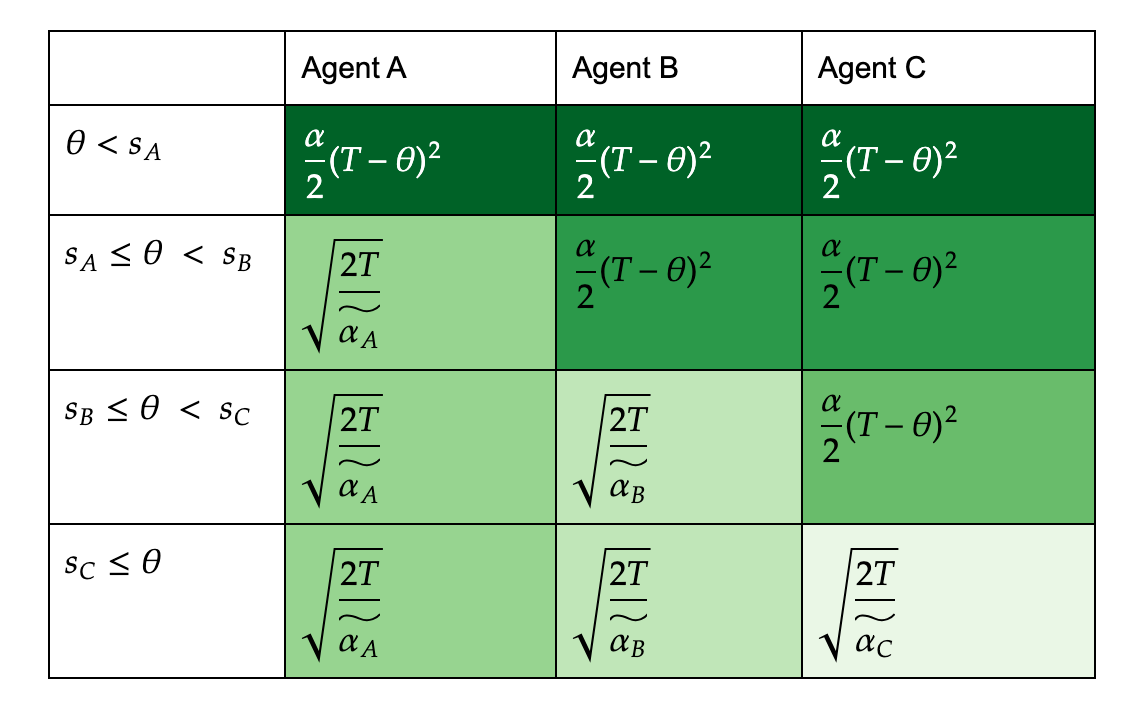}
    \caption{This table shows the reward the agents described in Section~\ref{subsec:optimism} receive if $\theta\,,$ the true threshold beyond which $f_2$ pays off, lies in different regions. The columns are increasing in grit from left to right. We can see that if $\theta$ is larger, more grit can actually result in less reward, since the agent switches back to the stable arm and accrues less reward there.}
    \label{fig:reward-visual}
\end{figure}

Observe that when the first agent switches, all agents have the same evidence about the striving arm. Likewise, when each agent switches, the remaining agents all have the same evidence, though different agents act differently, albeit all rationally, in response to this information. While it may at first seem disconcerting that people with access to the same evidence have {\em different} yet completely rational responses to it, this is is consistent with the philosophical thesis of Permissivism, which argues that ``some bodies of evidence permit more than one rational doxastic attitude toward a particular proposition'' \cite{lasonen-aarnio_permissivism_2023}. Thus, it is reasonable for different agents to have different beliefs about the underlying state of the world upon receiving a set of evidence (i.e., different agents have different beliefs about how long it is worth staying on an arm after receiving reward 0 for the first $s_A$ time). In fact, Morton argues that Permissivism exactly allows for grit to be conceived of as a factor in shifting people's behavior from the norm even when presented shared evidence. Indeed, in our model, we explicitly model this aspect of how beliefs + evidence $\rightarrow$ actions by having each agent hold different beliefs about the payoff slope $\alpha.$

This perspective also provides us a natural way in which to consider the optimal level of grit for this setting. In particular, if we can remove the uncertainty in the guess for $\alpha\,,$ then the only remaining uncertainty has to do with when the function will start improving. This suggests that agents who have guesses for $\alpha$ closer to the true $\alpha$ will fare better.

\subsection{Modelling Grit: Discomfort Tolerance} \label{subsec:discomfort}

In this setting, we study another relevant aspect of grit, namely discomfort tolerance. To do so, let us first introduce a cost to playing the striving arm.

\paragraph*{Cost to strive}

Let us consider a setting in which there is a cost to strive. The reward profile / bandit arms $f_1$ and $f_2$ look almost as before, with $\alpha = 1\,$:
$f_1(t) = 1 \, \forall \, t$ and $f_2(t) = \begin{cases}
    -1 & t < \theta \\
    t-\theta & t \ge \theta
\end{cases}\,.
$

The negative reward models the cost of striving, for instance financial debt or effort expenditure that can be offset by rewards from the stable arm to ensure the agent is not ``in the negative,'' as
the agent is not allowed have negative reward at any point in time. This corresponds to not being able to take out a loan. Thus, an agent who does not have any ``savings'' coming in must play $f_1$ before they ever play $f_2\,.$ We suppose that smallest piece of time an agent can split between the arms is 1 unit. Later, 
we consider what happens when the cost of striving is subsidized by a ``trust fund.''

\paragraph*{Comfort}
Now, further, let us suppose an agent always wants to have average reward at least $\gamma\,,$ that is, at time $t\,,$ the agent wants their accrued reward to be at least $\gamma \cdot t\,.$ They still aim to optimize the competitive ratio as before, except now they do so subject to the constraint that $\forall \, t\,, $ the reward accrued up to that point is at least $\gamma \, \cdot \, t\,.$

\begin{definition}
    We say an agent desires $\gamma$-comfort if at any time $t > 0\,,$ they require net reward at least $\gamma \cdot t\,.$
\end{definition}

The agent playing this game aims to solve:
$$
\max \qquad  \frac{ALG}{OPT} \quad \text{ s.t. } \quad \frac{ALG_t}{t} \ge \gamma\,.
$$

Since the agent can play an arm for fractional amounts of time, an agent who requires $\gamma$ comfort will play $\alpha_\gamma$ time on the stable arm and then $1-\alpha_\gamma$ time on the striving arm, alternating between the two as soon as possible, for $\alpha_\gamma \coloneqq (\gamma + 1)/2$\,, the time duration that guarantees the desired average reward:
\begin{align*}
    ALG_t &= \begin{cases}
        t & t \le \alpha_\gamma \\
        2\alpha_\gamma - t & \alpha_\gamma < t \le 1
    \end{cases} \\
    \qquad \Rightarrow  \qquad
    \frac{ALG_t}{t} &= \begin{cases}
        1 & t \le \alpha_\gamma \\
        \frac{2\alpha_\gamma}{t} - 1 & \alpha_\gamma < t \le 1
    \end{cases} \\
   \text{for } \alpha_\gamma < t \le 1\,,  \gamma &= 2 \alpha_\gamma - 1 \le \frac{2\alpha_\gamma}{t} - 1 \le 1\,.
\end{align*}

This is the most rational thing for them to do, since any additional steps on $f_1$ simply serve to offset future costs of $f_2$ but might prevent the agent from seeing the increase phase as quickly. We formalize this below and provide the proof in Appendix~\ref{appendix:proof-min-acc} (which proceeds via similar casework to the proof of Lemma~\ref{lemma:switchpt}).

\begin{definition}
    We call a strategy {\em minimally accumulating} for an agent who wants to be $\gamma$-comfortable if the average net reward at time $t\,,$ $ALG_t/t$ is exactly 1 until reaching time $\alpha_\gamma \coloneqq (\gamma + 1)/2$ when playing the stable arm and strictly decreasing until reaching {\em value} $\gamma$ when playing the striving arm. In other words, the agent plays the stable arm for $\alpha_\gamma$ time followed by the striving arm for $1-\alpha_\gamma$ time and repeats. 
\end{definition}
\begin{lemma} \label{lemma:min-acc-works}
    For each strategy that ``stockpiles'' reward along the way, there exists a minimally accumulating strategy that nets total reward at least as much as the stockpiling strategy. 
\end{lemma}

We present the result for the competitive ratio and reward for a $\gamma$-comfortable agent (proof in Appendix~\ref{appendix:proof-gamma-comfort}).

\begin{lemma} \label{lemma:gamma-comfort}
Suppose 
an agent who wants to be $\gamma$-comfortable plays $f_1$ for $\alpha_\gamma$ time followed by $f_2$ for $1-\alpha_\gamma$ time before reverting back to $f_1\,$ and continuing the process. Then, the agent wanting to maximize their competitive ratio subject to the constraint of the average reward always being at least $\gamma$ will switch after absolute time $T - \frac{\gamma}{2} - \frac12 \sqrt{\gamma^2 + 4T(2-\gamma)}$, achieving competitive ratio $\gamma + \frac{\gamma(1-\gamma)}{2T} + \frac{(1-\gamma) \sqrt{\gamma^2 - 4T (2 - \gamma)}}{2T}$. This corresponds to $\frac{1-\gamma}{2} \cdot \left(T - \frac{\gamma}{2} - \frac12 \sqrt{\gamma^2 + 4T(2-\gamma)} \right)$ time on the striving arm.
\end{lemma}

\begin{rmk}
    Let us consider a concrete numerical example: suppose $T = 150\,, \gamma = 0.5.$ Then, the agent will switch after about time 135. However, of that time, only about 34 would have been spent exploring, with the remaining 101 time spent on the stable arm. The competitive ratio achieved is 0.55.
\end{rmk}

\begin{rmk}
    
For insight, let us next consider the behavior for extreme values of $\gamma\,:$ if $\gamma = 0\,,$ the first two terms go away, and the competitive ratio is $\frac{\sqrt{2T}}{T} = \frac{2}{\sqrt{T}}\,.$ This exactly aligns with our computation before. On the other hand, if $\gamma \rightarrow 1\,,$ the competitive ratio actually nears 1! This is because the best possible thing to do for an agent who requires complete comfort is to always play the stable arm. Indeed, to better understand this outcome, let us also investigate the total amount of time spent exploring as a function of $\gamma\,.$ Taking the derivative of the expression for exploration time with respect to $\gamma$, we can see that it is negative for all $\gamma \in [0, 1]\,.$ 
This tells us that as an agent requires more ``comfort,'' they spend less time exploring on the striving arm, and so while their competitive ratio improves, their chance of witnessing the growth in the striving arm and benefiting from it is low.
\end{rmk}
\section{Financial Support} \label{sec:trust-fund}

A natural question following this discussion is how to incentivize a gritty agent to explore for longer. We have already seen that simply encouraging more grit could do more harm than good. We could encourage the agent to give up more comfort, but often a baseline level of comfort cannot be foregone -- for instance, one may have to pay rent, buy food, etc. For this, we study the setting where there is a cost to striving and the agent must have discomfort tolerance $\gamma > 0$. In this section, we consider an implementation of a ``trust fund,'' i.e., financial support we could provide an agent, and then show what conclusions we can draw about the strategy followed by an agent and their eventual reward. Here we consider the simplest possible implementation, which already has interesting outcomes, and we study a more nuanced implementation in the Appendix.

\subsection{No Safety Net} Now, since an agent without a safety net must alternate between $f_1$ and $f_2\,,$ the competitive ratio maximizing strategy is computed as follows: first, if $f_2$ never increases, the agent incurs the competitive ratio on the left below, and if $f_2$ increases right after the agent switches, they receive the competitive ratio on the right.

\begin{align}
    \frac{(1 - 1) \cdot \frac s2 + T - s}{T} &= \frac{(1 - 1) \cdot \frac s2 + T - s}{(1 - 1) \cdot \frac s2 + \frac{1}{2} (T-s)^2}\,. \\
    \Rightarrow
    s &= T  - \sqrt{2T}\,.
\end{align}

The duration of time after which the switch happens is $T-\sqrt{2T}\,,$ but the proportion of that time actually spend exploring the striving arm is half that, namely $\frac{T-\sqrt{2T}}{2}\,.$ Thus, if $\theta > \frac{T-\sqrt{2T}}{2}\,,$ this agent will not be able to reap the benefits of striving.

\subsection{Free Reimbursement} In this model, an agent with a support network gets reimbursed for free each time they taking a striving step. This is analogous to having a benefactor who financially supports the agent as much as needed to prevent their net reward from being negative. In this case, the effective arms for an agent with such unconditional support are now:

$$
\hat{f}_1(t) = 1 \, \forall \, t \qquad \hat{f}_2(t) = \begin{cases}
    0 & t < \theta \\
    t-\theta & t \ge \theta
\end{cases}\,.
$$

Thus, the analysis is as before (Lemma~\ref{lemma:alphat-switch} with $\alpha, \alphat=1$), and the agent will spend $T - \sqrt{2T}$ time on the striving arm before switching to the stable arm. Remarkably, the duration of time after which the agent without a safety net and the agent with unconditional support ``give up'' is the same! However, due to the safety net, the agent with it can explore the striving arm for twice as long. In particular, if $\theta \in [\frac{T-\sqrt{2T}}{2}, T -\sqrt{2T}]\,,$ then the latter agent gets $\frac 12 (T - \theta)^2 \ge \frac 12 \cdot 2T = T$ reward, while one without the safety net only gets $\sqrt{2T}$.

In the Appendix,
we extend this to a model where the benefactor only promises support for a fixed amount of time. In that case also, a similar qualitative result holds -- agents with and without support ``give up'' on striving at the same absolute time but the agent with the financial support gets to explore for a multiplicative factor longer. Note also that by mapping this onto the discomfort tolerance perspective, we can see that in a world where the agent must maintain a positive discomfort tolerance but the agent lacks financial support, they must spend less time exploring, whereas if they have financial support, they can explore longer.

\subsection{Which Wins? Trust Fund or Grit?}

In this section, we combine the pieces from the previous sections to understand the interplay between grit and financial support. We primarily present conclusions in this section; the calculations behind these conclusions are presented in detail in the Appendix.
Here, the rate of increase is unknown, and there is also a cost to striving.
Let us make two comparisons. First, let us investigate what happens when someone without a trust fund becomes more gritty. Then, we will study the effect of a trust fund on two people, one of whom is grittier than the other.

\noindent\textbf{\underline{No trust fund, increased grit.}} For this, let us compare the first two rows of the Table below.
Immediately, we see that increased grit, as before, leads to increased exploration time. In the last column, we present the reward that the agent receives if they don't witness the start of the payoff before switching, which we call ``stable reward'' for short. There, we can also see that the stable reward is {\em lower} when the agent is grittier. 

\noindent\textbf{\underline{Introduce trust fund, same grit.}} Now, let us compare the first and third rows of the table below.
In this case, we see that at the same grit level, the presence of a safety net allows for a much longer exploration horizon. Both in the presence of and in the absence of the safety net, the stable reward is the same. This shows that the presence of a safety net allows for essentially ``free'' exploration.

\subsubsection{Discussion} We can view these results from two perspectives. One perspective is descriptive: we observe that providing a safety net increases exploration time essentially ``for free.'' This happens since the agent does not have to split their time to ensure they are not in the negative. The other perspective is prescriptive: in settings where an external entity aims to encourage exploratory behavior, encouraging increased grittiness could lead to worse outcomes if the risk doesn't pay off. While we don't study this in this work, this could lead to future agents being discouraged from taking the grittier course of action. On the other hand, providing support to an agent who is already gritty encourages exploration without the prospect of bad outcomes in case the taken risk doesn't pay off. Thus, telling a gritty agent to be grittier is worse than giving them financial support to extend their exploration horizon. This corroborates and provides a simple mechanistic explanation for what \cite{morton2019grit, wooten_precarious_2022} observe.

    \begin{tabular}{c|c|c|c}
        grit level & safety net & exploration time & stable reward \\
        \hline 
        $\alphat_1$ & none & $\frac{T}{2} - \sqrt{\frac{T}{2\alphat_1}}$ & $\sqrt{\frac{2T}{\alphat_1}}$\\
        $\alphat_2 > \alphat_1$ & none & $\frac{T}{2} - \sqrt{\frac{T}{2\alphat_2}}$ & $\sqrt{\frac{2T}{\alphat_2}}$ \\
        $\alphat_1$ & yes & $T - \sqrt{\frac{2T}{\alphat_1}}$ & $\sqrt{\frac{2T}{\alphat_1}}$\\
    \end{tabular}
\section{Rationality as Being Bayesian} \label{sec:bayesian}

In this section, we explore a different notion of rationality. We define the decision-making process of agents who behave this way and then we explore the effect of uncertainty tolerance on outcomes.

\subsection{Rationality in This Model}

Until now, we have focused on a model of rationality that aims to minimize regret in hindsight, which we quantify through the competitive ratio. In this section, we instead consider a notion of rationality that is forward-looking, as it aims to directly quantify uncertainty about the future.  In particular, we suppose agents have a prior distribution $P$ that is supported on $\{1, 2, \dots, T-1, T\} \cup \{ N \},$ where $N$ represents ``never,'' and $P(x) \coloneqq \mathbb{P}\left[\text{increase occurs at time } x\right]\,.$

The agent updates their posterior as follows. Suppose at time $t-1$, the posterior is $P^{(t-1)}.$ If they play the arm for one time step and the increase has not yet occurred, then they update:

$$
P^{(t)}(x) \leftarrow \frac{P^{(t-1)}(x)}{1 - P^{(t-1)}(t)} \qquad \forall x > t\,.
$$

If the mass on the numerical elements of the support is 0, then the update places probability 1 onto the support element $N\,,$ meaning the agent's posterior suggests the arm will {\em never} payoff.

\subsection{Setting}
As before, there are two arms:
$$
f_1(t) = 1\, \forall \, t \qquad f_2(t) = \begin{cases} 0 & t<\theta \\ t-\theta & t\ge \theta
\end{cases}
\,.
$$

Unlike before, we assume the arm is played in discrete time steps of size 1, i.e, the agent pulling the arm increases the time by 1. This is so that the prior can be defined over a discrete domain.

Based on this and the posterior update described above, we have that the recursive formula for the expected reward is:
$$
Q(t) = \frac{1}{2}(T-t)^2 \cdot p_t + V(t+1) \cdot (1-p_t)\,,
$$
and reward of the policy of being on the striving arm given no increase so far and haven't switched so far is:
$$
V(t) = \max \{ T-t, Q(t)  \}
$$

where  $p_t = P(\text{ increase starts at time } t | \text{ increase has not started up until } t-1)\,.$ The boundary condition is $Q(T) = 0\,.$ The agent switches when the reward from switching exceeds the expected reward from staying.

\begin{lemma}
    Suppose $Q(t), V(t)$ are defined as above. Then, $V(t)$ computes the total reward accrued from time $t$ up to time $T$ by a policy that maximizes expected reward over its posterior.
\end{lemma}
\begin{proof}
We show this using induction. From the boundary condition, we have that $V(T) = \max\{0, 0 \} = 0.$ For the base case, we consider $t = T-1\,.$ There are two possible actions the policy could take. Suppose the policy plays the striving arm: the first possible outcome is that it pays off with probability $p_{T-1}$ as defined above, and the second possible outcome is that it doesn't, and the policy accrues reward $V(T)$. We have that $Q(T-1) = \frac 12 p_{T-1} + 0 (1-p_{T-1}) = \frac{p_{T-1}}{2}.$ If the policy does not play the striving arm and instead switches to the stable arm for good, it is guaranteed $T-(T-1) = 1$ reward. Thus, an expected reward maximizing policy will accrue $V(T-1) = \max \{1, \frac{p_{T-1}}{2}\} = 1\,$ reward at that time.

Next, consider the inductive assumption that for $t = t' < T\,, V(t')$ gives the correct total reward accrued from $t'$ to $T.$ Then, we must show that $V(t'-1)$ indeed is the correct total reward accrued from $t'-1$ to $T\,.$ If we play the stable arm from here onward (and we know from Lemma~\ref{lemma:min-acc-works} that once we switch to the stable arm, we have no reason to switch back to the striving arm), then we would get reward $T-(t'-1) = T - t' + 1.$ On the other hand, if we play the striving arm for one step, then if it pays off immediately (probability $p_{t'-1}$), we get reward $\frac 12 (T-t'+1)^2\,,$ and if not yet (probability $1-p_{t'-1}$), we get reward $V(t')\,.$ Since we know that $V(t')$ is the correct total reward accrued from $t'$ to $T$ by the inductive assumption, $Q(t'-1) = \frac 12 (T-t'+1)^2 \, p_{t'-1} + V(t') \, (1-p_{t'-1})$ is the correct total reward accrued from $t'-1$ to $T$ if the policy plays the striving arm. Finally, we do choose the action that maximizes expected reward, and so $V(t'-1) = \max \{ T-t'+1, Q(t'-1)  \}$ is indeed the correct total reward accrued from $t'$ to $T\,.$
\end{proof}

In this view, we may study the uncertainty tolerance aspect of grit. If two agents have similar priors, i.e., same family of distribution and same mean, then the variance of the prior reflects how tolerant they are to uncertainty about when the striving arm will pay off. We primarily study how this uncertainty tolerance affects the policy an agent follows. We can also then understand how the reward is affected as before.

\subsection{Results for Modelling Grit as Uncertainty Tolerance}

We consider the case where an agent has a Gaussian prior on when the striving arm will pay off. This corresponds to having some understanding of {\em when} the arm might start paying off but not being entirely sure that it will pay off then so allowing for some latitude. Thus, the variance of the Gaussian prior corresponds to the uncertainty tolerance of the agent.

We find that as the variance of the prior increases, the point $s$ at which the agent reverts to the stable arm also increases. Thus, as an agent becomes more gritty, there is a wider range in which if $\theta$ lies, they will witness it. At the same time, if $\theta$ lies above the blue curve in Figure~\ref{fig:gauss-bayes}, the agent would not witness the increase, and they would also collect less stable reward. From the asymptoting shape, we can conclude that beyond a certain point, uncertainty tolerance has limited benefits, as the stable reward decreases but the additional region is not increasing by much.

\begin{figure}
    \centering
    \includegraphics[width=0.5\linewidth]{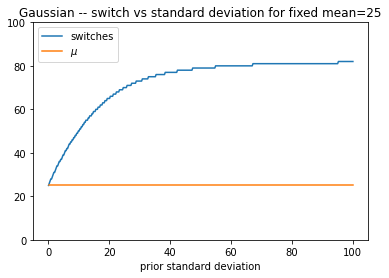}
    \caption{Switch point as a function of standard deviation $\sigma$ for prior of $\mathcal{N}(25, \sigma^2).$}
    \label{fig:gauss-bayes}
\end{figure}

Morton and Paul note that:
\begin{quote}
    ``Other things being equal, the gritty agent’s evidential threshold for updating her expectations of success will tend to be higher than the threshold an impartial observer would use. This is not because the perspective of the impartial observer is epistemically privileged, however; the Permissivist latitude applies to the policies of the agent and the observer alike. Rather, it is because the observer has no need to respond to the evidence in a way that guards against premature despair, and this should be reflected in his evidential policies.'' (p. 195 in \cite{morton2019grit})
\end{quote}

Our results corroborate this -- if we suppose the `impartial observer' switches if the striving arm has not yet paid off when $s$ reaches their expectation for when they payoff begins, then the gritty agent indeed has a higher threshold for updating their expectation of success.


\section{Discussion of Modelling Choices}

\textbf{Modelling Perspective} In this work, we provide a theoretical model for the following aspects of grit: we define a two-armed improving bandit instance that represents the decision-making problem agents are faced with. We further define two different objectives a decision-making agent might prioritize. Finally, we model different aspects of grit with careful study of variations of parameters in the bandit instance. While there are many folk notions of “grit”, the aspects we choose to model are based on the abstraction and formalization provided by \cite{morton2019grit}. We do not seek to evaluate whether or not their account is valid but rather to quantitatively formalize their notions and study the implications of their analysis. 
In doing so, we find additional confirmation for empirical work regarding the value of ``grants'' provided to people attempting ambitious choices. For instance, a study of grants given to entrepreneurs in Burkina Faso shows that even if immediate profits aren't improved, providing this financial support increases innovation and improves business practices, suggesting that those who get support get more time to explore and build a good foundation for their business \cite{grimm_short-term_2021}. Similarly, a study in Kenya found that intervening with grants to youth entrepreneurs at a time of crisis helped individuals maintain their businesses and produce more profits \cite{domenella_can_2021}. In a different vein, experiments in cognitive science and psychology show that children and adults alike are more likely to explore in the presence of caregivers \cite{DahmaniATG24}. These empirical studies suggest that if someone is inclined to take an ambitious action, then supporting them helps improve outcomes, exactly what is captured by our model (as summarized in the table). In the future, our model could be used to develop resource allocation schemes in such settings. It would also be interesting to study community-level effects when agents of varying levels of grit make decisions based on each other.

\noindent\textbf{Strengths} A strength of our model is that the same multi-armed bandits instance can be used to study many different facets of grit, and resulting outcomes are visible just through this two-armed bandits instance. Also, our modelling allows for a modular understanding of the effect of grit on first, behavior and second, outcome or reward associated with that behavior. This, then, allows us to study interventions that encourage similar behavior with less outcome risk. 

\noindent\textbf{Weaknesses} Our model is not without its shortcomings. We are limited to studying an agent's choice between two options in this framework. In real settings, the stable option might not have the same return for all time. Human rationality is rarely, if ever, executed exactly as competitive ratio maximization or expected reward maximization over a prior. However, these seem like natural simplifications that provide a starting point for quantitative analysis of this trait.

\section{Conclusion}
We introduced a quantitative model for studying the impact of grit and financial support on decision-making between a long-term, ambitious end and immediate-reward, stable end. We developed a two-armed bandit model in which to study this and formalized notions of rationality and grit that gave rise a family of strategies for this instance. Our modeling engages with several prior works in the social science literature, and we hope our approach and framework admit future work on quantitative study of grit. \par

\section{Discussion in Relation to Pessimsim Traps}

We now briefly reflect on the previous two chapters jointly. First, we consider implications from our work for the issues of striving and ambition raised by Morton in her series of works \cite{morton2019grit, morton2022resisting, morton_moving_2021}. We also briefly talk about next steps for our theoretical inquiry. 

In \cite{morton2019grit} and \cite{morton2022resisting}, Jennifer Morton studies two complementary aspects of pursuing ambition, especially when coming from a disadvantaged background. On the one hand, in \cite{morton2019grit}, she discusses what it means to be {\em gritty}, and we study this in Chapter~\ref{chap:grit}. 
On the other hand, in \cite{morton2022resisting}, she discusses evidential factors that arise from ones community, which we study in Chapter~\ref{chap:pess-traps}. Taken together, our analysis provides formalisms to study evidence propagation both within an individual's reasoning and between individuals in a community. We show that it can be equally rational for different individuals to be ambitious or not, despite having access to the same evidence. We show that {\em beliefs} about the state of the world can cause this difference. Similarly, in the social setting, we see that the influence of history can make different actions equally rational. In both settings, we should how financial intervention can provide an agent with additional trustworthy evidence that helps them make better decisions. \par 

Crucially, though Morton's description of pessimism traps indicates that the ``optimistic'' option should entail ambition and investment, our current information cascade-based model does not faithfully incorporate the investment aspect of the optimistic action. Likewise, our model for grit focuses on an individual's interaction with the decision by themself. A natural and important next step would combine the two into a model that allows us to simultaneously study social and behavioral influences. \par


\part{Reflections and Conclusion}
\chapter{Conclusion}

Having presented all the technical details and arguments of the thesis, we now draw conclusions. This thesis has presented novel theoretical results regarding the improving multi-armed bandits problem and a multi-community opinion dynamics model. We also study problems in social epistemology using techniques from theoretical computer science. We synthesize our learnings from working on this area in an argument about the value of theoretical models. \par

First, we will position the presented work within the unifying framework discussed in the introduction. Then, we will engage with the emergent theme of the thesis. Namely, we will track the progression in the thesis from abstract to concrete and reflect on what was gained and what was lost at each step. Finally, we will present problems left open in the thesis and research directions opened up by the thesis.

\section{Interplay between sequential and social decision-making}

Let us consider the original unifying framework introduced in Chapter~\ref{sec:unif-framework}. We discussed the idea that by fixing a setting of interest and formalizing an objective whose maximization is the desired outcome, we can search for optimal algorithms. We then suggested that we can, on the other hand, study the relationship between the setting and outcome for a fixed algorithm aimed at the objective. \par

In this thesis, Part~\ref{part:imab} focused on the former perspective, while Part~\ref{part:social-ep} focused on the second perspective. In Part~\ref{part:imab}, we extensively studied the improving multi-armed bandits problem. We found various algorithms to solve various reward maximization and best-arm identification objectives. In Part~\ref{part:social-ep}, we studied two problems. In Chapter~\ref{chap:pess-traps}, where we studied the pessimism traps problem, we assumed the decision-making heuristic of agents was fixed (perfect Bayesian updates), and we examined how changing the reward structure of the world through subsidies affected outcomes. In Chapter~\ref{chap:grit}, where we studied grit, we assumed the decision-making heuristic of agents was fixed (maximize competitive ratio or Bayesian updates) and explored how different payoff times affect different agents' outcomes. We then investigated how changing the reward of a bandit arm by providing a subsidy affected outcomes. \par

Through these complementary perspectives, we were able to characterize decision making in both general and specific settings. In the case of IMAB and grit, once we had a theoretically well-understood framework, we could apply it to study social problems. Likewise, we can take problems in the social setting and motivate new learning-theoretic or sequential decision-making models to study. Thus, these two perspectives work in conjunction to help us understand different aspects of decision-making.


\section{General vs Specific}
Next, let us consider how the thesis addressed the emergent theme of what we gain and what we lose by studying problems abstractly. The work presented in Chapter~\ref{chap:br25} is the most abstract in the thesis. Chapter~\ref{chap:alg-design-imab} shows how to make it more practical by adapting to structure, but overall the framework remains quite abstract. Then, the social epistemology chapters (Chapters~\ref{chap:pess-traps} and \ref{chap:grit}) study concrete problems and therefore look at relatively narrow settings. The methods of this thesis remain in the theoretical realm (with occasional experiments on stylized or offline real-world data), and so the thesis does not explore problems that are increasingly practical and nuanced, and it does not get to the full extent of real-world complexity. However, even at these three points along the continuum of abstract to real-world, we can learn several useful lessons. \par

First, the obvious point: nothing comes for free. As we get more concrete, we necessarily lose generality. As we get more abstract, we necessarily lose specificity. In the most general case, we could handle a rather strong adversary. Going from worst-case to beyond worst-case guarantees for IMAB, we were able to adapt to structure available in the setting, but we had to witness that structure, either via access to offline samples or through stronger assumptions. In going from these works to the work on grit, we could suddenly do much better than depending on the general algorithms. Further, that instantiation of the IMAB problem allowed us to evince the interplay between a very particular structure of the instance and very particular behaviors of agents. Thus, hearkening back to our argument in Chapter~\ref{chap:social-ep-intro}, it is crucial that we formalize our theoretical problems at the level of specificity at which we wish to solve them. It is often of theoretical interest to characterize as general a version of the problem as possible. It is often practically most useful to solve rather particular settings. \par

Related to that, this thesis also shows that fixing complexity in certain aspects of a problem can allow for exploring complexity in other aspects. In the first half of the thesis, we assume an agent that has the singular goal of optimizing their competitive ratio. Then, we are able to allow for rich complexity in the kinds of instances they might see, and we show how even if instances are chosen adversarially, the agent can succeed to a reasonable degree. On the other hand, in the second half, we assume that the agents have a relatively well-specified decision problem in front of them. In both cases, there are only two options they must decide between, and there is some uncertainty inherent to these options. Now, we are able to explore how complexity in social networks or in behavioral traits affect the decision between these two options. It would have been a lot more challenging to explore all these sources of complexity at once. While there is value to studying how sources of complexity interact, and there are certainly cases where compounding complexity behaves quite differently to the sum of their parts, there is significant epistemic value to deriving insight from simple, isolated models. We also argue this in Chapter~\ref{chap:social-ep-intro}.

\section{Open Problems and Future Directions}

Finally, let us talk about what remains open. Luckily for future researchers and myself, the answer is plenty. We first survey some concrete open problems. Then, we consider the prospects of our framework as a research agenda.

\subsection{Concrete Technical Questions}
\paragraph{Improving multi-armed bandits.} For the worst-case objective, we can consider a plethora of open problems that arise from modifying the nature of the improving reward function (instead of diminishing returns), adding in costs to playing different arms and/or switching between arms, and additional advice or feedback beyond bandit but less than full information. There are many related directions in the sequential decision-making literature, including related to sequential Pandora's box problems \cite{beyhaghicaipandorasurvey24, aouad2026pandorasboxproblemsequential, chawla_combinatorial_2026}. \par

Studying the IMAB problem in beyond-worst case settings is especially exciting due to relevance for practical problems such as hyperparameter tuning. In that vein, it is interesting to consider other beyond-worst case paradigms, such as warm starting or algorithms with predictions (which might relate to additional advice above). It is also interesting to consider partially or fully online models for data-driven algorithm design (so far, we study a setting where we have full offline access to past instances).

\paragraph{Opinion dynamics / social learning} Our work on pessimism traps provides an intervention in a particular opinion dynamics model. It would be interesting to consider to what degree this intervention is robust to exact  knowledge of the strength of signal parameter. Further, it would be interesting to identify other models where such an intervention breaks learning and frees agents from cascades. 

\subsection{Broad Future Directions}

The framework described in Section~\ref{sec:unif-framework} and instantiated throughout the thesis is quite broad. This means we can use to study a wide variety of problems. In general, as we study and characterize more models in the standard sequential decision-making framework, we have a better understanding of those models, and we can apply them to study particular social problems. Likewise, there are likely to be complex social settings that are {\em not} captured by existing theoretical models. Then, we can devise new theoretical models, and we can not only use them to understand the social setting but also study them in the standard theoretical way. Thus, this framework provides us with a way to seek out, frame, and solve an infinite set of problems. \par

In this thesis, we instantiated this framework to study decision-making when investment is required to witness payoff. There are plenty of other areas where it would be interesting to use this framework. A socially and politically relevant direction is opinion polarization, where there is already some amount of theoretical work. We can also hope to study human-AI collaboration in this framework. Overall, this framing encourages investigation of many timely and significant problems. \par

\section{Parting Thoughts}

Decision-making has long been studied by a wide range of scholars. Computer scientists are among the most recent entrants to the study of how we acquire, aggregate, and operationalize information. As more and more of our decisions are partially or even completely automated by algorithmic and AI systems, it is increasingly important that we apply our methods to study decision-making and use insights from such inquiry to proactively design algorithmic decision-augmenting or decision-making systems in a socially-beneficial way. I hope this thesis conveys not only the technical results contained therein but also the exhortation to be deliberate and thoughtful about how our work relates to the world we live in.




\bibliographystyle{alpha}
\bibliography{bibliography}

@INPROCEEDINGS{yao-principle,
  author={Yao, Andrew Chi-Chin},
  booktitle={18th Annual Symposium on Foundations of Computer Science (sfcs 1977)}, 
  title={Probabilistic computations: Toward a unified measure of complexity}, 
  year={1977},
  volume={},
  number={},
  pages={222-227},
  doi={10.1109/SFCS.1977.24}}

@inproceedings{heidari_tight_nodate,
author = {Heidari, Hoda and Kearns, Michael and Roth, Aaron},
title = {Tight policy regret bounds for improving and decaying bandits},
year = {2016},
isbn = {9781577357704},
publisher = {AAAI Press},
booktitle = {Proceedings of the Twenty-Fifth International Joint Conference on Artificial Intelligence},
pages = {1562–1570},
numpages = {9},
location = {New York, New York, USA},
series = {IJCAI'16}
}

@incollection{fiat_oil_2009,
	location = {Berlin, Heidelberg},
	title = {The Oil Searching Problem},
	volume = {5757},
	url = {http://link.springer.com/10.1007/978-3-642-04128-0_45},
	pages = {504--515},
	booktitle = {Algorithms - {ESA} 2009},
	publisher = {Springer Berlin Heidelberg},
	author = {{McGregor}, Andrew and Onak, Krzysztof and Panigrahy, Rina},
	editor = {Fiat, Amos and Sanders, Peter},
	urldate = {2024-01-22},
	year = {2009},
	langid = {english},
	doi = {10.1007/978-3-642-04128-0_45},
	note = {Series Title: Lecture Notes in Computer Science},
}

@article{thompson_likelihood_1933,
	title = {On the likelihood that one unknown probability exceeds another in view of the evidence of two samples.
        },
	volume = {25},
	issn = {0006-3444},
	url = {https://doi.org/10.1093/biomet/25.3-4.285},
	doi = {10.1093/biomet/25.3-4.285},
	pages = {285--294},
	number = {3},
	journal = {Biometrika},
	author = {{Thompson}, {William} R},
	date = {1933-12},
        year = {1933},
}

@incollection{gal_search_2011,
	title = {Search Games},
	rights = {Copyright © 2010 John Wiley \& Sons, Inc. All rights reserved.},
	isbn = {978-0-470-40053-1},
	url = {https://onlinelibrary.wiley.com/doi/abs/10.1002/9780470400531.eorms0912},
	booktitle = {Wiley Encyclopedia of Operations Research and Management Science},
	publisher = {John Wiley \& Sons, Ltd},
	author = {Gal, Shmuel},
	urldate = {2024-01-28},
	year = {2011},
	langid = {english},
	doi = {10.1002/9780470400531.eorms0912},
}

@book{noauthor_theory_2003,
        author = {Alpern, Steven and Gal, Shmuel},
	location = {Boston},
	title = {The Theory of Search Games and Rendezvous},
	volume = {55},
	isbn = {978-0-7923-7468-8},
	url = {http://link.springer.com/10.1007/b100809},
	series = {International Series in Operations Research \& Management Science},
	publisher = {Kluwer Academic Publishers},
	urldate = {2024-01-28},
	year = {2003},
	langid = {english},
	doi = {10.1007/b100809},
}

@misc{Riv23b,
   author =        { Ronald L. Rivest },
   title =         { Learning Learning Curves },
   date =          { 2023-11-03 },
   year =          {2023},
   note =          { (Unpublished) Slides from talk given at Rob Schapire's 60th birthday party. },
 url = {https://people.csail.mit.edu/rivest/pubs/Riv23b.pdf}
}

@book{lattimore_bandit_2020,
	edition = {1},
	title = {Bandit Algorithms},
	isbn = {978-1-108-57140-1 978-1-108-48682-8},
	url = {https://www.cambridge.org/core/product/identifier/9781108571401/type/book},
	publisher = {Cambridge University Press},
	author = {Lattimore, Tor and Szepesvári, Csaba},
	urldate = {2024-02-07},
	date = {2020-07-31},
	year = {2020},
	langid = {english},
	doi = {10.1017/9781108571401},
}

@misc{slivkins_introduction_2022,
	title = {Introduction to Multi-Armed Bandits},
	url = {http://arxiv.org/abs/1904.07272},
	number = {{arXiv}:1904.07272},
	publisher = {{arXiv}},
	author = {Slivkins, Aleksandrs},
	urldate = {2024-02-07},
	date = {2022-01-08},
	year = {2022},
	langid = {english},
	eprinttype = {arxiv},
	eprint = {1904.07272 [cs, stat]},
}

@article{hyperband_2017,
author = {Li, Lisha and Jamieson, Kevin and DeSalvo, Giulia and Rostamizadeh, Afshin and Talwalkar, Ameet},
title = {Hyperband: a novel bandit-based approach to hyperparameter optimization},
year = {2017},
issue_date = {January 2017},
publisher = {JMLR.org},
volume = {18},
number = {1},
issn = {1532-4435},
journal = {J. Mach. Learn. Res.},
month = {jan},
pages = {6765–6816},
numpages = {52}
}

@article{haussler_rigorous_1996,
	title = {Rigorous learning curve bounds from statistical mechanics},
	volume = {25},
	issn = {1573-0565},
	url = {https://doi.org/10.1007/BF00114010},
	doi = {10.1007/BF00114010},
	pages = {195--236},
	number = {2},
	journal = {Machine Learning},
	shortjournal = {Machine Learning},
	author = {Haussler, David and Kearns, Michael and Seung, H. Sebastian and Tishby, Naftali},
	date = {1996-11-01},
        year = {1996}
}

@inproceedings{li_efficient_2020,
	title = {Efficient Automatic {CASH} via Rising Bandits},
	volume = 34,
	issn = {2374-3468, 2159-5399},
	url = {http://arxiv.org/abs/2012.04371},
	doi = {10.1609/aaai.v34i04.5910},
	pages = {4763--4771},
	booktitle = {Proceedings of the {AAAI} Conference on Artificial Intelligence},
	shortjournal = {{AAAI}},
	author = {Li, Yang and Jiang, Jiawei and Gao, Jinyang and Shao, Yingxia and Zhang, Ce and Cui, Bin},
	urldate = {2024-03-20},
	date = {2020-04-03},
        year = {2020},
	eprinttype = {arxiv},
	eprint = {2012.04371 [cs, stat]}
}

@book{borodin_online_2005,
	title = {Online Computation and Competitive Analysis},
	isbn = {978-0-521-61946-2},
	pagetotal = {440},
	publisher = {Cambridge University Press},
	author = {Borodin, Allan and El-Yaniv, Ran},
	year = {2005},
	langid = {english},
}

@article{pathological,
 ISSN = {00129682, 14680262},
 URL = {http://www.jstor.org/stable/2999431},
 author = {Lones Smith and Peter Sørensen},
 journal = {Econometrica},
 number = {2},
 pages = {371--398},
 publisher = {[Wiley, Econometric Society]},
 title = {Pathological Outcomes of Observational Learning},
 urldate = {2023-10-09},
 volume = {68},
 year = {2000}
}

@article{coateLoury,
 ISSN = {00028282},
 URL = {http://www.jstor.org/stable/2117558},
 author = {Stephen Coate and Glenn C. Loury},
 journal = {The American Economic Review},
 number = {5},
 pages = {1220--1240},
 publisher = {American Economic Association},
 title = {Will Affirmative-Action Policies Eliminate Negative Stereotypes?},
 urldate = {2025-04-21},
 volume = {83},
 year = {1993}
}

@article{bikhchandani1992theory,
author = {Bikhchandani, Sushil and Hirshleifer, David and Welch, Ivo},
title = {A Theory of Fads, Fashion, Custom, and Cultural Change as Informational Cascades},
journal = {Journal of Political Economy},
volume = {100},
number = {5},
pages = {992-1026},
year = {1992},
doi = {10.1086/261849},

URL = { 
    
        https://doi.org/10.1086/261849
    
    

},
eprint = { 
    
        https://doi.org/10.1086/261849
    
    

}
}

@book{fricker2007epistemic,
  title={Epistemic Injustice: Power and the Ethics of Knowing},
  author={Fricker, Miranda},
  year={2007},
  publisher={Oxford University Press}
}

@article{Aronowitz2021-AROEBB-2,
	author = {Sara Aronowitz},
	doi = {10.1215/00318108-8998825},
	journal = {Philosophical Review},
	number = {3},
	pages = {339--383},
	title = {Exploring by Believing},
	volume = {130},
	year = {2021}
}

@book{chamley_rational_2004,
  author       = {Chamley, Christophe P.},
  title        = {Rational Herds: Economic Models of Social Learning},
  publisher    = {Cambridge University Press},
  address      = {Cambridge},
  year         = {2003},
  isbn         = {9780511616372},
  doi          = {10.1017/CBO9780511616372},
  note         = {Online publication January 2010}
}

@article{acemoglu_opinion_2011,
	title = {Opinion Dynamics and Learning in Social Networks},
	volume = {1},
	issn = {2153-0793},
	url = {https://doi.org/10.1007/s13235-010-0004-1},
	doi = {10.1007/s13235-010-0004-1},
	pages = {3--49},
	number = {1},
	journal = {Dynamic Games and Applications},
	shortjournal = {Dyn Games Appl},
	author = {Acemoglu, Daron and Ozdaglar, Asuman},
	urldate = {2024-08-15},
	date = {2011-03-01},
	langid = {english},
        year = {2011}
}

@incollection{sirbu_opinion_2017,
	location = {Cham},
	title = {Opinion Dynamics: Models, Extensions and External Effects},
	isbn = {978-3-319-25658-0},
	url = {https://doi.org/10.1007/978-3-319-25658-0_17},
	shorttitle = {Opinion Dynamics},
	pages = {363--401},
	booktitle = {Participatory Sensing, Opinions and Collective Awareness},
	publisher = {Springer International Publishing},
	author = {Sîrbu, Alina and Loreto, Vittorio and Servedio, Vito D. P. and Tria, Francesca},
	editor = {Loreto, Vittorio and Haklay, Muki and Hotho, Andreas and Servedio, Vito D.P. and Stumme, Gerd and Theunis, Jan and Tria, Francesca},
	urldate = {2024-08-15},
	year = {2017},
	langid = {english},
	doi = {10.1007/978-3-319-25658-0_17},
}

@inproceedings{das_modeling_2014,
	location = {New York, {NY}, {USA}},
	title = {Modeling opinion dynamics in social networks},
	isbn = {978-1-4503-2351-2},
	url = {https://dl.acm.org/doi/10.1145/2556195.2559896},
	doi = {10.1145/2556195.2559896},
	series = {{WSDM} '14},
	pages = {403--412},
	booktitle = {Proceedings of the 7th {ACM} international conference on Web search and data mining},
	publisher = {Association for Computing Machinery},
	author = {Das, Abhimanyu and Gollapudi, Sreenivas and Munagala, Kamesh},
	urldate = {2024-08-15},
	date = {2014-02-24},
        year = {2014}
}

@article{
cohen_2006_intervention,
author = {Geoffrey L. Cohen  and Julio Garcia  and Nancy Apfel  and Allison Master },
title = {Reducing the Racial Achievement Gap: A Social-Psychological Intervention},
journal = {Science},
volume = {313},
number = {5791},
pages = {1307-1310},
year = {2006},
doi = {10.1126/science.1128317},
URL = {https://www.science.org/doi/abs/10.1126/science.1128317},
eprint = {https://www.science.org/doi/pdf/10.1126/science.1128317},
}

@inproceedings{mossel1,
author = {Mossel, Elchanan and Mueller-Frank, Manuel and Sly, Allan and Tamuz, Omer},
title = {Social Learning Equilibria},
year = {2018},
isbn = {9781450358293},
publisher = {Association for Computing Machinery},
address = {New York, NY, USA},
url = {https://doi.org/10.1145/3219166.3219207},
doi = {10.1145/3219166.3219207},
booktitle = {Proceedings of the 2018 ACM Conference on Economics and Computation},
pages = {639},
numpages = {1},
location = {Ithaca, NY, USA},
series = {EC '18}
}

@inproceedings{Balcan2013CircumventingTP,
  title={Circumventing the Price of Anarchy: Leading Dynamics to Good Behavior},
  author={Maria-Florina Balcan and Avrim Blum and Y. Mansour},
  booktitle={SIAM journal on computing (Print)},
  year={2013},
  url={https://api.semanticscholar.org/CorpusID:10058683}
}

@article{banerjee1992simple,
  title={A simple model of herd behavior},
  author={Banerjee, Abhijit V},
  journal={The Quarterly Journal of Economics},
  volume={107},
  number={3},
  pages={797--817},
  year={1992},
  publisher={Oxford University Press}
}

@article{manski1993identification,
  title={Identification of Endogenous Social Effects: The Reflection Problem},
  author={Manski, Charles F.},
  journal={The Review of Economic Studies},
  volume={60},
  number={3},
  pages={531--542},
  year={1993},
  publisher={Oxford University Press}
}

@book{thaler2008nudge,
  title={Nudge: Improving Decisions About Health, Wealth, and Happiness},
  author={Thaler, Richard H. and Sunstein, Cass R.},
  year={2008},
  publisher={Yale University Press}
}

@article{kahneman1974judgment,
  title={Judgment under Uncertainty: Heuristics and Biases},
  author={Kahneman, Daniel and Tversky, Amos},
  journal={Science},
  volume={185},
  number={4157},
  pages={1124--1131},
  year={1974},
  publisher={American Association for the Advancement of Science}
}

@book{sunstein2001echo,
  title={Republic.com},
  author={Sunstein, Cass R.},
  year={2001},
  publisher={Princeton University Press}
}

@article{johnson2012beyond,
  title={Beyond nudges: Tools of a choice architecture},
  author={Johnson, Eric J. and Shu, Suzanne B. and Dellaert, Benedict G. C. and Fox, Craig and Goldstein, Daniel G. and Häubl, Gerald and Larrick, Richard P. and Payne, John W. and Peters, Ellen and Schkade, David and others},
  journal={Marketing Letters},
  volume={23},
  number={2},
  pages={487--504},
  year={2012},
  publisher={Springer}
}

@article{benartzi2017should,
  title={Should governments invest more in nudging?},
  author={Benartzi, Shlomo and Beshears, John and Milkman, Katherine L. and Sunstein, Cass R. and Thaler, Richard H. and Shankar, Maya and Tucker-Ray, Will and Congdon, William J. and Galing, Steven},
  journal={Psychological Science},
  volume={28},
  number={8},
  pages={1041--1055},
  year={2017},
  publisher={SAGE Publications Sage CA: Los Angeles, CA}
}

@article{anderson1997information,
  title={Information cascades in the laboratory},
  author={Anderson, Lisa R. and Holt, Charles A.},
  journal={American Economic Review},
  volume={87},
  number={5},
  pages={847--862},
  year={1997},
  publisher={American Economic Association}
}

@book{duckworth2016grit,
  title={Grit: The Power of Passion and Perseverance},
  author={Duckworth, Angela},
  year={2016},
  publisher={Scribner}
}

@article{morton2019grit,
  title={Grit},
  author={Morton, Jennifer M and Paul, Sarah K},
  journal={Ethics},
  volume={129},
  number={2},
  pages={175--203},
  year={2019},
  publisher={University of Chicago Press}
}

@article{morton2022resisting,
  title={Resisting pessimism traps: the limits of believing in oneself},
  author={Morton, Jennifer M},
  journal={Philosophy and Phenomenological Research},
  volume={104},
  number={3},
  pages={728--746},
  year={2022},
  publisher={Wiley Online Library}
}

@InProceedings{pessimism-traps-alg,
  author =  {Blum, Avrim and Diana, Emily and Ravichandran, Kavya and Tolbert, Alexander},
  title = {{Pessimism Traps and Algorithmic Interventions}},
  booktitle = {6th Symposium on Foundations of Responsible Computing (FORC 2025)},
  series =  {Leibniz International Proceedings in Informatics (LIPIcs)},
  year =  {2025},
  volume =  {329},
  editor =  {Bun, Mark},
  publisher = {Schloss Dagstuhl -- Leibniz-Zentrum f{\"u}r Informatik},
  address = {Dagstuhl, Germany},
  URL =   {https://arxiv.org/abs/2406.04462}
}

@phdthesis{wooten_precarious_2022,
	title = {Precarious Mobility: Trying and Failing to Get Ahead in the 21st Century},
	rights = {open},
	issn = {3737-2181},
	url = {https://dash.harvard.edu/handle/1/37372181},
	shorttitle = {Precarious Mobility},
	author = {Wooten, Tom},
	urldate = {2024-08-22},
	date = {2022-05-11},
	langid = {english},
	note = {Accepted: 2022-06-07T06:32:36Z},
        year = {2022},
        school = {Harvard University}
}

@inproceedings{patil_mitigating_2023,
	location = {Macau, {SAR} China},
	title = {Mitigating Disparity while Maximizing Reward: Tight Anytime Guarantee for Improving Bandits},
	isbn = {978-1-956792-03-4},
	url = {https://www.ijcai.org/proceedings/2023/456},
	doi = {10.24963/ijcai.2023/456},
	shorttitle = {Mitigating Disparity while Maximizing Reward},
	eventtitle = {Thirty-Second International Joint Conference on Artificial Intelligence \{{IJCAI}-23\}},
	pages = {4100--4108},
	booktitle = {Proceedings of the Thirty-Second International Joint Conference on Artificial Intelligence},
	publisher = {International Joint Conferences on Artificial Intelligence Organization},
	author = {Patil, Vishakha and Nair, Vineet and Ghalme, Ganesh and Khan, Arindam},
	urldate = {2024-01-22},
	date = {2023-08},
        year = {2023},
	langid = {english},
}

@inbook{lasonen-aarnio_permissivism_2023,
	location = {London},
	edition = {1},
	title = {Permissivism, Underdetermination, and Evidence},
	isbn = {978-1-315-67268-7},
	url = {https://www.taylorfrancis.com/books/9781315672687/chapters/10.4324/9781315672687-33},
	pages = {358--370},
	booktitle = {The Routledge Handbook of the Philosophy of Evidence},
	publisher = {Routledge},
	author = {Jackson, Elizabeth and {LaFore}, Greta},
	bookauthor = {Lasonen-Aarnio, Maria and Littlejohn, Clayton},
	urldate = {2025-02-13},
	date = {2023-11-24},
        year = {2023},
	langid = {english},
	doi = {10.4324/9781315672687-33},
}

@article{kamenetz_key_2015,
	title = {A Key Researcher Says 'Grit' Isn't Ready For High-Stakes Measures},
	url = {https://www.npr.org/sections/ed/2015/05/13/405891613/a-key-researcher-says-grit-isnt-ready-for-high-stakes-measures},
	journal = {{NPR}},
	author = {Kamenetz, Anya},
	urldate = {2025-02-13},
	date = {2015-05-13},
        year = {2015},
	langid = {english},
}

@article{Gibbon_seligman_2020,
	title = {Martin Seligman and the Rise of Positive Psychology},
	author = {Gibbon, Peter},
	url = {https://www.neh.gov/article/martin-seligman-and-rise-positive-psychology},
	titleaddon = {National Endowment for the Humanities},
	journal = {Humanities},
	year = {2020},
	volume = {41},
	number = {3},
	urldate = {2025-02-17},
	langid = {english},
}

@book{bernardosmith_bayesian_2000,
	title = {Bayesian Theory},
	author = {Bernardo, José M. and Smith, Adrian F. M. },
    publisher = {Wiley},
	url = {https://www.wiley.com/en-us/Bayesian+Theory-p-9780471494645},
	year = {2000},
	titleaddon = {Wiley.com},
	urldate = {2025-02-17},
	langid = {english},
}

@InProceedings{blum_nearly-tight_2024,
  title =    {Nearly-tight Approximation Guarantees for the Improving Multi-Armed Bandits Problem},
  author =       {Blum, Avrim and Ravichandran, Kavya},
  booktitle =    {Proceedings of The 36th International Conference on Algorithmic Learning Theory},
  pages =    {228--245},
  year =   {2025},
  editor =   {Kamath, Gautam and Loh, Po-Ling},
  volume =   {272},
  series =   {Proceedings of Machine Learning Research},
  month =    {24--27 Feb},
  publisher =    {PMLR},
  url =    {https://proceedings.mlr.press/v272/blum25a.html},
}

@book{seligman_learned_2006,
	title = {Learned Optimism: How to Change Your Mind and Your Life},
	publisher = {Vintage Books},
	author = {Seligman, Martin},
	year = {2006},
    isbn = {9780394579153}
}

@book{seligman_authentic_2002,
	title = {Authentic Happiness: Using the New Positive Psychology to Realize Your Potential for Lasting Fulfillment},
	isbn = {978-0-7432-4788-7},
	publisher = {Free Press},
	author = {Seligman, Martin},
	year = {2002},
}

@book{grimm_short-term_2021,
	title = {Short-Term Impacts of Targeted Cash Grants and Business Development Services: Experimental Evidence from Entrepreneurs in Burkina Faso},
	url = {https://hdl.handle.net/10986/36735},
	shorttitle = {Short-Term Impacts of Targeted Cash Grants and Business Development Services},
	publisher = {World Bank, Washington, {DC}},
	author = {Grimm, Michael and Soubeiga, Sidiki and Weber, Michael},
	urldate = {2025-05-01},
	date = {2021-12},
        year = {2021},
	langid = {english},
	doi = {10.1596/1813-9450-9877},
}

@article{domenella_can_2021,
	title = {Can Business Grants Mitigate a Crisis? Evidence from Youth Entrepreneurs in Kenya during {COVID}-19},
	url = {https://hdl.handle.net/10986/36693},
	shorttitle = {Can Business Grants Mitigate a Crisis?},
	journal = {World Bank, Washington, {DC}},
	author = {Domenella, Yanina and Jamison, Julian C. and Safir, Abla and Zia, Bilal},
	urldate = {2025-05-01},
	date = {2021-12},
        year = {2021},
	langid = {english},
	doi = {10.1596/1813-9450-9874},
}

@book{morton_moving_2021,
	title = {Moving Up Without Losing Your Way: The Ethical Costs of Upward Mobility},
	isbn = {978-0-691-21693-5},
	url = {https://books.google.com/books?id=ao4EEAAAQBAJ},
	publisher = {Princeton University Press},
	author = {Morton, J.M.},
	year = {2021},
	lccn = {2019937194},
}

@inproceedings{mussi2024best,
  title={Best arm identification for stochastic rising bandits},
  author={Mussi, Marco and Montenegro, Alessandro and Trov{\`o}, Francesco and Restelli, Marcello and Metelli, Alberto Maria},
  booktitle={Proceedings of the 41st International Conference on Machine Learning},
  pages={36953--36989},
  year={2024}
}

@inproceedings{metelli_stochastic_2022,
  title={Stochastic rising bandits},
  author={Metelli, Alberto Maria and Trovo, Francesco and Pirola, Matteo and Restelli, Marcello},
  booktitle={International Conference on Machine Learning},
  pages={15421--15457},
  year={2022},
  organization={PMLR}
}

@inproceedings{balcan2020data,
  title={{Data-Driven Algorithm Design} (Book Chapter)},
  author={Maria-Florina Balcan},
  booktitle={Beyond Worst-Case Analysis of Algorithms, Tim Roughgarden (Ed)},
  year={2020},
  publisher={{Cambridge University Press}}
}

@inproceedings{sharma2025offline,
author = {Sharma, Dravyansh and Suggala, Arun},
title = {Offline-to-online hyperparameter transfer for stochastic bandits},
year = {2025},
isbn = {978-1-57735-897-8},
publisher = {AAAI Press},
url = {https://arxiv.org/abs/2501.02926},
doi = {10.1609/aaai.v39i19.34243},
booktitle = {Proceedings of the Thirty-Ninth AAAI Conference on Artificial Intelligence and Thirty-Seventh Conference on Innovative Applications of Artificial Intelligence and Fifteenth Symposium on Educational Advances in Artificial Intelligence},
articleno = {2270},
numpages = {9},
series = {AAAI'25/IAAI'25/EAAI'25}
}

@inproceedings{agarwal2017corralling,
  title={Corralling a band of bandit algorithms},
  author={Agarwal, Alekh and Luo, Haipeng and Neyshabur, Behnam and Schapire, Robert E},
  booktitle={Conference on Learning Theory},
  pages={12--38},
  year={2017},
  organization={PMLR}
}

@inproceedings{luo2022corralling,
  title={Corralling a larger band of bandits: A case study on switching regret for linear bandits},
  author={Luo, Haipeng and Zhang, Mengxiao and Zhao, Peng and Zhou, Zhi-Hua},
  booktitle={Conference on Learning Theory},
  pages={3635--3684},
  year={2022},
  organization={PMLR}
}

@inproceedings{arora2021corralling,
  title={Corralling stochastic bandit algorithms},
  author={Arora, Raman and Marinov, Teodor Vanislavov and Mohri, Mehryar},
  booktitle={International Conference on Artificial Intelligence and Statistics},
  pages={2116--2124},
  year={2021},
  organization={PMLR}
}

@inproceedings{yanlcdb,
  title={LCDB 1.1: A Database Illustrating Learning Curves Are More Ill-Behaved Than Previously Thought},
  author={Yan, Cheng and Mohr, Felix and Viering, Tom Julian},
  booktitle={The Thirty-ninth Annual Conference on Neural Information Processing Systems Datasets and Benchmarks Track},
  year={2025}
}

@article{herbster2001tracking,
  title={Tracking the best linear predictor},
  author={Herbster, Mark and Warmuth, Manfred K},
  journal={Journal of Machine Learning Research},
  volume={1},
  number={Sep},
  pages={281--309},
  year={2001}
}

@inproceedings{sharma2020learning,
  title={Learning piecewise Lipschitz functions in changing environments},
  author={Sharma, Dravyansh and Balcan, Maria-Florina and Dick, Travis},
  booktitle={International Conference on Artificial Intelligence and Statistics},
  pages={3567--3577},
  year={2020},
  organization={PMLR}
}

@article{azevedo_neoliberal_2019,
	title = {Neoliberal Ideology and the Justification of Inequality in Capitalist Societies: Why Social and Economic Dimensions of Ideology Are Intertwined},
	volume = {75},
	rights = {© 2019 The Society for the Psychological Study of Social Issues},
	issn = {1540-4560},
	url = {https://onlinelibrary.wiley.com/doi/abs/10.1111/josi.12310},
	doi = {10.1111/josi.12310},
	shorttitle = {Neoliberal Ideology and the Justification of Inequality in Capitalist Societies},
	pages = {49--88},
	number = {1},
	journal = {Journal of Social Issues},
	author = {Azevedo, Flavio and Jost, John T. and Rothmund, Tobias and Sterling, Joanna},
	urldate = {2025-07-23},
	year = {2019},
	langid = {english},
}

@inproceedings{gupta2016pac,
  title={A {PAC} approach to application-specific algorithm selection},
  author={Gupta, Rishi and Roughgarden, Tim},
  booktitle={Proceedings of the 2016 ACM Conference on Innovations in Theoretical Computer Science},
  pages={123--134},
  year={2016}
}

@inproceedings{balcan2017learning,
  title={Learning-theoretic foundations of algorithm configuration for combinatorial partitioning problems},
  author={Balcan, Maria-Florina and Nagarajan, Vaishnavh and Vitercik, Ellen and White, Colin},
  booktitle={Conference on Learning Theory},
  pages={213--274},
  year={2017},
  organization={PMLR}
}

@inproceedings{balcan2018dispersion,
  title={Dispersion for data-driven algorithm design, online learning, and private optimization},
  author={Balcan, Maria-Florina and Dick, Travis and Vitercik, Ellen},
  booktitle={2018 IEEE 59th Annual Symposium on Foundations of Computer Science (FOCS)},
  pages={603--614},
  year={2018},
  organization={IEEE}
}

@inproceedings{blum2021learning,
  title={Learning complexity of simulated annealing},
  author={Blum, Avrim and Dan, Chen and Seddighin, Saeed},
  booktitle={{International Conference on Artificial Intelligence and Statistics (AISTATS)}},
  pages={1540--1548},
  year={2021},
  organization={PMLR}
}

@inproceedings{bartlett2022generalization,
  title={Generalization bounds for data-driven numerical linear algebra},
  author={Bartlett, Peter and Indyk, Piotr and Wagner, Tal},
  booktitle={Conference on Learning Theory (COLT)},
  pages={2013--2040},
  year={2022},
  organization={PMLR}
}

@inproceedings{jinsample,
  title={Sample Complexity of Posted Pricing for a Single Item},
  author={Jin, Billy and Kesselheim, Thomas and Ma, Will and Singla, Sahil},
  booktitle={Advances in Neural Information Processing Systems},
  year={2024}
}

@article{cheng2024learning,
  title={Learning cut generating functions for integer programming},
  author={Cheng, Hongyu and Basu, Amitabh},
  journal={Advances in Neural Information Processing Systems},
  volume={37},
  pages={61455--61480},
  year={2024}
}

@article{sakaue2024generalization,
  title={Generalization bound and learning methods for data-driven projections in linear programming},
  author={Sakaue, Shinsaku and Oki, Taihei},
  journal={Advances in Neural Information Processing Systems},
  volume={37},
  pages={12825--12846},
  year={2024}
}

@article{khodak2024learning,
  title={Learning to Relax: Setting Solver Parameters Across a Sequence of Linear System Instances},
  author={Khodak, Mikhail and Chow, Edmond and Balcan, Maria-Florina and Talwalkar, Ameet},
  year={2024},
  journal={The Twelfth International Conference on Learning Representations (ICLR)}
}

@inproceedings{balcan2024trees,
  title={Learning Accurate and Interpretable Decision Trees},
  author={Balcan, Maria-Florina and Sharma, Dravyansh},
  booktitle={Uncertainty in Artificial Intelligence},
  pages={288--307},
  year={2024},
  organization={PMLR}
}

@inproceedings{sharma2024no,
author = {Sharma, Dravyansh},
title = {No internal regret with non-convex loss functions},
year = {2024},
isbn = {978-1-57735-887-9},
publisher = {AAAI Press},
url = {https://ojs.aaai.org/index.php/AAAI/article/view/29412},
doi = {10.1609/aaai.v38i13.29412},
booktitle = {Proceedings of the Thirty-Eighth AAAI Conference on Artificial Intelligence and Thirty-Sixth Conference on Innovative Applications of Artificial Intelligence and Fourteenth Symposium on Educational Advances in Artificial Intelligence},
articleno = {1664},
numpages = {9},
series = {AAAI'24/IAAI'24/EAAI'24}
}

@article{balcanalgorithm,
  title={Algorithm Configuration for Structured Pfaffian Settings},
  author={Balcan, Maria Florina and Nguyen, Anh Tuan and Sharma, Dravyansh},
  journal={Transactions on Machine Learning Research},
  year={2025}
}

@inproceedings{chatziafratisaccelerating,
  title={Accelerating data-driven algorithm selection for combinatorial partitioning problems},
  author={Chatziafratis, Vaggos and Karmarkar, Ishani and Li, Yingxi and Vitercik, Ellen},
  booktitle={The Thirty-ninth Annual Conference on Neural Information Processing Systems},
  year={2025}
}

@article{khodak2023meta,
  title={Meta-learning adversarial bandit algorithms},
  author={Khodak, Mikhail and Osadchiy, Ilya and Harris, Keegan and Balcan, Maria-Florina and Levy, Kfir Y and Meir, Ron and Wu, Steven Z},
  journal={Advances in Neural Information Processing Systems},
  volume={36},
  pages={35441--35471},
  year={2023}
}

@article{balcan2024accelerating,
  title={Accelerating ERM for data-driven algorithm design using output-sensitive techniques},
  author={Balcan, Maria-Florina and Seiler, Christopher and Sharma, Dravyansh},
  journal={Advances in Neural Information Processing Systems},
  volume={37},
  pages={72648--72687},
  year={2024}
}

@inproceedings{
blum_theoretical_2025,
title={A Theoretical Model for Grit in Pursuing Ambitious Ends},
author={Blum, Avrim and Diana, Emily and Ravichandran, Kavya and Tolbert, Alexander},
booktitle={AAAI 2026 Artificial Intelligence for Social Impact Track},
year={2026},
url={http://arxiv.org/abs/2503.02952}
}

@inproceedings{
bgrs2025algorithm,
title={Algorithm design and sharper bounds for the improving multi-armed bandits problem},
author={Blum, Avrim and Garicano, Marten and Ravichandran, Kavya and Sharma, Dravyansh},
year={2026},
  booktitle =    {Proceedings of The 42th Conference on Uncertainty in Artifical Intelligence},
url={https://arxiv.org/abs/2511.10619}
}

@InCollection{sep-models-science,
	author       =	{Frigg, Roman and Hartmann, Stephan},
	title        =	{{Models in Science}},
	booktitle    =	{The {Stanford} Encyclopedia of Philosophy},
	editor       =	{Edward N. Zalta and Uri Nodelman},
	howpublished =	{\url{https://plato.stanford.edu/archives/win2025/entries/models-science/}},
	year         =	{2025},
	edition      =	{{W}inter 2025},
	publisher    =	{Metaphysics Research Lab, Stanford University}
}

@book{elgin2017,
    author = {Elgin, Catherine Z.},
    title = {True Enough},
    publisher = {The MIT Press},
    year = {2017},
    month = {10},
    isbn = {9780262036535},
    doi = {10.7551/mitpress/9780262036535.001.0001},
    url = {https://doi.org/10.7551/mitpress/9780262036535.001.0001},
}

@article{battermanrice2014,
 ISSN = {00318248, 1539767X},
 URL = {http://www.jstor.org/stable/10.1086/676677},
 author = {Robert W. Batterman and Collin C. Rice},
 journal = {Philosophy of Science},
 number = {3},
 pages = {349--376},
 publisher = {[The University of Chicago Press, Philosophy of Science Association]},
 title = {Minimal Model Explanations},
 urldate = {2026-05-08},
 volume = {81},
 year = {2014}
}

@article{valentini2012,
author = {Valentini, Laura},
title = {Ideal vs. Non-ideal Theory: A Conceptual Map},
journal = {Philosophy Compass},
volume = {7},
number = {9},
pages = {654-664},
doi = {https://doi.org/10.1111/j.1747-9991.2012.00500.x},
url = {https://compass.onlinelibrary.wiley.com/doi/abs/10.1111/j.1747-9991.2012.00500.x},
eprint = {https://compass.onlinelibrary.wiley.com/doi/pdf/10.1111/j.1747-9991.2012.00500.x},
year = {2012}
}

@book{rubinstein-fables,
  title     = "Economic Fables",
  author    = "Rubinstein, Ariel",
  year      = 2012,
  publisher = "Open Book Publishers",
}

@book{woodward2003,
    author = {Woodward, James},
    title = {Making Things Happen: A Theory of Causal Explanation},
    publisher = {Oxford University Press},
    year = {2004},
    month = {01},
    isbn = {9780195155273},
    doi = {10.1093/0195155270.001.0001},
    url = {https://doi.org/10.1093/0195155270.001.0001},
}

@book{klein-sourcesofpower,
author = {Klein, Gary},
year = {2001},
month = {01},
pages = {},
title = {Sources of Power: How People Make Decisions},
volume = {1},
journal = {Leadership and Management in Engineering},
doi = {10.1061/(ASCE)1532-6748(2001)1:1(21)},
publisher = {MIT Press}
}

@misc{hazan2023introductiononlineconvexoptimization,
      title={Introduction to Online Convex Optimization}, 
      author={Elad Hazan},
      year={2023},
      eprint={1909.05207},
      archivePrefix={arXiv},
      primaryClass={cs.LG},
      url={https://arxiv.org/abs/1909.05207}, 
}

@book{Roughgarden_2021, 
    author={Tim Roughgarden},
    place={Cambridge},
    title={Beyond the Worst-Case Analysis of Algorithms},
    publisher={Cambridge University Press}, year={2021}
    }

@inbook{chawla_combinatorial_2026,
author = {Shuchi Chawla and Dimitrios Christou and Amit Harlev and Ziv Scully},
title = {Combinatorial Selection with Costly Information},
booktitle = {Proceedings of the 2026 Annual ACM-SIAM Symposium on Discrete Algorithms (SODA)},
chapter = {},
pages = {664-717},
doi = {10.1137/1.9781611978971.26},
URL = {https://epubs.siam.org/doi/abs/10.1137/1.9781611978971.26},
eprint = {https://epubs.siam.org/doi/pdf/10.1137/1.9781611978971.26},
}

@article{beyhaghicaipandorasurvey24,
author = {Beyhaghi, Hedyeh and Cai, Linda},
title = {Recent Developments in Pandora's Box Problem: Variants and Applications},
year = {2024},
issue_date = {June 2023},
publisher = {Association for Computing Machinery},
address = {New York, NY, USA},
volume = {21},
number = {1},
url = {https://doi.org/10.1145/3699814.3699817},
doi = {10.1145/3699814.3699817},
journal = {SIGecom Exch.},
month = oct,
pages = {20–34},
numpages = {15}
}

@misc{aouad2026pandorasboxproblemsequential,
      title={The Pandora's Box Problem with Sequential Inspections}, 
      author={Ali Aouad and Jingwei Ji and Yaron Shaposhnik},
      year={2026},
      eprint={2507.07508},
      archivePrefix={arXiv},
      primaryClass={cs.CE},
      url={https://arxiv.org/abs/2507.07508}, 
}

@article{bbkg-powerofrandomization-94,
author = {Ben-David, Shai and Borodin, Allan and Karp, Richard and Tardos, Gábor and Wigderson, Avi},
year = {1994},
month = {01},
pages = {2-14},
title = {On the Power of Randomization in On-Line Algorithms.},
volume = {11},
journal = {Algorithmica},
doi = {10.1007/BF01294260}
}

@article{Parker_2020_adequacypurpose, title={Model Evaluation: An Adequacy-for-Purpose View}, volume={87}, DOI={10.1086/708691}, number={3}, journal={Philosophy of Science}, author={Parker, Wendy S.}, year={2020}, pages={457–477}}

@incollection{WOODWARD2007-causalmodelssocial,
title = {Causal models in the social sciences},
editor = {Stephen P. Turner and Mark W. Risjord},
booktitle = {Philosophy of Anthropology and Sociology},
publisher = {North-Holland},
address = {Amsterdam},
pages = {157-210},
year = {2007},
series = {Handbook of the Philosophy of Science},
issn = {18789846},
doi = {https://doi.org/10.1016/B978-044451542-1/50006-4},
url = {https://www.sciencedirect.com/science/article/pii/B9780444515421500064},
author = {James Woodward}
}

@inproceedings{Elliott-Graves2026-pointofprediction,
author="Elliott-Graves, Alkistis",
editor="Arabatzis, Theodore
and Arapostathis, Stathis
and Katsaloulis, Iraklis
and Tympas, Aristotle",
title="What is the Point of Prediction?",
bookTitle="The Perils and Promises of Prediction in the Natural Sciences: Historical and Epistemological Perspectives",
year="2026",
publisher="Springer Nature Switzerland",
address="Cham",
pages="9--25",
isbn="978-3-032-11706-9",
doi="10.1007/978-3-032-11706-9_2",
url="https://doi.org/10.1007/978-3-032-11706-9_2"
}

@article{eg-w-idealization,
author = {Elliott-Graves, Alkistis and Weisberg, Michael},
title = {Idealization},
journal = {Philosophy Compass},
volume = {9},
number = {3},
pages = {176-185},
doi = {https://doi.org/10.1111/phc3.12109},
url = {https://compass.onlinelibrary.wiley.com/doi/abs/10.1111/phc3.12109},
eprint = {https://compass.onlinelibrary.wiley.com/doi/pdf/10.1111/phc3.12109},
year = {2014}
}

@inproceedings{Weisberg2020AbstractionAR,
  title={Abstraction and Representational Capacity in Computational Structures},
  author={Michael Weisberg},
  year={2020},
  url={https://api.semanticscholar.org/CorpusID:213609770}
}

@article{Weisberg-whoismodeler,
author = {Michael Weisberg },
title = {Who is a Modeler?},
journal = {The British Journal for the Philosophy of Science},
volume = {58},
number = {2},
pages = {207-233},
year = {2007},
doi = {10.1093/bjps/axm011},

URL = {https://www.journals.uchicago.edu/doi/abs/10.1093/bjps/axm011},
eprint = {https://www.journals.uchicago.edu/doi/pdf/10.1093/bjps/axm011}
}

@inbook{Weisberg2018-validating-idealized,
 ISBN = {9781517905347},
 URL = {http://www.jstor.org/stable/10.5749/j.ctv5cg8vk.13},
 author = {Michael Weisberg},
 booktitle = {The Experimental Side of Modeling},
 pages = {240--264},
 publisher = {University of Minnesota Press},
 title = {Validating Idealized Models},
 urldate = {2026-05-14},
 year = {2018}
}

@article{Creel_2020, title={Transparency in Complex Computational Systems}, volume={87}, DOI={10.1086/709729}, number={4}, journal={Philosophy of Science}, author={Creel, Kathleen A.}, year={2020}, pages={568–589}}

@book{dray1964philosophy,
  title={Philosophy of history},
  author={Dray, William H},
  publisher={Prentice-Hall Inc.},
  year={1964}
}

@book{dray1979laws,
  title={Laws and explanation in history},
  author={Dray, William H},
  publisher={Oxford University Press},
  year={1979}
}

@article{FORBER-howpossible,
title = {Confirmation and explaining how possible},
journal = {Studies in History and Philosophy of Science Part C: Studies in History and Philosophy of Biological and Biomedical Sciences},
volume = {41},
number = {1},
pages = {32-40},
year = {2010},
issn = {1369-8486},
doi = {https://doi.org/10.1016/j.shpsc.2009.12.006},
url = {https://www.sciencedirect.com/science/article/pii/S1369848609000752},
author = {Patrick Forber},
}

@article{VERREAULTJULIEN-2019-howpossibly,
title = {How could models possibly provide how-possibly explanations?},
journal = {Studies in History and Philosophy of Science Part A},
volume = {73},
pages = {22-33},
year = {2019},
issn = {0039-3681},
doi = {https://doi.org/10.1016/j.shpsa.2018.06.008},
url = {https://www.sciencedirect.com/science/article/pii/S0039368117301231},
author = {Philippe Verreault-Julien},
}

@incollection{mohseni2024methods,
  title={Methods for modelers of science},
  author={Mohseni, Aydin},
  booktitle={Methods in the Philosophy of Science: A User's Guide},
  publisher={MIT Press},
  year={2025}
}

@article{Ylikoski02012014-hpe,
author = {Petri Ylikoski and N. Emrah Aydinonat},
title = {Understanding with theoretical models},
journal = {Journal of Economic Methodology},
volume = {21},
number = {1},
pages = {19--36},
year = {2014},
publisher = {Routledge},
doi = {10.1080/1350178X.2014.886470},
URL = { 
https://doi.org/10.1080/1350178X.2014.886470
},
eprint = { https://doi.org/10.1080/1350178X.2014.886470
}
}

@article{Schelling01071971,
author = {Thomas C. Schelling},
title = {Dynamic models of segregation},
journal = {The Journal of Mathematical Sociology},
volume = {1},
number = {2},
pages = {143--186},
year = {1971},
publisher = {Routledge},
doi = {10.1080/0022250X.1971.9989794},
URL = { https://doi.org/10.1080/0022250X.1971.9989794
},
eprint = {https://doi.org/10.1080/0022250X.1971.9989794
}
}

@book{potochnik-idealization-aims,
    author = {Potochnik, Angela},
    title = {Idealization and the Aims of Science},
    publisher = {University of Chicago Press},
    year = {2017},
    month = {11},
    isbn = {9780226507057},
    doi = {10.7208/chicago/9780226507194.001.0001},
    url = {https://doi.org/10.7208/chicago/9780226507194.001.0001},
}

@inproceedings{DahmaniATG24,
  title = {Caregiver presence promotes judgements of exploration},
  author = {Annya L. Dahmani and Dorsa Amir and Ashley J. Thomas and Alison Gopnik},
  year = {2024},
  url = {https://escholarship.org/uc/item/6s33x461},
  researchr = {https://researchr.org/publication/DahmaniATG24},
  cites = {0},
  citedby = {0},
  booktitle = {Proceedings of the 46th Annual Meeting of the Cognitive Science Society, CogSci 2024, Rotterdam, The Netherlands, July 24-27, 2024},
  editor = {Larissa K. Samuelson and Stefan Frank and Mariya Toneva and Allyson Mackey and Eliot Hazeltine},
  publisher = {cognitivesciencesociety.org},
}

\part*{Appendices}

\appendix
\chapter{Appendices for Improving Multi-Armed Bandits}
\section{Proof of Lemma~\ref{lemma:unknownT}}
\label{appendix:unknownTproof}

\begin{proof}
    Let $T'' = (2^{i} - 1)T_0$ be such that $T'' \le T \le 2 T''\,,$ which implies that the last $T'$ with which we we run Algorithms~\ref{alg:get-mhat} and \ref{alg:rrr} is $2^{i-1}T_0$. Note that $T' \le T'' \le  T \le 2 T'' \le 4 T' $
    Then, we consider what we gain when we spend $T'/2$ exploring and $T'/2$ exploiting. Our analysis of the reward from Algorithm~\ref{alg:rrr} will assume we start from 0 pulls in the exploitation phase, but in practice we are already higher than that.

    We show that the reward achieved running Algorithm~\ref{alg:rrr} for $T'/2$ steps is a constant fraction of $\OPT_T$.

Each arm has been pulled $T'/(2k)$ times in the exploration phase.   We know that $f^\star(T') \in [\frac12 \max_i \hat{m}_L^{(i)}, \max_i \hat{m}_U^{(i)}].$ As before, we actually run Algorithm~\ref{alg:rrr} with its parameter set to $T'/2 - k\,.$ Due to the diminishing returns property, we have that $\frac{f^\star(T')}{f^\star(T'/2 - k)}\frac{f(T'')}{f(T')}\frac{f^\star(T)}{f^\star(T'')} \le \frac{T'}{T'/2 - k} \frac{T''}{T'} \frac{T}{T''} \le 4 \cdot 2 \cdot 2 = 16\,.$ Thus, $f^\star(T) \le 2 f^\star(T'') \le 4 f^\star(T') \le 4 \max_i \hat{m}_U^{(i)}\,,$ and $c_2 = 16.$ 
Now, the extent of the interval is $128k\,,$ so the probability of choosing $\hat{m}$ that is within a constant factor of $f^\star(T'/2-k)$ is $\frac{1}{\log 128k}\,.$ 
  Now, we simply compute the reward achieved by the algorithm by applying Lemma~\ref{lemma:recur-value}. The algorithm plays for $T'/2 - k$ steps, while OPT plays for $T$ steps. In particular:

  \begin{align*}
      ALG_{T'/2} &= V\left( \frac{T'}{2} - k, k  \right) \ge \frac{\OPT_{T'/2-k}}{2c_2} \frac{1}{\gk} 
      \ge \left(\frac{\frac{T'}{2} - k}{T'}\right)^2 \frac{\OPT_{T'}}{2c_2} \frac{1}{\gk} \\
      &\ge \left(\frac{1}{2} - \frac{1}{4} \right)^2 \frac{\OPT_{T'}}{2c_2} \frac{1}{\gk} \ge \frac{1}{16} \frac{T'^2}{T^2} \mt \frac{1}{\gk} \\
      &\ge \frac{1}{16 \cdot 16} \mt \frac{1}{\gk}\,.
  \end{align*}

Plugging in the computed value for $c_2\,$ and considering the probability of choosing the correct $\hat{m}\,,$ we get $ALG_{T'/2} \ge \OPT/(8192 \cdot \gk \log(128\,k))\,.$
\end{proof}

\section{Maximum Reward Objective} \label{appendix:maxreward}

In certain situations, instead of trying to maximize the sum over rewards, we wish to just maximize the maximum reward achieved in a single step. For instance, if we are training different machine learning models, we care about the best performance the model can achieve, not the cumulative performance over training iterations. Our upper bound results from the main paper all hold up to constant factors for this objective, since we exploit the relationship between the maximum value of a function with diminishing returns and the area under it in many places. Our lower bound result also holds but requires a bit of care. In this section, we formally argue the small changes we make in order for the results in the main paper to translate to this different objective.

\subsection{Lower Bound}
We use the same construction and game as before but now directly argue about the pull with maximum reward rather than the sum. For this we have that the algorithm's expected reward can be upper bounded as follows:
\begin{align}
    ALG &= \mathbb{P}\left[E\right] \cdot \text{Reward under event } E + \mathbb{P}\left[E^c\right] \cdot \text{Reward under event } E^c \\
    &\le \frac{\sqrt{k}}{k} 1 + 1 \cdot \frac{1}{\sqrt{k}} = \frac{2}{\sqrt{k}} \\
    &= \frac{2 \, \OPT}{\sqrt{k}}\,.
\end{align}
So we have that the competitive ratio is at least $\Omega(\sqrt{k})$ as before.

\subsection{Upper Bound}

The following facts allow us to directly translate the theorem statements to equivalent ones under this new objective function.

\begin{fact}
    For a given function $f$ satisfying the diminishing returns property, we can relate the area under the function to its maximum value $m$ as follows:
    $$
    mT \ge \sum_{t = 1}^T f(t) \ge \frac{mT}{2}\,.
    $$
\end{fact}

\begin{fact}
    Out of a given set of functions $\mathcal{F}\,,$ each satisfying the diminishing returns property, define $f_1^\star \coloneqq \arg \max_{f \in \mathcal{F}} f(T)$ and $f_2^\star \coloneqq \arg \max_{f \in \mathcal{F}} \sum_{t = 1}^T f(t)\,.$ 
    Then, we have that $f_1^\star(T) \le 2 f_2^\star(T)\,.$
\end{fact}
\begin{proof}
    We have that:

    \begin{align}
        f_2^\star(T) \cdot T &\ge  \sum_{t = 1}^T f_2^\star(t) \\
        &\ge \sum_{t = 1}^T f_1^\star(t) \\
        &\ge \frac{f_1^\star(T) \cdot T}{2}\,.
    \end{align}
\end{proof}

Thus, we use that $\OPT_T  \ge  \frac{f_1^\star(T) \cdot T}{2}\,.$

\begin{fact}
    The maximum reward achieved in a single pull over the course of running the algorithm is at least the average reward per pull over the pulls of the algorithm, which is exactly the sum of rewards achieved by the algorithm, referred to as $ALG$ in the main body of the paper, divided by $T\,.$
\end{fact}

Putting these three facts together, we have:

\begin{enumerate}
    \item \textbf{Translating Theorem~\ref{thm:alg1approx}.} The maximum pull by the algorithm is at least $ALG/T\,,$ which by the theorem is at least $\OPT/(8 c_2 T \sqrt{k}) \ge f_1^\star(T)/(16 c_2 \sqrt{k})\,.$ Thus, the $O(\sqrt{k})$ competitive ratio still holds.
    \item \textbf{Translating Theorem~\ref{thm:generalresult}.} By the same argument above, we divide both sides of the result by $T$ and get that the maximum pull by the algorithm is at least $1/O(\sqrt{k} \log k)$ fraction of the maximum pull achievable.
    \item \textbf{Translating Lemma~\ref{lemma:unknownT}.} By the same argument as before, we divide both sides by $T$ to get that the maximum pull of the algorithm is at least $1 / (16384 \sqrt{k} \log(128 k))$ fraction of the maximum pull achievable.
\end{enumerate}

\section{Extension to noisy rewards}
\label{sec:noisyrewards}

\subsection{Extension of Algorithm~\ref{alg:rrr}}


Now, suppose that when we pull an arm, instead of getting the exact value of  $f_i(t_i)$, we get a reward value $\hat{f}_i(t_i) \in [(1-\epsilon) f_i(t_i), (1+\epsilon) f_i(t_i)]\,.$ We now wish to achieve cumulative reward that competes with the reward achieved by the best policy. The best policy is no longer necessarily a single arm, since the hat reward functions could be slightly non-monotone. However, it suffices to compare to the policy that plays the arm corresponding to the best pre-corruption arm $f^\star\,.$ This is for the following reason. Let $\pi$ be the policy that plays only $f^\star\,,$ and let $\hat{\pi}$ be the policy that achieves the optimal reward in the $\hat{f}$ problem. Define $V(\cdot)$ to be the value of the policy in the argument under the original problem and $\hat{V}(\cdot)$ be the value of the policy in the argument in the hat problem. Then:

\begin{align}
    \OPT_T = \hat{V}(\pi) &\ge (1-\epsilon) V(\pi) \quad \text{ reward at each step in hat problem at least } 1-\epsilon \text{ reward in original problem}\\
    &\ge (1-\epsilon) V(\hat{\pi}) \quad \text{ optimality of } \pi \text{ for } V\\
    &\ge \frac{1-\epsilon}{1+\epsilon} \hat{V}(\hat{\pi}) \quad \text{ reward at each step in original problem } \ge \frac{1}{1+\epsilon} \text{ of reward in hat problem} \label{eqn:valhat-pihat}
\end{align}

Thus, in the rest of this section until the end, we compare to $\hat{V}(\pi) = \OPT_T\,,$ and then finally we can lower bound that by the quantity in Eqn.~\ref{eqn:valhat-pihat}.

We show that the same algorithm works with a minor modification. In particular, instead of switching arms away from arm $i$ when $f_i(t_i) < m \cdot t_i/T\,,$ we now switch away from arm $i$ when $\hat{f}_i(t_i) < (1-\epsilon) m t_i / T\,.$ Note that the algorithm therefore must have knowledge of the value of $\epsilon\,.$

We must show two things: first, that as before, we never switch off the best arm; second, that similar to before, playing an arm for time $t_i$ accrues at least a constant fraction of $\lbt{t_i-1}\,.$ The first can be seen as follows: for an arm $f_i\,,$ as long as $f_i(t_i) \ge t_i \, m/T\,,$ we have $\hat{f}_i(t_i) \ge (1-\epsilon) f_i(t_i) \ge (1-\epsilon) t_i \, m/T\,,$ so we will continue playing that arm. In particular, this holds for the best arm.


For the second, we prove a variant of Claim~\ref{cl:armreward}:
\newcommand{\epsfactor}{\frac{1-\epsilon}{1+\epsilon}}
\begin{claim}
    Suppose $m \in [\frac{1}{c_2} f^\star(T), f^\star(T)]\,.$ In the noisy case, if arm $i$ is played by Algorithm~\ref{alg:rrr} with the adaptation mentioned above for $t_i$ steps, then the algorithm receives total reward at least $\lbt{t_i - 1} \cdot \epsfactor$ in those steps.
\end{claim}

\begin{proof}
    We know that arm $i$ is played until $\hat{f}_i(t_i) < (1-\epsilon) m t_i / T\,.$ This means the total reward from this arm is:

\begin{align}
    \sum_{\tau = 1}^{t_i - 1} \hat{f}_i(t_i) &\ge \frac{(1-\epsilon) \, m}{T} \sum_{\tau = 1}^{t_i - 1} \tau \\
    &\ge \frac{(1-\epsilon) \, m}{T} \frac{(t_i - 1)^2}{2}\,.
\end{align}

Since $\OPT_T \le \hat{f}^\star(T) \, T \le (1+\epsilon)f^\star(T)\,T \le (1+\epsilon) c_2 \, m \, T\,,$ 
the right hand side above is at least $\frac{1-\epsilon}{1+\epsilon} \, \lbt{t_i - 1}\,.$
\end{proof}

Finally, we must incorporate this into the recursion of Lemma~\ref{lem:recurrence} and then change Lemma~\ref{lemma:recur-value} slightly. 
In particular:

\begin{align}
        V(T', k') 
        &\ge \frac {1}{k'} \OPT_{T'+k'} + \left( 1 - \frac{1}{k'} \right) \min_{0 \le t \le T'+k'-1} \left\{ \epsfactor \lbt{t} + V(T'-t, k'-1)  \right\}
        \label{eqn:noisyrecurrence}
    \end{align}
where we define $V(T',k')=0$ for $T' \leq 0\,, \forall \, k'$. 

Finally, we aim now to show that $V(T', k') \ge \epsfactor \lbt{T'} \, \frac{1}{\gka{k'}}\,.$ The base cases go through as before, since $\epsfactor$ times the lower bound is only smaller than the lower bound for which we have shown it. Further, let the inductive assumption be $V(T'', k'') \ge \epsfactor \lbt{T''} \, \frac{1}{\gka{k''}}\,, \forall \, T'' < T'\,, k'' < k'\,.$ Now, we can lower bound the right hand side of Eqn.~\ref{eqn:noisyrecurrence} by:
$$
\epsfactor \left( \frac{1}{k'} \OPT_{T'} + \left( 1 - \frac{1}{k'} \right) \min_t \left\{ \lbt{t} + \frac{1}{\gka{k'}} \lbt{T'-t}\right\}  \right)\,.
$$

Now, we can analyze everything multiplied by $(1-\epsilon)$ exactly as before, and we would prove the following statement:

\begin{theorem}
    Suppose that when we pull an arm $i$ for the $t_i$th time, instead of getting value $f_i(t_i)\,,$ we get $\hat{f}_i(t_i) \in (1\pm \epsilon) f_i(t_i)\,.$ Suppose we run the adapted version of Algorithm~\ref{alg:rrr} as discussed above with a parameter $m \in [\frac{1}{c_2} f^*(T), f^*(T)]$.
    Then the recurrence given in Eqn.~\ref{eqn:noisyrecurrence} evaluates as follows: for all $T'\,, V(T', k') \ge \epsfactor \lbt{T'} \cdot \frac{1}{\gka{k'}}\,.$ In particular, $V(T, k) \ge \epsfactor \frac{\OPT_T}{2c_2 \cdot \gk} \ge \left(\epsfactor\right)^2 \frac{\hat{V}(\hat{\pi})}{2c_2 \cdot \gk}\,.$ 
\end{theorem}

Finally, we can incorporate the extra switching steps to get a theorem like before with the same $\left(\epsfactor\right)^2$ multiplicative factor.

\subsection{Extension of Algorithm~\ref{alg:get-mhat}}

Now, we define:
$$
\hat{m}_L = \frac{\hat{f}\left
(\frac{T}{2k}\right)}{1+\epsilon}\, \qquad \hat{m}_U = 2k \, \cdot \, \frac{\hat{f}\left
(\frac{T}{2k}\right)}{1-\epsilon}\,.
$$  

It is clear by our assumptions on $\hat{f}$ and diminishing returns that $f(T)$ lies between $\hat{m}_L$ and $\hat{m}_U\,.$ Further, it is clear that $\hat{m}_U \le 2k \, \frac{1+\epsilon}{1-\epsilon} \hat{m}_L\,.$ The rest of the arguments go through, and now the approximation factor is $O\left(\left( \epsfactor\right)^2\sqrt{k} \log\left(  \frac{1+\epsilon}{1-\epsilon} \, k\right)\right)\,.$

Note that this definition of $\hat{m}_U$ could have also been used in the noiseless case with the same guarantee. However, in practice we may get a better interval if we do what we originally had in the algorithm.

\chapter{Appendices for Algorithm Design for Improving Multi-Armed Bandits}
\section{Additional Related Work}\label{appendix:additional-related}
\textit{Improving (Rising) Bandits.} Initially, \cite{heidari_tight_nodate} formulated the improving (and decaying) bandits problem where payoffs increase the more the arm is played, and obtain sub-linear policy regret. \cite{patil_mitigating_2023} show that a linear regret is unavoidable in the worst-case and obtain a tight $\Theta(k)$ on the competitive ratio of deterministic algorithms. 
Then, \cite{blum_nearly-tight_2024} extend the development on this problem to randomized algorithms, 
achieving an $O(\sqrt{k} \log k)$ upper bound, nearly matching their $\Omega(\sqrt{k})$ lower bound. A related line of work is that on (deterministic, rested) {\em rising} bandits \cite{metelli_stochastic_2022,mussi2024best}. This line of work bounds the instance-dependent policy regret on benign instances but otherwise considers similar increasing reward functions. We note that policy regret is a stronger notion of regret than {\em external regret}, where an algorithm is compared to the best action on {\em the same} set of rewards. In policy regret (and in the competitive ratios we consider), an algorithm is compared to the best policy for the sequence of rewards generated by {\em playing that policy}.\looseness-1

\textit{Data-Driven Algorithm Design.}  Data-driven algorithm design is emerging as a powerful approach for using machine learning to design algorithms that have strong performance on “typical” input instances, as opposed to worst-case or average-case analysis~\cite{gupta2016pac,balcan2017learning,balcan2018dispersion}. The approach is known to be effective in both statistical and online learning settings \cite{balcan2020data}, and, under this paradigm, a growing line of research has successfully developed techniques for designing several fundamental  algorithms in machine learning and beyond (e.g.~\cite{blum2021learning,bartlett2022generalization,jinsample,khodak2024learning,balcan2024trees,cheng2024learning,sakaue2024generalization,sharma2024no,balcanalgorithm}). Recent work \cite{sharma2025offline} has shown how to tune hyperparameters in standard multi-armed bandit algorithms like UCB, LinUCB and GP-UCB under this paradigm. While they focus on stochastic bandits, we extend their techniques to tune hyperparameters in the non-stochastic improving bandits setting. Another related work \cite{khodak2023meta} shows how to tune hyperparameters for adversarial bandit algorithms in an online-with-online meta-learning setting. While we focus on statistical complexity of tuning our bandit algorithms, an interesting future direction is to give computationally efficient algorithms~\cite{balcan2024accelerating,chatziafratisaccelerating}.\looseness-1

\section{Full Proofs for Sharper Competitive Ratio}
\label{appendix:competitive-ratio}
In this Appendix, we give full proofs of our optimal sharper competitive ratio under for each value of the Concavity Envelope Exponent $\beta_I$ satisfied by the instance $I$, along with a formal proof of the ``doubling trick'' applied in our context for completeness. Our analysis extends the techniques due to \cite{blum_nearly-tight_2024}.
\subsection{Upper Bound} \label{appendix:upperbd-proof}
Suppose an instance has Concavity Envelope Exponent $\beta_I \in (0,1)$,  and fix $\alpha \in (\beta_I, 1)$. Let $\gamma := \frac{\alpha}{\alpha+1}$.
For $\tau'\ge0$ and $k'\ge1$, let $V(\tau',k')$ denote the expected reward earned in the next $\tau'+k'$ pulls when $k'$ arms are still untouched. Let $V(\tau'',\cdots)=0$ for all $\tau''\le0$.

We analyze $\textit{PTRR}_\alpha$ through a recurrence that trades off two properties. First, the optimal arm is never abandoned when $\alpha > \beta_I$ (Lemma~\ref{lem:never-drop}). Second, abandoning any non‑optimal arm after $t$ pulls yields at least the “area under the envelope” up to $t$ (Lemma~\ref{lem:area}). These two facts give the value recurrence in Lemma~\ref{lem:rec}: with probability $1/k'$ we hit $f^*$ and earn $\mathrm{OPT}_{\tau'+k'}$; otherwise we bank the area from time $t$ and recurse from $(\tau'-t,k'-1)$ with the minimizer attained on $[0,\tau']$. We then solve the recurrence by normalizing $t$ to $y=t/\tau'$ and minimizing a one‑variable convex form $u\,y^{\alpha+1}+v(1-y)^{\alpha+1}$ exactly (Lemma~\ref{lem:exact-min}). A clean lower bound on this minimum (Lemma~\ref{lem:balance}) closes the induction and yields
$$
V(\tau',k') \ge \frac{m}{(\alpha+1)\tau^\alpha}\cdot \frac{(\tau')^{\alpha+1}}{2\,(k'+1)^\gamma}
$$
(Lemma~\ref{lem:solve}). Finally, substituting $(\tau',k')=(T-k,k)$ and $\mathrm{OPT}_T$’s comparison to $m$ gives Theorem~\ref{thm:upper}. A more detailed overview of the proof is included after the theorem.

Our analysis of $\mathit{PTRR}_\alpha$ holds for any parameter $m$ that satisfies
$$\frac{1}{c_2} f^*(T-k) \leq m \leq f^*(T) \left( \frac{T-k}{T} \right)^\alpha$$ for some constant $c_2 \geq 1$. In particular, setting $m := \frac{T - k}{T} f^{\star}(T)$ gives $c_2 = \frac{T}{T - k} \in [1,2]$ when $T \ge 2k$. Since $\frac{T-k}{T} \in (0,1]$ and $\alpha \in (0,1)$, we also know that $f^{\star}(T) \frac{T - k}{T} \leq f^*(T) \left(\frac{T-k}{T}\right)^\alpha$, and therefore that $\frac{1}{2} f^{\star}(T - k) \le m \le f^*(T) \left( \frac{T-k}{T} \right)^\alpha$ in this case. 

\begin{lemma}[Optimal arm is never dropped]\label{lem:never-drop}
$\textit{PTRR}_\alpha$ will never switch away from the optimal arm $f^*$. 
\end{lemma}
\begin{proof}
By definition of the CEE, we know that
$
f_{\star}(t) \ge\ f_{\star}(T)\left( \frac{t}{T} \right)^{\beta_I}
$
for all $0\le t\le T$. Moreover, recall that $m \leq f^*(T) \left( \frac{\tau}{T}\right)^\alpha$. Since $\frac{\tau}{T} \in (0,1]$ and $\alpha > \beta_I$, it follows that
$$
f^{\star}(t) \ge f^{\star}(T) \left( \frac{t}{T} \right)^{\beta_I}
\ge f^{\star}(T) \left( \frac{t}{T} \right)^\alpha = f^{\star}(T) \left( \frac{\tau}{T} \right)^{\alpha} \left( \frac{t}{\tau} \right)^{\alpha}
\ge  m \left( \frac{t}{\tau} \right)^{\alpha}.
$$
for all $t \le T$, and therefore that the test never fails once the optimal arm is reached.
\end{proof}

\begin{lemma}[Area before abandonment]\label{lem:area}
If the keep-test holds for $t_i-1$ pulls on arm $i$ and fails at $t_i$, we obtain a reward of at least
$$
 \frac{m}{(\alpha+1)\tau^\alpha}\,(t_i-1)^{\alpha+1}.
$$
\end{lemma}
\begin{proof}
For $1\le s<t_i$, we know that $f_i(s)\geq m(s/\tau)^\alpha$. Summing and using the monotonicity of $x^\alpha$ on $[0,\infty)$ gives
$$
 \sum_{s=1}^{t_i-1} f_i(s)\ \ge\ m\tau^{-\alpha}\sum_{s=1}^{t_i-1}s^\alpha \ \ge\ m\tau^{-\alpha}\int_{0}^{t_i-1} x^\alpha\,dx \ \geq \ \frac{m}{(\alpha+1)\tau^\alpha}\,(t_i-1)^{\alpha+1}.
$$
\end{proof}

\noindent With $k'$ untouched arms, we either hit the best arm now (probability $1/k'$) and then take the optimum path, or we spend $t$ pulls on a non-best arm, bank the area from Lemma \ref{lem:area}, and recurse on $(\tau' - t, k' - 1)$. To ensure a worst-case guarantee, we take the minimum over all feasible abandonment times $t$.

\begin{lemma}[Value recurrence]\label{lem:rec}
For all $\tau'\ge0$, $k'\ge1$,
\begin{equation}
V(\tau',k')\ \ge\ \frac{1}{k'}\,\mathrm{OPT}_{\tau'+k'} +\Bigl(1-\frac{1}{k'}\Bigr)\min_{0\leq t\leq \tau'+k'-1}\!\left[\frac{m}{(\alpha+1)\tau^\alpha}\,t^{\alpha+1}+V(\tau'-t,k'-1)\right]\!, \label{eq:rec}
\end{equation}
and the minimum is attained on $[0,\tau']$.
\end{lemma}
\begin{proof}
With probability $1/k'$ the next arm is $f^*$ and, by Lemma \ref{lem:never-drop}, the algorithm stays on it and earns $\mathrm{OPT}_{\tau'+k'}$. Else it plays a non-optimal arm. If it abandons this arm at time $t$, Lemma~\ref{lem:area} gives the earned area and the process recurses on $(\tau'-t,k'-1)$. If $t\geq \tau'$, then $V(\tau'-t,k'-1)=0$ while the area term increases in $t$, which implies that the minimizer lies in $[0,\tau']$.
\end{proof}

\noindent Below, we prove two technical results that aid our analysis of the recurrence. First, the recurrence reduces to minimizing $u\, y^p + v(1 - y)^p$ over $y \in [0,1]$. This convex function has a unique minimizer balancing the two terms. We write that minimum in closed form to keep constants explicit.

\begin{lemma}[One-variable minimum]\label{lem:exact-min}
For $p>1$ and $u,v>0$,
$$
 \min_{y\in[0,1]}\ \{\,u\,y^p+v(1-y)^p\,\} \ =\ \bigl(u^{-1/(p-1)}+v^{-1/(p-1)}\bigr)^{-(p-1)}.
$$
\end{lemma}
\begin{proof}
Let $f(y):=u\,y^p+v(1-y)^p$ on $[0,1]$. For $p>1$, we have
$$
 f''(y)=u\,p(p\!-\!1)y^{p-2}+v\,p(p\!-\!1)(1-y)^{p-2}>0\quad(y\in(0,1)),
$$
so $f$ is strictly convex. Moreover, note that
$$
 f'(y)=u p\,y^{p-1}-v p\,(1-y)^{p-1},\quad f'(0^+)=-vp<0,\ \ f'(1^-)=up>0,
$$
which implies that $f'$ has a unique global minimizer $y^\star\in(0,1)$. Solving $u\,{y^\star}^{p-1}=v\,(1-y^\star)^{p-1}$ gives
$$
 y^\star=\frac{v^{1/(p-1)}}{u^{1/(p-1)}+v^{1/(p-1)}}.
$$
Letting $a:=u^{1/(p-1)}$ and $b:=v^{1/(p-1)}$, it follows that
$$
 \min_{y\in[0,1]} f(y)=f(y^\star)=\frac{a^{p-1}b^p+b^{p-1}a^p}{(a+b)^p} =\frac{(ab)^{p-1}}{(a+b)^{p-1}} =\bigl(u^{-1/(p-1)}+v^{-1/(p-1)}\bigr)^{-(p-1)}.
$$
\end{proof}

\noindent We now lower-bound the exact minimum with a simple balancing inequality that cleanly shows the dependence on $k'$.

\begin{lemma}[Balancing inequality]\label{lem:balance}
Let $\alpha\in(0,1]$ and $\gamma:=\frac{\alpha}{\alpha+1}$. For all integers $k'\ge1$,
\begin{equation}\label{eq:balance-half}
 \frac{1}{k'}+\Bigl(1-\frac{1}{k'}\Bigr)\Bigl(1+\bigl(2\,k'^{\gamma}\bigr)^{1/\alpha}\Bigr)^{-\alpha} \ \ge\ \frac{1}{2\,(k'+1)^{\gamma}}.
\end{equation}
\end{lemma}
\begin{proof}
Let $A:=(2\,k'^{\gamma})^{1/\alpha}\ge0$. Note that the function $u\mapsto (1+u)^{-\alpha}$ is convex and decreasing on $[0,\infty)$ for every $\alpha>0$, which implies that
$$
 (1+A)^{-\alpha} =A^{-\alpha}\Bigl(1+\frac{1}{A}\Bigr)^{-\alpha} \ \ge\ A^{-\alpha}\Bigl(1-\frac{\alpha}{A}\Bigr) =\frac{1}{2\,k'^{\gamma}}-\frac{\alpha}{2^{1+1/\alpha}\,k'}
$$
for $A>0$. Substituting this lower bound into the left-hand side gives
$$
 \begin{aligned}
 \frac{1}{k'}+\Bigl(1-\frac{1}{k'}\Bigr)(1+A)^{-\alpha} &\geq \frac{1}{k'}+\Bigl(1-\frac{1}{k'}\Bigr)\!\left(\frac{1}{2\,k'^{\gamma}}-\frac{\alpha}{2^{1+1/\alpha}\,k'}\right)\\
 &= \frac{1}{2\,k'^{\gamma}}+\frac{1}{k'}-\frac{1}{2\,k'^{\gamma+1}} -\frac{\alpha}{2^{1+1/\alpha}\,k'}+\frac{\alpha}{2^{1+1/\alpha}\,k'^2}.
 \end{aligned}
$$
Since $k'^{\gamma}\ge1$, we know that $-\frac{1}{2\,k'^{\gamma+1}}\geq -\frac{1}{2k'}$ and $\frac{\alpha}{2^{1+1/\alpha}\,k'^2}\ge0$. Moreover, $2^{-1/\alpha}\le\frac12$ for $\alpha\in(0,1]$, which implies that $\frac{\alpha}{2^{1+1/\alpha}}\le\frac{\alpha}{2}$, and therefore that
$$
 \frac{1}{k'}-\frac{1}{2\,k'^{\gamma+1}}-\frac{\alpha}{2^{1+1/\alpha}\,k'} \ \ge\ \frac{1}{k'}-\frac{1}{2k'}-\frac{\alpha}{2k'} \ =\ \frac{1-\alpha}{2k'}\ \ge\ 0
$$
(with equality only when $\alpha=1$). Combining these estimates gives
$$
 \frac{1}{k'}+\Bigl(1-\frac{1}{k'}\Bigr)(1+A)^{-\alpha} \ \ge\ \frac{1}{2\,k'^{\gamma}}+\frac{1-\alpha}{2k'}\ \ge\ \frac{1}{2\,k'^{\gamma}}.
$$
Finally, $(k'+1)^\gamma\geq k'^\gamma$ implies that $\frac{1}{2\,k'^{\gamma}}\geq \frac{1}{2\,(k'+1)^\gamma}$, proving \eqref{eq:balance-half}.
\end{proof}

\noindent We can now prove the master lower bound on $V(\tau', k')$ by induction on $(k', \tau')$. The base cases are immediate. The inductive step plugs the one-variable minimum into the recurrence and then uses the balancing inequality.

\begin{lemma}[Solving the recurrence]\label{lem:solve}
For all $\tau'\ge0$ and $k'\ge1$,
\begin{equation}\label{eq:IH-half}
 V(\tau',k')\ \ge\ \frac{m}{(\alpha+1)\tau^\alpha}\ \frac{(\tau')^{\alpha+1}}{2\,(k'+1)^{\gamma}}.
\end{equation}
\end{lemma}
\begin{proof}
We prove this by induction on $(k',\tau')$. \textit{Base cases.} If $\tau'=0$ there is nothing to show. If $k'=1$, then we have
$$
 V(\tau',1)=\mathrm{OPT}_{\tau'}\ \ge\ \sum_{s=1}^{\tau'} m\Bigl(\frac{s}{\tau}\Bigr)^\alpha \ \ge\ \frac{m}{(\alpha+1)\tau^\alpha}(\tau')^{\alpha+1} \ \ge\ \frac{m}{(\alpha+1)\tau^\alpha}\ \frac{(\tau')^{\alpha+1}}{2\cdot 2^\gamma},
$$
as $2\cdot 2^\gamma\geq 1$. \textit{Inductive step.} Now assume \eqref{eq:IH-half} holds for all $(k''\!,\tau'')$ with $k''<k'$ and $0 \leq \tau'' \leq \tau'$. From \eqref{eq:rec}, we know that
$$
 V(\tau',k')\ \ge\ \frac{1}{k'}\,\mathrm{OPT}_{\tau'+k'}+\Bigl(1-\frac{1}{k'}\Bigr)\min_{0\leq t\leq \tau'}\!\left[\frac{m}{(\alpha+1)\tau^\alpha}\,t^{\alpha+1}+V(\tau'-t,k'-1)\right]\!.
$$

Using CEE and $m \leq f^\star(T)(\tau/T)^\alpha$, recall that we have
$$
f^\star(s) \geq f^\star(T) \left( \frac{s}{T} \right)^{\beta_I} \geq f^\star(T) \left( \frac{s}{T} \right)^\alpha = f^\star(T) \left( \frac{\tau}{T} \right)^\alpha \left( \frac{s}{\tau} \right)^\alpha \geq m \left( \frac{s}{\tau} \right)^\alpha
$$
for all $s \leq T$, which implies that
$$
 \frac{1}{k'}\,\mathrm{OPT}_{\tau'+k'}\ \ge\ \frac{1}{k'}\sum_{s=1}^{\tau'} m\Bigl(\frac{s}{\tau}\Bigr)^\alpha\ \geq \ \frac{m}{(\alpha+1)\tau^\alpha}\,\frac{(\tau')^{\alpha+1}}{k'}.
$$
Applying the inductive hypothesis to $V(\tau'-t,k'-1)$ and letting $y:=t/\tau'\in[0,1]$ gives
$$
 \begin{aligned}
 V(\tau',k')&\geq \frac{m}{(\alpha+1)\tau^\alpha}\,(\tau')^{\alpha+1}\left[\frac{1}{k'}+\Bigl(1-\frac{1}{k'}\Bigr)\min_{y\in[0,1]}\Bigl\{y^{\alpha+1}+\frac{1}{2\,k'^{\gamma}}(1-y)^{\alpha+1}\Bigr\}\right]\!.
 \end{aligned}
$$
By Lemma~\ref{lem:exact-min}, the minimum equals $\bigl(1+(2k'^{\gamma})^{1/\alpha}\bigr)^{-\alpha}$. Lemma~\ref{lem:balance} then yields \eqref{eq:IH-half}.
\end{proof}

\noindent Finally, we apply Lemma \ref{lem:solve} with $(\tau', k') = (T - k, k)$ to obtain the desired competitive ratio.

\begin{theorem}[Upper bound]\label{thm:upper}
Assume $T\geq 2k$. If we run $PTRR_\alpha$ with $\tau = T-k$ and $m := \frac{T - k}{T} f^{\star}(T)$, we obtain
$$
 \mathbb{E}[\mathrm{reward}] \ge \frac{1}{2^{\alpha + 3}(\alpha + 1)} \cdot \frac{\mathrm{OPT}_T}{(k + 1)^{\gamma}},
$$ where $\gamma=\frac{\alpha}{\alpha+1}$. 
The competitive ratio is $O(k^{\alpha / (\alpha + 1)})$.
\end{theorem}
\begin{proof} ({\bf Overview.})
Two simple properties drive our analysis. First, the optimal arm is never abandoned when $\alpha > \beta_I$. Once we encounter $f^*$, we stay on it forever. Second, if we ever abandon a non‑optimal arm at time $t$, the cumulative reward collected on that arm is at least the “area” under the threshold up to $t$
$$
 \sum_{s=1}^{t-1} f_i(s) \ge \frac{m}{(\alpha + 1) \tau^\alpha}(t - 1)^{\alpha + 1}.
$$
\noindent These two facts yield a value recurrence. Let $V(\tau', k')$ be the expected reward from a state with $k'$ untouched arms and $\tau'$ `free'' pulls left. With probability $1/k'$ the next random pick is $f^*$, after which the algorithm earns $\mathrm{OPT}_{\tau' + k'}$. Otherwise it spends $t$ pulls on a bad arm, banks the area above, and recurses on $(\tau' - t, k' - 1)$. Minimizing over feasible $t$ gives
$$
 V(\tau', k') \ge \frac{1}{k'} \mathrm{OPT}_{\tau' + k'} + \left(1 - \frac{1}{k'}\right) \min_{t \in [0, \tau']} \left\{ \frac{m}{(\alpha + 1) \tau^\alpha} t^{\alpha + 1} + V(\tau' - t, k' - 1) \right\}.
$$
Induct on $(\tau',k')$. The recurrence reduces to a one-variable balance
$$
 \min_{y \in [0,1]} \left\{ u\, y^{\alpha+1} + v(1 - y)^{\alpha+1} \right\},
$$
which yields
$$
 V(\tau',k') \geq \frac{m}{(\alpha+1)\tau^{\alpha}} \cdot \frac{(\tau')^{\alpha+1}}{2(k'+1)^{\alpha/(\alpha+1)}}.
$$
Finally, plug in $(\tau, k)$ with $\tau=T-k$, use $T \geq 2k$ so $\tau \ge T/2$, and upper-bound $\text{OPT}_T \le c_2 \,m\,(T/\tau)^{\alpha}T$ to obtain the desired result.

({\bf Full proof of final step.}) Applying Lemma~\ref{lem:solve} with $(\tau',k')=(T-k,k)$ (where the $k$ extra time accounts for the ``switching'' steps when the keep-test first fails) and $T \geq 2k$ gives
$$
 \mathbb{E}[\mathrm{reward}]\ \ge\ \frac{m}{(\alpha+1)\tau^\alpha}\cdot \frac{\tau^{\alpha+1}}{2\,(k+1)^\gamma} \ =\ \frac{m\,\tau}{2(\alpha+1)}\cdot \frac{1}{(k+1)^\gamma}.
$$
Since $\mathrm{OPT}_T\leq f^*(T) \cdot T \le c_2\,m\,(T/\tau)^{\beta_I}T \leq c_2\,m\,(T/\tau)^{\alpha}T$ (by the CEE), we know that $m \tau \ge \frac{(\tau / T)^{1 + \alpha}}{c_2} \, \text{OPT}_T$, and therefore that
$$
 \mathbb{E}[\mathrm{reward}] \ \ge\ \frac{1}{2(\alpha+1)c_2}\left(\frac{\tau}{T}\right)^{1+\alpha}\cdot \frac{\mathrm{OPT}_T}{(k+1)^\gamma} \geq \frac{1}{2^{\alpha + 3}(\alpha + 1)} \cdot \frac{\mathrm{OPT}_T}{(k + 1)^{\gamma}},
$$
as desired.
\end{proof}

\subsection{Lower Bound}
\label{appendix:lowerbd-proof}
To prove a matching lower bound, we construct hard a family of instances with $\beta_I = \beta$ on which every algorithm has
expected approximation factor at least on the order of $k^{\beta/(\beta+1)}$. The idea is straightforward and mirrors the construction in \cite{blum_nearly-tight_2024}: one arm $g$ is defined as $m (t/T)^\beta$,
while the others all match it up to a time $s$ and then flatten so that no deterministic sequence of pulls can safely discard one before investing $s$ pulls. We let $\mathcal{D}_s$ denote the distribution that picks one arm uniformly at random to be $g$. Lemma~\ref{lem:PF-index} shows every instance drawn from $\mathcal{D}_s$ has $\beta_I=\beta$. For any deterministic schedule, Lemma~\ref{lem:generous} upper-bounds the expected reward by balancing $\mathrm{OPT}_T$ with the flat value $Tg(s)$. Lemma~\ref{lem:opt-lb} gives $\mathrm{OPT}_T\ge mT/(\beta+1)$. Combining yields Lemma~\ref{lem:ratio}: with $x:=s/T$, 
$$
\frac{\mathbb{E}[\mathrm{ALG}_T]}{\mathrm{OPT}_T}\ \le\ h(x):=\frac{1}{k x}+(\beta+1)x^\beta.
$$
Lemma~\ref{lem:xstar} minimizes $h$ at $x^\star=[k\beta(\beta+1)]^{-1/(\beta+1)}$ and gives $h(x^\star)=\Theta\!\big(k^{-\beta/(\beta+1)}\big)$. Setting $s=\lfloor x^\star T\rfloor$ and assuming $T\ge 2/x^\star$, Lemma~\ref{lem:round} controls discretization within a constant factor. This proves Theorem~\ref{thm:LB} for deterministic schedules. Yao’s principle extends it to randomized algorithms with the same exponent. As before, a more detailed overview of the proof is included after the theorem.

\paragraph{Hard family.}
Fix $k\ge2$, $T\ge1$, and $m>0$. For $s\in\{1,\dots,T\}$ define a good arm as
$$ 
g(t):=m(t/T)^\beta,\quad t=1,\dots,T.
$$
Define for this $s$ a bad arm by
$$
f_{\mathrm{bad}}(t)=
\begin{cases}
g(t),&t\le s,\\
g(s),&t>s.
\end{cases}
$$

\noindent Let $\mathcal{D}_s$ be the distribution that chooses one arm uniformly at random to be $g$ and
sets all other arms to $f_{\mathrm{bad}}$.

\begin{lemma}[Membership and Concavity Envelope Exponent]\label{lem:PF-index}
Every instance drawn from $\mathcal{D}_s$ has $\beta_I=\beta$.
\end{lemma}

\begin{proof}
For $g$, we know that LE$(\beta)$ holds with equality. Since $$f_{\mathrm{bad}}(T)=g(s)\le g(T)=m,$$
we likewise know that $$f_{\mathrm{bad}}(t)\ge g(s)(t/T)^\beta=f_{\mathrm{bad}}(T)(t/T)^\beta$$ for all $t\le T$ and every $f_{\mathrm{bad}}$.
It follows that LE$(\beta)$ holds for each arm, and therefore that $\beta_I\le\beta$. 

Now suppose for the sake of contradiction that $\beta'<\beta$. Note that
$$m(t/T)^\beta< m(t/T)^{\beta'}=g(T)(t/T)^{\beta'},$$ which implies that $g$ violates LE$(\beta')$. It follows that $\beta_I=\beta$.
\end{proof}

To show that no algorithm can outperform $\textit{PTRR}_\beta$ on this distribution, we will consider the following `generous game.' Fix a deterministic sequence $S$ of $T$ arm pulls. Let $A$ denote the number of distinct arms that receive at least $s$ pulls under $S$, and note that $A \leq \lfloor\frac{T}{s}\rfloor.$ Draw an arbitrary instance from $\mathcal D_s$, play $S$ on this instance, and only afterwards reveal which arm was good. If $S$ gave arm $g$ at least $s$ pulls, pay out a reward of $\mathrm{OPT}_T$. Else provide the actual reward that $S$ obtained. Note that paying $\mathrm{OPT}_T$ in certain cases can only increase the payoff of the sequence, which implies that the payoff of this game upper-bounds the expected reward $\mathbb{E}_{\mathcal{D}_s} [\mathrm{ALG}_T]$.

\begin{lemma}[Generous evaluation]\label{lem:generous}
For any deterministic algorithm $\mathrm{ALG}$, let 
$S$ be the sequence played by $\mathrm{ALG}$ when all arms are $f_\text{bad}$, and let $A$ be the number of arms that receive at least $s$ pulls under $S$.
Then 
\begin{equation}\label{eq:gen}
\mathbb{E}_{\mathcal{D}_s}[\mathrm{ALG}_T]
\ \le\ \frac{A}{k}\,\mathrm{OPT}_T\ +\ \Bigl(1-\frac{A}{k}\Bigr)\,T\,g(s)
\ \le\ \frac{T}{k s}\,\mathrm{OPT}_T\ +\ T\,m\Bigl(\frac{s}{T}\Bigr)^\beta.
\end{equation}
\end{lemma}

\begin{proof}
Draw an instance from $\mathcal{D}_s$ and run $S$. With probability $A/k$, the good arm lies among those $A$ arms, and the reward is at most $\mathrm{OPT}_T$. Otherwise the good arm is pulled $<s$ times, which implies that any pull is worth at most $g(s)$, and therefore that the realized reward is at most $T \cdot g(s) = T \cdot m(s/T)^\beta$. Taking expectations gives the desired reward.
\end{proof}

\noindent We now bound $\text{OPT}_T$.
\begin{lemma}[Lower bound on the benchmark]\label{lem:opt-lb}
$$\mathrm{OPT}_T\ \ge\ \sum_{t=1}^{T} g(t)\ \ge\ \dfrac{mT}{\beta+1}.$$
\end{lemma}

\begin{proof}
Note that $$\mathrm{OPT}_T = \sum_{t=1}^{T} g(t)= m\sum_{t=1}^{T}(t/T)^{\beta} \ge mT^{-\beta}\!\int_{0}^{T} x^{\beta}\,dx = mT/(\beta+1),$$
as $x^{\beta}$ is non-decreasing on \([0,T]\) for \(\beta>0\).
\end{proof}

\noindent Combining the earlier generous bound with this lower bound gives a clean formula for the approximation ratio as a function of $x=s/T.$

\begin{lemma}[Ratio at a given $s$]\label{lem:ratio}
For any deterministic algorithm $\mathrm{ALG}$ and $x:=s/T\in\{1/T,\dots,1\}$,
$$
\frac{\mathbb{E}_{\mathcal{D}_s}[\mathrm{ALG}_T]}{\mathrm{OPT}_T}
\ \le\ h(x):=\frac{1}{k x}+(\beta+1)\,x^\beta.
$$
\end{lemma}

\begin{proof}
Divide \eqref{eq:gen} by Lemma~\ref{lem:opt-lb}.
\end{proof}

\noindent Simple calculus gives a unique minimizer $x^*$ and the exact value $h(x^*)$.

\begin{lemma}[Continuous minimizer]\label{lem:xstar}
$h$ has a unique minimizer on $(0,1]$ at
$$
x^\star\ :=\ \bigl[k\,\beta(\beta+1)\bigr]^{-1/(\beta+1)}\,,
$$
and
$$
h(x^\star)\ =\ (\beta+1)^2\,[\beta(\beta+1)]^{-\beta/(\beta+1)}\,k^{-\beta/(\beta+1)}.
$$
\end{lemma}

\begin{proof}
First, note that $$h'(x)=-1/(k x^2)+(\beta+1)\beta x^{\beta-1} = \frac{(\beta+1)\beta\,x^{\beta+1}-\tfrac1k}{x^{2}}$$ is strictly increasing in $x>0$, which implies that it has a unique minimum $$
(\beta+1)\beta\,(x^\star)^{\beta+1}=\frac1k
\quad\Longrightarrow\quad
x^\star=\big[k\,\beta(\beta+1)\big]^{-1/(\beta+1)}.
$$

\noindent At \(x^\star\), we use the first‑order condition to get
$$
h(x^\star)=\frac{1}{k x^\star}+(\beta+1)(x^\star)^\beta
=(\beta+1)\beta(x^\star)^\beta+(\beta+1)(x^\star)^\beta
=(\beta+1)^2(x^\star)^\beta.
$$
Since \((x^\star)^\beta=[k\,\beta(\beta+1)]^{-\beta/(\beta+1)}\), it follows that
\[
h(x^\star)=(\beta+1)^2\,[\beta(\beta+1)]^{-\beta/(\beta+1)}\,k^{-\beta/(\beta+1)}.
\]

\end{proof}

\noindent Recall that $s$ must be an integer. Following \cite{blum_nearly-tight_2024}, we take $s = \lfloor x^* T \rfloor$ and require $x^* T \geq 2$. This simple condition ensures $s/T$ lies between $x^*/2$ and $x^*$, inflating $h$ by at most a constant factor.
\begin{lemma}[Rounding]\label{lem:round}
Let $s:=\lfloor x^\star T\rfloor$. If $x^\star T\ge2$, then $s/T\in[x^\star/2,\,x^\star]$ and
$$
h\!\Bigl(\frac{s}{T}\Bigr)\ \le\ \frac{3}{2}\,h(x^\star).
$$
\end{lemma}

\begin{proof}
Since $x^\star T\ge2$, we know that $s/T\ge (x^\star T-1)/T\ge x^\star/2$, and trivially $s/T\le x^\star$.
Let $s/T=a x^\star$ with $a\in[1/2,1]$. Using $1/(k x^\star)=(\beta+1)\beta (x^\star)^\beta$, it follows that
$$
\frac{h(a x^\star)}{h(x^\star)}
=\frac{\beta/a + a^\beta}{\beta+1}
\ \le\ \frac{2\beta + 1}{\beta+1}
\ \le\ \frac{3}{2},
$$
since $1/a\le2$ and $a^\beta\le1$ for $a\in[1/2,1]$, $\beta\in(0,1]$.
\end{proof}

\noindent We now plug our choice of $s$ into the ratio bound. Because the bound holds for every deterministic schedule against $D_s$, it also holds for any randomized algorithm by linearity of expectation.
\begin{theorem}[Lower Bound]\label{thm:LB}
Fix $\beta\in(0,1]$ and $k\ge2$. Suppose
$
T\ \ge\ \frac{2}{x^\star}\ =\ 2\,[\beta(\beta+1)]^{\frac{1}{\beta+1}}\,k^{\frac{1}{\beta+1}}.
$
Then there exists a distribution on instances with $\beta_I=\beta$ such that for every (possibly randomized) algorithm,
$$
\frac{\mathbb{E}[\mathrm{ALG}_T]}{\mathrm{OPT}_T}
\ \le\ \frac{3}{2}\,(\beta+1)^2\,[\beta(\beta+1)]^{-\beta/(\beta+1)}\,k^{-\beta/(\beta+1)}.
$$

\end{theorem}

\begin{proof} ({\bf Overview.})
We construct a similar hard family to \cite{blum_nearly-tight_2024}. Define a ``good'' arm as the power curve $g(t) = m(t/T)^\beta$. Let the other $k - 1$ arms copy $g$ up to a breakpoint $s$ and then flatten. Every arm satisfies $\text{LE}(\beta)$, and the good arm violates any stricter floor, so $\beta_I = \beta$. Let $\mathcal{D}_s$ denote the distribution that picks one arm uniformly at random to be $g$.

Against any fixed schedule $S$ of length $T$, let $A$ denote the number of distinct arms that receive at least $s$ pulls under $S$. Note that a ``generous'' evaluation that pays $\text{OPT}_T$ if the good arm lies among those $A$ and otherwise pays at most $Tg(s)$ provides an upper-bound for the expected reward of $S$ under a random draw from $D_s$. Since $A \le \lfloor T/s \rfloor$ and $\mathrm{OPT}_T \ge mT/(\beta + 1)$, it follows that
$$
\frac{\mathbb{E}[\mathrm{ALG}_T]}{\mathrm{OPT}_T} \le \frac{1}{kx} + (\beta + 1)x^\beta \quad \text{where } x := s/T \in (0,1].
$$
Call the right-hand side $h(x)$. Simple calculus gives a unique minimizer
$$
x^* = \left[k\beta(\beta + 1)\right]^{-1/(\beta+1)}, \quad \text{ which evaluates to } h(x^*) = (\beta + 1)^2 \left[\beta(\beta + 1)\right]^{-\beta/(\beta+1)} k^{-\beta/(\beta+1)}.
$$
Let $s = \lfloor x^\star T \rfloor$ and $T \ge 2/x^\star$. Using $\frac{1}{k x^*} = \beta (\beta + 1) (x^*)^{\beta}$, we get $h(s/T) \leq \tfrac{3}{2} h(x^*),$ which implies that
$$
\frac{\mathbb{E}[\mathrm{ALG}_T]}{\mathrm{OPT}_T}\ \le\ \tfrac{3}{2} h(x^\star) = \frac{3}{2}\,(\beta+1)^2\,[\beta(\beta+1)]^{-\beta/(\beta+1)}\,k^{-\beta/(\beta+1)}.
$$
By Yao’s principle, the same bound holds for randomized algorithms.

({\bf Full proof of final step.}) Let $s=\lfloor x^\star T\rfloor$, and note that the condition on $T$ ensures $x^\star T\ge2$. Combine Lemmas~\ref{lem:ratio}, \ref{lem:xstar}, and \ref{lem:round} to bound the ratio for any deterministic schedule against $\mathcal{D}_s$. Since this bound holds for every deterministic schedule, it also holds for any randomized algorithm by linearity of expectation (by Yao’s principle). Lemma~\ref{lem:PF-index} guarantees $\beta_I=\beta$ for the constructed family.
\end{proof}

\subsection{Doubling Trick for Unknown T}\label{appendix:unknownT}
We remove the need to know $T$ by using the doubling schedule and exploration procedure of \cite{blum_nearly-tight_2024}. At a high level, we start with a guess $T_0 = 4k$, pretend this is the horizon, and spend $T'/2$ steps using the same adaptation of \textbf{Algorithm 2} of \cite{blum_nearly-tight_2024} that they use to estimate $\widehat{m}$, the reward of the best arm, for the relevant time scale. After this, we spend $T'/2$ steps exploiting (running $PTRR_\alpha$) with $\tau' := \frac{T'}{2} - k$ and $m=\frac{\widehat{m}}{2\cdot16^\alpha}$ (we shrink $\hat{m}$ to ensure we don't discard the best arm). 
If time remains after $T'$ steps, we double $T'$ and repeat. This yields the below result:

\begin{theorem}[Unknown-$T$ guarantee]\label{thm:unknownT-final}
Assume $\beta_I \in (0,1)$ and fix $\alpha \in (\beta_I, 1)$, $\gamma = \alpha / (\alpha + 1)$. If $T > 4k$, then the doubling trick above achieves
$$
\mathbb{E}[\mathrm{ALG}_T] \ge \frac{1}{2048 \cdot 16^\alpha (\alpha + 1) \log(128k)} \cdot \frac{\mathrm{OPT}_T}{(k + 1)^\gamma}.
$$
\end{theorem}
\begin{proof}
Consider $i$ such that $T' = 2^i \, T_0$ is the last iteration for which an explore / exploit cycle was  completed. Define $T'' \coloneqq \sum_{j = 0}^{i} 2^j T_0 = (2^{i+1} - 1) T_0 $. Then $T'' < T \le 2 T''\,.$ We will argue two things: first, we will show that spending $T'/2$ time on the procedure for estimating $m$ provides a sufficiently good estimate for $m$ for our purposes; second, we will quantify the gap between spending $T'/2$ collecting reward from the instance based on our estimated $m$ and $T$ time spent on the optimal arm.

Let $\tau := T'/2 - k$. Let $\hat{m}$ denote the estimated maximum value at horizon $\tau + k$ from running the estimation procedure for time $T'/2$ (from time $T''-T' $ to $T''-T'/2$). From the analysis in \cite{blum_nearly-tight_2024}, we know that $\frac{1}{2} \widehat{m} \le f^{\star}(\tau) \le 2\widehat{m}$ with probability at least $\frac{1}{\log(128k)}$, and that
$$
    \frac{T}{\tau}= \frac{T}{\frac{T'}{2}-k} \le \frac{T}{\frac{T'}{4}} = 4 \, \frac{T}{T''} \frac{T''}{T'} \le 4 \cdot 2 \cdot 2 = 16\,.$$
Let $m := \frac{\hat m}{2 \cdot 16^\alpha}$, and run $\textit{PTRR}_\alpha$ with $m$, $\tau$ for $\frac{T'}{2}$ steps.

Since $\tau/T \ge 1/16$ and $\widehat{m} \le 2 f^{\star}(\tau) \le 2 f^{\star}(T)$, we know that
$$
m \le \frac{2 f^{\star}(T)}{2 \cdot 16^\alpha} \le f^{\star}(T) \left( \frac{\tau}{T} \right)^{\alpha},
$$
and therefore that 
$$
m \left( \frac{t}{\tau} \right)^\alpha \le f^{\star}(T) \left( \frac{t}{T} \right)^\alpha \le f^{\star}(T) \left( \frac{t}{T} \right)^{\beta_I} \le f^{\star}(t)
$$ for all $t \le T$. It follows that $\textit{PTRR}_\alpha$ again does not switch away from the best arm $f^*$. 

Now, we proceed to the second part of our argument.
Applying Lemma~\ref{lem:solve} with $(\tau', k') = (\tau, k)$ gives
$$
\mathbb{E}[\text{exploit}] \ge \frac{m \tau}{2(\alpha + 1)(k + 1)^\gamma}, \qquad \gamma = \frac{\alpha}{\alpha + 1}.
$$
Since $m = \widehat{m}/(2 \cdot 16^\alpha)$ and $\tau / T \ge 1/16$, it follows that
$$
\mathbb{E}[\text{exploit}] \ge \frac{\widehat{m} \tau}{4(\alpha + 1) 16^\alpha (k + 1)^\gamma} \ge \frac{\widehat{m} T}{64(\alpha + 1) 16^\alpha (k + 1)^\gamma}.
$$

Since $f^{\star}(T) \le 16 f^{\star}(\tau) \le 32 \widehat{m}$, it further follows that $\mathrm{OPT}_T \le f^{\star}(T) T \le 32 \widehat{m} T$, and therefore that 
$$
\frac{\mathbb{E}[\text{exploit}]}{\mathrm{OPT}_T} \ge \frac{1}{2048 (\alpha + 1) 16^\alpha} \cdot \frac{1}{(k + 1)^\gamma}.
$$
Multiplying by the $1 / \log(128k)$ selection probability gives
$$
\mathbb{E}[\mathrm{ALG}_T] \ge \frac{1}{2048 (\alpha + 1) 16^\alpha \log(128k)} \cdot \frac{\mathrm{OPT}_T}{(k + 1)^\gamma}.
$$
\end{proof}

\section{Details for Sample Complexity Analysis}
We now provide some background from prior work and full proofs for the sample complexity of tuning the $\alpha$ parameter in our algorithm family $PTRR_\alpha$.

\subsection{Results from Prior Work} 
\label{appendix:ss25results}
\begin{definition}[Definition 1 in Arxiv verson of \cite{sharma2025offline}]
    Suppose the derandomized dual function $l_T^{I, \pmb{z}}(\rho)$ is a piecewise constant function. The derandomized dual complexity of $\mathcal{D}$ w.r.t. instance I is given by $\mathcal{Q}_\mathcal{D} \coloneqq \E{I \sim \mathcal{D}}{\E{z}{q\left(l_T^{I, \pmb{z}}(\cdot)\right)}}\,,$ where $q(\cdot)$ is the number of pieces over which the function is piecewise constant.
\end{definition}

\noindent Having defined this complexity measure that is capable of characterizing the complexity of an algorithm family of interest, we now present the main theorem we apply from \cite{sharma2025offline}, a uniform convergence guarantee on learning the best value of the parameter $\alpha$:
\begin{theorem}[Theorem 6.1 in Arxiv Version of \cite{sharma2025offline}] \label{thm:ss25main}
    Consider the Hyperparameter Transfer setup for any arbitrary $\mathcal{D}$\footnote{To reiterate, we now consider $\mathcal{D}$ supported over $\mathcal{I} \times \Pi_k\,,$ where $\Pi_k = \{ \pi_k\}$ the set of permutations of $[k]$.} and suppose the derandomized dual function $l_T^{I, \pmb{z}}(\alpha)$ is a piecewise constant function. For any $\epsilon, \delta > 0, N$ problems $\{I_i\}_{i = 1}^N$ sampled from $\mathcal{D}$ with corresponding random coins $\{\pmb{z}_i\}_{i = 1}^N$ such that $N = O\left( \left(\frac{H}{\epsilon}\right)^2 (\log \mathcal{Q}_\mathcal{D} + \log \frac 1 \delta )  \right)$ are sufficient to ensure that with probability at least $1-\delta,$ for all $\pmb{\alpha} \in \mathcal{P},$ we have that:
    
    \begin{equation*}        
    \left| \frac 1N \sum_{i = 1}^N l_T^{P_i, \pmb{z}_i}(\pmb{\alpha}) - \E{P \sim \mathcal{D}}{l_T^P(\pmb{\alpha})}    \right| < \epsilon
    \end{equation*}

\end{theorem}

Note that by the argument below, this guarantee solves the Hyperparameter Transfer Setting described in Definition~\ref{defn:hypertransf}.

\subsection{Uniform Convergence Implies Population Loss Near-Optimality}
\label{appendix:ucpoploss}
We recall the standard argument below for completeness. 

\begin{lemma}
    [Uniform convergence implies near-optimality in population loss]
    Suppose $\hat{\alpha}$ is the minimizer of $\frac 1N \sum_{i = 1}^N l(P_i, \alpha)$ for $N$ large enough to satisfy uniform convergence with error $\epsilon$ and $\alpha^\star \coloneqq \arg \min_{\alpha} \E{P \sim \mathcal{D}}{l_T(P, \alpha)}.$ Then, 
    $$
    \left |\E{P \sim \mathcal{D}}{l_T(P, \hat{\alpha})} - \E{P \sim \mathcal{D}}{l_T(P, \alpha^\star)} \right | < 2\epsilon\,.
    $$
\end{lemma}
\newcommand{\emploss}[1]{\frac 1N \sum_{i = 1}^N l_T(P_i, #1)}

\begin{proof}
    We can see this by adding and subtracting empirical losses:
    \begin{align}
        \left |\E{P \sim \mathcal{D}}{l_T(P, \hat{\alpha})} - \E{P \sim \mathcal{D}}{l_T(P, \alpha^\star)} \right | &= 
        \bigg |\E{P \sim \mathcal{D}}{l_T(P, \hat{\alpha})} - \emploss{\hat{\alpha}}\\ &+ \emploss{\hat{\alpha}} - \emploss{\alpha^\star} \label{eqn:neg}\\&+ \emploss{\alpha^\star} - \E{P \sim \mathcal{D}}{l_T(P, \alpha^\star)} \bigg | \\
        &\le \bigg |\E{P \sim \mathcal{D}}{l_T(P, \hat{\alpha})} - \emploss{\hat{\alpha}} \bigg | + \\
        &+ \bigg| \emploss{\alpha^\star} - \E{P \sim \mathcal{D}}{l_T(P, \alpha^\star)} \bigg | \\
        &\le 2\epsilon
    \end{align}
    Note that the term in Eqn.~\ref{eqn:neg} is negative, since $\hat{\alpha}$ is the minimizer of the empirical loss, and the last inequality follows from the uniform convergence guarantee. 
\end{proof}

\subsection{Average Regret} \label{appendix:avgregret}

Now, let us consider specific instantiations for an $H$-bounded loss function.

\begin{definition} \label{defn:avgregret}
    Suppose at time $t$, the reward collected by the algorithm is $r_t.$ Define the {\em average regret} as:

    $$
    R(T) \coloneqq\frac{1}{T} \left( \max_i \sum_{t = 1}^T f_i(t) -  \sum_{t = 1}^T r_t \right)\,.
    $$
    That is, the regret in hindsight is in reference to the best fixed arm.
\end{definition}

\begin{fact}
    The average regret is $H$-bounded for $H = m\,.$
\end{fact}

\noindent Thus, if we are interested in solving the problem with respect to average regret, we get the sample complexity below:
\begin{corollary}
        For the Hyperparameter Transfer setting for the improving multi-armed bandits problem optimizing for averaged regret, $N = O\left( \left(\frac{m}{\epsilon}\right)^2 (\log kT + \log \frac 1 \delta )  \right)$ instances drawn from $\mathcal{D}$ suffice to get the uniform convergence guarantee in Theorem~\ref{thm:ss25main}.
\end{corollary}

Note that the guarantee of this corollary is that the parameters found will produce a sequence of pulls that has regret that is close to the regret of the sequence of pulls induced by the {\em best set of parameters}.

\subsection{Computational Complexity of ERM}
\label{appendix:erm-alg}
In Section~3.3 we analyze the sample complexity of selecting $\alpha$ by ERM over
the family $\mathcal{A}=\{\text{PTRR}_\alpha:\alpha\in(0,1]\}$ in the
Hyperparameter Transfer Setting (Definition~2.6), using the uniform convergence
results from Appendix~B.1 and
Appendix~B.2. Here we add to that analysis by showing that, given offline
access to the full learning curves, the ERM minimizer over the
continuous domain $\mathcal{P}=(0,1]$ can be computed exactly. Our implementation is efficient for small $k$, and we leave open the question of efficient exact or approximate implementation for large $k$.

As in Section~3.3, we derandomize Algorithm~1 by augmenting each instance with a
permutation $z\in\Pi_k$ that fixes the order in which arms are sampled from $S$.
We therefore work with i.i.d.\ samples $(I_1,z_1),\dots,(I_N,z_N)\sim (D')^N$. For
fixed $(I,z)$ and $\alpha$, the execution of $\text{PTRR}_\alpha$ is
deterministic. Let $l^{I,z}_T(\alpha)$ denote the corresponding (derandomized)
dual loss value at horizon $T$ (Appendix~B.1). 

If $S=\emptyset$ while $t<T$, we halt (this does not occur for appropriate alpha, but ensures that the algorithm is well-defined for
all $\alpha$ and does not affect the arguments below).

\paragraph{Finite critical set for fixed $(I,z)$}

Fix an augmented instance $(I,z)$ with reward curves
$\{f_i(t)\}_{i\in[k],\,t\in\{0,1,\dots,T\}}$, and recall that Algorithm~1 makes $\alpha$-dependent decisions through the keep-test
\begin{equation}
\label{eq:ptrr_keep_test_app}
f_i(t_i)\ \ge\ m\Big(\frac{t_i}{\tau}\Big)^\alpha,
\qquad t_i\in\{0,1,\dots,T\}.
\end{equation}
For each pair $(i,t)$ with $i\in[k]$ and $t\in\{1,\dots,T\}\setminus\{\tau\}$,
define $c_{i,t}\in(0,1]$ to be the unique solution (if it exists) to the equality
\begin{equation}
\label{eq:critical_point_def_app}
f_i(t)\ =\ m\Big(\frac{t}{\tau}\Big)^\alpha,
\qquad \alpha\in(0,1].
\end{equation}
Uniqueness holds because for fixed $t\neq\tau$, the map $\alpha\mapsto m(t/\tau)^\alpha$
is strictly monotone. Let the critical set be
$$
\mathcal{C}(I)
\ :=\
\big\{c_{i,t}\in(0,1] : i\in[k],\ t\in\{1,\dots,T\}\setminus\{\tau\},
\text{ and \eqref{eq:critical_point_def_app} has a solution in }(0,1]\big\}.
$$
Then clearly $|\mathcal{C}(I)|\le kT$.

\begin{lemma}[Constancy between critical values]
\label{lem:constancy_between_critical}
Fix $(I,z)$. The function $\alpha\mapsto l^{I,z}_T(\alpha)$ is constant on each
connected component of $(0,1]\setminus \mathcal{C}(I)$.
Moreover, any change in $l^{I,z}_T(\alpha)$
can occur only at $\alpha\in\mathcal{C}(I)$.
\end{lemma}

\begin{proof}
For each $i\in[k]$ and $t\in\{0,1,\dots,T\}$, define the predicate
$$
\text{Test}_{i,t}(\alpha)
\ :=\
\mathbf{1}\!\left[\, f_i(t)\ \ge\ m\Big(\frac{t}{\tau}\Big)^\alpha \,\right].
$$
When $t=\tau$, the right-hand side equals $m$ and is independent of $\alpha$.
When $t\neq\tau$, strict monotonicity of $\alpha\mapsto m(t/\tau)^\alpha$ implies that
$\mathsf{Test}_{i,t}(\alpha)$ can change value only at the unique equality solution
$c_{i,t}$ (if it exists). Therefore, on any connected component
$U\subset(0,1]\setminus\mathcal{C}(I)$, every predicate $\mathsf{Test}_{i,t}(\alpha)$
is constant for all $\alpha\in U$.

Now fix $\alpha,\alpha'\in U$ and couple the executions of $\text{PTRR}_\alpha$ and
$\text{PTRR}_{\alpha'}$ on the same augmented instance $(I,z)$.
Because $z$ fixes the order of arms sampled from $S$, the only remaining branching
in Algorithm~1 is the outcome of the keep-test \eqref{eq:ptrr_keep_test_app} at the
current arm and current pull-count. Each such branch is determined by some
$\mathsf{Test}_{i,t}(\cdot)$, and all of these predicates agree on $U$.
Therefore both executions take identical branches at every step and therefore generate
the same pull sequence and the same accumulated reward, which implies that $l^{I,z}_T(\alpha)=
l^{I,z}_T(\alpha')$. This proves constancy on $U$ and that trace changes can occur
only at $\alpha\in\mathcal{C}(I)$.
\end{proof}

Because Algorithm~1 uses a non-strict inequality in \eqref{eq:ptrr_keep_test_app},
the value of $l^{I,z}_T(\alpha)$ at $\alpha\in\mathcal{C}(I)$ can differ from
the constant values on the adjacent open intervals. Therefore, any exact ERM
procedure must treat critical points as candidates.

\paragraph{Exact ERM over $\alpha\in(0,1]$}

Given i.i.d.\ samples $(I_1,z_1),\dots,(I_N,z_N)\sim(D')^N$, define the empirical
objective
\[
\widehat{L}_N(\alpha)
\ :=\
\frac{1}{N}\sum_{j=1}^N l^{I_j,z_j}_T(\alpha),
\qquad \alpha\in(0,1].
\]
We show how to compute an exact minimizer
$\widehat{\alpha}\in\arg\min_{\alpha\in(0,1]} \widehat{L}_N(\alpha)$. For each sample $j$, form $\mathcal{C}(I_j)$ and sort its distinct elements as
$$
0 < c_{j,1} < \cdots < c_{j,q_j} \le 1,
\qquad q_j \le kT.
$$
Define the open intervals induced by these breakpoints:
\[
U_{j,0}:=(0,c_{j,1}),\quad
U_{j,r}:=(c_{j,r},c_{j,r+1})\ (r=1,\dots,q_j-1),\quad
U_{j,q_j}:=(c_{j,q_j},1),
\]
(with the convention that if $q_j=0$ then $U_{j,0}=(0,1)$).
By Lemma~\ref{lem:constancy_between_critical}, we know that $l^{I_j,z_j}_T(\alpha)$ is constant
on each $U_{j,r}$, but may take a different value at each $\alpha=c_{j,r}$. Define the global critical set
$$
\mathcal{C}_{\mathrm{all}}
\ :=\
\bigcup_{j=1}^N \mathcal{C}(I_j),
\qquad |\mathcal{C}_{\mathrm{all}}|\le NkT,
$$
and let its distinct elements be
$0 < b_1 < \cdots < b_M < 1$ (possibly $M=0$).

\begin{theorem}[Running time for implementing exact ERM for $\text{PTRR}_\alpha$ over the derandomized loss function]
\label{thm:exact_erm_ptrr}
Given offline access to full reward curves for each training instance, an exact
empirical minimizer
$\widehat{\alpha}\in\arg\min_{\alpha\in(0,1]} \widehat{L}_N(\alpha)$
can be computed in time
$$
O\!\left(NkT^2\ +\ NkT\log(NkT)\right).
$$
\end{theorem}

\begin{proof}
We first show that it suffices to minimize $\widehat{L}_N(\alpha)$ over a finite
candidate set. By Lemma~\ref{lem:constancy_between_critical},
for each $j$, the function $l^{I_j,z_j}_T(\alpha)$ is constant on each open interval
of $(0,1]\setminus\mathcal{C}(I_j)$. Therefore the empirical average
$\widehat{L}_N(\alpha)$ is constant on each open interval of
$(0,1]\setminus\mathcal{C}_{\mathrm{all}}$. The only points at which
$\widehat{L}_N(\alpha)$ can differ from these open-interval constants are the
breakpoints in $\mathcal{C}_{\mathrm{all}}$ and the endpoint $\alpha=1$, which implies that an exact minimizer is attained by some $\alpha$ in the finite set
$$
\mathcal{A}_{\mathrm{cand}}
\ :=\
\Big(\{b_1,\dots,b_M,1\}\Big)
\ \cup\
\Big\{\text{one representative }\tilde\alpha_r\in(b_r,b_{r+1})
:\ r=0,\dots,M\Big\}
$$
(where we set $b_0:=0$ and $b_{M+1}:=1$).

We now describe an evaluation procedure for $\mathcal{A}_{\mathrm{cand}}$ and bound
its runtime. For each sample $j$, we compute
(i) the constant value of $l^{I_j,z_j}_T(\alpha)$ on each local interval $U_{j,r}$
by simulating $\text{PTRR}_\alpha$ at one representative point in $U_{j,r}$, and
(ii) the value at each local breakpoint $c_{j,r}$ by simulating at $\alpha=c_{j,r}$,
and also simulate once at $\alpha=1$.
Each simulation performs at most $T$ pulls and therefore runs in $O(T)$ time.
Since each sample has at most $q_j\le kT$ breakpoints and $q_j+1\le kT+1$ open
intervals, this requires $O((2q_j+2)\cdot T)=O(kT^2)$ time per sample and
$O(NkT^2)$ time total.

Next, we sort the global breakpoint set $\mathcal{C}_{\mathrm{all}}$ in
$O(NkT\log(NkT))$ time. Sweep $\alpha$ from left to right across the sorted list.
On each global open interval $(b_r,b_{r+1})$, compute $\widehat{L}_N(\tilde\alpha_r)$
using the current local-interval value for each sample. At each breakpoint $b_r$,
compute $\widehat{L}_N(b_r)$ by using the precomputed breakpoint value for those
samples that have a local breakpoint at $b_r$, and otherwise using the current
local-interval value. Track the best value encountered across all candidates in
$\mathcal{A}_{\mathrm{cand}}$, including $\alpha=1$. This yields an exact minimizer
over $(0,1]$.

The sweep itself is linear in the number of breakpoint events, $O(NkT)$, and is
dominated by the $O(NkT^2)$ simulation cost and the $O(NkT\log(NkT))$ sorting cost.
\end{proof}

This is a purely computational result, which shows that the ERM
minimizer assumed in Appendix~B.2 can be computed exactly over the continuous
parameter space $\alpha\in(0,1]$ from offline learning curves in time
$$
O\!\left(NkT^2\ +\ NkT\log(NkT)\right).
$$

\noindent We can use this result for implementing exact ERM over the derandomized dual loss function to implement exact ERM over the true dual loss function (which involves expectation over the randomization of PTRR$_\alpha$) by taking an average over $k!$ permutations of the arms.

\subsection{An Algorithm for Finding a Suitable Value of the Parameter}
\label{appendix:approx-learning-alpha}

In this section, we provide a natural algorithm for finding a suitable value of the parameter $\alpha$ for the PTRR algorithm based on access to instances. We then analyze the sample complexity, runtime complexity, and performance of the algorithm. \par

At a high level, the algorithm starts by (approximately) identifying the underlying value of the CEE, $\beta_i$ for each instance $i$ in the training set. Then, we aggregate estimated values $\hat{\beta}_i$ are aggregated by taking the maximum and that aggregated value (or 1, if it is bigger than 1) is returned as a value of $\alpha$ to use in future deployments. We now present a very general version of the result.

\begin{algorithm}
\caption{Approximate Learner}\label{alg:learn-alpha}
\begin{algorithmic}[1]
\State \textbf{Require:} $n$ samples from distribution $\mathcal{D}$
\For{each instance $I_i$}
    \State $\hat{\beta}_i \leftarrow \texttt{CEE\_Oracle}(I_i)$
\EndFor
\State $\beta \gets \max_i  \hat{\beta}_i $ \\
\Return $\beta$
\end{algorithmic}
\end{algorithm}




For both steps, we can adjust the sophistication of the protocols. For the first step, we start by assuming we have an oracle that returns $CEE(I)$ for any instance $I\,.$ We can implement an approximate oracle by dividing the interval $(0, 1]$ into subintervals of size $\epsilon$ and binary searching over that discrete set for the smallest satisfying value. Similarly, for the second step, it is easiest to analyze the aggregation protocol that simply takes the maximum. However, this has the awkward property that with more samples, the predicted parameter increases toward the supremum of the support of the induced distribution over $\beta$s. Instead, we ought to use a $1-\delta$ order statistic that concentrates around its mean. This implies discarding some probability mass as part of the failure probability.

At a high level, in order to argue that this algorithm provides interesting, non-vacuous guarantees on the performance of PTRR with the learned parameter on a new instance drawn from the same distribution, we argue that the learned value of the parameter is good for a good fraction of the support of the distribution. To argue about this, we first define the induced distribution over $\beta$:

\begin{definition}
    For each instance $I_i$ drawn from the distribution $\mathcal{D}\,,$ suppose we compute $\beta_i = CEE(I).$ We call the distribution over the $\beta_i$ the {\em induced distribution over } $\beta$s.
\end{definition}

Next, we define the optimal parameter $\alpha^\star$ for the distribution. 

\begin{definition}
    Define $\alpha^\star$ as the maximizer of the expected competitive ratio over the distribution of instances. In particular, $\alpha^\star \coloneqq \arg \max_{\alpha \in (0, 1]} \E{I_\text{test} \sim \mathcal{D}}{\frac{Rew(PTRR_\alpha, I_\text{test})}{OPT(I_\text{test})}}\,.$
\end{definition}

Now, assuming the induced distribution over $\beta$ for a distribution $\mathcal{D}$ has some density and cumulative density, we can show that, for a fixed number of samples, with quantifiable probability (approaching 1 as the number of samples goes to infinity), we can bound the reward ratio in terms of the learned parameter. We formalize this in the theorem below.

\begin{theorem}
    Suppose the induced distribution over $\beta$s has density $f(\beta)$ and cumulative density $F(\beta)\,.$ Suppose we run Algorithm~\ref{alg:learn-alpha} with a fixed number of samples $n > 10$ and receive $\hat{\alpha}$ as the learned value of the parameter. If $\hat{\alpha} \ge1\,,$ we run $PTRR_1$ and achieve $1/\sqrt{k}$ fraction of the reward. Otherwise, 
    with probability $\left( \frac{1}{n+1} \right)^{1/n}\left( \frac{n}{n+1}\right)\,$ (which approaches 1 as $n \rightarrow \infty)$ over the test sample and the training samples, the reward of running $PTRR_{\hat{\alpha}}$ on the test sample is at least $1/k^{\hat{\alpha}/(\hat{\alpha}+1)}$ fraction of running $PTRR_{\alpha^\star}\,$ on that instance. That is:
    \begin{equation} \label{eqn:reward-ratio}
    \frac{Rew(PTRR_{\hat{\alpha}}, I_\text{test})}{Rew(PTRR_{\alpha^\star}, I_\text{test})} \ge \Omega \left(\frac{1}{k^{\hat{\alpha}/(\hat{\alpha}+1)}} \right)\,.
    \end{equation}
\end{theorem}


\begin{rmk}
    Note that this is different to the PAC guarantee given by ERM in two important ways. First, we cannot choose arbitrarily small failure probability and error values and draw a number of samples that is a function of those. Secondly, we are providing a guarantee for a single fixed test instance drawn from the distribution, not for the ``average'' test instance. In particular, we are using the fact that PTRR with the best value of the parameter for the distribution cannot get more reward on the instance than the optimal reward achievable on that instance. 
\end{rmk}

\begin{proof}

In order to prove this result, at a high-level, we will show that even if we resign ourselves to failure on a small fraction of the distribution (i.e., bad approximation ratio), we can still achieve a good approximation with high probability on the rest. To show this, we proceed in three steps:
\begin{enumerate}
    \item First, we define two events that are crucial to our guarantee: first, that the learned $\hat{\alpha}$ is sufficiently large; second, that the test example is sufficiently nice.
    \item We argue that for any test example for which the CEE is at most $\tau\,,$ running $PTRR_{\hat{\alpha}}$ obeys the reward ratio in Eqn.~\ref{eqn:reward-ratio} as long as $\hat{\alpha} \ge \tau$.
    \item We argue that with good probability over the training sample, $\hat{\alpha} \ge \tau\,,$ and with good probability over the test instance, $CEE(I_\text{test}) \le \tau\,.$ Thus, we can combine the probabilities of the relevant events to show that with the stated probability, the reward ratio guarantee holds.
\end{enumerate}

\paragraph{Relevant Events} At a high level, our strategy will be to resign ourselves to failure on a small fraction of the support of the induced distribution over $\beta$s in the interest of guaranteeing an approximation on the rest of the support. To that end, we define the following two events:
\begin{align}
    \mathcal{E}_1 &\coloneqq \{  \hat{\alpha} \ge \tau  \} \\
    \mathcal{E}_2 &\coloneqq \{  CEE(I_\text{test}) \le \tau \}
\end{align}

We know from the analysis in Section~\ref{sec:intro3} that for an instance with CEE $\beta$\,, running $PTRR_\alpha$ for any $\alpha \ge \beta$ will achieve $1/k^{\alpha/(\alpha+1)}$ fraction of the optimal reward. Thus, on a new test example, provided the value of the PTRR parameter we use is larger than the CEE of the instance, we will accrue sufficient reward. We formalize this subsequently.

\paragraph{Reward Ratio Holds} Now, recall from Theorem~\ref{thm:ptrr-alpha} that for an instance with CEE $\beta_I\,,$ $PTRR_\alpha$ for any $\alpha > \beta_I$ accrues $O(OPT/k^{\alpha/(\alpha+1)})$ reward. Thus, if $\hat{\alpha} > \tau > \beta_\text{test} \coloneqq CEE(I_\text{test})\,,$ the $Rew(PTRR_{\hat{\alpha}}, I_\text{test}) \ge O(OPT/k^{\alpha/(\alpha+1)})\,.$ Finally, we observe that by the definition of $OPT\,,$ we have that $OPT \ge Rew(PTRR_\alpha, I_\text{test}) \, \forall \alpha\,,$ so in particular this holds for $\alpha = \alpha^\star\,,$ the value of $\alpha$ for which the expected competitive ratio is maximized. 
Thus, we have that if $\hat{\alpha} \ge \tau$ and $CEE(I_\text{test}) \le \tau\,,$ then:
$$
\frac{Rew(PTRR_{\hat{\alpha}}, I_\text{test})}{Rew(PTRR_{\alpha^\star}, I_\text{test})} \ge O \left(\frac{1}{k^{\hat{\alpha}/(\hat{\alpha}+1)}} \right)\,.
$$

Thus, it remains to argue that the conditions in the previous statement occur with good probability.

\paragraph{Probability} First, let us consider event $\Pr{}{\mathcal{E}_2}\,.$ This is simply the cumulative density at $\beta=\tau\,, F(\tau)\,.$ Next, we analyze $\Pr{}{\mathcal{E}_1}:$
\begin{align}
    \Pr{}{\mathcal{E}_1} &= \Pr{}{\hat{\alpha} \ge \tau} \\
    &= \Pr{}{\max_{i}\hat{\beta}_i \ge \tau} \\
    \text{ under oracle } &= 1 - \Pr{}{\forall \, i \, \beta_i \le \tau} \\
    &= 1 - F(\tau)^n\,.
\end{align}
The probability of success is the probability of the intersection of these two events, so the probability of success is $F(\tau)(1-F(\tau)^n)\,.$ Now, it suffices to show that we can pick a $\tau$ that gives us a good probability of success. We can therefore pick the $\tau$ that maximizes the probability of success, namely $\arg \max_{\tau} F(\tau)(1-F(\tau)^n)\,.$
\begin{align}
    \max_{\tau} F(\tau)(1-F(\tau)^n) \Leftrightarrow f(\tau) - (n+1) F(\tau)^n f(\tau) &= 0 \\
    f(\tau)\left( 1 - (n+1) F(\tau)^n \right) &= 0 \\
    \frac{1}{n+1} &= F(\tau)^n \\
    \tau = F^{-1}\left(  \left(   \frac{1}{n+1} \right)^{1/n}  \right)\,.
\end{align}

Plugging this back into the probability of success, we get the stated result.

\end{proof}

\subsection{Full Empirical Evaluation}
\label{appendix:empirical}
The main phenomenon highlighted by the theory is that the parameter $\alpha$ controls the exploration--abandonment tradeoff, and therefore that the best $\alpha$ can depend on the underlying instance. The purpose of this empirical section is to provide evidence on real learning-curve data that different instances prefer different $\alpha$. We use sample-wise learning curves from the Learning Curve Database (LCDB~1.1), specifically the CC-18 benchmarks.
LCDB stores error-rate curves at multiple training-set sizes (``anchors'') for a fixed set of learners, together with repeated evaluations under a nested cross-validation protocol. 

Each dataset $d$ defines one improving-bandit instance.
Arms correspond to learners, and the time index corresponds to LCDB anchor points.
We convert error rates to rewards through
$
r_{i,d}(t) \;=\; 1 - \mathrm{err}_{i,d}(t),
$
so that larger rewards correspond to better performance.
LCDB stores multiple repetitions per (dataset, learner, anchor) due to nested cross-validation.
We average across these repetitions (ignoring missing values) to obtain a single mean learning curve per (dataset, learner).
This yields deterministic reward functions $r_{i,d}(\cdot)$, matching our setting.

LCDB curves can terminate early for some dataset--learner pairs (in the stored tensors this appears as NaNs).
For a fixed horizon $T$, we restrict to datasets for which all included learners have finite values for anchors $t=1,\dots,T$.
This ensures that each dataset corresponds to a well-defined bandit instance with a common horizon.

For each usable dataset $d$ and each $\alpha$ on a fixed grid, we run $\mathrm{PTRR}_\alpha$ for a horizon of $T$ pulls.
The only randomness in our implementation is the random ordering in which arms are first considered. We average results over 200 random seeds. We evaluate performance using the normalized cumulative reward
$$
\frac{\mathbb{E}[\mathrm{ALG}(d;\alpha)]}{\mathrm{OPT}(d)}
\qquad\text{where}\qquad
\mathrm{OPT}(d)=\max_i \sum_{t=1}^T r_{i,d}(t),
$$
which is the reciprocal of the competitive ratio $\mathrm{OPT}/\mathbb{E}[\mathrm{ALG}]$ from Definition~2.2. Note that $\mathrm{OPT}(d)$, the best fixed-arm policy on $d$ in hindsight, is not necessarily the global best policy here due to the absence of monotonicity.

\paragraph{Main experiment: $T=44$, $k=22$.}
Our primary experiment uses horizon $T=44$ and a learner set of size $k=22$ (we exclude two learners with $\geq 50 \%$ ill-behavior in the CC-18 file).
Under this choice, we obtain 27 datasets with complete prefixes of length $T$ across all $k$ learners.


\begin{figure}[H]
    \centering
    \begin{subfigure}[t]{0.49\linewidth}
        \centering
        \includegraphics[width=\linewidth]{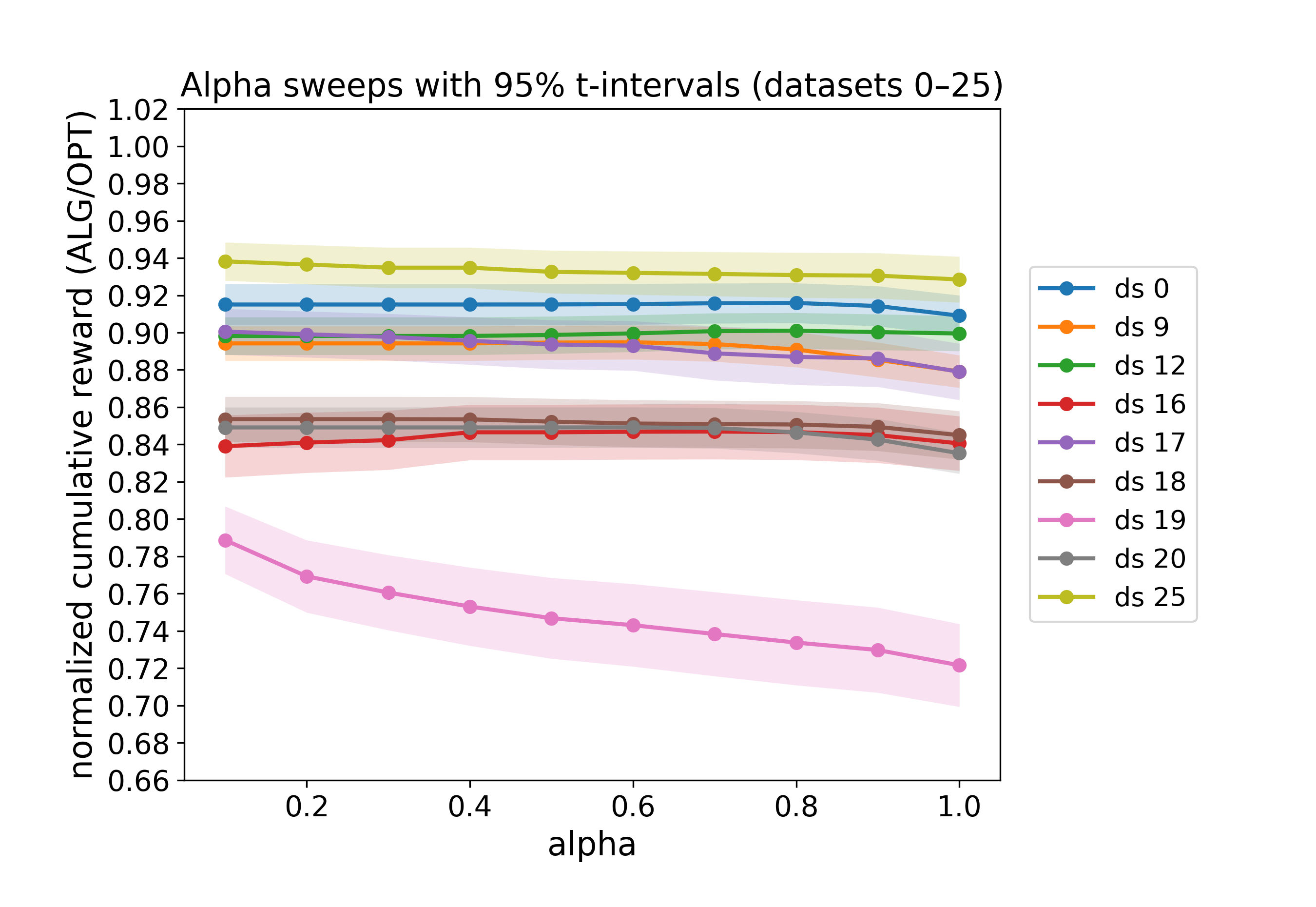}
        \label{fig:alpha_sweep_block1_ci_bigfont}
    \end{subfigure}\hfill
    \begin{subfigure}[t]{0.49\linewidth}
        \centering
        \includegraphics[width=\linewidth]{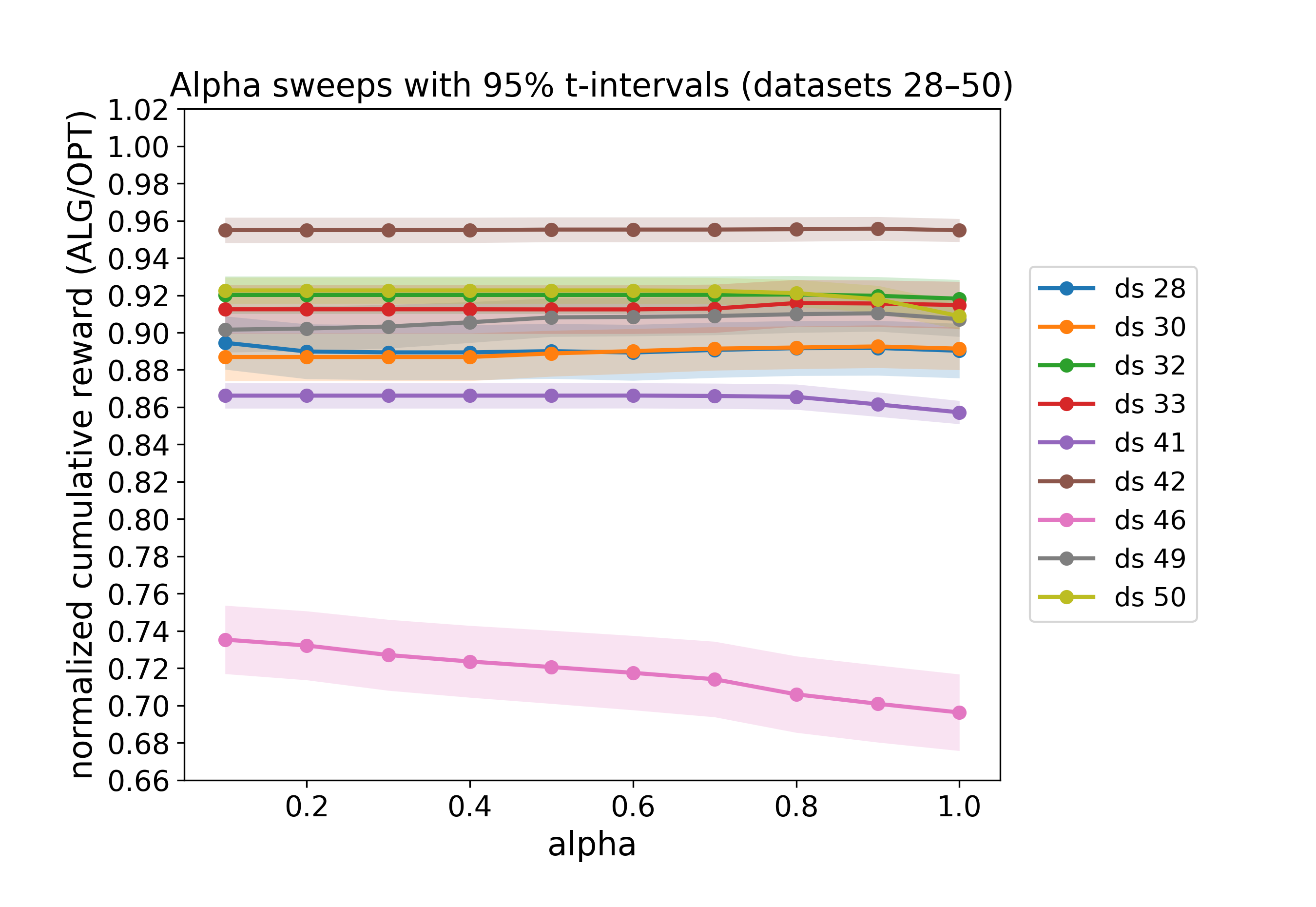}
        \label{fig:alpha_sweep_block2_ci_bigfont}
    \end{subfigure}

    \vspace{0.6em}

    \begin{subfigure}[t]{0.49\linewidth}
        \centering
        \includegraphics[width=\linewidth]{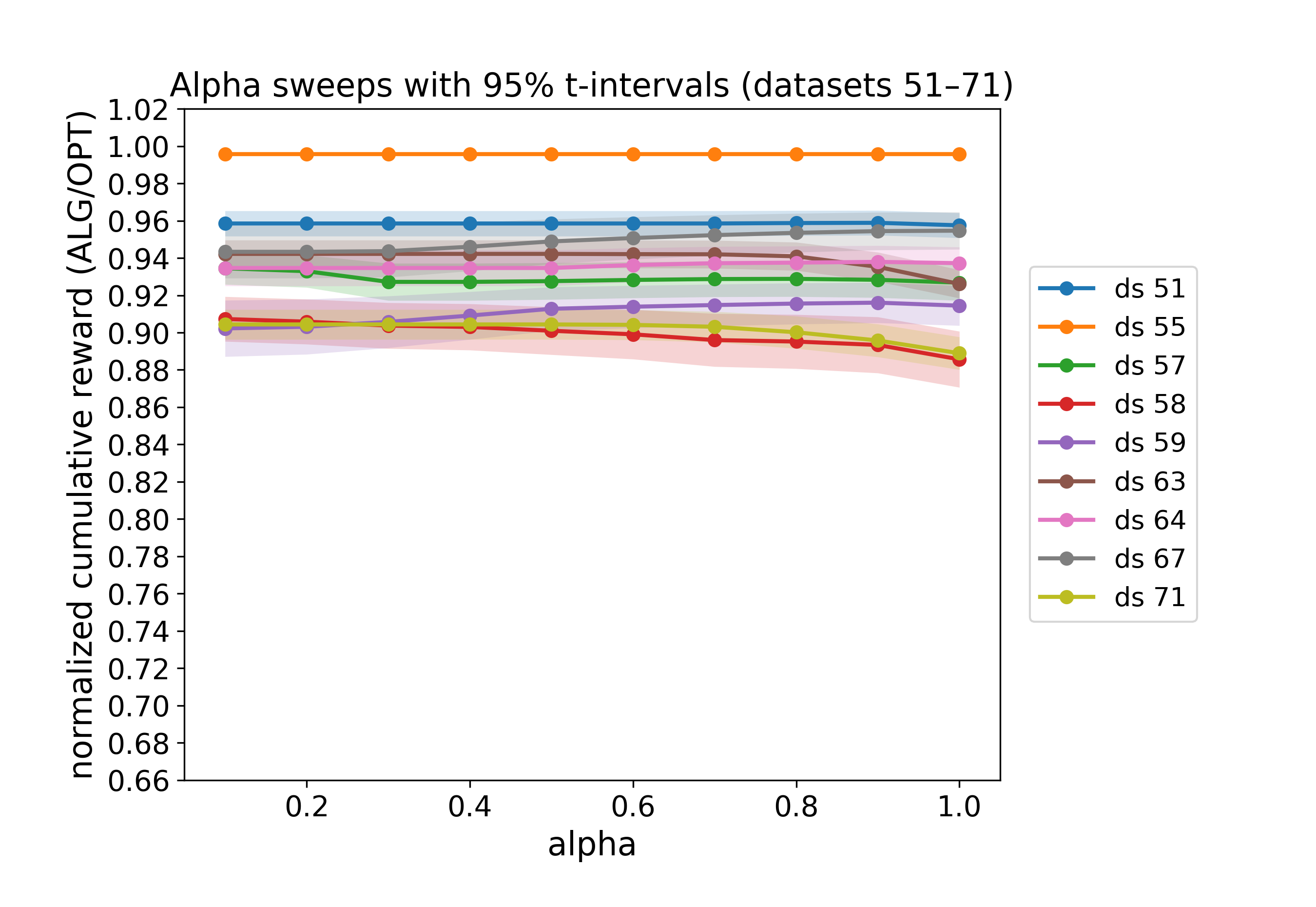}
        \label{fig:alpha_sweep_block3_ci_bigfont}
    \end{subfigure}

    \caption{\textbf{Sensitivity of $\mathrm{PTRR}_\alpha$ to $\alpha$ on all LCDB instances ($T=44$, $k=22$).} Each curve corresponds to one CC-18 dataset $d$ from LCDB~1.1 and reports the normalized cumulative reward $\mathbb{E}[\mathrm{PTRR}_\alpha(d)]/\mathrm{OPT}(d)$ as a function of $\alpha\in\{0.1,0.2,\dots,1.0\}$, where $\mathrm{OPT}(d)=\max_i\sum_{t=1}^T r_{i,d}(t)$ is the cumulative reward of the best fixed-arm policy in hindsight under the same horizon (which does not necessarily correspond to the global optimal policy due to the absence of monotonicity) 
and $r_{i,d}(t)=1-\mathrm{err}_{i,d}(t)$ is the mean (over cross-validation) reward at anchor $t$ for arm $i$. For each $(d,\alpha)$, $\mathbb{E}[\mathrm{PTRR}_\alpha(d)]$ is estimated by averaging over $200$ random arm orderings. The shaded regions are pointwise 95\% Student-$t$ confidence intervals across the 200 runs (mean $\pm\, t_{0.975,199}\cdot \mathrm{sd}/\sqrt{200}$). For most datasets, performance differences across $\alpha$ are small relative to the confidence intervals, while a minority show a significant trend across $\alpha$ on this grid.}
    \label{fig:alpha_sweep_all_blocks_ci_bigfont}
\end{figure}

We sweep $\alpha \in \{0.1,0.2,\dots,1.0\}$.
Figure \ref{fig:alpha_sweep_all_blocks_ci_bigfont} shows all of the per-dataset performance curves $\alpha \mapsto \mathbb{E}[\mathrm{ALG}]/\mathrm{OPT}$, demonstrating that the maximizer varies across certain instances. Across the 27 datasets, the best $\alpha$ is widely distributed: 11 datasets select $\alpha=0.1$, while many others select $\alpha$ in the range $[0.8,1.0]$ (it is worth noting that 4 of the 11 cases are actually flat across several small $\alpha$ values). Mechanistically, smaller $\alpha$ corresponds to more aggressive early abandonment in $\mathrm{PTRR}_\alpha$, so these datasets are those for which aggressive abandonment is (on this grid) empirically most favorable.

\begin{figure}[H]
    \centering
    \includegraphics[width=1\linewidth]{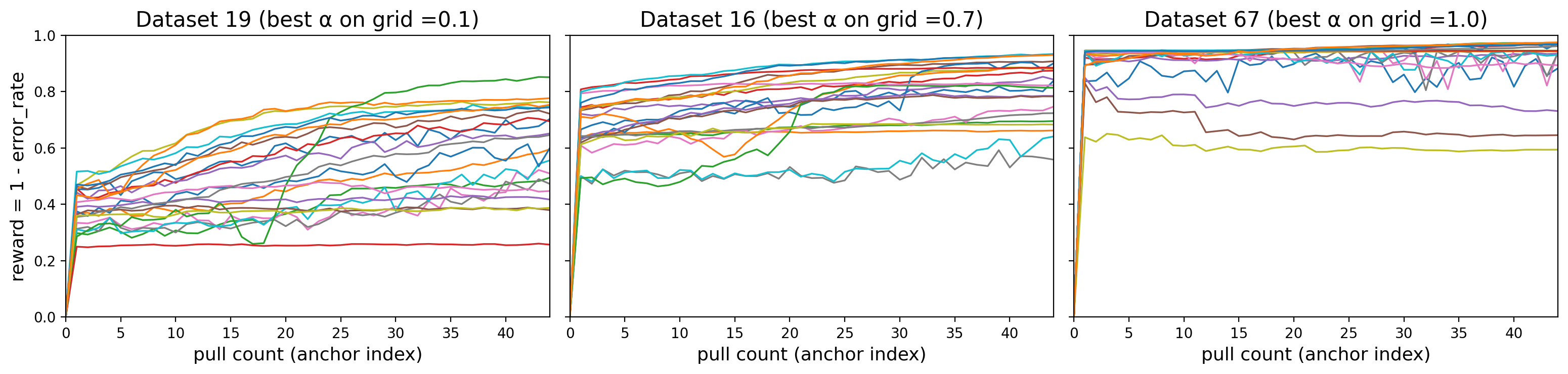}
    \caption{\textbf{Mean LCDB reward curves for three datasets with distinct best $\alpha$ values on the grid.} Each panel overlays the mean reward curves $r_{i,d}(t)=1-\mathrm{err}_{i,d}(t)$ across anchors $t$ for all $k=22$ arms on a single CC-18 dataset $d$. The title of each panel reports the value of $\alpha\in\{0.1,0.2,\dots,1.0\}$ that maximizes the estimated normalized cumulative reward $\mathbb{E}[\mathrm{PTRR}_\alpha(d)]/\mathrm{OPT}(d)$ at horizon $T=44$ on that dataset. 
    We selected these datasets to illustrate the reward dynamics underlying distinct best $\alpha$ values on the grid. \looseness-1
    \label{fig:learning curves appendix}}
\end{figure}

Naturally, real learning curves in LCDB are not guaranteed to satisfy the monotonicity/concavity assumptions used in this work (and largely don't). To connect $\alpha$ to concrete learning dynamics, we plot mean learning curves with fixed axes for three datasets with distinct best $\alpha$'s.
Figure \ref{fig:learning curves appendix} displays all learners for dataset 19 (which selects $\alpha=0.1$), dataset 16 (which selects $\alpha = 0.7)$ and index 67 (which selects $\alpha=1.0$). Qualitatively, these plots suggest a mechanism consistent with the influence of $\alpha$ on $\mathrm{PTRR}_\alpha$, where datasets preferring smaller $\alpha$ tend to exhibit early separation between `good' and `bad' arms (making aggressive abandonment beneficial), while datasets preferring larger $\alpha$ exhibit closer early performance among many learners (making aggressive abandonment riskier). We emphasize that this interpretation is qualitative and based directly on the plotted mean curves; the purpose of these plots is merely contextualize the observed instance dependence.

\section{Best-of-both-worlds for Maximizing Cumulative Reward}

In this section, we consider the problem of getting the best policy regret possible if it is sublinear and reverting to optimal competitive ratio if it is not. In the main body, we considered a similar formulation for best-arm identification. Here, we instead hybridize between an algorithm known to get good policy regret on good instances from \cite{metelli_stochastic_2022} and an algorithm known to get optimal competitive ratio on worst-case instances from \cite{blum_nearly-tight_2024}. We take a data-driven approach to identifying exactly how to hybridize, i.e., exactly when to switch from the policy-regret-optimizing algorithm to the competitive-ratio-optimizing algorithm. We note that as a result of this approach, we do {\em not} get per-instance worst-case guarantees. This is an interesting direction for future work.

\subsection{Motivating Examples}
\cite{blum_nearly-tight_2024} give a randomized algorithm for improving bandits (also known as deterministic rested rising multi-armed bandits) that achieves the optimal $\sqrt{k}$ competitive ratio on worst-case instances. Although it is well known that a sublinear regret is not attainable for worst-case instances of improving bandits, under some niceness assumptions, UCB-style algorithms~\cite{metelli_stochastic_2022} can be shown to obtain sublinear policy regret (regret w.r.t.\ the best single arm pulled for the entire time horizon). Here we present examples that show that both the above algorithms from the literature may be sub-optimal and there are instances where each is dominated by the other.

\begin{example}
{\it The randomized robin algorithm of \cite{blum_nearly-tight_2024} may suffer linear policy regret on instances where the UCB-based algorithm of ~\cite{metelli_stochastic_2022} achieves sublinear regret.} We set the reward function for the best arm as $f_{i^*}(t)=1$ for all $t$, and for any other arm $i\ne i^*$ as $f_{i}(t)=t/T$, where $T$ is the time horizon. Now the randomized robin algorithm selects the optimal arm first with probability $1/k$, and otherwise, it keeps playing a sub-optimal arm. The expected total reward is $T\cdot\frac{1}{k}+\frac{T}{2}\cdot\frac{k-1}{k}=\frac{T}{2}\left(1+\frac{1}{k}\right)$, which corresponds to  $\Omega(T)$ regret. On the other hand, the UCB-based algorithm has a logarithmic upper bound on its regret. Indeed, after $O(\log T)$ exploratory pulls of any sub-optimal arm, it will always prefer the optimal arm, implying an upper bound of $O(k\log T)$ on the policy regret.
\end{example}

\begin{example}
{\it The UCB-based algorithm of ~\cite{metelli_stochastic_2022} may have $\Omega(k)$ competitive ratio on some instances.} We use the example used in the lower bound construction of \cite{blum_nearly-tight_2024}. Set $f_{i^*}(t)=t/T$ for all $t$ for the optimal arm $i^*$, and for any other arm $i\ne i^*$ as $f_{i}(t)=\min\left\{t/T,\frac{1}{\sqrt{k}}\right\}$. Due to the exploration term, while the arm rewards are identical, each arm gets pulled an equal number of times. By time $T$, each arm gets pulled $T/k$ times for a total reward of $T/2k$, sub-optimal by a factor of $k$.
\end{example}

\subsection{Data-driven Hybrid Approach}

Our goal, therefore, is to take a step toward devising algorithms that can achieve sublinear regret for nice instances but fall back to optimal competitive ratio for general instances.
To this end, we describe an algorithm that interpolates between a UCB-based algorithm by \cite{metelli_stochastic_2022} that gets sublinear regret on nice instances and the algorithm of \cite{blum_nearly-tight_2024} which achieves the optimal worst-case competitive ratio.


\begin{algorithm}
\caption{Regret-optimizing Hybrid Algorithm, i.e. \texttt{Regret-Hybrid}$_B$} \label{alg:reg-hybr}
\begin{algorithmic}
    \State Run Algorithm 1 (R-ed-UCB) of \cite{metelli_stochastic_2022} for B time steps 
    \State Run PTRR for T-B time steps
\end{algorithmic}
\end{algorithm}


We define a family of algorithms parameterized by $B$:

\begin{definition} \label{defn:regret-hybrid}
    Define the family of algorithms \texttt{Regret-Hybrid} $\coloneqq \{ \texttt{Regret-Hybrid}_B : B \in [T] \} \,, $ where $\texttt{Regret-Hybrid}_B$ is Algorithm~\ref{alg:reg-hybr}.
\end{definition}

Now, as before, we analyze $Q_D$ and then instantiate the result for a loss function of interest.

\begin{lemma}
    For the family $\texttt{Regret-Hybrid}$ defined in Defn.~\ref{defn:regret-hybrid}, the improving multi-armed bandits problem, and any piecewise constant loss function, $Q_D \le k\,T^2\,.$
\end{lemma}
\begin{proof}
    $B$ takes on $T$ discrete values, and for each fixed $B\,,$ by Lemma~\ref{lemma:bdqd}, there are at most $k\,T$ behaviors for the PTRR family on an instance.
\end{proof}

In order to study cumulative reward via regret, we can use the average regret loss function defined in Defn.~\ref{defn:avgregret}. Then, extending Theorem~\ref{thm:hybrid-alg-sc}, we get the following corollary:

\begin{corollary}
        For the Hyperparameter Transfer setting for the improving multi-armed bandits problem optimizing for averaged regret over the algorithm family described in Defn.~\ref{def:fullyhybridfam}, $N = O\left( \left(\frac{m}{\epsilon}\right)^2 (\log kT + \log \frac 1 \delta )  \right)$ instances drawn from $\mathcal{D}$ suffice to get the uniform convergence guarantee in Theorem~\ref{thm:ss25main}.
\end{corollary}

Note that this result provides an algorithm that is competitive with the best possible algorithm in the family on instances from distribution $\mathcal{D}.$ Thus, that algorithm may not provide the worst-case fallback guarantee on a {\em fixed} instance. As mentioned earlier, this is an interesting consideration for future work.

\section{BAI Comparison with Prior Work and Proof Details}\label{appendix:PR}

We provide below a comparison with prior work on best arm identification along with complete proof details for Section \ref{sec:bothworlds}.

\subsection{Comparison With Prior Work}

First, let us compare the BAI task in \cite{mussi2024best} to ours. In their work, they study the task of identifying the arm with the greatest single pull in the instance (this equals the pull of that arm at time $T$ due to reward monotonicity). In our work, we study the task of identifying the arm with the greatest cumulative reward in the instance. While these tasks are slightly different, we note that both are important and relevant tasks, and the best arm for one task will be a 2-approximation (that is, not be more than a factor of two suboptimal) for the other task. 
On the one hand, consider a setting in which each arm is a technology, and investing in developing the technology increases its utility. If we are interested in identifying one out of many technologies that will have the highest utility after the investment period, we are interested in identifying the arm with the highest single pull, corresponding to \cite{mussi2024best}'s setting. On the other hand, consider a setting in which each arm is an advertisement, and the reward associated with a pull is the amount of money a user spends as a result of that advertisement. A corporation would care to identify the arm that has the highest cumulative reward and then use that arm for future users. Thus, both are interesting and valid goals albeit slightly different from each other. \par

Having defined the tasks, we will now see more clearly that the difference in the ``niceness'' conditions relates to this, as well. We consider the deterministic variant of the setting which is the focus of this work (that is, set the stochasticity in the instances to zero). Now, we study the conditions and terms in Theorem 4.1 of \cite{mussi2024best}. Note that setting the stochasticity to zero implies $\epsilon, \sigma = 0\,.$ Then, $y$ is no longer a function of $a\,,$ and thus Eqn. 4 is satisfied for {\em all} $a\,,$ so we can achieve BAI with probability 1 if the conditions are met. This is the same guarantee we achieve, and so we are ready to compare the conditions more carefully. \par

In \cite{mussi2024best}, they study $y_i\,,$ the number of pulls of arm $i$ until the first difference is smaller than the average gap between the best single pull over all arms and the best single pull of this arm. This quantity corresponds to our quantity $h_i(\epsilon)\,,$ which is the number of pulls of arm $i$ after which the amount of cumulative reward we could achieve if we continued growing with the same slope is bounded by $\epsilon.$ Thus, both $y_i$ and $h_i$ study how quickly arms ``converge'' to their final value, $y_i$ directly in the value (of individual pull) sense and $h_i$ in the cumulative sense. \par

Finally, we note that \cite{mussi2024best} compares the sum of these ``convergence times'' to $(1-\iota)T\,,$ whereas we compare to a parameter $B\,.$ In both cases, the smaller the comparison value, the more value we can extract from the best arm in the remaining $\iota T$ or $T-B$ time. Thus, the overall flavor of the conditions in both works is the same, namely that if arms ``converge quickly,'' then BAI is possible. The exact notions of the words in quotes is what differs.

\subsection{Difficulty in Corralling Improving Bandit Algorithms}\label{appendix:corralling}

A natural approach towards obtaining best-of-both-worlds guarantees in the above examples would be to design a meta-algorithm that can potentially switch between \cite{blum_nearly-tight_2024} and \cite{mussi2024best}. We examine whether it is possible in the improving bandits setting to perform corralling~\cite{agarwal2017corralling,arora2021corralling,luo2022corralling}, a recently developed paradigm for meta-learning bandit algorithms in a fully online setting. Our main result here is an impossibility result which rules out the ability to use corralling to achieve best-of-both-worlds in a fully online improving bandits setting. We show that in the improving bandits setting, no matter what the meta-algorithm is, it is possible to suffer linear regret relative to best base algorithm.

In the meta-learning setting, we have a collection of $M$ {\it base} algorithms $\mathcal{B}=\{B_1,\dots,B_M\}$ for the IMAB problem and the goal is to (nearly) recover the guarantees of the best algorithm on any given instance. Each base algorithm can give its own prediction for which arm to play next, given the history of arm pull and rewards so far and the  meta-learner can decide which arm to actually pull based on these predictions. For best arm identification, we define the sub-optimality ratio of the meta-algorithm as

$$\overline{R}_T(\mathcal{M},\mu,\mathcal{B}):=\left[\max_{j}R_T^{B_j}(\mu)\right]/R_T^{\mathcal{M}}(\mu),$$
\noindent where $\mu$ denotes the IMAB instance, $R_T^{B_j}$ denotes the cumulative reward of the best arm selected by base algorithm $B_j$, and $R_T^{\mathcal{M}}$ denotes the cumulative reward of the best arm selected by the corralling meta-algorithm $\mathcal{M}$.

\begin{theorem}[Lower Bound for Best Arm Identification in Corralling Improving Bandits]
Consider deterministic multi-arm rested bandits with rising concave reward sequences. Let $\mathcal{B}=\{B_1,\dots,B_M\}$ be any finite class of (possibly randomized) base algorithms. A meta-algorithm $\mathcal M$ selects at each round one base algorithm whose recommended arm is executed.
Then 
\[
\inf_{\mathcal M}
\sup_{\mu,\mathcal{B}}
\overline{R}_T(\mathcal{M},\mu,\mathcal{B})
\;\ge\;
2.
\]
\label{theorem:corralling-bai}
\end{theorem}

\begin{proof}
Let $k=3$. Fix horizon $T$ and let $L=T/2$. Choose $\Delta>0$ sufficiently small. Construct three deterministic concave rising environments $E^j$ as follows.

For $s \le L$, define
\[
\mu_1(s)=\mu_2(s)=\mu_3(s)=s\Delta.
\]
That is, the first $L$ pulls of each arm yield identical rewards in all environments.

For $s>L$, define:

\medskip
\noindent
\textbf{Environment $E^j$:}
\[
\mu_j(s)=L\Delta+(s-L)\Delta,
\qquad
\mu_{i}(s)=L\Delta, \text{ for }i\ne j.
\]

\medskip
All reward sequences are nondecreasing and concave.
Observe that the  environments are identical until some arm is pulled more than $L$ times. In $E^j$, arm $j$ is uniquely optimal after the divergence. Set the base algorithm class $\mathcal{B}=\{B_1,B_2,B_3\}$, where $B_j$ always pulls arm $j$ and outputs it as the best arm.

Consider any meta-algorithm $\mathcal M$. Note that the meta-algorithm can pull at most one base algorithm more than $L$ times, say $B_1$ (WLOG). Since rewards are identical for the first $L$ pulls of each arm, $\mathcal M$ cannot distinguish between environments $E^2$ and $E^3$. So, no matter what best arm is selected by $\mathcal M$, we can always select an environment where it picks a sub-optimal arm. In any environment the optimal arm gets cumulative reward twice as much as any sub-optimal arm. 
Therefore,
\[
\sup_{\mu\in\{E^j\}}
\overline{R}_T(\mathcal{M},\mu,\mathcal{B})
\ge
2.
\]
\end{proof}

\noindent For cumulative reward maximization, we seek to bound the meta-regret with respect to the best algorithm among the base algorithms


$$\tilde{R}_T(\mathcal{M},\mu,\mathcal{B}):=\max_{j}R_T^{B_j}(\mu) - R_T^{\mathcal{M}}(\mu),$$

\noindent where $\mu$ denotes the IMAB instance, $R_T^{B_j}$ denotes the cumulative reward if algorithm $B_j$ is used exclusively to decide the arm pulls for $T$ rounds, and $R_T^{\mathcal{M}}$ denotes the cumulative reward collected by the corralling meta-algorithm $\mathcal{M}$.

\begin{theorem}[Minimax Impossibility of Corralling Improving Bandits]
Consider deterministic two-arm rested bandits with rising concave reward sequences and horizon $T$. Let $\mathcal{B}=\{B_1,\dots,B_M\}$ be any finite class of (possibly randomized) base algorithms. A meta-algorithm $\mathcal M$ selects at each round one base algorithm whose recommended arm is executed.
Then there exists a universal constant $c>0$ such that\looseness-1
\[
\inf_{\mathcal M}
\sup_{\mu,\mathcal{B}}
\tilde{R}_T(\mathcal{M},\mu,\mathcal{B})
\;\ge\;
cT.
\]
\label{theorem:corralling}
\end{theorem}

\begin{proof}[Proof Sketch]
    We construct two environments that are identical for the first $T/2$ pulls of each arm but diverge thereafter, with opposite optimal arms. Any meta-algorithm attempting to compete with the best run-alone base must effectively determine which environment it faces before the divergence point. However, since the two instances are indistinguishable during the prefix, identifying the correct environment requires linear exploration. Consequently, linear meta-regret is unavoidable on at least one of the two instances. A full proof is located in Appendix \ref{appendix:corralling}.
\end{proof}

\noindent This theorem establishes a  minimax barrier for corralling in improving bandits.  The obstruction is information-theoretic: the two environments are indistinguishable for a linear prefix yet demand opposite commitments thereafter. This motivates the need to consider alternative approaches towards achieving best-of-both-worlds guarantees.

We provide below a complete proof for Theorem \ref{theorem:corralling}.

\begin{proof}
Fix horizon $T$ and let $L=T/2$. Choose $\Delta>0$ sufficiently small. Construct two deterministic concave rising environments $E^+$ and $E^-$ as follows.

For $s \le L$, define
\[
\mu_1(s)=\mu_2(s)=s\Delta.
\]
That is, the first $L$ pulls of either arm yield identical rewards in both environments.

For $s>L$, define:

\medskip
\noindent
\textbf{Environment $E^+$:}
\[
\mu_1(s)=L\Delta,
\qquad
\mu_2(s)=L\Delta+(s-L)\Delta.
\]

\medskip
\noindent
\textbf{Environment $E^-$:}
\[
\mu_2(s)=L\Delta,
\qquad
\mu_1(s)=L\Delta+(s-L)\Delta.
\]

\medskip
Both reward sequences are nondecreasing and concave.
Observe that the two environments are identical until some arm is pulled more than $L$ times. In $E^+$, arm $2$ is uniquely optimal after the divergence; in $E^-$, arm $1$ is uniquely optimal. Set the base algorithm class $\mathcal{B}=\{B_1,B_2\}$, where $B_j$ always pulls arm $j$.

Consider any meta-algorithm $\mathcal M$. Since rewards are identical for the first $L$ pulls of each arm, the distribution over the first $L$ actions of $\mathcal M$ is identical under $E^+$ and $E^-$. Let $N_1$ be the expected number of pulls of arm $1$ by round $L$.
Without loss of generality, suppose $N_1 \le L/2$ (otherwise swap the roles of the arms).

Consider environment $E^-$, where arm $1$ becomes uniquely optimal after the divergence. Let $j^*$ be an index maximizing $R_T^{B_j}(E^-)$. Base algorithm $B_1$ must obtain $\Theta(T\Delta)$ additional reward by exploiting arm $1$ after round $L$.

Because $\mathcal M$ allocates at most $L/2$ pulls to arm $1$ during the prefix, it under-invests in the arm that later becomes uniquely optimal. Due to concavity, the cumulative post-divergence reward is linear in the number of additional pulls. Thus, there exists a constant $c>0$ such that
\[
R_T^{B_{1}}(E^-)
-
R_T^{\mathcal M}(E^-)
\ge
cT.
\]

If instead $N_1>L/2$, the same argument applied to $E^+$ yields an identical bound. Therefore,
\[
\sup_{\mu\in\{E^+,E^-\}}
\tilde{R}_T(\mathcal{M},\mu,\mathcal{B})
\ge
cT.
\]
\end{proof}

\subsection{Proofs for results in Section~\ref{sec:hybridguarantees}} \label{appendix:hybridproofs}
We prove the lemma used in  Section~\ref{sec:hybridguarantees} to justify the quantities $L_i, U_i$ used in Algorithm~\ref{alg:hybrid}.

\begin{lemma}\label{lem:envelope} 
For every arm $i$ and every $t \leq T$, we have $L_i(t) \leq f_i(T) \leq U_i(t)$.
\end{lemma}
\begin{proof}
By concavity, we know that $\gamma_i(s)$ is non-increasing, and therefore that for any $x \ge t$, we have
$$
f_i(x) - f_i(t) = \sum_{s=t}^{x-1} \left(f_i(s+1) - f_i(s)\right) \le \sum_{s=t}^{x-1} \gamma_i(t) = (x - t)\gamma_i(t).
$$
Setting $x = T$ gives $f_i(T) \le f_i(t) + (T - t)\gamma_i(t) = U_i(t).$ By monotonicity, we likewise know that $L_i(t) = f_i(t)\leq F_i(T)$.  
\end{proof}

We include below a complete proof for Theorem \ref{thm:hybrid-single2}.

\noindent{\it Proof (of Theorem \ref{thm:hybrid-single2}.)}

\begin{proof}[Proof of 1. (certificate correctness)]
Suppose an instance $I$ satisfies \(\mathrm{GCC}(\theta)\). Since $\triangle_i(t)$ is non-increasing for all $i$ and
$\sum_{i=1}^k h_i(\Delta_I/3) \le \theta$, we know that there are at most $\theta$ total pulls (across arms) such that $\triangle_i(t_i) > \Delta_I/3$. Note that we can never have $L_{i}(t_i)  > \max_{j \ne i} U_j(t_j)$ for some $i \neq i^*$, as we know by concavity and monotonicity that $L_i(t) \le f_i(T) \le U_i(t)$, so this would imply $f_{i^*} (T) \leq U_{i^*}(t_{i^*}) < L_i (t_i) \leq f_{i}(T)$ (a contradiction). Since $B>\theta$ and Stage $1$ always pulls the arm $i$ that maximizes $(U_i - L_i)$, it follows that the algorithm will reach some point (namely $t = \theta$) where $\triangle_i(t_i) \leq \Delta_I/3$ for all $i$. At this point, for each $j \ne i^\star$, we have
$$
U_j(t_j) \le f_j(T) + \frac{\Delta_I}{3} \le f^*(T) - \Delta_I + \frac{\Delta_I}{3} = f^*(T)  - \frac{2\Delta_I}{3}.
$$
Moreover, $i^*$ satisfies
$$
L_{i^\star}(t_{i^*})  \ge f^*(T) - \frac{\Delta_I}{3}.
$$ It follows that $L_{i^\star}(t_{i^*})  > \max_{j \ne i^\star} U_j(t_j)$, and therefore that the algorithm returns $i^*$ in Stage $1$.
\end{proof}

\begin{proof}[Proof of 2. (approximation fallback)]
Write $g^\star(h):=\max_i g_i(h)=\max_i f_i(t_i+h)$.
Let $T_\text{rem} := T-B$, and let $\tau' = T_\text{rem}-k$. Suppose $m' := (\tau'/T)f^{\star}(T)$. Since 
$g^{\star}(T) = \max_i f_i(T) \ge f^{\star}(T)$, we know that
$$
m' = \frac{\tau'}{T} f^{\star}(T) \le f^{\star}(T) \left( \frac{\tau'}{T} \right)^{\alpha} \le g^{\star}(T) \left( \frac{\tau'}{T} \right)^{\alpha}.
$$
Using monotonicity and the fact that $g^{\star}(T) \le 2 f^{\star}(T)$, we also know that $g^{\star}(\tau') \le 2 f^{\star}(T) = 2(T/\tau') m'$, and therefore that \ $m' \ge (\tau'/(2T)) g^{\star}(\tau')$. Since 
$B \le T/2$ and $T \ge 4k$, it follows that
$$
\frac{1}{8} g^{\star}(\tau') \le m' \le g^{\star}(T) \left( \frac{\tau'}{T} \right)^{\alpha}.
$$
Run \textit{PTRR}\(_\alpha\) for $T_\text{rem}$ steps with parameters $m'$ and $\tau'$. By Theorem \ref{thm:ptrr-alpha}, we know that
$$
\mathbb{E}\big[\mathrm{ALG}_{T_\text{rem}} \big]\ \ge\ 
\frac{\mathrm{OPT}^{\mathrm{res}}_{T_\text{rem}}}{C_\alpha c_2(k+1)^{\alpha/(1+\alpha)}},
$$
where $\mathrm{OPT}^{\mathrm{res}}_{T_{\mathrm{rem}}}$ denotes the optimal cumulative reward for $\{g_i\}$ over $T_{\mathrm{rem}}$ rounds and $C_\alpha = 2^{\alpha+2}(\alpha+1)$.

Now note that $\mathrm{OPT}^{\mathrm{res}}_{T_{\mathrm{rem}}}\ge \tfrac12g^\star(T_{\mathrm{rem}})T_{\mathrm{rem}}$, and that the algorithm’s maximum single‑pull reward dominates its average reward (Facts B.1 and B.3 in the appendix of \cite{blum_nearly-tight_2024}). Let $\hat i$ denote the arm that achieves this maximum single-pull reward, and note that

$$
\mathbb{E}\Big[\max_{t\le T_{\mathrm{rem}}}\text{reward}_t\Big]\ \ge\ \frac{1}{T_{\mathrm{rem}}}\mathbb{E}\big[\mathrm{ALG}_{T_{\mathrm{rem}}}\big]\ \ge\ \frac{g^\star(T_{\mathrm{rem}})}{2C_\alpha c_2(k+1)^{\alpha/(1+\alpha)}}.
$$

Since $f_{\hat i}(T)\ge \max_{t\le T_{\mathrm{rem}}}\text{reward}_t$ by monotonicity, we know that taking expectations gives
$$
\mathbb{E}\big[f_{\hat i}(T)\big]\ \ge\ \frac{g^\star(T_{\mathrm{rem}})}{2C_\alpha c_2(k+1)^{\alpha/(1+\alpha)}}.
$$
Monotonicity and concavity give $g^\star(T_{\mathrm{rem}})\ge (T_{\mathrm{rem}}/T)f^\star(T)\ge \tfrac12 f^\star(T)$, as $T_{\mathrm{rem}}\ge T/2$. Combining with $c_2\le 8$ from Step~1, it follows that
$$
\mathbb{E}\big[f_{\hat i}(T)\big]\ \ge\ \frac{1}{2C_\alpha c_2(k+1)^{\alpha/(1+\alpha)}}\cdot \frac{1}{2}f^\star(T)\ \ge\ \frac{1}{2^{\alpha+7}(\alpha+1)}(k+1)^{-\alpha/(1+\alpha)}f^\star(T),
$$
as desired.
\end{proof}

\subsection{Proof of Lemma \ref{lemma:bdqd-hybrid}}

\begin{proof}
    We argue this by applying the proof of Lemma~\ref{lemma:bdqd}. For a fixed (augmented) instance (i.e., fixed instance and fixed random permutation), we are interested in computing the number of different behaviors as we vary $B, \alpha\,.$ To understand this, let us start by fixing $B$. Then, by Lemma~\ref{lemma:bdqd}, we know that there are at most $kT$ possible behaviors as we vary $\alpha\,.$ Now, for a fixed value of $B,$ for either algorithm, the sequence of pulls is determined, which exactly determines the loss in Stage 1. Thus, each value of $B$ corresponds to at most 1 new value of the loss. This implies that there are at most $kT$ possible behaviors in both stages for a fixed value of $B.$ Since there are at most $T$ values of $B,$ we have that there are at most $kT^2$ possible behaviors.
\end{proof}

\subsection{Cumulative Reward Best Arm Identification} \label{appendix:cum-reward-bai}
In the below sections, we provide comparable BAI guarantees to Section~\ref{sec:hybridguarantees} for cumulative reward instead of maximum reward. As before, we work in the standard improving‑bandits setting with $k$ arms, a known horizon $T$, and non‑decreasing concave reward functions $f_i$. Each algorithm Hybrid$_{\alpha,B}$ has two stages. Stage 1 uses a UCB-style envelope: At each step, the algorithm computes a lower bound $L_i(n)$ and terminal upper bound $U_i(n)$ on the final accumulated reward of every arm and pulls the arm with the largest optimistic estimate $U_i$. If the lower bound of one arm dominates the terminal upper bound of every other arm, Hybrid$_{\alpha,B}$ commits to this arm. 
If no commit occurs by time $B$, Stage 2 runs PTRR$_\alpha$ and finds an arm whose expected terminal reward is at least a substantial fraction of the best arm’s.\looseness-1 \par

For the terminal envelope, we define $L_i(t) := F_i(t) + (T - t) f_i(t),
\triangle_i(t) := \frac{(T - t)(T - t + 1)}{2} \, \gamma_i(t - 1),$ and $
U_i(t) := L_i(t) + \triangle_i(t), 
$
where $F_i(T) \coloneqq \sum_{t = 1}^T f_i(t)$ and $\gamma_i (t-1) := f_i(t) - f_i (t-1)$. We set $U_i(0) : = \infty$ to ensure first pulls. Using concavity and monotonicity, it is again straightforward to prove that $L_i(t) \leq F_i(T) \leq U_i(t)$ for all $i,t$.

\begin{algorithm}[h]
\caption{$\textit{Cumulative Hybrid}_{\alpha, B}$}\label{alg:cumhybrid}
\begin{algorithmic}[1]
\State \textbf{Require:} $m$
\State \textbf{Stage 1}: $t\gets0$ 
\For{each arm $i$}
\State $t_i\gets0$, $F_i\gets0$, $L_i\gets0$, $U_i\gets+\infty$
\EndFor
\While{$t<B$}
  \For{each $i$ with $t_i\ge1$}
     \State $L_i \gets F_i + (T-t_i)\,f_i(t_i)$ \hfill 
     \State $\gamma_i \gets f_i(t_i)-f_i(t_i-1)$ \hfill 
     \State $U_i \gets L_i + \frac{(T-t_i)(T-t_i+1)}{2}\,\cdot \gamma_i$ 
  \EndFor
  \State $\hat i \gets \arg\max_i L_i$,\quad $U_{\mathrm{next}}\gets \max_{j\ne \hat i} U_j$
  \If{$L_{\hat i} > U_{\mathrm{next}}$} 
     \State \textbf{return} $\hat i$.
  \EndIf
  \State $i' \gets \arg\max_i (U_i - L_i)$
  \State \textbf{pull} $i'$;\; $t_{i'}\gets t_{i'}+1$, $t\gets t+1$, $F_{i'}\gets F_{i'}+f_{i'}(t_{i'})$
\EndWhile
\State \textbf{Stage 2}:
 $\tau' \gets (T - B) - k$, $m' \gets \left(\frac{\tau'}{T}\right) \cdot m$
\For{each $i$}
\State $g_i(s)\gets f_i(t_i + s)$
\EndFor \\
\Return $\hat i \gets \text{arm returned by } \textit{PTRR}_\alpha$ with parameters $(m', \tau')$ on $\{g_i\}$ for $T-B$ steps.

\end{algorithmic}
\end{algorithm}


\subsection{
Best-of-Both-Worlds Best arm identification guarantees}
In this section, we show that \textit{Cumulative Hybrid} contains algorithms that simultaneously (i) guarantee best arm identification on sufficiently benign instances, and (ii) preserve tight (up to constants) multiplicative bounds for approximating the best arm
on adversarial instances.\footnote{With additional information about stronger concavity, we achieve sharper bounds by using the corresponding version of PTRR.}

We start by defining a class of `sufficiently benign' instances and prove that our algorithm is guaranteed to return the best arm on all members of this class. If an instance is not in this class, Stage~2 pursues {\em approximate} BAI and runs $PTRR_\alpha$ with $\alpha$ dependent on the strength of concavity ($\alpha = 1$ works for {\em all} instances). Having reverted to an approximate goal, we identify an arm $\hat{i}$ whose final reward satisfies
$\mathbb{E}[f_{\hat{i}}(T)] \ge \Omega\left(k^{\frac{-\alpha}{1+\alpha}}\right) f^{\star}(T)$.  

As before, we assume for simplicity that both $T$ and $f^*(T)$ are known to the algorithm, which we use to set $\tau' = (T-B)-k$ and $m' = \frac{\tau'}{T} \cdot f^*(T)$. As in Section~\ref{sec:sharper CR}, these assumptions can be removed with only $O(\log k)$ overhead (see also Appendix \ref{appendix:unknownT}). 

We will now define a condition under which we can guarantee best arm identification.

\begin{definition}[Per-arm terminal budget, $h_i$]
For any arm $i$ and $\epsilon>0$, the terminal budget $h_i$ of arm $i$ is defined as
$
h_i(\epsilon) := \min\{\,n \in \{2,\ldots,T\} : \triangle_i(n) \le \epsilon\,\},
$
where $\triangle_i(t) := \frac{(T - t)(T - t + 1)}{2} \, \gamma_i(t - 1)$. 
\end{definition}

\begin{definition}[Best Arm Gap, $\Delta_I$]
For any instance $I$, its Best Arm Gap $\Delta_I$ is defined as
 $\Delta_I := \text{OPT}_T - \max_{j \ne i^*} F_j(T).$
\end{definition}

\begin{definition}[Gap Clearance Condition, $GCC(B)$]
For any $B\le T$, we say that an instance satisfies the Gap Clearance Condition condition $GCC(B)$ if $\Delta_I > 0$ and
$
\sum_{i=1}^K h_i(\Delta_I / 3) \;\le\; B.
$
\end{definition}

Under concavity, $\triangle_i(n)$ denotes the worst-case remaining terminal mass for arm $i$ after $n$ pulls: it is the most reward an adversary can ``hide in the tail'' by continuing with the last observed slope. 
The per-arm budget $h_i(\epsilon)$ is therefore the minimal number of pulls needed to ensure its optimistic terminal value is close to current lower envelope. 
Taking $\epsilon = \Delta_I/3$, once a suboptimal arm has been pulled this many times, its best possible continuation still loses to the best arm’s lower envelope. 
Thus $\mathrm{GCC}(B)$ states that the total work needed to certify the best arm fits within the mid-horizon budget $B$. 
If the sum exceeds $B$, concavity allows at least one suboptimal arm to remain plausibly optimal by time $B$, so no sound mid-horizon certificate can be guaranteed.

\begin{theorem}[best-of-both-worlds guarantees] \label{thm:hybrid-single}
Suppose an instance $I\in\mathcal{I}$ has Concavity Envelope Exponent $\beta_I\in(0,1]$. Algorithm \ref{alg:cumhybrid} with $\alpha \in(\beta_I, 1]$  satisfies the following properties:
\begin{enumerate}
\item If the instance further satisfies $\Delta_I>0$, and $GCC(\theta)$ holds for $\theta\le T/2$, then the algorithm with $B\in (\theta,T/2]$  identifies and commits to the best arm $i^\star$ in Stage~1.
    \item If Stage~1 does not certify a best arm by time $B$, then Stage~2 finds an approximate best arm $\hat i$ such that 
    $\mathbb{E}\big[F_{\hat i}(T)\big]\ \ge\ \Omega{\left(k^{\frac{-\alpha}{1+\alpha}}\right)}\; F^\star(T),$ where the expectation is  over the randomness of the algorithm. 
\end{enumerate}
\end{theorem}

\begin{proof}[Proof Sketch]
We argue (1) and (2) separately. (1) $\mathrm{GCC}(\theta)$ implies that the per–arm budgets $h_i(\epsilon)$ sum to at most $\theta$. As long as the certificate fails, Stage~1 pulls an arm with slack $>\Delta_I/3$, so these budgets are met within $B\ge\theta$ pulls. Then every suboptimal arm has $U_j\le \mathrm{OPT}_T- \frac{2\Delta_I}{3}$ and the best arm has $L_{i^\star}\ge \mathrm{OPT}_T- \frac{\Delta_I}{3}$, which implies that $L_{i^\star}\ge\max_{j\ne i^\star}U_j$. We commit to $i^\star$.
(2) If no certificate fires by $B\le T/2$, running $\textit{PTRR}_\alpha$ for $T_{\mathrm{rem}}$ steps yields an average reward within a $k^{\alpha/(1+\alpha)}$ factor of the residual optimum (up to constants). Using the fact that the best single pull is at least the average, and $\mathrm{OPT}^{\mathrm{res}}$ is at least a constant times $g^\star(T_{\mathrm{rem}})\,T_{\mathrm{rem}}$ by concavity, we get $\mathbb{E}[f_{\hat i}(T)]\ge \Omega\!\big(k^{-\alpha/(1+\alpha)}\big)\,f^\star(T)$. 
\end{proof}

\begin{proof}[Proof of 1. (certificate correctness)]
Suppose an instance $I$ satisfies \(\mathrm{GCC}(\theta)\). Since $\triangle_i(t)$ is non-increasing for all $i$ and
$\sum_{i=1}^k h_i(\Delta_I/3) \le \theta$, we know that there are at most $\theta$ total pulls (across arms) such that $\triangle_i(t_i) > \Delta_I/3$. Note that we can never have $L_{i}(t_i)  > \max_{j \ne i} U_j(t_j)$ for some $i \neq i^*$, as we know that $L_i(t) \le F_i(T) \le U_i(t)$, so this would imply $F_{i^*} (T) \leq U_{i^*}(t_{i^*}) < L_i (t_i) \leq F_{i}(T)$ (a contradiction). Since $B>\theta$ and Stage $1$ always pulls the arm $i$ that maximizes $(U_i - L_i)$, it follows that the algorithm will reach some point (namely $t = \theta$) where $\triangle_i(t_i) \leq \Delta_I/3$ for all $i$. At this point, for each $j \ne i^\star$, we have
$$
U_j(t_j) \le F_j(T) + \frac{\Delta_I}{3} \le \mathrm{OPT}_T - \Delta_I + \frac{\Delta_I}{3} = \mathrm{OPT}_T - \frac{2\Delta_I}{3}.
$$
Moreover, $i^*$ satisfies
$$
L_{i^\star}(t_{i^*})  \ge F_{i^\star}(T) - \Delta_I/3 = \mathrm{OPT}_T - \Delta_I/3.
$$ It follows that $L_{i^\star}(t_{i^*})  > \max_{j \ne i^\star} U_j(t_j)$, and therefore that the algorithm returns $i^*$ in Stage $1$.
\end{proof}

\begin{proof}[Proof of 2. (approximation fallback)]
Write $g^\star(h):=\max_i g_i(h)=\max_i f_i(t_i+h)$.
Let $T_\text{rem} := T-B$, and let $\tau' = T_\text{rem}-k$. Suppose $m' := (\tau'/T)f^{\star}(T)$. Since 
$g^{\star}(T) = \max_i f_i(T) \ge f^{\star}(T)$, we know that
$$
m' = \frac{\tau'}{T} f^{\star}(T) \le f^{\star}(T) \left( \frac{\tau'}{T} \right)^{\alpha} \le g^{\star}(T) \left( \frac{\tau'}{T} \right)^{\alpha}.
$$
Using monotonicity and the fact that $g^{\star}(T) \le 2 f^{\star}(T)$, we also know that $g^{\star}(\tau') \le 2 f^{\star}(T) = 2(T/\tau') m'$, and therefore that \ $m' \ge (\tau'/(2T)) g^{\star}(\tau')$. Since 
$B \le T/2$ and $T \ge 4k$, it follows that
$$
\frac{1}{8} g^{\star}(\tau') \le m' \le g^{\star}(T) \left( \frac{\tau'}{T} \right)^{\alpha}.
$$
Run \textit{PTRR}\(_\alpha\) for $T_\text{rem}$ steps with parameters $m'$ and $\tau'$. From the analysis of \textit{PTRR}\(_\alpha\) (Theorem \ref{thm:ptrr-alpha}), we know that
$$
\mathbb{E}\big[\mathrm{ALG}_{T_\text{rem}} \big]\ \ge\ 
\frac{\mathrm{OPT}^{\mathrm{res}}_{T_\text{rem}}}{C_\alpha c_2(k+1)^{\alpha/(1+\alpha)}},
$$
where $\mathrm{OPT}^{\mathrm{res}}_{T_{\mathrm{rem}}}$ denotes the optimal cumulative reward for $\{g_i\}$ over $T_{\mathrm{rem}}$ rounds and $C_\alpha = 2^{\alpha+2}(\alpha+1)$.

Now note that $\mathrm{OPT}^{\mathrm{res}}_{T_{\mathrm{rem}}}\ge \tfrac12g^\star(T_{\mathrm{rem}})T_{\mathrm{rem}}$, and that the algorithm’s maximum single‑pull reward dominates its average reward (Facts B.1 and B.3 in the appendix of \cite{blum_nearly-tight_2024}). Let $\hat i$ denote the arm that achieves this maximum single-pull reward, and note that

$$
\mathbb{E}\Big[\max_{t\le T_{\mathrm{rem}}}\text{reward}_t\Big]\ \ge\ \frac{1}{T_{\mathrm{rem}}}\mathbb{E}\big[\mathrm{ALG}_{T_{\mathrm{rem}}}\big]\ \ge\ \frac{g^\star(T_{\mathrm{rem}})}{2C_\alpha c_2(k+1)^{\alpha/(1+\alpha)}}.
$$

Since $f_{\hat i}(T)\ge \max_{t\le T_{\mathrm{rem}}}\text{reward}_t$ by monotonicity, we know that taking expectations gives
$$
\mathbb{E}\big[f_{\hat i}(T)\big]\ \ge\ \frac{g^\star(T_{\mathrm{rem}})}{2C_\alpha c_2(k+1)^{\alpha/(1+\alpha)}}.
$$
Monotonicity and concavity give $g^\star(T_{\mathrm{rem}})\ge (T_{\mathrm{rem}}/T)f^\star(T)\ge \tfrac12 f^\star(T)$, as $T_{\mathrm{rem}}\ge T/2$. Combining with $c_2\le 8$ from Step~1, it follows that
$$
\mathbb{E}\big[f_{\hat i}(T)\big]\ \ge\ \frac{1}{2C_\alpha c_2(k+1)^{\alpha/(1+\alpha)}}\cdot \frac{1}{2}f^\star(T)\ \ge\ \frac{1}{2^{\alpha+7}(\alpha+1)}(k+1)^{-\alpha/(1+\alpha)}f^\star(T),
$$
as desired.

Finally, in terms of cumulative reward of the chosen vs. best arms:
$$
\E{}{F_{\hat{i}}(T)} \ge \frac T2 \E{}{f_{\hat{i}}(T)} \ge \frac T2 \frac{1}{2^{\alpha+7}(\alpha+1)}(k+1)^{-\alpha/(1+\alpha)}f^\star(T) \ge \mathrm{OPT}_T \frac{1}{2^{\alpha+8}(\alpha+1)}(k+1)^{-\alpha/(1+\alpha)}
$$
\end{proof}

\chapter{Appendices for Pessimism Traps}
\section{Key features of pessimism traps}
\label{appendix:features}
Morton's epistemic characterization of pessimism traps builds upon several empirical characterizations from scholars studying the phenomenon in the field of education. Morton elucidates pessimism traps via the following key distinguishing features [pgs. 732-3 of \cite{morton2022resisting}, paraphrased here]:
\begin{enumerate}
    \item Making the ambitious choice is an investment, involving a long duration before payoff and hard work.
    \item There is a feasible alternative that still has a reasonable payoff.
    \item There is risk involved in the ambitious choice, enough that it would affect the choice of a risk-averse agent.
    \item There is strong evidence that succeeding in the ambitious end when coming from the group of interest is of low likelihood.
    \item Not pursuing the ambitious end will not change the pessimistic view they hold about it.
\end{enumerate}

Importantly, the choices in question aren't even necessarily the ``right'' and ``wrong'' ones, but rather they are ``ambitious'' and ``moderate.'' 

 \section{Related Works} \label{appendix:related-work}
 
Our model is motivated by several strands of research across several different areas. The work of \cite{bikhchandani1992theory} on information cascades examines how rational agents often ignore their private signals to imitate others' actions. Similar analyses oflooks at rational herding behavior are presented by \cite{banerjee1992simple}.

The classic models of information cascades and herding behavior have been expanded through various theoretical and empirical studies to better understand the nuances and complexities of these phenomena in different contexts and under varying assumptions. \cite{bikhchandani1992theory} initially framed the classic information cascade model, demonstrating how individuals, despite possessing private information, often conform to the erroneous actions of predecessors due to the strong influence of prior actions. This model has served as a baseline for exploring various dimensions of information processing within groups. Building on this foundational work,  \cite{anderson1997information} conducted laboratory experiments to observe cascade behavior in controlled settings, adding empirical evidence to the theoretical predictions. Their work highlights how real-world decision-making often reflects the tendencies predicted by cascade models but with notable variations depending on the specifics of the information and context.

Another relevant literature that studies consensus formation is that of opinion dynamics and social learning, which we briefly survey here. In this vein, \cite{sirbu_opinion_2017} have a review paper and \cite{chamley_rational_2004} has a book which both review both simple and more complex models for how people’s beliefs vary in relation to the beliefs of those they interact with. \cite{acemoglu_opinion_2011} study both Bayesian and non-Bayesian models for how agents update their beliefs, investigating consensus and asymptotic learning of state (i.e., what are the true beliefs in the world). Finally, \cite{das_modeling_2014} present a new model in which to study opinion dynamics.

From a philosophical perspective, the concepts of epistemic and hermeneutical injustice by \cite{fricker2007epistemic} provide a philosophical perspective for understanding how marginalized individuals may discount their knowledge due to systemic societal biases. \cite{fricker2007epistemic} explains how societal biases can influence individual knowledge recognition and acquisition, impacting decision-making processes in marginalized communities. This philosophical perspective may help understand the dynamics of pessimism traps, as further detailed by \cite{morton2022resisting} in their examination of individuals' ethical and epistemic dilemmas in such traps. \cite{Aronowitz2021-AROEBB-2} also extends the notion of an exploration and exploitation trade-off, common in the computer science literature, to the act of believing, ultimately arguing that ``epistemic rationality fundamentally concerns time.''

The contributions of \cite{manski1993identification} to social learning expand our understanding of how macro-level patterns, like those seen in economic models of cascades, are influenced by micro-level decisions. In behavioral economics, \cite{thaler2008nudge} present a concept of ``nudges'' that aligns with our discussions on how minor policy adjustments can realign individual decision-making with optimal outcomes, supporting our proposed interventions to counteract the effects of pessimism traps. \cite{thaler2008nudge}'s foundational work on ``nudging" provides an insightful lens through which to view subtle policy interventions to improve decision-making processes. In their book, \textit{Nudge: Improving Decisions About Health, Wealth, and Happiness} \cite{thaler2008nudge}, they introduce the concept of choice architecture—a method of influencing choice by organizing the context in which people make decisions. This approach does not restrict freedom of choice or significantly alter economic incentives; instead, it facilitates better decision-making through subtle environmental modifications. This is particularly relevant to our model, where nudges could be strategically employed to counter pessimism traps by subtly realigning perceived options with more optimistic or beneficial outcomes. Similarly, work in theoretical computer science and game theory, such as \cite{Balcan2013CircumventingTP}, study ``nudging'' specifically in the context of equilibria, where the goal is to redirect a population from a less desirable to a more desirable equilibrium.

Additionally, empirical studies have supported the application of nudges in various domains. For instance, \cite{johnson2012beyond} extends the discussion on the efficacy of nudges in real-world settings, demonstrating their potential to lead to significant changes in behavior across health, financial, and environmental domains. These interventions, while minimal, leverage psychological and behavioral insights to encourage choices that individuals are likely to benefit from without heavy-handed enforcement. Integrating nudging into public policy, particularly in health and environmental strategies, as discussed by  \cite{benartzi2017should}, shows how these concepts have moved beyond theoretical discussions to practical implementations. Their analysis highlights how small changes in policy design can have outsized effects on behavior, aligning well with our analysis of interventions designed to mitigate the effects of information cascades and decision-making under uncertainty.

Finally, \cite{coateLoury}’s model of statistical discrimination examines addresses  how how perceived barriers and individuals' response the responses of individuals to these barriers can perpetuate cycles of disadvantage, reinforcing the relevance of our model in discussing issues faced by marginalized groups. The psychological perspective provided by \cite{kahneman1974judgment} adds depth to our understanding of why individuals might disregard their private information, supporting the economic models of information cascades with cognitive theories on biases and heuristics. Lastly, \cite{sunstein2001echo} examines analysis on group polarization and echo chambers provides socio-psychological insights into how information cascades can exacerbate group divisions, emphasizing the multidisciplinary nature of studying these phenomena.

\section{Proof for Posterior Update (Eqn.~\ref{eq:post})}

\begin{proof}

First, applying Bayes' theorem and given the conditional independence of $s_t$ and $\barH{t-1}$ when conditioned on the best action, we have:
\begin{align*}
&\Pr{}{\world{A}|s_t, \barH{t-1}} = \frac{\Pr{}{s_t|\world{A}}\Pr{}{\barH{t-1}|\world{A}}\Pr{}{\world{A}}}{\Pr{}{s_t, \barH{t-1}}}
\end{align*}

Expanding the denominator using the law of total probability:
\begin{align*}
\Pr{}{s_t, \barH{t-1}} &= \Pr{}{\world{A}} \Pr{}{s_t, \barH{t-1}|\world{A}} + \Pr{}{\world{B}} \Pr{}{s_t, \barH{t-1}|\world{B}} \\
\text{cond. indep.}\Rightarrow \quad  &= \Pr{}{\world{A}}\Pr{}{s_t|\world{A}} \Pr{}{\barH{t-1}|\world{A}} + \Pr{}{\world{B}}\Pr{}{s_t|\world{B}} \Pr{}{\barH{t-1}|\world{B}}
\end{align*}

Finally, because we assume a uniform prior over which action is correct, we have $\Pr{}{\world{A}} = \Pr{}{\world{B}}$ and can thus cancel out all of those terms.

\end{proof}

\section{Formulation as a Random Walk} \label{appendix:random-walk}
Calculating the posterior can be difficult depending on the tie-breaking rule, as agents have to reverse engineer previous individuals' thought processes. However, with the tie-breaking rule we employ and our eventual subsidy, we can reformulate this process as a one-dimensional random walk on the integers. We begin with a toy example for intuition.

\begin{lemma}[Asymmetric Simple Random Walk in -1 to 1]
\label{lem:srw}
While the number of choices for action $A$ and action $B$ in the history $\barH{t-1}$ differ by no more than 1, agent $t$ will follow their signal. Thus, the process is an asymmetric simple random walk in this interval.
\end{lemma}

\begin{proof}
We will prove this through induction. Our inductive hypothesis is that if the previous $t-1$ agents acted according to their signal, and if the number of choices of action $A$ and $B$ differ by at most one, then agent $t$ will follow their signal. To begin, we establish the base case. Consider the first agent. Note that the hypothesis that the previous $t-1$ agents acted according to their signal is vacuously true. Now, let us look at the two signal options for the first agent to verify that they indeed follow their signal in each case. 
If the first agent receives signal $s_1=A$, they will update their posterior, according to Eqn.~\ref{eq:post} as:

\begin{align*}
\Pr{}{\world{A}|s_1=A, H_0 = \{\}} = \frac{\Pr{}{s_1=A|\world{A}}\Pr{}{H_0=\{\}|\world{A}}}{\sum_{w \in \{A,B\}}\Pr{}{s_1=A|\world{w}}\Pr{}{H_0=\{\}|\world{w}}}
\end{align*}

Now, because $H_0$ is always $\{\}$, we can set $\Pr{}{H_0=\{\}|\world{A}} = \Pr{}{H_0=\{\}|\world{B}} = 1$ and simplify the above expression to:

\begin{align*}
\Pr{}{\world{A}|s_1=A, H_0} &= \frac{\Pr{}{s_1=A|\world{A}}}{\Pr{}{s_1=A|\world{A}}+ \Pr{}{s_1=A|\world{B}}} = \frac{p}{p + (1-p)} = p
\end{align*}

Because $p > \frac{1}{2}$, the first agent will choose action $A$ in this case, thus following their signal.

If, instead the agent receives signal $B$, they will update their posterior as:
\begin{align*}
\Pr{}{\world{A}|s_1=B, H_0} &= \frac{\Pr{}{s_1=B|\world{A}}}{\Pr{}{s_1=B|\world{A}}+ \Pr{}{s_1=B|\world{B}}}
= \frac{1-p}{1-p + p}  = 1-p
\end{align*}

In this case, they will choose action $B$, again following their signal. From this, we can say that the first agent's action is identical to the action of their signal and so is distributed as a Bernoulli random variable with success parameter $p$.

 This satisfies the base case. Now, we proceed by induction. Suppose that the previous $t-1$ agents acted according to their signal, i.e., $H_{t-1} = \barH{t-1}$. Let $|A|$ indicate the number of choices for $A$ observed among the first $t-1$ agents, and let $|B|$ be the number of choices for $B$ among those agents. Then, if the $t^{th}$ agent gets signal $A$, their posterior is:

\begin{align*}
\Pr{}{\world{A}|s_t=A, H_{t-1}} = &\frac{p^{|A|+1}(1-p)^{|B|}}{p^{|A|+1}(1-p)^{|B|} + p^{|B|}(1-p)^{|A|+1}}
\end{align*}

If $|A| = |B|$, then the posterior is equal to $p$, and the agent follows their signal.
If $|A| = |B| - 1$, the posterior is equal to $\frac{1}{2}$, and the agent breaks the tie by following their signal. Finally, if $|A| = |B| + 1$, the posterior is $\frac{p^2}{p^2 + (1-p)^2}$, which is greater than $\frac{1}{2}$ for $p> \frac{1}{2}$.

Equivalently, if they get the signal $B$, the posterior is:

\begin{align*}
\Pr{}{\world{A}|s_t=B, H_{t-1}} = &
\frac{p^{|A|}(1-p)^{|B|+1}}{p^{|A|}(1-p)^{|B|+1} + p^{|B|+1}(1-p)^{|A|}}
\end{align*}

\end{proof}

\begin{lemma}[Stopping Points at -2 or 2]
\label{lem:stop}
A cascade will begin if the number of choices for action $A$ and action $B$ in the history $H_{t-1}$ differ by 2.
\end{lemma}

\begin{proof}
To see this, assume that at time $t$,  $|A| - |B| \geq 2$ or $|A| - |B| \leq -2$. Consider the first case. If agent $t$ gets signal $A$, their posterior is

\[
\frac{p}{1-p}^{|A|-|B|+1} \geq \frac{p}{1-p}^3 \geq 1 
\]

and so they take action $A$. If they get signal $B$, their posterior is  
\[
\frac{p}{1-p}^{|A|-|B|-1}\geq \frac{p}{1-p} \geq 1 
\]

So, regardless of their signal, they choose action $A$. Thus, a cascade begins, because this condition will be maintained for each subsequent agent. An equivalent analysis applies to the second case.
\end{proof}

\begin{lemma}[Probability of Wrong Cascade] \label{lemma:prcascade}
The probability of an incorrect cascade for $p>\frac{1}{2}$ is  $$\frac{\left( \frac{1-p}{p} \right)^2 + \left( \frac{1-p}{p} \right)^3}{1 + \left( \frac{1-p}{p} \right) + \left( \frac{1-p}{p} \right)^2 + \left( \frac{1-p}{p} \right)^3}\,.$$
\end{lemma}
\begin{proof}
We can compute this probability via a recurrence. Consider a random walk on the number line. Define $X_i$ as the probability a walk starting at $i$ reaches 2 before it reaches -2. Then, we have that:
\begin{align*}
    X_i = p \, X_{i+1} + (1-p) \, X_{i -1} \quad ; \quad 
    X_2 = 1 \quad ; \quad
    X_{-2} = 0
\end{align*}

We can solve this system as follows:
\begin{align*}
    Y_i &\coloneqq X_i - X_{i-1} \\
   p \, X_i + (1-p) \, X_i &=  p \, X_{i+1} + (1-p) \, X_{i -1} \Leftrightarrow (1-p) \, Y_i = p \, Y_{i+1} \\
   Y_i &= Y_0 \, \left(  \frac{1-p}{p} \right)^i \\
   \sum_{i = -1}^2 Y_i = X_2 - X_{-2} = 1 \quad \Rightarrow \quad
   Y_0  &= \frac{1}{\sum_{i = -1}^2 \left( \frac{1-p}{p} \right)^i} \\
   X_i = X_{-2} + \sum_{j = -1}^i Y_j &= \frac{1}{\sum_{j = -1}^2 \left( \frac{1-p}{p} \right)^j} \sum_{k = -1}^i \left(  \frac{1-p}{p} \right)^k \,.
\end{align*}
Simplifying this somewhat, we have that starting at 0:
\begin{align*}
    \Pr{}{\text{up cascade}} &= \frac{1 + \left( \frac{1-p}{p} \right)}{1 + \left( \frac{1-p}{p} \right) + \left( \frac{1-p}{p} \right)^2 + \left( \frac{1-p}{p} \right)^3} \\
    \Pr{}{\text{down cascade}} &= \frac{\left( \frac{1-p}{p} \right)^2 + \left( \frac{1-p}{p} \right)^3}{1 + \left( \frac{1-p}{p} \right) + \left( \frac{1-p}{p} \right)^2 + \left( \frac{1-p}{p} \right)^3} \\
\end{align*}

\end{proof}
Therefore, there is a substantial probability of an incorrect cascade, increasing as $p$ gets closer to $\frac{1}{2}$.

\begin{lemma}[Expected Starting Time] \label{lemma:estart} 
In expectation, it will take $\frac{2}{1-2p + 2p^2}$ agents for a cascade to begin.
\end{lemma}

\begin{proof}

To compute this, let us consider the same random walk defined in Lemma~\ref{lemma:prcascade}. With respect to this random walk, let us now define $Z_i = \E{\text{time to } \pm 2 \text{ starting from } i}\,.$ That is, we are interested in the expected time it takes to converge to a cascade. We can write the following equations and initial conditions ($Z_2 = 0 ;
    Z_{-2} = 0$) and solve manually:

\begin{align*}
    Z_i &= p\,(Z_{i+1} + 1) + (1-p)\, (Z_{i -1} + 1) = 1 + p\,(Z_{i+1}) + (1-p)\, (Z_{i -1})\\
    Z_0 &= 1 + p\,(Z_{1}) + (1-p)\, (Z_{-1}) \quad ; \quad
    Z_1 = 1 + p \, Z_2 + (1-p)\, Z_0 = 1 + (1-p) Z_0 \\
    Z_{-1} &= 1 + p \, Z_0 + (1-p)\, Z_{-2} = 1 + p \, Z_0 \\
    Z_0 &= 1 + p\,(1 + (1-p) Z_0) + (1-p)\, (1 + p \, Z_0) = 2 + 2\, p \, (1-p) \, Z_0 = \frac{2}{1-2\,p\,(1-p)}\,.
\end{align*}

\end{proof}

\section{Time-Varying Subsidy Proofs}


\subsection{Proof of Theorem~\ref{thm:subsidy}}

\begin{proof}
Consider the scenario where the net expected reward for choosing action $A$ includes both the base reward $R$ and a potential subsidy $r_t$:
\begin{align*}
\left(R + r_t \right) \Pr{\world{A} \mid \barH{t-1}, s_t} + r_t \cdot \Pr{}{\world{B} \mid \barH{t-1}, s_t} = r_t + R \cdot \Pr{}{\world{A} \mid H_{t-1}, S_t}.
\end{align*}

Similarly, the reward for choosing action $B$ is:
\[
R \cdot \Pr{}{\world{B} \barH{t-1}, s_t}
\]

We seek to determine $r_t$ such that it ensures action $A$ is at least as preferable when $S_t = A$ and less preferable when $S_t = B$. This leads to the following conditions:
\begin{align*} \text{(1) } r_t + R \cdot \Pr{}{\world{A} \mid \barH{t-1}, s_t = A} &\geq  R \cdot \Pr{}{\world{B} \mid \barH{t-1}, s_t = A} \\ \text{(2) } r_t + R \cdot \Pr{}{\world{A} \mid \barH{t-1}, s_t = B} &\leq R \cdot \Pr{}{\world{B} \mid \barH{t-1}, s_t = B}
\end{align*}

Simplifying these inequalities, we find:
\begin{align*}
    r_t &\geq R \left( \Pr{}{\world{B} \mid \barH{t-1}, s_t = A}- \Pr{}{\world{A} \mid \barH{t-1}, s_t = A }\right), \\
    r_t &\leq R \left( \Pr{}{\world{B} \mid \barH{t-1}, s_t = B} - \Pr{}{\world{A} \mid \barH{t-1}, s_t = B}\right).
\end{align*}

Now, let us rewrite the probabilities by expanding the posterior in terms of the evidence (and recalling that the prior over best actions is uniform)
\begin{align*}
    \Pr{}{\world{A} \rvert \barH{t-1}, s_t = A} &= \frac{\Pr{}{\barH{t-1}, s_t = A \rvert \world{A}} }{\Pr{}{\barH{t-1}, s_t = A \rvert \world{A}} + \Pr{}{ \barH{t-1}, s_t = A \rvert \world{B}} } \\
    &= \frac{p \cdot \Pr{}{\barH{t-1} \rvert \world{A}} }{p \cdot \Pr{}{\barH{t-1} \rvert \world{A}}  + (1- p) \cdot \Pr{}{\barH{t-1} \rvert \world{B}}}
\end{align*}

\begin{align*}
    \Pr{}{\world{A} \rvert \barH{t-1}, s_t = B} &= \frac{\Pr{}{\barH{t-1}, s_t = B \rvert \world{A}}}{\Pr{}{\barH{t-1}, s_t = B \rvert \world{A}}+ \Pr{}{\barH{t-1}, S_t = B \rvert \world{B}} } \\
    &= \frac{(1-p) \cdot \Pr{}{\barH{t-1} \rvert \world{A}}}{p \cdot \Pr{}{\barH{t-1}\rvert \world{A}}  + (1- p) \cdot \Pr{}{\barH{t-1} \rvert \world{B}} }
\end{align*}

\begin{align*}
    \Pr{}{\world{B} \rvert \barH{t-1}, s_t = A} &= \frac{\Pr{}{\barH{t-1}, s_t = A \rvert \world{B}} }{\Pr{}{\barH{t-1}, s_t = A \rvert \world{A}} + \Pr{}{ \barH{t-1}, s_t = A \rvert \world{B}} } \\
    &= \frac{p \cdot \Pr{}{\barH{t-1} \rvert \world{B}} }{p \cdot \Pr{}{\barH{t-1} \rvert \world{A}}  + (1- p) \cdot \Pr{}{\barH{t-1} \rvert \world{B}}}
\end{align*}

\begin{align*}
    \Pr{}{\world{B} \rvert \barH{t-1}, s_t = B} &= \frac{\Pr{}{\barH{t-1}, s_t = B \rvert \world{B}}}{\Pr{}{\barH{t-1}, s_t = B \rvert \world{B}}+ \Pr{}{\barH{t-1}, S_t = B \rvert \world{B}} } \\
    &= \frac{(1-p) \cdot \Pr{}{\barH{t-1} \rvert \world{B}}}{p \cdot \Pr{}{\barH{t-1}\rvert \world{A}}  + (1- p) \cdot \Pr{}{\barH{t-1} \rvert \world{B}} }
\end{align*}

Suppose $|A|$ people have decided on $A$ in the history $\barH{t-1}$ (excluding any decisions made during a cascade). We define:
\begin{align*}
\gamma_t \coloneqq \frac{\Pr{}{\barH{t-1} \mid \world{B}}}{\Pr{}{\barH{t-1} \mid \world{A}}} = \frac{(1-p)^{|A|} p^{t-1-|A|}}{p^{|A|} (1-p)^{t-1-|A|}} = \left( \frac{1-p}{p} \right)^{2|A|-t} \frac{1-p}{p}.
\end{align*}

Then, we can re-write the above expressions as 
\begin{align*}
    \Pr{}{\world{A} \rvert \barH{t-1}, s_t = A} = \frac{1}{1  + \frac{1- p}{p} \gamma_t}
 \quad ; \quad
    \Pr{}{\world{B} \rvert \barH{t-1}, s_t = A} = \frac{\gamma_t }{1 + \frac{1- p}{p} \gamma_t}
\end{align*}

Similarly, we can re-write the expressions in our second condition as 
\begin{align*}
    \Pr{}{\world{A} \rvert \barH{t-1}, s_t = B} = \frac{1}{\frac{p}{1-p} + \gamma_t}
 \quad ; \quad
    \Pr{}{\world{B} \rvert \barH{t-1}, s_t = B} =  \frac{\gamma_t}{\frac{p}{1-p}  + \gamma_t}
\end{align*}

This implies that $R \left( \frac{\gamma_t -1 }{1 + \frac{1- p}{p} \gamma_t}\right) \leq r_t \leq R \left( \frac{\gamma_t - 1}{\frac{p}{1-p} + \gamma_t }\right)$

Finally, we show that there is always a non-negative value for the subsidy in this interval. Although this is not strictly required, we want to avoid a situation where an external entity is actually taking money from the agents. To do this, it is sufficient to see that $\gamma_t \geq 1$ in the region where the subsidy is needed to cause agents to act according to their signal and when the correct cascade has not yet been reached. By Lemmas \ref{lem:srw} and \ref{lem:stop}, this is the region where $|B| - |A| \geq 2$. Then,

\begin{align*}
\gamma_t = \frac{(1-p)^{|A|} p^{t-1-|A|}}{p^{|A|} (1-p)^{t-1-|A|}}  = \frac{(1-p)^{|A|} p^{|B|}}{p^{|A|} (1-p)^{|B|}} = \frac{p}{1-p}^{|B|-|A|}  \geq 1
\end{align*}

\end{proof}

\subsection{Proof of Theorem~\ref{thm:subsidy_budget}}
\label{appendix:proof-subsidylen}

\begin{proof}
We apply Wald's first equation to upper bound the amount of time for which the subsidy must be released. This time, we imagine that we must move from a state where there are two more choices for $B$ than $A$ to a state where there are two more choices for $A$ than $B$. This reduces to the problem of a random walk with probability $p$ of moving right hitting 2 when starting from -2 (with no left stopping point). Let $N$ be the stopping time at which this event occurs. Using Wald's equation, we have $\mathbb{E}[S_N]=4$ and $\mathbb{E}[X_1]= p \cdot 1 + (1-p) \cdot -1 = 2p-1$. Then,
$\mathbb{E}[N] = \frac{\mathbb{E}[S_N]}{\mathbb{E}[X_1]} = \frac{4}{2p-1}$. The corresponding subsidy calculation considers the worst-case scenario where each round necessitates the maximum subsidy $R$, leading to:
\begin{align*}
\mathbb{E}[\text{Total subsidy}] = R \times E(N) = R \times \frac{4}{2p-1}.
\end{align*}
\end{proof}

\section{Details for extension to multiple groups} \label{appendix:k-group}

\subsection{Defining The Setting}

We know that there is some way in which to subsidize the agent that incentivizes them to reveal their signal.
\begin{fact} \label{lemma:value-exists}
    For a group with signal strength $p$, for each value $x = |A|-|B|\,,$ there exists a value $v_{x, p}$ such that if the government provides a subsidy to the agent in play when there are $x$ more $A$ actions than $B$ actions of $v_{x, p}$ (i.e., subsidizes action $A$ by $v_{x, p}$ if $v_{x, p} > 0$ and action $B$ by $|v_{x, p}|$ otherwise), then the agent acts in such a way that their signal is revealed. The subsidy value $v_{x, p}$ is a function of $x\,,$ and $p\,,$ the probability that the signal is aligned with the correct action.
\end{fact}

\begin{proof}
    The existence of this subsidy is guaranteed by the fact that we construct such a subsidy in the previous section. In particular, see Algorithm~\ref{alg:subsidy} and Theorem~\ref{thm:subsidy}.
\end{proof}

\subsection{Analyzing the Subsidy Scheme}

We can model the state of a fixed group receiving a randomly chosen subsidy from the distribution $\mathcal{D}$ using a random walk, as before. In Lemma~\ref{lemma:random-walk-gvt}, we describe that random walk, following which in Definition~\ref{defn:rw-aug} we define a related walk. 

\subsubsection{Proof of Lemma~\ref{lemma:random-walk-gvt}}

\begin{proof}

    First, as before, we can formalize this process as a random walk, where the location on the random walk depends on the net difference between the number of people taking the ``positive'' action for their group and the number taking the ``negative'' action. It is clear that if an agent reveals their signal (which is ``right'' (both in terms of alignment with world and in direction on the walk) with probability $p$) they take a step in the ``right'' direction. Now, if the agent does not reveal their signal, there is no step taken on the random walk, since the location on the random walk is given by the number of revealed $A$ signals less the number of revealed $B$ signals. If the agent does not reveal their signal, then the action they take is ascribed by an agent to the subsidy they received and therefore does not affect the location on the random walk.
    We formalize this in the lemma before.
    
    \begin{lemma} \label{lemma:rw-iff-signal}
     A step of size 1 is taken on the random walk if and only if the agent revealed their signal. 
    \end{lemma}
    
    If (and only if, per the lemma above) the agent playing at a given time reveals their signal as a result of the government subsidy, they take a step on the random walk. 
    Since there could be a range of subsidies that are signal-revealing at a given location on the random walk for a value of $p\,,$ there may be multiple values the government could pick that would cause signal-revealing. Let these values be $\{ v_j \}_{j = 1}^m\,.$ Then, $\alpha_i \coloneqq \sum_{j} \Pr{v_j}\,.$ Now, with probability $\alpha_i\,,$ the signal is revealed and a step is taken, and with probability $1-\alpha_i\,,$ the signal is not revealed and the walk stays in the same state. If the signal is revealed, then as before, the walk proceeds from $i$ to $i+1$ with probability $p$ and from $i$ to $i-1$ with probability $1-p\,.$
    
\end{proof}

\begin{rmk}
    If a group for whom the correct action is A / right reaches $+L > 2\,,$ then they are in the correct cascade for them. Similarly, if B / left is the correct action for a group and they reach $-L < -2$ in the random walk, then they are in the correct cascade for them.
\end{rmk}

    We showed in the previous section in Lemma~\ref{lem:stop} that $L=2$ actually suffices for pushing people into the cascade. However, in this setting, because the government provides the signal-revealing subsidy even after one group reaches $L=2\,,$ it may be possible to get a ``bad'' string of signals that push a group back toward the pessimism trap. Thus, we need to sustain it for longer. We show in subsequent lemmas exactly for how much longer we need to apply it.

Next, we analyze the simplified random walk $\mathcal{R}$ (Lemmas~\ref{lemma:prl-minusl}, \ref{lemma:whptrw}) and then show that we can use the analysis of $\mathcal{R}$ to study $\mathcal{R}_\text{gvt}$ (Lemma~\ref{lemma:analyse-rw-aug}).

\begin{lemma} \label{lemma:prl-minusl}
    In random walk $\mathcal{R}$ as defined in Definition~\ref{defn:rw-aug}, the probability that when starting at location $i\,,$ the walk hits $\Lrw$ before hitting $-\Lrw$ is given by:
    \begin{align*}
    x_i &= \Pr{}{\text{walk hits } \Lrw \text{ before hitting } -\Lrw \text{ when starting at } i}  \\
    & = \frac{2p-1}{(1-p)\, \left( \alpp^{-L} - \alpp^{L}  \right)} \sum_{j = -\Lrw + 1}^{i} \alpp^j\,.
    \end{align*}
    Thus, in order for $x_0 \ge 1-\delta\,,$ i.e., for the probability of hitting the correct cascade before the incorrect one when the walk starts from 0 to be high, we need $L = \frac{\log(1/\delta - 1)}{\log(p/(1-p))}\,.$ 
\end{lemma}

\begin{proof}
    We can analyze this by means of a recurrence. For $x_i$ defined as in the lemma statement, we have that:
    \begin{align*}
        x_i &= (1-p_\text{min}) \, x_i + p_\text{min} \, p \, x_{i + 1} + p_\text{min} \, (1-p) \, x_{i - 1} \\
        \Leftrightarrow x_i &=  p \, x_{i + 1} +  (1-p) \, x_{i - 1} \\
        \Leftrightarrow p \, ( x_i - x_{i + 1}) &= (1-p) \, (x_{i -1} - x_{i}) \\
    \end{align*}
    Now, we write down the boundary conditions: $x_{-\Lrw} = 0$ and $x_{\Lrw} = 1\,.$ Next, we solve the dynamics:
    \begin{align}
        \text{Define } y_i \coloneqq x_{i} - x_{i-1}
        &\Rightarrow (1-p) \, y_{i} = p \, y_{i+1}\\
        \text{Solution } \qquad \quad \qquad y_i &= c \left(\frac{1-p}{p}\right)^i \\
        x_\Lrw - x_{-\Lrw} = 1 &= \sum_{i = -\Lrw + 1}^{\Lrw} y_i = \sum_{i = -\Lrw + 1}^{\Lrw} c \alpp^i \\
        \Leftrightarrow c &= \frac{1}{\sum_{i = -\Lrw + 1}^{\Lrw} \alpp^i} = \frac{1 - \frac{1-p}{p}}{\alpp^{-L + 1} \left(1 - \alpp^{2L + 2} \right) } \\
        &= \frac{\frac{2p-1}{p}}{\alpp\, \left( \alpp^{-L} - \alpp^{L}  \right)}
    \end{align}

    Plugging this back into the definition of $y_i\,,$ we get:
    \begin{align}
        x_i = x_{i - 1} + \frac{2p-1}{(1-p)\, \left( \alpp^{-L} - \alpp^{L}  \right)} \, \alpp^i
    \end{align}

    Finally, let us incorporate the initial condition $x_{-\Lrw} = 0\,.$ Then, $$x_{-\Lrw + 1} = \frac{2p-1}{(1-p)\, \left( \alpp^{-L} - \alpp^{L}  \right)} \, \alpp^{-\Lrw + 1}\,,$$ and in general:
    $$
    x_i = \frac{2p-1}{(1-p)\, \left( \alpp^{-\Lrw} - \alpp^{\Lrw}  \right)} \sum_{j = -\Lrw + 1}^{i} \alpp^j\,.
    $$
    Note that when $ i = \Lrw\,,$ indeed $x_\Lrw = 1\,.$

    Finally, we compute $x_0 = \frac{2p-1}{(1-p)\, \left( \alpp^{-\Lrw} - \alpp^{\Lrw}  \right)} \sum_{j = -\Lrw + 1}^{0} \alpp^j\,$ and set this to be at least $1-\delta$ so we can determine what the appropriate threshold $L$ is. In particular:
    \begin{align}
        x_0 &= \frac{2p-1}{(1-p)\, \left( \alpp^{-\Lrw} - \alpp^{\Lrw}  \right)} \left( \frac{\alpp^{-\Lrw + 1} - \alpp}{1 - \alpp}  \right) \\
        &= \frac{\alpp^{-\Lrw} - 1}{\alpp^{-\Lrw} - \alpp^\Lrw} = \frac{1}{1 + \alpp^{\Lrw}} \\
        \text{want } \quad &\ge 1-\delta\,.
    \end{align}
    Solving, we get $L \ge \frac{\log(1/\delta - 1)}{\log(p/(1-p))}\,.$

\end{proof}

\begin{lemma} \label{lemma:whptrw}
    After $T = \frac{2 L}{\pmin \, (2p - 1)} + \frac{2 \log(1/\delta)}{\pmin^2 \, (2p-1)^2}$ steps of this random walk, with probability at least $1-\delta$, the walk will have moved at least $L\,$ steps net to the right.
\end{lemma}

\begin{proof}
    Let us define the outcome of step $i$ in the following way:
    $$
    X_i \coloneqq \begin{cases} +1 & \text{ step right, i.e., w.p. } \, p_\text{min} \, p \\
    0 & \text{ step right, i.e., w.p. } \, 1-p_\text{min} \\
    -1 & \text{ step right, i.e., w.p. } \, p_\text{min} \, (1-p)   \end{cases}\,.
    $$
    We are interested in upper bounding $\Pr{}{\sum_{i = 1}^T X_i \le \Lrw}\,.$ To do so, we will apply a Hoeffding bound. In particular, we have that $\mathbb{E}\left[ X_i  \right] = - \pmin + \pmin \, p + \pmin \, p = \pmin (2 \, p - 1)\,$ and so $\mathbb{E}\left[ \sum_{i = 1}^T X_i  \right] = \pmin \, T \, (2\, p - 1)\,.$ Applying the Hoeffding bound:
    \begin{align}
        \Pr{}{\sum_{i = 1}^T X_i \le \mathbb{E}\left[ \sum_{i = 1}^T X_i  \right] - \left(\mathbb{E}\left[ \sum_{i = 1}^T X_i  \right] - \Lrw \right)} &\le \exp \left( \frac{-2 \left( \Lrw -  \pmin \, T \, (2\, p - 1)  \right)^2}{T \cdot 4}    \right) \\
        &= \exp \left( \frac{-2 \left( \Lrw -  \pmin \, T \, (2\, p - 1)  \right)^2}{4 T}    \right)\,.
    \end{align}
    Finally, we can solve the following quadratic in $T$ to show that the stated value of $T$ suffices:
    \begin{align}
            4 \log(1/\delta)\,T &\le 2 \Lrw^2 - 4 \Lrw \, \pmin \, (2p -1) \, T \,+ 2\,\pmin^2\, T^2 \, (2p - 1)^2 \\
            T &\ge  \max \bigg \{   \frac{-(- 4 \Lrw \, \pmin \, (2p -1) - 4 \log(1/\delta))}{4\,\pmin^2\,(2p - 1)^2} \pm \\
            & \qquad \frac{\sqrt{(- 4 \Lrw \, \pmin \, (2p -1) - 4 \log(1/\delta))^2 - 4(2\,\pmin^2\,(2p - 1)^2)(2 \Lrw^2)}}{4\,\pmin^2\,(2p - 1)^2}   \bigg \} \\
            \Leftarrow T &\ge \frac{2 \left( 4 \Lrw \, \pmin \, (2p -1) + 4 \log(1/\delta) \right) }{4\,\pmin^2\,(2p - 1)^2}   \\
            &= \frac{2L}{\pmin \, (2p - 1)} + \frac{2 \log(1/\delta)}{\pmin^2 \, (2p-1)^2}
    \end{align}

\end{proof}

We show that we can analyze $\mathcal{R}_\text{gvt}$ by analyzing the simpler-to-analyze $\mathcal{R}\,.$

\begin{lemma} \label{lemma:analyse-rw-aug}
    We can analyze the random walk $\mathcal{R}_\text{gvt}$ described in Lemma~\ref{lemma:random-walk-gvt}, i.e., the one reflecting the dynamics of the game in the presence of the described government subsidy, by instead analyzing the random walk $\mathcal{R}\,$ given in Definition~\ref{defn:rw-aug}. In particular, the probability $x_i = \Pr{}{\text{walk hits } \Lrw \text{ before hitting } -\Lrw \text{ when starting at } i}$ is the same in both random walks,  $\mathcal{R}\,, \mathcal{R}_\text{gvt}\,.$ Further, let $T_\mathcal{R}$ be the number of steps needed to be taken in $\mathcal{R}$ for the walk to hit $L\,$ with high probability. Suppose $T_{\mathcal{R}_\text{gvt}}$ is the number of steps needed to achieve the same in $\mathcal{R}_\text{gvt}$. Then, $T_{\mathcal{R}_\text{gvt}} \le T_\mathcal{R}\,.$
\end{lemma}

\begin{proof}
    We consider the two parts separately.

    \textbf{Probability} Let us write the recurrence equations for each of the random walks. First, for $\mathcal{R}_{\text{gvt}}$:
    $$
    x_i = (1-\alpha_i)\, x_i + \alpha_i\, (1-p)\, x_{i - 1} + \alpha_i\, p\, x_{i + 1}\,.
    $$
    And for $\mathcal{R}\,:$
    $$
    x_i = (1-p_\text{min})\, x_i + p_\text{min} \,(1-p)\, x_{i - 1} + p_\text{min}\, p\, x_{i + 1}\,.
    $$
    Now, observe that subtracting the first term on the right hand side from both sides we get, respectively:
    \begin{align*}
        \alpha_i\, x_i &= \alpha_i\, (1-p) \,x_{i - 1} + \alpha_i\, p\, x_{i + 1} \\
        p_\text{min}\, x_i &= p_\text{min}\, (1-p)\, x_{i - 1} + p_\text{min}\, p\, x_{i + 1}\,.
    \end{align*}
    This can be simplified to:
    \begin{align*}
        x_i &= (1-p)\, x_{i - 1} +  p \, x_{i + 1} \\
         x_i &= (1-p)\, x_{i - 1} + p \, x_{i + 1}\,.
    \end{align*}
    These equations are the same, meaning that the dynamics are the same. If the boundary conditions are the same, then the values will also be the same.

    \textbf{Time Required To Converge} For this, we simply observe that the relationship between $\alpha_i$ in $\mathcal{R}_\text{gvt}$ and $\pmin$ in $\mathcal{R}$ is $\alpha_i \ge \pmin \, \forall i\,.$ Let us consider the instances where a step is taken in the original walk $\mathcal{R}_\text{gvt}$ as compared to where steps are taken in the analyzed walk $\mathcal{R}\,.$ Define $D_i \coloneqq \mathbb{I}\left[ \text{ step taken in } \mathcal{R}  \right]\,.$
    Define:
    $$
    \tilde{D}_i = \begin{cases}
    1 & \text{if }\,D_i = 1 \\
    1 & \text{ w.p. } \frac{\alpha_i - \pmin}{1-\pmin} \text{ if } \, D_i = 0 \\ 
    0 & \text{ otherwise.}
    \end{cases}
    $$
    
    Note that $\tilde{D}_i$ is $1$ with probability $\alpha_i\,$ and 0 with probability $1-\alpha_i\,.$ 
    Additionally, due to the coupling, $T_\alpha \coloneqq \sum_{i = 1}^T \tilde{D}_i \ge \sum_{i = 1}^T D_i \eqqcolon T_{\pmin} \,.$ 
    As a result, there are strictly more steps taken in $\mathcal{R}_\text{gvt}$ than in the $\mathcal{R}\,.$ Conditioned on taking a step, the probabilities of right and left steps are the same.  
    Thus, for a threshold of interest $Q\,,$ the probability of not exceeding it after $T_\alpha$ steps is smaller than the probability of not exceeding it after $T_{\pmin}$ steps.
   
\end{proof}

\subsubsection{Proof of Theorem~\ref{thm:kgroup}}
Finally, we present the proof of the main theorem, Theorem~\ref{thm:kgroup}.

\begin{proof}

\newcommand{\green}[1]{\textcolor{green}{#1}}

We start by fixing a group to look at. We show that the subsidy behaves, as before, as a random walk that group takes on the number line and show how long it takes to achieve with high probability a sufficient condition for the walk to finish in an up cascade. We then union bound over the failure probability and appropriately scale the time required to get the stated result.

Thus, let us start by fixing a group. Suppose for this group that $A$ is the correct choice (``right'' is the direction for an up cascade). All of our subseequent arguments hold symmetrically in the case that $B$ is the correct choice (``left'' is the direction for an up cascade). By Lemma~\ref{lemma:random-walk-gvt}, we have the description of a random walk $\mathcal{R}_\text{gvt}$ that models the setting in Definition~\ref{defn:multigroup-game} with the government subsidy described in Definition~\ref{defn:government-subsidy}. Further, Lemma~\ref{lemma:analyse-rw-aug} establishes that for the quantities of our interest, we can study $\mathcal{R}$ defined in Definition~\ref{defn:rw-aug} instead of $\mathcal{R}_\text{gvt}\,.$ 

With this having been established, let us define a tolerance parameter $\delta_0 \coloneqq \delta/(3k)$ which represents the probability of a single failure for a single group. If there is probability at most $\delta_0$ that the walk does not net $L$ steps and probability at most $\delta_0$ that even if the walk nets at least $L$ steps, it hit $-L$ first, then the total failure probability for this walk is at most $2\delta_0 = \delta/k\,.$ We want the group to take sufficiently many steps in the random walk that it exceeds a threshold $L\,,$ sending it into an up cascade. In the past, we considered $L = 2$ because the government knew the correct action and only subsidized it, but now that we're subsidizing both actions, we need to protect against hitting a bad cascade before hitting the correct one. By Lemma~\ref{lemma:whptrw}, we have that $T = \frac{2L}{\pmin \, (2p - 1)} + \frac{2 \log(1/\delta_0)}{\pmin^2 (2p-1)^2}\,$ steps suffice for netting $L$ steps to the right with probability at least $1-\delta_0 = 1-\delta/(3k)\,.$ 

Now, we do want to make sure the walk did not hit a down cascade before it hits the up cascade. For this, we analyze the probability of hitting the down cascade first in Lemma~\ref{lemma:prl-minusl}. As before, we want the probability of failure to be at most $\delta_0\,,$ and so applying Lemma~\ref{lemma:prl-minusl}, $L \ge \frac{\log(1/\delta_0 - 1)}{\log(p/(1-p))}\,.$

\newcommand{\nagentsneed}{\frac{ 2 \frac{\log(1/\delta_0 - 1)}{\log(p/(1-p))}}{\pmin \, (2p - 1)} + \frac{2 \log(1/\delta_0)}{\pmin^2 (2p-1)^2}}

Thus, with probability at least $1-2\delta_0 = 1-2\delta/(3k)\,,$ this group will end in an up cascade after $\nagentsneed$ steps (i.e., agents from this group are seen) without any negative side effects. 

Next, we analyze how long it takes to see enough agents from this group with high probability. An agent is from group $j$ with probability at least $g_\text{min}\,,$ and so with probability at least $1-\delta_0\,,$ after $\frac{2T^\star + 2\log(1/\delta_0)}{g_\text{min}}$ agents, we have seen agents from that group at least $T^\star$ times. Thus, we have that with probability at least $1-\delta/k\,,$ after $\frac{2\left( \nagentsneed \right) + 2\log(3k/\delta)}{g_\text{min}}$ steps, with high probability, for a fixed group, we have (1) seen enough agents from that group to (2) see enough ``correct'' signals for that group and (3) not enter a bad cascade first. Finally, we union bound over the $k$ groups to get that with probability at least $1-\delta\,,$ $k$ times this number of steps suffices for \textit{all} groups to reach positive cascades.

\end{proof}

\section{Extra Experiments}

In this section, we provide additional plots from our experiments. 

\subsection{Proportion Correct Cascades}
\begin{figure}[ht!]
    \centering
    \includegraphics[width=0.75\textwidth]{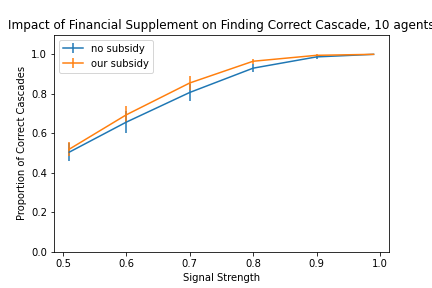}
    \caption{Proportion Correct Cascades for 10 Agents}
    \label{fig:subsidy_10_agents}
\end{figure}

\begin{figure}[ht!]
    \centering
    \includegraphics[width=0.75\textwidth]{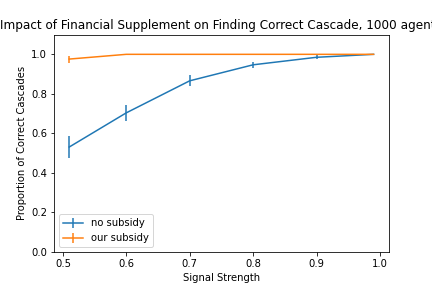}
    \caption{Proportion Correct Cascades for 1000 Agents}
    \label{fig:subsidy_1000_agents}
\end{figure}

\subsection{Subsidy Size}

\begin{figure}[ht!]
    \centering
    \includegraphics[width=0.75\textwidth]{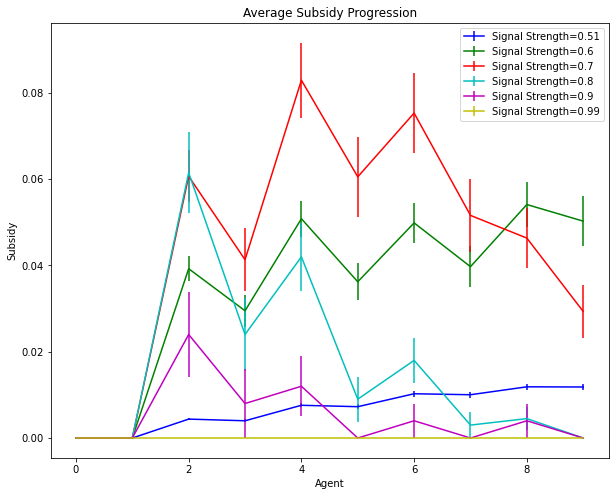}
    \caption{Average Subsidy Progression for 10 Agents}
    \label{fig:subsidy_10_agents_prog}
\end{figure}

\begin{figure}[ht!]
    \centering
    \includegraphics[width=0.75\textwidth]{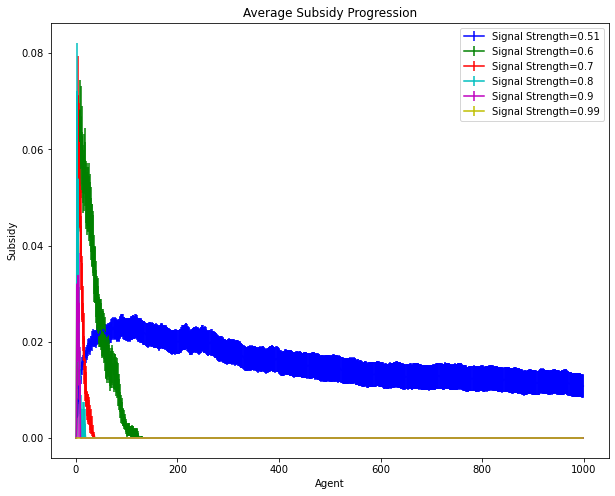}
    \caption{Average Subsidy Progression for 1000 Agents}
    \label{fig:subsidy_1000_agents_prog}
\end{figure}

\chapter{Appendices for Grit}

\section{Related Work} \label{appendix:grit-related-work}

We survey two categories of work related to ours. First, we discuss work from the social sciences that observes grit in people and analyzes its role and impact in society. We also include work that abstracts and formalizes what differentiates grit from other related traits. Next, we focus on relevant computer science literature related to the formal frameworks we use in our analysis.

One of the most popular studies of grit is psychologist Angela Duckworth's book \cite{duckworth2016grit}. In it, she discusses how across fields and circumstances, the trait of grit distinguishes people who succeed from people who succumb to their circumstances. This work had a significant impact on popular understanding of the impact of hard work on success, suggesting that grit is often a more important factor than talent or proclivity. Relatedly, psychologist Martin Seligman has a large body of work encouraging positive attitudes as a path to success and good outcomes \cite{seligman_authentic_2002, seligman_learned_2006}. School systems like the Knowledge is Power Program (KIPP) Charter Schools have operationalized this somewhat academic research (including by bringing Seligman on as a consultant \cite{Gibbon_seligman_2020}) to build their signature ``no excuses'' philosophy of education, emphasizing personal responsibility and grit as paths to success. 

More recently, sociologist Tom Wooten studied the impact of the ``no excuses'' educational approach in his dissertation \cite{wooten_precarious_2022}, using ethnographic methods to study outcomes for six students who attended schools with this educational philosophy, particularly exploring the mechanisms by which these systems perpetuate poverty. In particular, he argues that upward mobility in the United States is fundamentally precarious, with seemingly small disruptions destabilizing those coming from impoverished backgrounds much more than those who have more financial stability. Wooten's findings are the impetus for us to investigate the impacts of financial support on the behavior of gritty agents in Section~\ref{sec:trust-fund}. He also reflects on the pitfalls of an educational system that overemphasizes grit, writing that, ``In lieu of such large, fundamental changes, my research also points to smaller policy shifts that could make a big difference. One shift is that primary and secondary schools should carefully examine the explicit and implicit lessons they teach their students about
`grit,' altering their curricula to emphasize balance and pacing.'' \par

Abstracting findings from several such observational studies, \cite{morton2019grit} develop a philosophical theory of grit, arguing that it is possible due to a philosophical concept called \textit{Permissivism} for two agents to interpret the same body of evidence in completely different ways, and in fact grit is a consequence of Permissivism. We build heavily on the abstractions derived in this work. \par

Finally, we survey some computer science literature that we draw on in order to quantitatively formalize and study the question of decision-making with and without grit. The perspectives we take on rationality are inspired by objectives that are commonly optimized in the computer science literature, including the competitive ratio studied in online learning \cite{borodin_online_2005} and the Bayesian approach to uncertainty quantification  \cite{bernardosmith_bayesian_2000}. The framework we use to represent the decision problem is an instance of the multi-armed bandits problem, a well-studied framework for making decisions when the payoff is unknown \cite{slivkins_introduction_2022}. More particularly, we suppose the bandit arms have structured reward, and the structure of interest is improving, first formalized by \cite{heidari_tight_nodate} and studied in general by \cite{patil_mitigating_2023, blum_nearly-tight_2024}.
\section{Why This Instance and Not Something Else?} \label{subsubsec:whynotflat}

The instance defined in Equation~\ref{eqn:main-instance} reflects settings where grit pays off eventually but even then not immediately, i.e., there is only a gradual increase of reward even after $t = \theta\,.$ This captures certain situations we may be interested in -- for instance, consider a computer science theory PhD program, where the first several years are spent taking classes to bolster one's mathematical skills, following which there is still a ramp up period as the student gets accustomed to doing research. Thus, it is natural to suppose the early years of PhD candidacy yield some research progress, while the later years yield significantly more. On the other hand, there are many natural cases in which once the payoff starts, it reaches its complete potential immediately; a natural example of this is when well-digging or searching for oil. In these cases, a bandit arm formulation more like the following may be more reasonable:
$$
f_1(t) = 1 \, \forall \, t \qquad f_2(t) = \begin{cases}
    0 & t < \theta \\
    m & t \ge \theta
\end{cases}\,.
$$

If we aim to maximize the competitive ratio, we consider two extreme worst cases: first, one in which the agent spent $s$ time on the striving arm but instead should have spent all their time on the stable arm; second, one in which the agent spent $s$ time on the striving arm to no avail but would've witnessed the increase on the striving arm if they'd stuck around infinitesimally longer. In this case, the former is $(T-s)/T\,,$ which is decreasing with $s\,,$ while the latter is $(T-s)/(m(T-s)) = 1/m\,,$ which is non-increasing. This implies that as long as the agent switches to the stable arm before $s = (1-1/m)T\,$, they achieve competitive ratio at least $1/m\,.$ 
Thus, achieving the optimal competitive ratio is trivial and there is no strategy to the game. Owing to this, we study a slightly more complicated setting -- the instance in which $f_2$ increases linearly once it starts paying off.

In general, the non-trivial instances are ones in which the first competitive ratio extreme case is decreasing with $s$ and the second competitive ratio extreme case is increasing with $s\,.$ If we define $F_2 \coloneqq \int f_2\,,$ we find $F_2$ needs to be increasing with $s$ in order to find a sensible switch point that balances the worse cases, i.e.:
$$
\frac{T-s}{T} = \frac{T-s}{F_2(T-s)} \Rightarrow s = T - F_2^{-1}(T) \Rightarrow CR = \frac{F_2^{-1}(T)}{T}\,.
$$

We also note that the interleaving results are specific to a deterministic, noise-free $f_1\,;$ if the stable arm had noisy payoffs with an unknown mean value, then the agent would want to perform a small amount of interleaving in order to better understand the tradeoff.

\section{Proofs of Results}

\subsection{Proof of Lemma~\ref{lemma:alphat-switch}} \label{appendix:proof-alphat-switch}
\begin{proof}

Recall our previous argument that all deterministic strategies boil down to playing $f_2$ for $s$ steps and then switching to $f_1$. With that in mind, there are two extremes of what could happen to the competitive ratio, one of which decreases with the longer spent on $f_2$, and the other of which increases with time spent on $f_2\,.$ Thus, we compute them and equalize:
\begin{align}
    \text{competitive ratio if never increase } &= \frac{T-s}{T} \\
    \text{compeititve ratio if increase right after switch } &= \frac{T-s}{\frac{\alphat}{2} (T-s)^2}\,. \\
    \text{equalizing, } \frac{T-s}{T} &= \frac{T-s}{\frac{\alphat}{2} (T-s)^2} \\
    s &= T - \sqrt{\frac{2T}{\alphat}}\,.
\end{align}

\end{proof}


\subsection{Proof of Lemma~\ref{lemma:min-acc-works}} \label{appendix:proof-min-acc}

\begin{proof}
    For a strategy, let $\{t_i\}_{i = 0}^n$ be the set of times at which the agent switches from one arm to the other with $t_0 = 0$. Then, $[t_i, t_{i+1}]$ for even $i$ are the intervals played on the stable arm, and the remaining intervals are played on the striving arm.
    We argue that we can convert any strategy that is not-minimally accumulating into a minimally-accumulating strategy that accrues at least as much reward as the original non-minimally-accumulating strategy. 

    Suppose the total amount of time spent on the stable arm $f_1$ is $T_1 \coloneqq \sum_{i \text{ even}} (t_{i+1} - t_i)\,.$  and the total time spent on striving arm $f_2$ is $T_2 \coloneqq \sum_{i \text{ odd}} (t_{i+1} - t_i)\,,$ such that $T_1 + T_2 = T\,.$ Recall $\theta\,,$ the threshold until which one must play $f_2$ before it pays off. If $T_2 < \theta\,,$ then redefining the intervals such that the strategy becomes minimally-accumulating does not change the total accumulated reward, since the reward accrued from an arm is only a function of the time {\em on that particular arm}. Thus, the total reward is $T_1$ regardless of the order in which it is accrued.

    Let us now consider the second case, $T_2 > \theta\,.$ Now, let us break up the time spent on arm $f_1$ into the $\frac{1+\gamma}{1-\gamma} \, \theta$ time that must happen prior to playing the striving arm for $\theta$ time and the remaining $T_1 - \frac{1+\gamma}{1-\gamma} \, \theta$ time. (Due to the comfort restriction, we know $T_1 - \frac{1+\gamma}{1-\gamma} \, \theta \ge 0\,.$) As argued previously, we can rearrange interval endpoints for the first $\frac{1+\gamma}{1-\gamma} \, \theta$ time spent on $f_1$ and $\theta$ time spent on $f_2$ with no changes in overall reward. In particular, we can rearrange it to be minimally accumulating. Now, for the remaining $T_1 - \frac{1+\gamma}{1-\gamma} \, \theta$ time, if we play it on the stable arm (at any time), we accrue $R = T_1 - \frac{1+\gamma}{1-\gamma} \, \theta$ reward. However, having satisfied the comfort requirements, if we instead use it at the end of the game and on arm $f_2\,,$ we instead gain reward $(T_2 - \theta - R/2)R\,.$ If $R/2 + T_2 - \theta > 1\,,$ then this is strictly better. If not, we can still play this time on the stable arm. Thus, we have constructed a minimally-accumulating strategy that accrues at least as much reward as the original strategy.

\end{proof}

\subsection{Proof of Lemma~\ref{lemma:gamma-comfort}} \label{appendix:proof-gamma-comfort}

\begin{proof}
    As before, we compute the competitive ratio in two cases: one in which the stable arm played the whole time would be optimal and one in which the striving arm pays off as soon as the agent switches for good. Since the agent is alternating between $\alpha_\gamma \coloneqq \frac{\gamma+1}{2}$ time on $f_1$ and $1-\alpha_\gamma$ time on $f_2\,,$ we must account for this in the computation:

    $$
    \frac{\alpha_\gamma \cdot s - (1-\alpha_\gamma)s + T - s}{T} \, = \, \frac{\alpha_\gamma \cdot s - (1-\alpha_\gamma)s + T - s}{\frac12 (T-s)^2 + \alpha_\gamma \cdot s - (1-\alpha_\gamma)s}\,.
    $$

Solving, we get:
\begin{align}
    s^2 - 2sT + T^2 + s(2\alpha_\gamma-1) - 2 T &= 0 \\    
    s &=  T - \alpha + \frac12 -\frac12 \sqrt{(2\alpha_\gamma - 1)^2 + 4T(3-2\alpha)} \\
     &= T - \frac{\gamma}{2} - \frac{\sqrt{\gamma^2 + 4T(2 - \gamma)}}{2}\,.
\end{align}

We can plug this back in to get the competitive ratio:
$$
\frac{(2\alpha_\gamma - 1)s + T - s}{T} = \gamma + \frac{\gamma(1-\gamma)}{2T} + \frac{(1-\gamma) \sqrt{\gamma^2+ 4T (2-\gamma)}}{2T} \,.
$$

Finally, the amount of time spent exploring on the striving arm is $1-\alpha_\gamma$ fraction of the total time pre-switch, which is:
$$
\frac{1-\gamma}{2} \cdot \left(T - \frac{\gamma}{2} - \frac{\sqrt{\gamma^2 + 4T(2 - \gamma)}}{2} \right)\,.
$$

\end{proof}

\section{Fixed Time Financial Support}
\label{appendix:fixed-time-support}
Now, suppose the unconditionality of support is a little weaker. Instead of ``as much as needed,'' the benefactor only promises $R$ units of support. This could correspond to a parent committing to financially support a child until 18, or 21, or 25, etc. Now, the calculus is a little different. First we consider the case where $R$ is large, say like $T\,.$
\begin{align}
    \text{competitive ratio if never increase } &= \frac{R - s + T - s}{R + T} \\
    \text{competitive ratio if increase right after switch } &= \frac{R - s + T-s}{R - s + \frac{1}{2} (T-s)^2}\,. \\
    \text{equalizing, } \frac{R - s + T - s}{R + T} &= \frac{R - s + T-s}{R - s + \frac{1}{2} (T-s)^2} \\
    s = T + 1 - \sqrt{4T + 1}\,.
\end{align}

Interestingly, in this model the agent switches to the stable arm a little sooner than in the previous case -- which makes sense as the agent knows not to expect unbounded financial support. However, the agent can still explore for that full time owing to the financial support. Thus, when compared to an agent without financial support, this agent can explore for a longer time by a factor of $\frac{T+1 - \sqrt{4T + 1}}{\frac{T - \sqrt{2T}}{2}} \ge \frac{2(T - \sqrt{5T})}{T - \sqrt{2T}}\,,$ which is at least 1.5 for large enough $T$ ($T \ge 23$).

These results allow us to conclude that even when all agents have nearly identical ``outward'' behavior, i.e., the point at which they finally give up on striving and switch to the stable arm, the presence of a trust fund allow agents to explore for longer, i.e., spend more time on the striving arm. Our result provides a quantitative mechanism by which we can explain why agents who seemingly spend similar amounts of time on ambitious goals can see significant differences in outcome when they come from different kinds of resource backgrounds.

\section{Calculations For Interplay of Grit and Trust Fund} \label{appendix:combine}

Now, we look at what happens when the rate of increase of the arm is unknown so some amount of grit plays into the decision of when to switch, but there is also a cost to striving and net reward is never allowed to be negative. Formally, at each time step, the agent chooses between:

$$
f_1(t) = 1 \, \forall \, t \qquad f_2(t) = \begin{cases}
    -1 & t < \theta \\
    \alpha (t-\theta) & t \ge \theta
\end{cases}\,.
$$

As before, an agent has a guess $\alphat$ of how fast they think the increase will happen. Let us first consider someone without a trust fund and guess $\alphat$. For them, the factors to balance are:
\begin{align}
\frac{(1 - 1) \cdot s + T - s}{T} &= \frac{(1 - 1) \cdot s + T - s}{(1 - 1) \cdot s + \frac{\alphat}{2} (T-s)^2} \\
\frac{T - s}{T} &= \frac{T - s}{\frac{\alphat}{2} (T-s)^2} 
\end{align}

which gives switching point $s = T - \sqrt{\frac{2T}{\alphat}}\,$ and $\frac{T}{2} - \sqrt{\frac{T}{2\alphat}}$ time spent on the striving arm. However, an agent with a fixed large safety net (say, $R = T$) can explore for longer, trading off:

\begin{align}
    \frac{R - s + T - s}{R + T} &= \frac{R - s + T-s}{R - s + \frac{\alphat}{2} (T-s)^2} \\
    s &= T + \frac1\alphat - \sqrt{\frac{4T}{\alphat} + \frac 1\alphat}\,.
\end{align}

In this case, the switching point is as above, and the time spent on the striving arm is the same. 
\singlespacing

\end{document}